\documentclass[11pt]{report}

\usepackage{s01_ew}
\usepackage{gg}
\usepackage{ff}
\usepackage{4f_s01}
\usepackage{gc}
\usepackage{mw}

\usepackage[english]{babel}
\usepackage{graphicx,rotating}
\usepackage{a4p,here}
\usepackage{cite,epsfig}

\renewcommand{\Huge}{\huge}
\newcommand{\updates}[1]%
 {\fbox{\parbox{\linewidth}{\textbf{Updates with respect to summer 2000:}\\#1}}}

\input{rotate}

\begin{document}
\flushbottom
\begin{titlepage}
\begin{center}
\Large {EUROPEAN ORGANIZATION FOR NUCLEAR RESEARCH\\}
\end{center}
\vspace*{0.2cm}
\begin{flushright}
       CERN-EP/2001-098 \\
       LEPEWWG/2001-02  \\
       ALEPH 2001-078 PHYS 2001-028 \\
       DELPHI 2001-131 PHYS 906 \\
       L3 Note 2723   \\
       OPAL PR 350    \\
       hep-ex/0112021 \\
       {\bf  December 17th, 2001} \\
\end{flushright}

\vspace*{0.5cm}

\begin{center}
\boldmath
\Huge {\bf A Combination of Preliminary \\
               Electroweak Measurements and \\
            Constraints on the Standard Model\\[.5cm]

}
\unboldmath

\vspace*{1.0cm}
\Large {\bf
The LEP Collaborations\footnote{The LEP Collaborations each take
responsibility for the preliminary results of their own.}
 ALEPH, DELPHI, L3, OPAL,\\
    the LEP Electroweak Working Group\footnote{%
WWW access at {\tt http://www.cern.ch/LEPEWWG}

The members of the 
LEP Electroweak Working Group 
who contributed significantly to this
note are: \\
D.~Abbaneo,         %
J.~Alcaraz,         %
P.~Antilogus,       %
A.~Bajo-Vaquero,    %
P.~Bambade,         %
E.~Barberio,        %
M.~Biglietti        %
A.~Blondel,         %
S.~Blyth,           %
D.~Bourilkov,       %
P.~Casado,          %
D.G.~Charlton,      %
P.~Checchia,        %
R.~Chierici,        %
R.~Clare,           %
B.~de~la~Cruz,      %
M.~Elsing,          %
P.~Garcia-Abia,     %
M.W.~Gr\"unewald,   %
A.~Gurtu,           %
J.B.~Hansen,        %
P.~Hansen,          %
R.~Hawkings,        %
J.~Holt,            %
R.W.L.~Jones,       %
B.~Kersevan,        %
N.~Kjaer,           %
E.~Lan{\c c}on,     %
L.~Malgeri,         %
C.~Mariotti,        %
M.~Martinez,        %
F.~Matorras,        %
C.~Matteuzzi,       %
S.~Mele,            %
E.~Migliore,        %
M.N.~Minard,        %
K.~M\"onig,         %
A.~Oh,              %
C.~Parkes,          %
U.~Parzefall,       %
Ch.~Paus,           %
M.~Pepe-Altarelli,  %
B.~Pietrzyk,        %
O.~Pooth,           %
G.~Quast,           %
P.~Renton,          %
H.~Rick,            %
S.~Riemann,         %
J.M.~Roney,         %
H.~Ruiz,            %
K.~Sachs,           %
S.~Spagnolo,        %
A.~Straessner,      %
D.~Strom,           %
R.~Tenchini,        %
F.~Teubert,         %
E.~Tournefier,      %
A.~Valassi,         %
S.~Villa,           %
H.~Voss,            %
C.P.~Ward,          %
N.K.~Watson,        %
P.S.~Wells.         %
}\\
and the SLD Heavy Flavour and Electroweak Groups\footnote{%
N.~de Groot,        %
P.C.~Rowson,        %
V.~Serbo,           %
D.~Su.              %
}\\
}
\vskip 0.5cm
\large\textbf{Prepared from Contributions of the LEP and SLD
  Experiments \\
to the 2001 Summer Conferences.}\\
\end{center}
\vfill
\begin{abstract}
  This note presents a combination of published and preliminary
  electroweak results from the four LEP collaborations and the SLD
  collaboration which were prepared for the 2001 summer conferences.
  Averages from $\Zzero$ resonance results are derived for hadronic
  and leptonic cross sections, the leptonic forward-backward
  asymmetries, the $\tau$ polarisation asymmetries, the $\bb$ and
  $\cc$ partial widths and forward-backward asymmetries and the $\qq$
  charge asymmetry.  Above the $\Zzero$ resonance, averages are
  derived for di--fermion cross sections and forward-backward
  asymmetries, W--pair, Z--pair and single--W production cross
  section, electroweak gauge boson couplings, W mass and width and W
  decay branching ratios. For the first time, total and differential
  cross sections for di--photon production are combined.
  
  The main changes with respect to the experimental results presented
  in summer 2000 are updates to the Z-pole heavy flavour results from
  SLD and LEP and to the W mass from LEP.  The results are compared
  with precise electroweak measurements from other experiments.  Using
  a new evaluation of the hadronic vacuum polarisation, the parameters
  of the Standard Model are evaluated, first using the combined LEP
  electroweak measurements, and then using the full set of electroweak
  results.
\end{abstract}
\end{titlepage}
\setcounter{page}{2}
\renewcommand{\thefootnote}{\arabic{footnote}}
\setcounter{footnote}{0}

\chapter{Introduction}
\label{sec-Intro}

This paper presents an update of combined results on electroweak
parameters by the four LEP experiments and SLD using published and
preliminary measurements, superseding previous
analyses\cite{bib-EWEP-00}.  Results derived from the $\Zzero$
resonance are based on data recorded until the end of 1995 for the LEP
experiments and 1998 for SLD.
Since 1996 LEP has run at energies above the W-pair production
threshold.  In 2000, the final year of data taking at LEP, the total
delivered luminosity was as high as in 1999; the maximum
centre-of-mass energy attained was close to 209~\GeV\ although most of
the data taken in 1999 was collected at 205 and 207~\GeV.  By the end
of $\LEPII$ operation, a total integrated luminosity of approximately
700\pb\ per experiment has been recorded above the Z resonance.

The $\LEPI$ (1990-1995) $\Zzero$-pole measurements consist of the
hadronic and leptonic cross sections, the leptonic forward-backward
asymmetries, the $\tau$ polarisation asymmetries, the $\bb$ and $\cc$
partial widths and forward-backward asymmetries and the $\qq$ charge
asymmetry.  The measurements of the left-right cross section
asymmetry, the $\bb$ and $\cc$ partial widths and
left-right-forward-backward asymmetries for b and c quarks from SLD
are treated consistently with the LEP data.  Many technical aspects of
their combination are described in References~\citen{LEPLS},
\citen{ref:lephf} and references therein.

The $\LEPII$ (1996-2000) measurements are di--fermion cross sections
and forward-backward asymmetries; di--photon production, W--pair,
Z--pair and single--W production cross sections, and electroweak gauge
boson self couplings.  W boson properties, like mass, width and decay
branching ratios are also measured.

Several measurements included in the combinations are still
preliminary. 

This note is organised as follows:
\begin{description}
\item [Chapter~\ref{sec-LS}] $\Zzero$ line shape and leptonic
  forward-backward asymmetries;
\item [Chapter~\ref{sec-TP}] $\tau$ polarisation;
\item [Chapter~\ref{sec-ALR}] Measurement of polarised asymmetries at SLD;
\item [Chapter~\ref{sec-HF}] Heavy flavour analyses;
\item [Chapter~\ref{sec-QFB}] Inclusive hadronic charge asymmetry;
\item [Chapter~\ref{sec-GG}] Photon-pair production at energies above the Z;
\item [Chapter~\ref{sec-FF}] Fermion-pair production at energies above the Z;
\item [Chapter~\ref{sec-4F}] W and four-fermion production;
\item [Chapter~\ref{sec-GC}] Electroweak gauge boson self couplings;
\item [Chapter~\ref{sec-MW}] W-boson mass and width;
\item [Chapter~\ref{sec-eff}] Interpretation of the Z-pole results
  in terms of effective couplings of the neutral weak current;
\item [Chapter~\ref{sec-MSM}] Interpretation of all results, also
  including results from neutrino interaction and atomic parity
  violation experiments as well as from CDF and D\O\  in terms of
  constraints on the Standard Model
\item [Chapter~\ref{sec-Conc}] Conclusions including prospects for the future.
\end{description}
To allow a quick assessment, a box highlighting the updates is given
at the beginning of each section.

\boldmath
\chapter{$\Zzero$ Lineshape and Lepton Forward-Backward Asymmetries}\label{sec-LS-SM}
\label{sec-LS}
\unboldmath

\updates{ Unchanged w.r.t. summer 2000: All experiments have 
  published final results which enter in the combination. 
  The final combination procedure is used. }

\noindent
The results presented here are based on the full \LEPI{} data set.
This includes the data taken during the
energy scans in 1990 and 1991 in the range\footnote{In this note
  $\hbar=c=1$.}  $|\roots-\MZ|<3$~\GeV{}, the data collected at the
$\Zzero$ peak in 1992 and 1994 and the precise energy scans in
1993 and 1995 ($|\roots-\MZ|<1.8$~\GeV{}).  
The total event statistics are given in Table~\ref{tab-LSstat}.
Details of the individual analyses can be found in 
References~\citen{ALEPHLS,DELPHILS,L3LS,OPALLS}. 

\begin{table}[hbtp]
\begin{center}\begin{tabular}{lr} %
\begin{minipage}[b]{0.49\textwidth}
\begin{center}\begin{tabular}{r||rrrr||r}
  \multicolumn{6}{c}{$\qq$}  \\
\hline
   year & A &   D  &   L  &  O  & all \\
\hline
'90/91  & 433 &  357 &  416 &  454 &  1660\\
'92     & 633 &  697 &  678 &  733 &  2741\\
'93     & 630 &  682 &  646 &  649 &  2607\\
'94     &1640 & 1310 & 1359 & 1601 &  5910\\
'95     & 735 &  659 &  526 &  659 &  2579\\
\hline
 total  & 4071 & 3705 & 3625 & 4096 & 15497\\
\end{tabular}\end{center}
\end{minipage}
   &
\begin{minipage}[b]{0.49\textwidth}
\begin{center}\begin{tabular}{r||rrrr||r}
  \multicolumn{6}{c}{$\leptlept$} \\
\hline
   year & A &   D  &   L  &  O  & all \\
\hline
'90/91  &  53 &  36 &  39  &  58  &  186 \\
'92     &  77 &  70 &  59  &  88  &  294 \\
'93     &  78 &  75 &  64  &  79  &  296 \\
'94     & 202 & 137 & 127  & 191  &  657 \\
'95     &  90 &  66 &  54  &  81  &  291 \\
\hline
total   & 500 & 384 & 343  & 497  & 1724 \\
\end{tabular}\end{center}
\end{minipage} \\
\end{tabular} \end{center}
\caption[Recorded event statistics]{
The $\qq$ and $\leptlept$ event statistics, in units of $10^3$, used
for the analysis of the $\Zzero$ line shape and lepton forward-backward
asymmetries by the experiments ALEPH (A), DELPHI (D), L3
(L) and OPAL (O).
}
\label{tab-LSstat}
\end{table}

For the averaging of results the LEP experiments provide a standard
set of 9 parameters describing the information contained in hadronic
and leptonic cross sections and leptonic forward-backward
asymmetries.  These parameters are
convenient for fitting and averaging since they have small
correlations. They are:
\begin{itemize}
\item The mass $\MZ$ and total width $\GZ$ of the Z boson, where
  the definition is based on the Breit-Wigner denominator
  $(s-\MZ^2+is\GZ/\MZ)$ with $s$-dependent width~\cite{ref:QEDCONV}.
\item The hadronic pole cross section of Z exchange:
\begin{equation}
\shad\equiv{12\pi\over\MZ^2}{\Gee\Ghad\over\GZ^2}\,.
\end{equation}
Here $\Gee$ and $\Ghad$ are the partial widths of the $\Zzero$ for
decays into electrons and hadrons.

\item The ratios:
\begin{equation}\label{eqn-sighad}
 \Ree\equiv\Ghad/\Gee, \;\; \Rmu\equiv\Ghad/\Gmumu \;\mbox{and}\;
\Rtau\equiv\Ghad/\Gtautau.
\end{equation}
Here $\Gmumu$ and $\Gtautau$ are the partial widths of the $\Zzero$
for the decays $\Ztomumu$ and $\Ztotautau$.  Due to the mass of
the $\tau$ lepton, a difference of 0.2\% is expected between the
values for $\Ree$ and $\Rmu$, and the value for $\Rtau$, even under
the assumption of lepton universality~\cite{ref:consoli}.
\item The pole asymmetries, $\Afbze$, $\Afbzm$ and $\Afbzt$, for the
  processes $\eeee$, $\eemumu$ and $\eetautau$. In terms of the real
  parts of the effective vector and axial-vector neutral current
  couplings of fermions, $\gvf$ and $\gaf$, the pole asymmetries
  are expressed as
\begin{equation}
\label{eqn-apol}
\Afbzf \equiv {3\over 4} \cAe\cAf
\end{equation}
with
\begin{equation}
\label{eqn-cAf}
\cAf\equiv\frac{2\gvf \gaf} {\gvf^{2}+\gaf^{2}}\ = 2 \frac{\gvf/\gaf} {1+(\gvf/\gaf)^{2}}\,.
\end{equation}
\end{itemize}
The imaginary parts of the vector and axial-vector coupling constants
as well as real and imaginary parts of the photon vacuum polarisation
are taken into account explicitly in the fitting formulae and are fixed to
their Standard Model values.
The fitting procedure takes into account the effects of initial-state
radiation~\cite{ref:QEDCONV} to 
${\cal O}(\alpha^3)$~\cite{ref:Jadach91,ref:Skrzypek92,ref:Montagna96}, as
well as the $t$-channel and  the $s$-$t$ interference contributions in the case 
of $\ee$ final states.

The set of 9 parameters does not describe hadron and
lepton-pair production completely, because it does not include the
interference of the $s$-channel $\Zzero$ exchange with the $s$-channel
$\gamma$ exchange.  For the results presented in this section and used
in the rest of the note, the $\gamma$-exchange contributions and the
hadronic $\gamma\Zzero$ interference terms are fixed to their Standard
Model values.  The leptonic $\gamma\Zzero$ interference terms are
expressed in terms of the effective couplings.

\begin{table}[tp] \begin{center}{\small
\begin {tabular} {lr|@{\,}r@{\,}r@{\,}r@{\,}r@{\,}r@{\,}r@{\,}r@{\,}r@{\,}r}
\hline %
\multicolumn{2}{c|}{~}& \multicolumn{9}{c}{correlations} \\
\multicolumn{2}{c|}{~} & $\MZ$ & $\GZ$ & $\shad$ &
     $\Ree$ &$\Rmu$ & $\Rtau$ & $\Afbze$ & $\Afbzm$ & $\Afbzt$ \\
\hline %
\multicolumn{2}{l}{ $\pzz \chi^2/N_{\rm df}\,=\, 169/176$}& 
                                      \multicolumn{9}{c}{ALEPH} \\
\hline %
 $\MZ$\,[\GeV{}]\hspace*{-.5pc} & 91.1891 $\pm$ 0.0031     &   
  1.00 \\
 $\GZ$\,[\GeV]\hspace*{-2pc}  &  2.4959 $\pm$ 0.0043     & 
  .038 & ~1.00 \\ 
 $\shad$\,[nb]\hspace*{-2pc}  &  41.558 $\pm$ 0.057$\pz$ &   
 $-$.091 & $-$.383 & ~1.00 \\
 $\Ree$        &  20.690 $\pm$ 0.075$\pz$ &   
  .102  &~.004 & ~.134 & ~1.00 \\
 $\Rmu$        &  20.801 $\pm$ 0.056$\pz$ &   
 $-$.003 & ~.012 & ~.167 & ~.083 & ~1.00 \\ 
 $\Rtau$       &  20.708 $\pm$ 0.062$\pz$ &   
 $-$.003 & ~.004 & ~.152 & ~.067 & ~.093 & ~1.00 \\ 
 $\Afbze$      &  0.0184 $\pm$ 0.0034     &   
 $-$.047 & ~.000 & $-$.003 & $-$.388 & ~.000 & ~.000 & ~1.00 \\ 
 $\Afbzm$      &  0.0172 $\pm$ 0.0024     &   
 .072 & ~.002 & ~.002 & ~.019 & ~.013 & ~.000 & $-$.008 & ~1.00 \\ 
 $\Afbzt$      &  0.0170 $\pm$ 0.0028     &   
 .061 & ~.002 & ~.002 & ~.017 & ~.000 & ~.011 & $-$.007 & ~.016 & ~1.00 \\
    ~          & \multicolumn{2}{c}{~}                \\[-0.5pc]
\hline %
\multicolumn{2}{l}{$\pzz \chi^2/N_{\rm df}\,=\, 177/168$} & 
                                        \multicolumn{9}{c}{DELPHI} \\
\hline %
 $\MZ$\,[\GeV{}]\hspace*{-.5pc}   &  91.1864 $\pm$ 0.0028    &
 ~1.00 \\ 
 $\GZ$\,[\GeV]\hspace*{-2pc}   &  2.4876 $\pm$ 0.0041     &
 ~.047 & ~1.00 \\ 
 $\shad$\,[nb]\hspace*{-2pc}   &  41.578 $\pm$ 0.069$\pz$ &
 $-$.070 & $-$.270 & ~1.00 \\ 
 $\Ree$        &  20.88  $\pm$ 0.12$\pzz$ &
 ~.063 & ~.000 & ~.120 & ~1.00 \\ 
 $\Rmu$        &  20.650 $\pm$ 0.076$\pz$ &
 $-$.003 & $-$.007 & ~.191 & ~.054 & ~1.00 \\ 
 $\Rtau$       &  20.84  $\pm$ 0.13$\pzz$ &
 ~.001 & $-$.001 & ~.113 & ~.033 & ~.051 & ~1.00 \\ 
 $\Afbze$      &  0.0171 $\pm$ 0.0049     &
 ~.057 & ~.001 & $-$.006 & $-$.106 & ~.000 & $-$.001 & ~1.00 \\ 
 $\Afbzm$      &  0.0165 $\pm$ 0.0025    &
 ~.064 & ~.006 & $-$.002 & ~.025 & ~.008 & ~.000 & $-$.016 & ~1.00 \\ 
 $\Afbzt$      &  0.0241 $\pm$ 0.0037     & 
 ~.043 & ~.003 & $-$.002 & ~.015 & ~.000 & ~.012 & $-$.015 & ~.014 & ~1.00 \\
    ~          & \multicolumn{2}{c}{~}                \\[-0.5pc]
\hline %
\multicolumn{2}{l}{$\pzz \chi^2/N_{\rm df}\,=\, 158/166 $}  & 
                                    \multicolumn{9}{c}{L3} \\
\hline %
 $\MZ$\,[\GeV{}]\hspace*{-.5pc}   &  91.1897 $\pm$ 0.0030      & 
 ~1.00 \\ 
 $\GZ$\,[\GeV]\hspace*{-2pc}   &   2.5025 $\pm$ 0.0041      & 
 ~.065 & ~1.00 \\ 
 $\shad$\,[nb]\hspace*{-2pc}   &   41.535 $\pm$ 0.054$\pz$  & 
 ~.009 & $-$.343 & ~1.00 \\ 
 $\Ree$        &   20.815  $\pm$ 0.089$\pz$ & 
 ~.108 & $-$.007 & ~.075 & ~1.00 \\ 
 $\Rmu$        &   20.861  $\pm$ 0.097$\pz$ & 
 $-$.001 & ~.002 & ~.077 & ~.030 & ~1.00 \\ 
 $\Rtau$       &   20.79 $\pz\pm$ 0.13$\pzz$& 
 ~.002 & ~.005 & ~.053 & ~.024 & ~.020 & ~1.00 \\ 
 $\Afbze$      &   0.0107 $\pm$ 0.0058      & 
 $-$.045 & ~.055 & $-$.006 & $-$.146 & $-$.001 & $-$.003 & ~1.00 \\ 
 $\Afbzm$      &   0.0188 $\pm$ 0.0033      & 
 ~.052 & ~.004 & ~.005 & ~.017 & ~.005 & ~.000 & ~.011 & ~1.00 \\ 
 $\Afbzt$      &   0.0260 $\pm$ 0.0047      & 
 ~.034 & ~.004 & ~.003 & ~.012 & ~.000 & ~.007 & $-$.008 & ~.006 & ~1.00 \\
    ~          & \multicolumn{2}{c}{~}                \\[-0.5pc]
\hline %
\multicolumn{2}{l}{$\pzz \chi^2/N_{\rm df}\,=\, 155/194 $} & 
                                      \multicolumn{9}{c}{OPAL} \\
\hline %
 $\MZ$\,[\GeV{}]\hspace*{-.5pc}   & 91.1858 $\pm$ 0.0030 &
 ~1.00 \\ 
 $\GZ$\,[\GeV]\hspace*{-2pc}   & 2.4948  $\pm$ 0.0041 & 
 ~.049 & ~1.00 \\ 
 $\shad$\,[nb]\hspace*{-2pc}   & 41.501 $\pm$ 0.055$\pz$ & 
 ~.031 & $-$.352 & ~1.00 \\ 
 $\Ree$        & 20.901 $\pm$ 0.084$\pz$ &
 ~.108 & ~.011 & ~.155 & ~1.00 \\ 
 $\Rmu$        & 20.811 $\pm$ 0.058$\pz$ &
 ~.001 & ~.020 & ~.222 & ~.093 & ~1.00 \\ 
 $\Rtau$       & 20.832 $\pm$ 0.091$\pz$ &
 ~.001 & ~.013 & ~.137 & ~.039 & ~.051 & ~1.00 \\ 
 $\Afbze$      & 0.0089 $\pm$ 0.0045 &
 $-$.053 & $-$.005 & ~.011 & $-$.222 & $-$.001 & ~.005 & ~1.00 \\ 
 $\Afbzm$     & 0.0159 $\pm$ 0.0023 &
 ~.077 & $-$.002 & ~.011 & ~.031 & ~.018 & ~.004 & $-$.012 & ~1.00 \\ 
 $\Afbzt$      & 0.0145 $\pm$ 0.0030 & 
 ~.059 & $-$.003 & ~.003 & ~.015 & $-$.010 & ~.007 & $-$.010 & ~.013 & ~1.00 \\
\hline %
\end{tabular}}%
\caption[Nine parameter results]{\label{tab-ninepar}
Line Shape and asymmetry parameters from fits to the data of the four
LEP experiments and their correlation coefficients. }
\end{center}
\end{table} 

The four sets of nine parameters provided by the LEP experiments are
presented in Table~\ref{tab-ninepar}.  For performing the average over
these four sets of nine parameters, the overall covariance matrix is
constructed from the covariance matrices of the individual LEP
experiments and taking into account common systematic
errors~\cite{LEPLS}.  The common systematic errors include theoretical
errors as well as errors arising from the uncertainty in the LEP beam
energy.  The beam energy uncertainty contributes an uncertainty of
$\pm1.7~\MeV$ to $\MZ$ and $\pm1.2~\MeV$ to $\GZ$. In addition, the
uncertainty in the centre-of-mass energy spread of about $\pm1~\MeV$
contributes $\pm0.2~\MeV$ to $\GZ$.  The theoretical error on
calculations of the small-angle Bhabha cross section is
$\pm$0.054\,\%\cite{bib-lumthopal} for OPAL and
$\pm$0.061\,\%\cite{bib-lumth99} for all other experiments, and
results in the largest common systematic uncertainty on $\shad$.  QED
radiation, dominated by photon radiation from the initial state
electrons, contributes a common uncertainty of $\pm$0.02\,\% on
$\shad$, of $\pm0.3$~\MeV{} on $\MZ$ and of $\pm0.2$~\MeV{} on $\GZ$.
The contribution of $t$-channel diagrams and the $s$-$t$ interference
in $\Zzero\ra\ee$ leads to an additional theoretical uncertainty
estimated to be $\pm0.024$ on $\Ree$ and $\pm0.0014$ on $\Afbze$,
which are fully anti--correlated.  Uncertainties from the
model-independent parameterisation of the energy dependence of the
cross section are almost negligible, if the definitions of
Reference\,\cite{bib-PCP99} are applied. Through unavoidable remaining
Standard Model assumptions, dominated by the need to fix the
$\gamma$-$\Zzero$ interference contribution in the $\qq$ channel,
there is some small dependence of $\pm 0.2$ \MeV{} of $\MZ$ on the
Higgs mass, $\MH$ (in the range 100 \GeV{} to 1000 \GeV{}) and the
value of the electromagnetic coupling constant. Such ``parametric''
errors are negligible for the other results. The combined
parameter set and its correlation matrix are given in
Table~\ref{tab-zparavg}.

\begin{table}[htb]\begin{center}
\begin {tabular} {lr|r@{\,}r@{\,}r@{\,}r@{\,}r@{\,}r@{\,}r@{\,}r@{\,}r}
\hline %
\multicolumn{2}{c|} {without lepton universality} & 
                                    \multicolumn{9}{l}{~~~correlations} \\
\hline %
\multicolumn{2}{c|}{$\pzz \chi^2/N_{\rm df}\,=\,32.6/27 $} &
   $\MZ$ & $\GZ$ & $\shad$ &
     $\Ree$ &$\Rmu$ & $\Rtau$ & $\Afbze$ & $\Afbzm$ & $\Afbzt$ \\
\hline %
 $\MZ$ [\GeV{}]  & 91.1876$\pm$ 0.0021 &
 ~1.00 \\
 $\GZ$ [\GeV]  & 2.4952 $\pm$ 0.0023 &
 $-$.024 & ~1.00 \\ 
 $\shad$ [nb]  & 41.541 $\pm$ 0.037$\pz$ &
 $-$.044 & $-$.297 & ~1.00 \\ 
 $\Ree$        & 20.804 $\pm$ 0.050$\pz$ &
 ~.078 & $-$.011 & ~.105 & ~1.00 \\ 
 $\Rmu$        & 20.785 $\pm$ 0.033$\pz$ & 
 ~.000 & ~.008 & ~.131 & ~.069 & ~1.00 \\ 
 $\Rtau$       & 20.764 $\pm$ 0.045$\pz$ &  
 ~.002 & ~.006 & ~.092 & ~.046 & ~.069 & ~1.00 \\ 
 $\Afbze$      & 0.0145 $\pm$ 0.0025 &
 $-$.014 & ~.007 & ~.001 & $-$.371 & ~.001 & ~.003 & ~1.00 \\ 
 $\Afbzm$      & 0.0169 $\pm$ 0.0013 &
 ~.046 & ~.002 & ~.003 & ~.020 & ~.012 & ~.001 & $-$.024 & ~1.00 \\ 
 $\Afbzt$      & 0.0188 $\pm$ 0.0017 &
 ~.035 & ~.001 & ~.002 & ~.013 & $-$.003 & ~.009 & $-$.020 & ~.046 & ~1.00 \\ 
\multicolumn{3}{c}{~}\\[-0.5pc]
\multicolumn{2}{c} {with lepton universality} \\
\hline %
\multicolumn{2}{c|}{$\pzz \chi^2/N_{\rm df}\,=\,36.5/31 $}  & 
   $\MZ$ & $\GZ$ & $\shad$ & $\Rl$ &$\Afbzl$ \\
\hline %
 $\MZ$ [\GeV{}]  & 91.1875$\pm$ 0.0021$\pz$    &
 ~1.00 \\ 
 $\GZ$ [\GeV]  & 2.4952 $\pm$ 0.0023$\pz$    &
 $-$.023  & ~1.00 \\ 
 $\shad$ [nb]  & 41.540 $\pm$ 0.037$\pzz$ &
 $-$.045 & $-$.297 &  ~1.00 \\ 
 $\Rl$         & 20.767 $\pm$ 0.025$\pzz$ &
 ~.033 & ~.004 & ~.183 & ~1.00 \\ 
 $\Afbzl$      & 0.0171 $\pm$ 0.0010   & 
 ~.055 & ~.003 & ~.006 & $-$.056 &  ~1.00 \\ 
\hline %
\end{tabular} 
\caption[]{
  Average line shape and asymmetry parameters from the data of the
  four LEP experiments,  without and with the
  assumption of lepton universality. }
\label{tab-zparavg}
\end{center}
\end{table}

\begin{figure}[p]
\vspace*{-0.6cm}
\begin{center}
  \mbox{\includegraphics[width=0.9\linewidth]{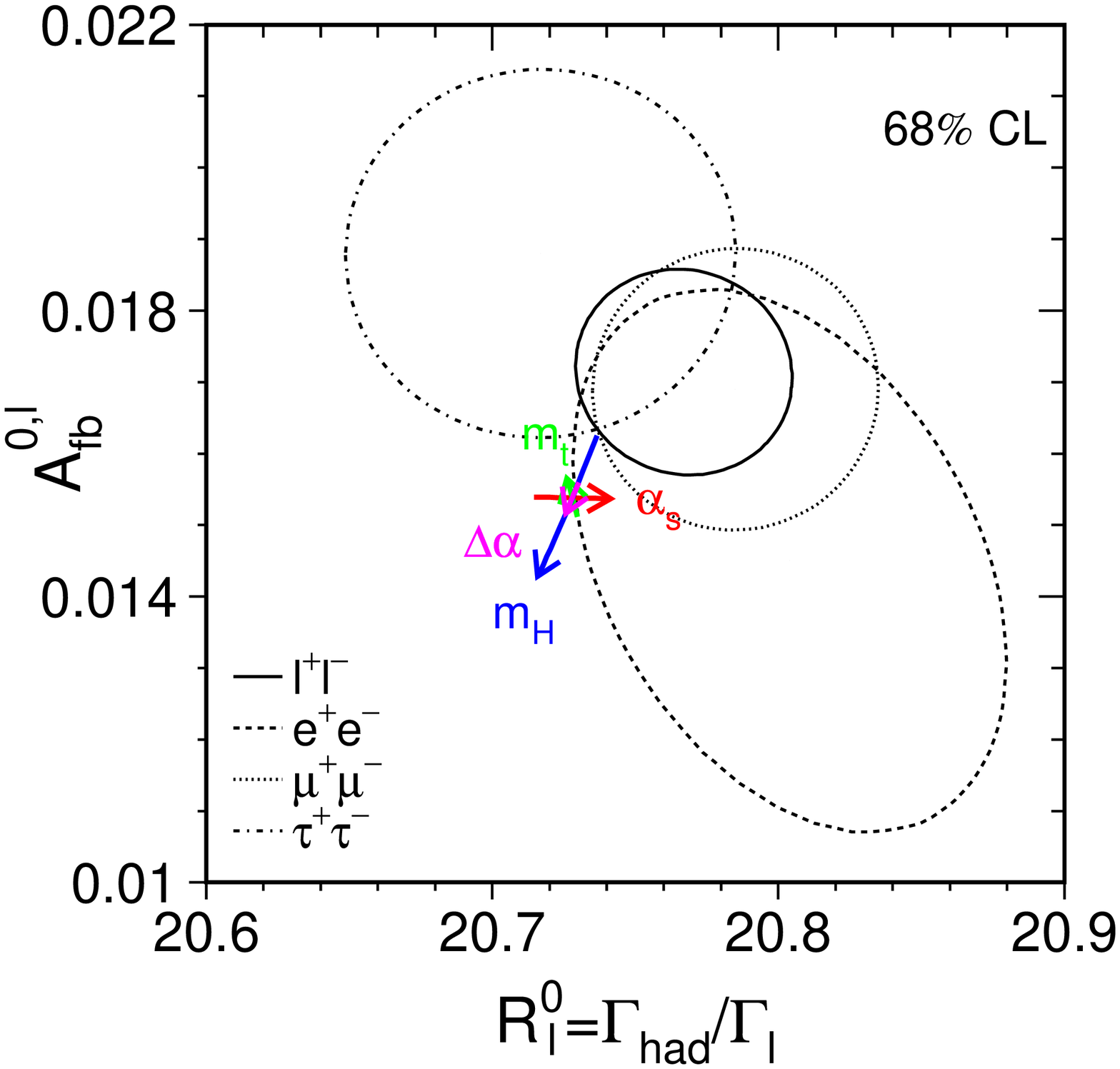}}
\end{center}
\caption[]{
  Contours of 68\% probability in the $\RZ$-$\Afbzl$ plane. For better
  comparison the results for the $\tau$ lepton are corrected to
  correspond to the massless case.  The $\SM$ prediction for
  $\MZ=91.1875$~\GeV{}, $\Mt=174.3$~\GeV{}, $\MH=300$~\GeV{}, and
  $\alfmz=0.118$ is also shown. The lines with arrows correspond to
  the variation of the $\SM$ prediction when $\Mt$, $\MH$, $\alfmz$
  and $\Delta\alpha_{\mathrm{had}}^{(5)}(\MZ^2)$ are varied in the
  intervals $\Mt=174.3\pm5.1$~\GeV{}, $\MH=300^{+700}_{-186}~\GeV$,
  $\alfmz=0.118\pm0.002$ and
  $\Delta\alpha_{\mathrm{had}}^{(5)}(\MZ^2)=0.02761\pm0.00036$,
  respectively. The arrows point in the direction of increasing values
  of $\Mt$, $\MH$, $\alfas$ and
  $\Delta\alpha_{\mathrm{had}}^{(5)}(\MZ^2)$.  }
\label{fig-LU} 
\end{figure}

If lepton universality is assumed, the set of 9 parameters  
is reduced to a set of 5 parameters. $\RZ$ is defined as 
$\RZ\equiv\Ghad/\Gll$, where $\Gll$ refers to the partial 
$\Zzero$ width for the decay into a pair of massless charged 
leptons.  The data of each of the four LEP experiments are 
consistent with lepton universality (the difference in $\chi^2$ 
over the difference in d.o.f.{} with and without the assumption 
of lepton universality is 3/4, 6/4, 5/4 and 3/4 for ALEPH, DELPHI, 
L3 and OPAL, respectively). The lower part of Table~\ref{tab-zparavg} 
gives the combined result and the corresponding correlation matrix.  
Figure~\ref{fig-LU} shows, for each lepton species and for the 
combination assuming lepton universality, the resulting 68\% 
probability contours in the $\RZ$-$\Afbzl$ plane. Good agreement
is observed. 

For completeness the partial decay widths of the $\Zzero$ boson 
are listed in Table~\ref{tab-widths}, although 
they are more correlated than the ratios given in 
Table~\ref{tab-zparavg}. The leptonic pole cross-section, 
$\sll$, defined as
\begin{eqnarray}
\sll & \equiv & {12\pi\over\MZ^2}{\Gll^2\over\GZ^2} \, ,
\end{eqnarray}
in analogy to $\shad$, is shown in the last line of the Table.
Because QCD final state corrections appear twice in the 
denominator via $\GZ$, $\sll$ has a higher sensitivity to $\alpha_s$ 
than $\shad$ or $\Rl$, where the dependence on QCD corrections is 
only linear.

\begin{table}[hbtp] \begin{center} 
\begin{tabular} {lr|r@{\,}r@{\,}r@{\,}r}
\hline %
 \multicolumn{2}{c|}{without lepton universality} &
                            \multicolumn{4}{l}{~~~correlations} \\
 & & $\Ghad$ & $\Gee$ & $\Gmumu$ & $\Gtautau$ \\
\hline %
$\Ghad$ [\MeV]     & 1745.8$\pz\pm$2.7$\pz\pzz$ & ~1.00 \\
$\Gee$ [\MeV]      & 83.92$\pm$0.12$\pzz$& $-$0.29 & ~1.00 \\ 
$\Gmumu$ [\MeV]    & 83.99$\pm$0.18$\pzz$& ~0.66 & $-$0.20 & ~1.00 \\
$\Gtautau$ [\MeV]  & 84.08$\pm$0.22$\pzz$&  0.54 & $-$0.17 & ~0.39 & ~1.00 \\  
\hline %
\multicolumn{6}{c}{~} \\[-0.5pc]
\hline %
 \multicolumn{2}{c|}{with    lepton universality} &
                            \multicolumn{4}{l}{~~~correlations} \\
 & & $\Ginv$ & $\Ghad$ & $\Gll$ &  \\
\hline %
$\Ginv$ [\MeV]     & $\pz$499.0$\pzz\pm$1.5$\pz\pzz$    & ~1.00 \\
$\Ghad$ [\MeV]     & 1744.4$\pzz\pm$2.0$\pz\pzz$   & $-$0.29 & ~1.00 \\
$\Gll$ [\MeV]       & 83.984$\pm$0.086$\pz$  & ~0.49 & ~0.39 & ~1.00 \\
\hline %
$\Ginv/\Gll$        & {$\pz$5.942\pz$\pm$0.016$\pz$} &    \\
\hline %
$\sll$ [nb]      & {2.0003$\pm$0.0027} &  \\
\hline %
\end{tabular} 
\caption[]{
  Partial decay widths of the $\Zzero$ boson, derived from the results of the
  9-parameter averages in Table~\ref{tab-zparavg}. In the
  case of lepton universality, $\Gll$ refers to the partial $\Zzero$ width for
  the decay into a pair of massless charged leptons.  }
\label{tab-widths}
\end{center}
\end{table}
\boldmath
\section{Number of Neutrino Species}
\label{sec-Nnu}
\unboldmath

An important aspect of our measurement concerns the information
related to $\Zzero$ decays into invisible channels. Using the results
of Table~\ref{tab-zparavg}, the ratio of
the $\Zzero$ decay width into invisible particles and the leptonic
decay width is determined:
\begin{eqnarray}
\Ginv / \Gll & = & 5.942\pm 0.016\,.
\end{eqnarray}
The Standard Model value for the ratio of the partial widths to
neutrinos and charged leptons is:
\begin{eqnarray}
(\Gnn / \Gll)_{\mathrm{SM}} & = & 1.9912\pm 0.0012\,.
\end{eqnarray}
The central value is evaluated for $\MZ=91.1875$~\GeV{} 
and the error quoted accounts for
a variation of $\Mt$ in the range $\Mt=174.3\pm5.1~\GeV$ and a
variation of $\MH$ in the range $100~\GeV \le \MH \le 1000~\GeV$.  
The number of light neutrino
species is given by the ratio of the two expressions listed above:
\begin{eqnarray}
\Nnu & = & 2.9841\pm 0.0083,
\end{eqnarray}
which is two standard deviations below the value of 3 expected from 3
observed fermion families.

Alternatively, one can assume 3 neutrino species and determine the
width from additional invisible decays of the Z.  This yields
\begin{eqnarray}
  \Delta\Ginv & = & -2.7 \pm 1.6\ \MeV.
\end{eqnarray}
The measured total width
is below the Standard Model expectation.  
If a conservative approach is taken to limit the result to only
positive values of $\Delta\Ginv$ and renormalising the probability for
$\Delta\Ginv\ge0$ to be unity, then the resulting 95\% CL upper limit on
additional invisible decays of the Z is
\begin{eqnarray}
  \Delta\Ginv & < & 2.0\ \MeV.
\end{eqnarray}

The theoretical error on the luminosity\cite{bib-lumth99} constitutes
a large part of the uncertainties on $\Nnu$ and $\Delta\Ginv$.

\boldmath
\chapter{The $\tau$ Polarisation}
\label{sec-TP}
\unboldmath

\updates{ OPAL has finalised their results. While all results are now
  final, the combination procedure itself is still preliminary. }

\noindent
The longitudinal $\tau$ polarisation $\cal {P}_{\tau}$ of $\tau$
pairs produced 
in $\Zzero$ decays is defined as
\begin{eqnarray}
{\cal P}_{\tau} & \equiv & 
\frac{\sigma_{\mathrm{R}} - \sigma_{\mathrm{L}}}
{\sigma_{\mathrm{R}} + \sigma_{\mathrm{L}}} \, ,
\end{eqnarray}
where $\sigma_{\mathrm{R}}$ and $\sigma_{\mathrm{L}}$ are the $\tau$-pair cross sections for
the production of a right-handed and left-handed $\tau^-$,
respectively. The distribution of $\ptau$ as a function of the polar
scattering angle $\theta$ between the $\mathrm{e}^-$ and the $\tau^-$,
at $\roots = \MZ$, is given by
\begin{eqnarray}
\label{eqn-taupol}
{\cal P}_{\tau}(\cos\theta) & = &
 - \frac{\cAt(1+\cos^2\theta) + 2    \cAe\cos\theta}
             {1+\cos^2\theta  + 2\cAt\cAe\cos\theta} \, ,
\end{eqnarray}
with $\cAe$ and $\cAt$ as defined in Equation~(\ref{eqn-cAf}).
Equation~(\ref{eqn-taupol}) is valid for pure Z exchange.  The effects of
$\gamma$ exchange, $\gamma$-$\Zzero$ interference and electromagnetic
radiative corrections in the initial and final states
are taken into account in the experimental analyses.  In
particular, these corrections account for the $\sqrt{s}$ dependence of
the $\tau$ polarisation, which is important because
the off-peak data are included in the event samples for all
experiments.
When averaged over all production angles $\cal {P}_{\tau}$ is a
measurement of $\cAt$. As a function of $\cos\theta$, $\cal
{P}_{\tau}(\cos\theta)$ provides nearly independent determinations of
both $\cAt$ and $\cAe$, thus allowing a test of the universality of
the couplings of the $\Zzero$ to $\mathrm{e}$ and $\tau$.

Each experiment makes separate $\ptau$ measurements using the five
$\tau$ decay modes e$\nu \overline{\nu}$, $\mu\nu \overline{\nu}$,
$\pi\nu$, $\rho\nu$ and
$a_{1}\nu$\cite{bib-ALEPHTAU,bib-DELPHITAUnew,bib-L3TAUfin,bib-OPALTAU}.
The $\rho\nu$ and $\pi\nu$ are the most sensitive channels,
contributing weights of about $40\%$ each in the average.  DELPHI and
L3 also use an inclusive hadronic analysis. The combination is made
using the results from each experiment already averaged over the
$\tau$ decay modes.

\section{Results}

Tables~\ref{tab-tau1} and~\ref{tab-tau2} show the most recent results
for $\cAt$ and $\cAe$ obtained by the four LEP
collaborations\cite{bib-ALEPHTAU,bib-DELPHITAUnew,bib-L3TAUfin,bib-OPALTAU}
and their combination.  Although the size of the event samples used by
the four experiments are roughly equal, smaller errors are quoted by
ALEPH. This is largely associated with the higher angular granularity
of the ALEPH electromagnetic calorimeter.  Common systematic errors
arise from uncertainties in radiative corrections (decay radiation) in
the $\pi\nu$ and $\rho\nu$ channels, and in the modelling of the
$a_{1}$ decays\cite{bib-EWPPE187}.  These errors and their
correlations need further investigation, but are already taken into
account in the combination (see also Reference~\citen{bib-L3TAUfin}).
The statistical
correlation between the extracted values of $\cAt$ and $\cAe$ is small
($\le$ 5\%). %

The average values for $\cAt$ and $\cAe$:
\begin{eqnarray}
  \cAt & = & 0.1439 \pm 0.0043 \\
  \cAe & = & 0.1498 \pm 0.0049 \,,
\end{eqnarray}
with a correlation of 0.013, are compatible, in good agreement with
neutral-current lepton universality. Assuming $\mathrm{e}$-$\tau$
universality, the values for $\cAt$ and $\cAe$ can be combined. This
combination is performed including the small common systematic errors
between $\cAt$ and $\cAe$ within each experiment and between
experiments.  The combined result of $\cAt$ and $\cAe$ is:
\begin{eqnarray}
  \cAl & = & 0.1465 \pm 0.0033 \,,
\end{eqnarray}
where the error includes a systematic component of 0.0016.

\begin{table}[htbp]
\renewcommand{\arraystretch}{1.15}
\begin{center}
\begin{tabular}{|ll||c|}
\hline
Experiment & & $\cAt$ \\
\hline
\hline
ALEPH  &(90 - 95), final       & $0.1451\pm0.0052\pm0.0029$  \\
DELPHI &(90 - 95), final       & $0.1359\pm0.0079\pm0.0055$  \\
L3     &(90 - 95), final       & $0.1476\pm0.0088\pm0.0062$  \\
OPAL   &(90 - 95), final       & $0.1456\pm0.0076\pm0.0057$  \\
\hline
\hline
LEP Average &  preliminary     & $0.1439\pm0.0043$           \\
\hline
\end{tabular}
\end{center}
\caption[]{
  LEP results for $\cAt$.  The
  first error is statistical and the second systematic. In the LEP average,
  statistical and systematic errors are combined in quadrature. The systematic
  component of the error 
  is $\pm0.0026$.}
\label{tab-tau1}
\end{table}
\begin{table}[htbp]
\renewcommand{\arraystretch}{1.15}
\begin{center}
\begin{tabular}{|ll||c|}
\hline
Experiment & & $\cAe$ \\
\hline
\hline
ALEPH   &(90 - 95), final       & $0.1504\pm0.0068\pm0.0008$  \\
DELPHI  &(90 - 95), final       & $0.1382\pm0.0116\pm0.0005$  \\
L3      &(90 - 95), final       & $0.1678\pm0.0127\pm0.0030$  \\
OPAL    &(90 - 95), final       & $0.1454\pm0.0108\pm0.0036$  \\
\hline
\hline
LEP Average &  preliminary      & $0.1498\pm0.0049$  \\
\hline
\end{tabular}
\end{center}
\caption[]{
  LEP results for $\cAe$.  The
  first error is statistical and the second systematic. In the LEP average,
  statistical and systematic errors are combined in quadrature. The systematic
  component of the error 
  is $\pm 0.0009$.}
\label{tab-tau2}
\end{table}

\boldmath
\chapter{Measurement of polarised lepton asymmetries at SLC}
\unboldmath
\label{sec-ALR}

\updates{ Unchanged w.r.t. summer 2000: SLD has final results for
  \ALR{} and the leptonic left-right forward-backward asymmetries.  }

\noindent
The measurement of the left-right cross section asymmetry ($\ALR$) by
SLD\cite{ref:sld-s00} at the SLC provides a systematically precise,
statistics-dominated determination of the coupling $\cAe$, and is
presently the most precise single measurement, with the smallest
systematic error, 
of this quantity.  In principle the analysis is
straightforward: one counts the numbers of Z bosons produced by left
and right longitudinally polarised electrons, forms an asymmetry, and
then divides by the luminosity-weighted e$^-$ beam polarisation
magnitude (the e$^+$ beam is not polarised):
\begin{equation}
  \label{eq:ALR}
  \ALR = \frac{N_{\mathrm{L}} - N_{\mathrm{R}}}%
              {N_{\mathrm{L}} + N_{\mathrm{R}}}%
         \frac{1}{P_{\mathrm{e}}}.
\end{equation}
Since the advent of high polarisation ``strained lattice'' GaAs
photo-cathodes (1994), the average electron
polarisation at the interaction point has been in the range 73\% to 77\%.
The method requires no
detailed final state event identification ($\ee$ final state events
are removed, as are non-Z backgrounds) and is insensitive to all
acceptance and efficiency effects.  The small total systematic error
of  0.64\% relative 
is
dominated by the 0.50\%
relative 
systematic error in the determination of the e$^-$ polarisation.
The 
relative 
statistical error on $\ALR$ is about 1.3\%.

The precision Compton polarimeter detects beam electrons
that are scattered by photons from a circularly polarised laser.
Two additional polarimeters that are sensitive to the Compton-scattered
photons and which are operated in the absence of positron beam, 
have verified the precision polarimeter result and are used to set a
calibration uncertainty of 0.4\% relative.
In 1998, a dedicated experiment was performed in order to test directly
the expectation that accidental polarisation of the positron beam was
negligible; the e$^+$ polarisation was found to be consistent with 
zero ($-0.02\pm 0.07$)\%.

The $\ALR$ analysis includes several very small corrections. The
polarimeter result is corrected for higher order QED and
accelerator related effects, a total of
($-0.22\pm0.15$)\% relative for 1997/98 data. 
The event asymmetry is
corrected for backgrounds and accelerator asymmetries, a total of
($+0.15\pm0.07$)\% relative, for 1997/98 data.

The translation of the $\ALR$ result to a ``pole'' value is a ($-2.5\pm0.4$)\%
relative shift, where the uncertainty arises from the precision of the
centre-of-mass
energy
determination.  
This small error due to the beam energy measurement 
reflects the results
of a scan of the Z peak used to calibrate the energy spectrometers 
to $\MZ$ from LEP data. 
The pole value, $\ALRz$, is equivalent to a measurement
of $\cAe$.

The 2000  result is included in a running average of all
of the SLD $\ALR$ measurements (1992, 1993, 1994/1995, 1996, 1997 and 1998).
This updated result for $\ALRz$ ($\cAe$)
is $0.1514 \pm 0.0022$.
In addition, the left-right forward-backward asymmetries for leptonic
final states are measured\cite{ref:sld-asym}.  From these, the parameters $\cAe$,
$\cAm$ and $\cAt$ can be determined.  The results are
$\cAe = 0.1544 \pm 0.0060$, $\cAm = 0.142 \pm 0.015$ and $\cAt =
0.136 \pm 0.015$. 
The lepton-based result for $\cAe$ can be combined 
with the $\ALRz$ result to yield
$\cAe = 0.1516 \pm 0.0021$, 
including small correlations in the systematic errors.
The correlation of this measurement with  
$\cAm$ and $\cAt$ is 
indicated in Table~\ref{tab:corr-As}.

Assuming lepton universality, the $\ALR$ result and the results on the
leptonic left-right forward-backward asymmetries 
can be combined, while accounting
for small
correlated systematic errors, yielding
\begin{equation}
 \cAl = 0.1513 \pm 0.0021.
\end{equation}

\begin{table}[h]
\centering
\begin{tabular}{c|ccc} 
       & $\cAe$ & $\cAm$ & $\cAt$ \\
\hline
$\cAe$ &  1.000  \\
$\cAm$ &  0.038 & 1.000 \\
$\cAt$ &  0.033 & 0.007  & 1.000 \\
\end{tabular}
\caption{Correlation coefficients  between $\cAe$,
$\cAm$ and $\cAt$}
\label{tab:corr-As}
\end{table}

\boldmath
\chapter{Results from b and c Quarks}
\label{sec-HF}
\unboldmath


\updates{ ALEPH has updated their $\Abb$ jet-charge measurement using
  a neural net charge tag
  
  DELPHI has presented new measurements of $\Abb$ using a neural net
  charge tag
  
  SLD has updated $\Rc$ and most $\cAb$ and $\cAc$ measurements.  }

\section{Introduction}

The relevant quantities in the heavy quark sector at \LEPI/SLD which are
currently determined by the combination procedure are:
\begin{itemize}
\item The ratios
  of the b and c quark partial widths of the Z to its total hadronic
  partial width: $\Rbz \equiv \Gbb / \Ghad$ and $\Rcz \equiv \Gcc /
  \Ghad$. (The symbols \Rb, \Rc{} are used to denote the
  experimentally measured ratios of event rates or cross sections.)
\item The forward-backward asymmetries, \Abb{} and \Acc.
\item The final state coupling parameters $\cAb,\,\cAc$ obtained from the
  left-right-forward-backward asymmetry at SLD.
\item The semileptonic branching ratios, $\Brbl$, $\Brbclp$ and $\Brcl$, and
  the average time-integrated $\Bzero\Bzerob$ mixing parameter, $\chiM$. 
  These are often determined at the same time or with similar methods
  as the asymmetries.
  Including them in the combination greatly reduces the errors.
  For example the measurements of $\chiM$ act as an effective measurement of 
  the   charge tagging efficiency, so that all errors coming from the mixture of
  different lepton sources in $\bb$ events cancel in the asymmetries.

\item The probability that a c quark produces a $\Dplus$, $\Ds$, 
 $\Dstarp$ meson\footnote{%
   Actually the product $\PcDst$ is fitted because this quantity is
   needed and measured by the LEP experiments.}  or a charmed baryon.
 The probability that a c quark fragments into a $\Dzero$ is
 calculated from the constraint that the probabilities for the weakly
 decaying charmed hadrons add up to one.  
\end{itemize}
A full description of the averaging procedure is published in \cite{ref:lephf};
the main motivations for the procedure are outlined here.  
Several analyses measure
more than one parameter simultaneously, for example the asymmetry measurements
with leptons or D mesons.
Some of the measurements of electroweak parameters depend explicitly
on the values of other parameters, for example \Rb{} depends on \Rc.
The common tagging and analysis techniques lead to common sources of
systematic uncertainty, in particular for the double-tag measurements
of \Rb.  The starting point for the combination is to ensure that all
the analyses use a common set of assumptions for input parameters
which give rise to systematic uncertainties.  
The input parameters are updated
and extended \cite{ref:lephfnew} to accommodate new analyses
and more recent measurements.  The correlations and interdependencies
of the input measurements are then taken into account in a $\chi^2$
minimisation which results in the combined electroweak parameters and
their correlation matrix.

\section{Summary of Measurements and Averaging Procedure}

All measurements are presented by the LEP and SLD collaborations in
a consistent manner for the purpose of combination.
The tables prepared by the experiments include a detailed breakdown of
the systematic error of each measurement and its dependence on other
electroweak parameters. Where necessary, the experiments apply small
corrections to their results in order to use agreed values and ranges
for the input parameters to calculate systematic errors.  The
measurements, corrected where necessary, are summarised in
Appendix~\ref{app-HF-tab} in Tables~\ref{tab:Rbinp}--\ref{tab:RcPcDstinp},
where the statistical and systematic errors are quoted separately.
The correlated systematic entries are from physics sources shared with one or
more other results in the tables and are derived from the full
breakdown of common systematic uncertainties. The uncorrelated
systematic entries come from the remaining sources.


\subsection{Averaging Procedure}

A $\chi^2$ minimisation procedure is used to derive the values of the
heavy-flavour electroweak parameters,  following the procedure
described in 
Reference~\citen{ref:lephf}.  The full statistical and systematic
covariance matrix for all measurements is calculated.  This
correlation matrix takes into account correlations between different measurements
of one experiment and between different experiments.  The
explicit dependence of each measurement on the other parameters is
also accounted for.  

Since c-quark events form the main background in the \Rb{} analyses,
the value of \Rb{} depends on the value of \Rc. If \Rb{} and \Rc{} are
measured in the same analysis, this is reflected in the correlation
matrix for the results.  However the analyses do not determine \Rb\ 
and \Rc\ simultaneously but instead measure \Rb\ for an assumed value
of \Rc. In this case the dependence is parameterised as
\begin{eqnarray}
 \Rb & = & 
 \Rb^{\rm{meas}} + a(\Rc) \frac {(\Rc - \Rc^{\rm{used}} )} {\Rc}.
\label{eq:rbrc}
\end{eqnarray}
In this expression, $\Rb^{\rm{meas}}$ is the result of the analysis
assuming a value of $\Rc = \Rc^{\rm{used}}$. The values of
$\Rc^{\rm{used}}$ and the coefficients $a(\Rc)$ are given in
Table~\ref{tab:Rbinp} where appropriate. The dependence of all other
measurements on other electroweak parameters is treated in the same
way, with coefficients $a(x)$ describing the dependence on parameter
$x$.

\subsection{Partial Width Measurements}

The measurements of \Rb{} and \Rc{} fall into two categories. In the
first, called a single-tag measurement, a method to select b or c
events is devised, and the number of tagged events is counted. This
number must then be corrected for backgrounds from other flavours and
for the tagging efficiency to calculate the true fraction of hadronic
\Zzero{} decays of that flavour. The dominant systematic errors come
from understanding the branching ratios and detection efficiencies
which give the overall tagging efficiency. For the second technique,
called a double-tag measurement, each event is divided into two
hemispheres.  With $N_t$ being the number of tagged hemispheres,
$N_{tt}$ the number of events with both hemispheres tagged and
$N_{\rm{had}}$ the total number of hadronic \Zzero{} decays one has
\begin{eqnarray}
   \frac{N_t}{2N_{\rm{had}}} &=& \effb \Rb
                        + \effc  \Rc +
                        \effuds ( 1 - \Rb - \Rc ) ,\\
   \frac{N_{tt}}{N_{\rm{had}}} &=& \Cb \effb^2 \Rb
                +    \Cc \effc^2 \Rc +
                          {\cal C}_{\mathrm{uds}} \effuds^2 ( 1 - \Rb - \Rc ) ,
\end{eqnarray}
where $\effb$, $\effc$ and $\effuds$ are the tagging efficiencies per
hemisphere for b, c and light-quark events, and $\Cq \ne 1$ accounts
for the fact that the tagging efficiencies between the hemispheres may
be correlated.  In the case of \Rb{} one has $\effb\gg\effc\gg\effuds$,
$\Cb \approx 1$. The correlations for the other flavours can be
neglected. These equations can be solved to give \Rb{} and $\effb$.
Neglecting the c and uds backgrounds and the correlations, they are
approximately given by
\begin{eqnarray}
\effb &\approx& 2 N_{tt} / N_t  , \\
\Rb &\approx& N_t^2 / (4N_{tt}N_{\rm{had}}).
\end{eqnarray}
The double-tagging method has the advantage that the b tagging
efficiency is derived 
from the data, reducing the systematic
error. The residual background of other flavours in the sample, and
the evaluation of the correlation between the tagging efficiencies in
the two hemispheres of the event are the main sources of systematic
uncertainty in such an analysis.

This method can be enhanced by including more tags. All additional 
efficiencies can be determined from the data, reducing the statistical 
uncertainties without adding new systematic uncertainties.

Small corrections must be applied to the results to obtain the partial
width ratios \Rbz{} and \Rcz{} from the cross section ratios \Rb{} and \Rc{}.
These corrections depend slightly on the 
invariant mass cutoff of the simulations used by the experiments;
they are applied by the collaborations before the combination.

The partial width measurements included are:
\begin{itemize}
\item Lifetime (and lepton) double-tag measurements for \Rb{} from
  ALEPH\cite{ref:alife}, DELPHI\cite{ref:drb}, L3\cite{ref:lrbmixed},
  OPAL\cite{ref:omixed} and SLD\cite{ref:SLD_R_B}.  These are the most
  precise determinations of \Rb.
  Since they completely dominate the combined result, no other \Rb{}
  measurements are used at present.
  The basic features of the double-tag technique are discussed above.
  In the ALEPH, DELPHI, OPAL and SLD measurements the charm rejection is
  enhanced by using the invariant mass information. DELPHI, OPAL and SLD
  also add kinematic information from the particles at the
  secondary vertex.
  The ALEPH and DELPHI measurements make use of several different
  tags; this improves the statistical accuracy and reduces the
  systematic errors due to hemisphere correlations and charm
  contamination, compared with the simple single/double tag.
\item Analyses with D/$\Dstarpm$ mesons to measure \Rc{} from
  ALEPH, DELPHI and OPAL.
  All measurements are constructed in such a way that no assumptions on the
  charm fragmentation are necessary as these are determined from the
  \LEPI\ data.  The
  available measurements can be divided into three groups:
\begin{itemize}
\item inclusive/exclusive double tag (ALEPH\cite{ref:arcd}, 
  DELPHI\cite{ref:drcd,ref:drcc}, OPAL\cite{ref:orcd}): In a first
  step $\Dstarpm$ mesons are reconstructed in several decay channels
  and their production rate is measured, which depends on the product
  $\Rc \times \PcDst$.  This sample of $\cc$ (and $\bb$) events is
  then used to measure $\PcDst$ using a slow pion tag in the opposite
  hemisphere.  In the ALEPH measurement \Rc{} is unfolded internally
  in the analysis so that no explicit $\PcDst$ is available. 
\item exclusive double tag (ALEPH\cite{ref:arcd}): 
  This analysis uses exclusively
  reconstructed $\Dstarp$, $\Dzero$ and $\Dplus$ mesons in different
  decay channels. It has lower statistics but better purity than the
  inclusive analyses.
\item reconstruction of all weakly decaying charmed states
  (ALEPH\cite{ref:arcc},  DELPHI\cite{ref:drcc}, OPAL\cite{ref:orcc}): 
  These analyses make the assumption that the production fractions
  of $\Dzero$, $\Dplus$, $\Ds$ and $\Lc$ 
  in c-quark jets of $\cc$ events add up to one with small corrections
  due to unmeasured charm strange baryons.
  This is a single tag measurement, relying only on knowing
  the decay branching ratios of the charm hadrons.  
  These analyses are also used to measure the c hadron production
  ratios which are needed for the \Rb{} analyses.  
\end{itemize}
\item A lifetime plus mass double tag from SLD to measure
  \Rc\cite{ref:SLD_R_C}.  This analysis uses the same tagging
  algorithm as the SLD \Rb{} analysis, but with the neural net tuned to
  tag charm. Although the
  charm tag has a purity of about 84\%, most of the background is from
  b which can be measured with high precision from the b/c mixed tag rate.
\item A measurement of \Rc{} using single leptons assuming $\Brcl$ from
  ALEPH \cite{ref:arcd}.
\end{itemize}
To avoid effects from nonlinearities in the fit, for the inclusive/exclusive
single/double tag and for the charm-counting analyses, the products
\RcPcDst, \RcfDz, \RcfDp, \RcfDs{} and \RcfLc that are actually
measured in the analyses are directly used as inputs to the fit.
The measurements of the production rates of weakly decaying charmed
hadrons, especially \RcfDs{} and \RcfLc{} have substantial errors due
to the uncertainties in the branching ratios of the decay mode used. 
Since these errors are
relative, there is a potential bias towards lower measurements.
To avoid this bias, for the production rates of weakly decaying charmed 
hadrons the logarithm of the production rates instead of the rates themselves
are input to the fit. For \RcfDz{} and \RcfDp{} the difference between
the results using the logarithm or the value itself is negligible. For
\RcfDs{} and \RcfLc{} the difference in the extracted value of $\Rc$ 
is about one
tenth of a standard deviation.

\subsection{Asymmetry Measurements}
\label{sec:asycorrections}
All b and c asymmetries given by the experiments correspond to full
acceptance.

The QCD corrections to the forward-backward asymmetries depend
strongly on the experimental analyses.  For this reason the numbers
given by the collaborations are also corrected for QCD effects. A
detailed description of the procedure can be found
in \cite{ref:afbqcd} with updates reported in \cite{ref:lephfnew}.

For the 12- and 14-parameter fits described above, the LEP peak and
off-peak asymmetries are corrected to $ \sqrt {s} = 91.26$ \GeV{}
using the predicted dependence from ZFITTER\cite{ref:ZFITTER}. The
slope of the asymmetry around $\MZ$ depends only on the axial coupling
and the charge of the initial and final state fermions and is thus
independent of the value of the asymmetry itself, i.e., the effective
electroweak mixing angle.

After calculating the overall averages, the quark pole asymmetries
$\Afbzq$, defined in terms of effective couplings, are derived from
the measured asymmetries by applying corrections as listed in
Table~\ref{tab:aqqcor}. These corrections are due to the energy shift
from 91.26 \GeV{} to $\MZ$, initial state radiation, $\gamma$ exchange
and $\gamma$-$\Zzero$ interference.  A very small correction due to
the nonzero value of the b quark mass is included in the last
correction.  All corrections are calculated using ZFITTER.

\begin{table}[bhtb]
\begin{center}
\begin{tabular}{|l||l|l|}
\hline
Source   & $\delta A_{\mathrm{FB}}^{\mathrm{b}}$
         & $\delta A_{\mathrm{FB}}^{\mathrm{c}}$ \\
\hline
\hline
$\sqrt{s} = \MZ $       & $ -0.0013 $  & $ -0.0034$  \\
QED corrections         & $ +0.0041 $  & $ +0.0104$  \\
$\gamma$, $\gamma$-$\Zzero$, mass & $ -0.0003 $  & $ -0.0008$  \\
\hline
\hline
Total                   & $ +0.0025 $  & $ +0.0062$  \\
\hline
\end{tabular}
\end{center}
\caption[]{%
  Corrections to be applied to the quark asymmetries as 
   $A_{\mathrm{FB}}^0 = A_{\mathrm{FB}}^{\mathrm{meas}}
  + \delta A_{\mathrm{FB}}$.}
\label{tab:aqqcor}
\end{table}

The SLD left-right-forward-backward asymmetries are also corrected for all
radiative effects and are directly presented in terms of $\cAb$ and $\cAc$.

The measurements used are:
\begin{itemize}
\item Measurements of \Abb{} and \Acc{} using leptons from 
  ALEPH\cite{ref:alasy}, DELPHI\cite{ref:dlasy}, L3\cite{ref:llasy} and
  OPAL\cite{ref:olasy}.  
  These analyses measure either \Abb{} only from a high $p_t$ lepton
  sample or they obtain \Abb{} and \Acc{} from a fit to the lepton
  spectra. In the case of OPAL the lepton information is combined
  with hadronic variables in a neural net. DELPHI uses in addition lifetime
  information and jet-charge in the hemisphere opposite to the lepton to
  separate the different lepton sources.
  Some asymmetry analyses also measure $\chiM$.
\item Measurements of \Abb{} based on lifetime tagged events with a
  hemisphere charge measurement from ALEPH\cite{ref:ajet}, 
  DELPHI\cite{ref:djasy,ref:dnnasy}, L3\cite{ref:ljet} and OPAL\cite{ref:ojet}.
  These measurements contribute roughly the same weight to the
  combined result as the lepton fits.  
\item Analyses with D mesons to measure \Acc{} from
  ALEPH\cite{ref:adsac} or \Acc{} and \Abb{} from
  DELPHI\cite{ref:ddasy} and OPAL\cite{ref:odsac}.
\item Measurements of \cAb{} and \cAc{} from SLD.
  These results include measurements using 
  lepton \cite{ref:SLD_AQL}, 
  D meson \cite{ref:SLD_ACD} and 
  vertex mass plus hemisphere charge \cite{ref:SLD_ABJ} 
  tags, which have similar sources of
  systematic errors as the LEP asymmetry measurements. 
  SLD also uses vertex mass for bottom or charm tagging in conjunction
  with a kaon tag or a vertex charge tag for both $\cAb$ and $\cAc$ 
  measurements \cite{ref:SLD_ABK,ref:SLD_vtxasy,ref:SLD_ACV}.
\end{itemize}

\subsection{Other Measurements}

The measurements of the charmed hadron fractions $\PcDst$, $\fDp$, $\fDs$
and $\fcb$ are included in the \Rc{} measurements and are described there.

ALEPH\cite{ref:abl}, DELPHI\cite{ref:dbl}, L3\cite{ref:lbl,ref:lrbmixed} and 
OPAL\cite{ref:obl} measure $\Brbl$, $\Brbclp$ and $\chiM$ or a subset of them
from a sample of leptons opposite to a b-tagged hemisphere and from a
double lepton sample. 
DELPHI\cite{ref:drcd} and OPAL\cite{ref:ocl} measure $\Brcl$ from a sample
opposite to a high energy $\Dstarpm$.
%
\section{Results}\label{sec-HFSUM}

\begin{figure}[htbp]
\vspace*{-0.6cm}
\begin{center}
  \mbox{\includegraphics[width=0.9\linewidth]{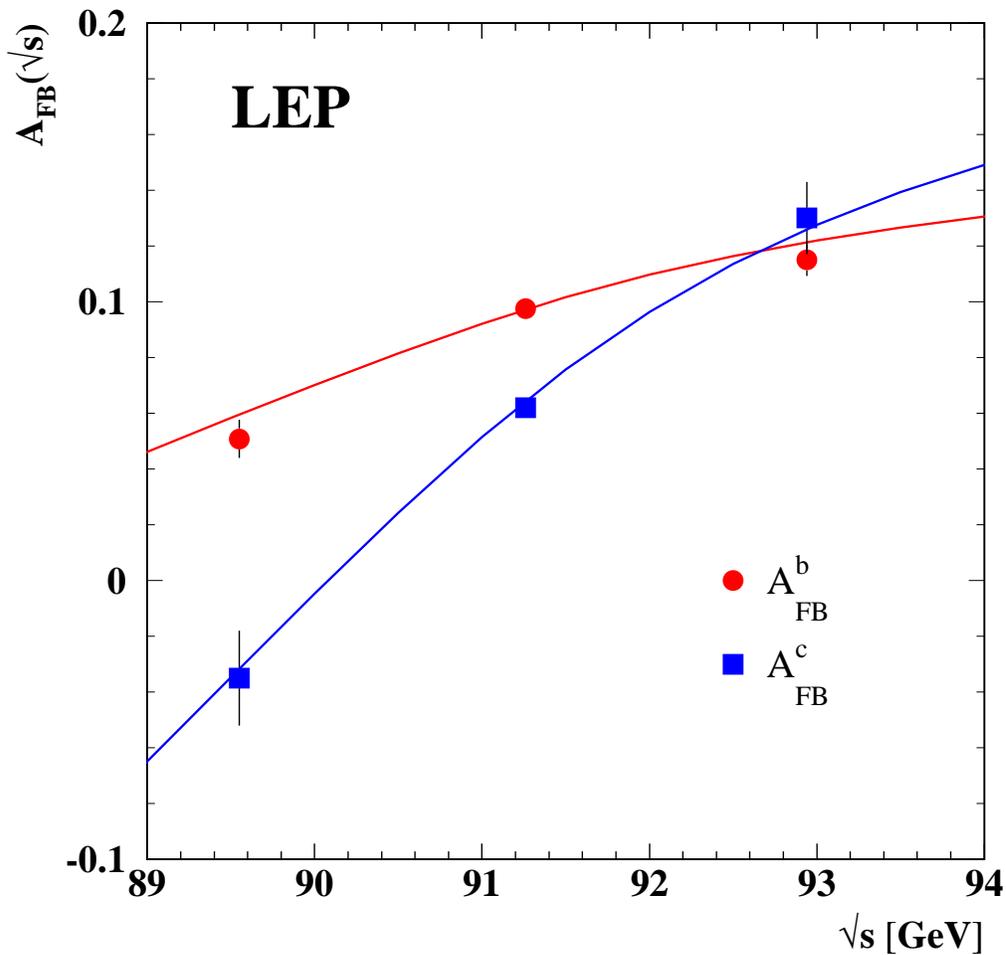}}
\end{center}
\caption[]{%
  Measured asymmetries for b and c quark final states as a function of
  the centre-of-mass energy. }
\label{fig-afbene}
\end{figure}

In a first fit the asymmetry measurements on peak, above peak and
below peak are corrected to three common centre-of-mass energies and
are then  combined at each energy point. The results of
this fit, including the SLD results, are given in
Appendix~\ref{app-HF-fit}.  The dependence of the average asymmetries on
centre-of-mass energy agrees with the prediction of the Standard
Model, as shown in Figure~\ref{fig-afbene}.
A second fit is made to derive the pole asymmetries $\Afbzq$ from the
measured quark asymmetries, in which all the off-peak asymmetry
measurements are corrected to the peak energy before combining. This fit
determines a total of 14 parameters:
the two partial widths, two LEP asymmetries, 
two coupling parameters from SLD,
three semileptonic branching ratios, the average mixing parameter and the
probabilities for c quark to fragment into a $\Dplus$, a $\Ds$, a
$\Dstarp$, or a charmed baryon.
If the SLD measurements are excluded from the fit there are 12 parameters to
be determined.
Results for
the non-electroweak parameters are independent of the
treatment of the off-peak asymmetries and the SLD data.

\subsection{Results of the 12-Parameter Fit to the LEP Data}
\label{sec-HFSUM-LEP}

Using the full averaging procedure gives the following combined
results for the electroweak parameters:
\begin{eqnarray}
  \label{eqn-hf4}
  \Rbz    &=& 0.21651  \pm  0.00072  \\
  \Rcz    &=& 0.1689   \pm  0.0047   \nonumber \\
  \Afbzb  &=& 0.0990   \pm  0.0017   \nonumber \\
  \Afbzc  &=& 0.0684   \pm  0.0035   \,,\nonumber
\end{eqnarray}
where all corrections to the asymmetries and partial widths are
applied.  The $\chi^2/$d.o.f.{} is $44/(90-12)$. The corresponding
correlation matrix is given in Table~\ref{tab:12parcor}.

\begin{table}[htbp]
\begin{center}
\begin{tabular}{|l||rrrr|}
\hline
&\makebox[1.2cm]{\Rbz}
&\makebox[1.2cm]{\Rcz}
&\makebox[1.2cm]{$\Afbzb$}
&\makebox[1.2cm]{$\Afbzc$}\\
\hline
\hline
\Rbz      & $  1.00$&$ -0.17$&$ -0.09$&$  0.02$ \\
\Rcz      & $ -0.17$&$  1.00$&$  0.07$&$ -0.01$ \\
$\Afbzb$  & $ -0.09$&$  0.07$&$  1.00$&$  0.15$ \\
$\Afbzc$  & $  0.02$&$ -0.01$&$  0.15$&$  1.00$ \\
\hline
\end{tabular}
\end{center}
\caption[]{
  The correlation matrix for the four electroweak parameters from the
  12-parameter fit.}
\label{tab:12parcor}
\end{table}

\subsection{Results of the 14-Parameter Fit to LEP and SLD Data}
\label{sec-HFSUM-LEP-SLD}

Including the SLD results for \Rb, \Rc, \cAb{} and \cAc{} into the fit the
following results are obtained:
\begin{eqnarray}
  \label{eqn-hf6}
  \Rbz    &=&  0.21646 \pm 0.00065  \,,\\
  \Rcz    &=&  0.1719  \pm 0.0031   \nonumber\,,\\
  \Afbzb  &=&  0.0990  \pm 0.0017   \nonumber\,,\\
  \Afbzc  &=&  0.0685  \pm 0.0034   \nonumber\,,\\
  \cAb    &=&  0.922   \pm 0.020    \nonumber\,,\\
  \cAc    &=&  0.670   \pm 0.026    \nonumber\,,
\end{eqnarray}
with a $\chi^2/$d.o.f.{} of $47/(99-14)$. The corresponding
correlation matrix is given in Table~\ref{tab:14parcor}
and the largest errors for the electroweak parameters are listed in Table
\ref{tab:hferrbk}.

\begin{figure}[htbp]
\vspace*{-0.6cm}
\begin{center}
  \mbox{\includegraphics[width=0.9\linewidth]{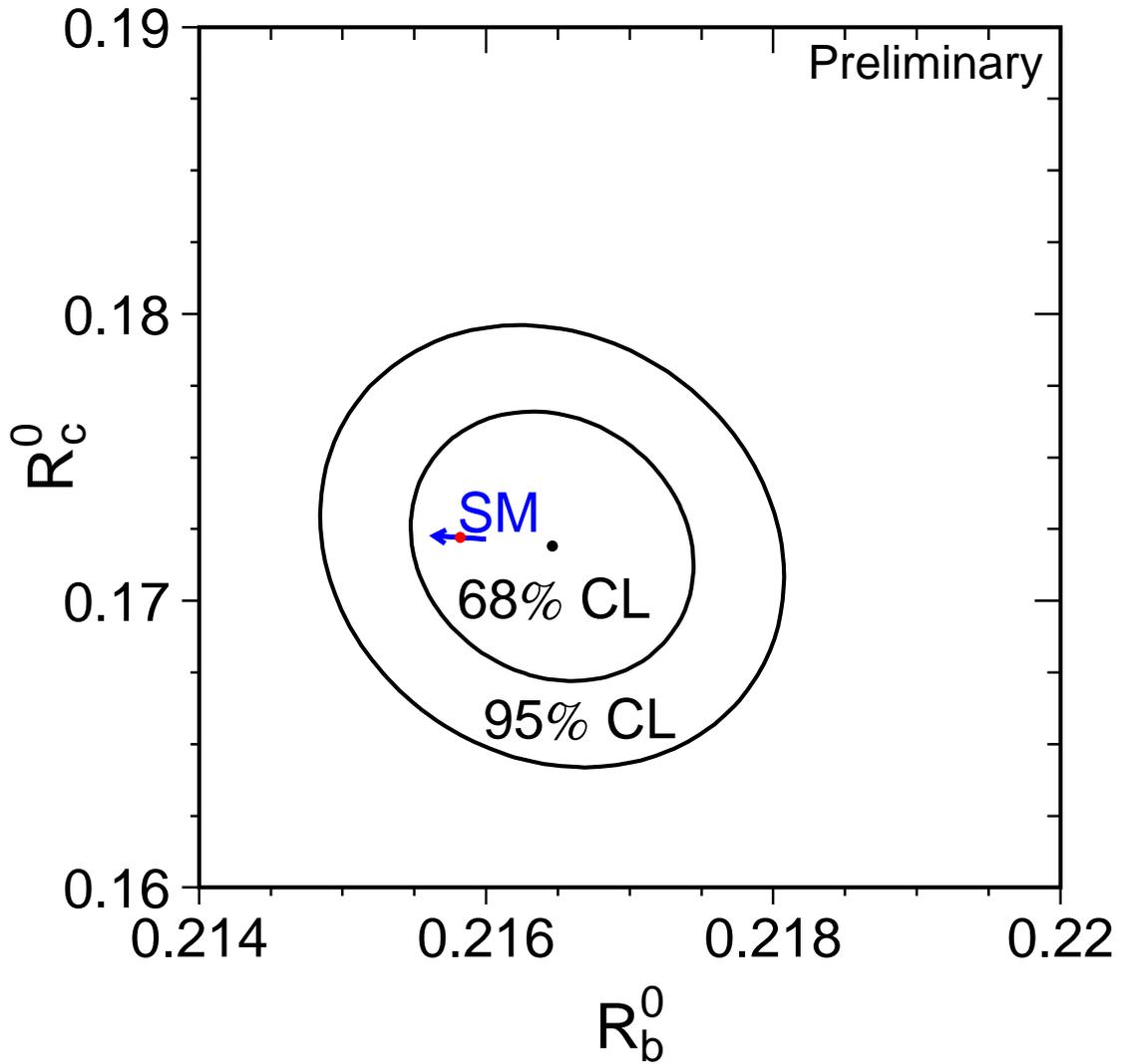}}
\end{center}
\caption[]{%
  Contours in the ($\Rbz$,$\Rcz$) plane derived from the LEP+SLD
  data, corresponding to 68\% and 95\% confidence levels assuming
  Gaussian systematic errors. The Standard Model prediction for
  $\Mt=174.3 \pm 5.1$~\GeV{} is also shown. The arrow points in the
  direction of increasing values of $\Mt$.  }
\label{fig-RbRc}
\end{figure}

In deriving
these results the parameters $\cAb$ and $\cAc$ are treated as
independent of the forward-backward asymmetries $\Afbzb$ and $\Afbzc$
(but see Section~\ref{sec-AF} for a joint analysis). In
Figure~\ref{fig-RbRc} the results for $\Rbz$ and $\Rcz$ are shown
compared with the Standard Model expectation.

\begin{table}[htbp]
\begin{center}
\begin{tabular}{|l||rrrrrr|}
\hline
&\makebox[1.2cm]{\Rbz}
&\makebox[1.2cm]{\Rcz}
&\makebox[1.2cm]{$\Afbzb$}
&\makebox[1.2cm]{$\Afbzc$}
&\makebox[0.9cm]{\cAb}
&\makebox[0.9cm]{\cAc}\\
\hline
\hline
\Rbz     & $  1.00$&$ -0.14$&$ -0.08$&$  0.01$&$ -0.08$&$  0.04$   \\
\Rcz     & $ -0.14$&$  1.00$&$  0.04$&$ -0.01$&$  0.03$&$ -0.05$   \\  
$\Afbzb$ & $ -0.08$&$  0.04$&$  1.00$&$  0.15$&$  0.02$&$  0.00$   \\
$\Afbzc$ & $  0.01$&$ -0.01$&$  0.15$&$  1.00$&$  0.00$&$  0.01$   \\
\cAb     & $ -0.08$&$  0.03$&$  0.02$&$  0.00$&$  1.00$&$  0.13$   \\
\cAc     & $  0.04$&$ -0.05$&$  0.00$&$  0.01$&$  0.13$&$  1.00$   \\
\hline
\end{tabular}
\end{center}
\caption[]{
  The correlation matrix for the six electroweak parameters from the
  14-parameter fit.  }
\label{tab:14parcor}
\end{table}

\begin{table}[htbp]
\begin{center}
\begin{tabular}{|c|c|c|c|c|c|c|}
\hline
&\makebox[1.2cm]{\Rbz}
&\makebox[1.2cm]{\Rcz}
&\makebox[1.2cm]{$\Afbzb$}
&\makebox[1.2cm]{$\Afbzc$}
&\makebox[0.9cm]{\cAb}
&\makebox[0.9cm]{\cAc}\\
 & $(10^{-3})$ & $(10^{-3})$ & $(10^{-3})$ & $(10^{-3})$ 
 & $(10^{-2})$ & $(10^{-2})$ \\
\hline
statistics & 
$0.43$ & $2.3$ & $1.6$ & $3.0$ & $1.5$ & $2.1$ \\
internal systematics &
$0.29$ & $1.4$ & $0.6$ & $1.4$ & $1.2$ & $1.5$ \\
QCD effects &
$0.18$ & $0.1$ & $0.3$ & $0.1$ & $0.3$ & $0.2$ \\
BR(D $\rightarrow$ neut.)&
$0.14$ & $0.3$ & $0$   & $0$ & $0$ & $0$ \\
D decay multiplicity &
$0.13$ & $0.3$ & $0  $ & $0  $ & $0$ & $0$ \\
BR(D$^+ \rightarrow$ K$^- \pi^+ \pi^+) $&
$0.09$ & $0.2$ & $0  $ & $0$ & $0$ & $0  $ \\
BR($\Ds \rightarrow \phi \pi^+) $&
$0.03$ & $0.5$ & $0  $ & $0$ & $0$ & $0$ \\
BR($\Lambda_{\mathrm{c}} \rightarrow $p K$^- \pi^+) $&
$0.06$ & $0.5$ & $0  $ & $0.1$ & $0$ & $0  $ \\
D lifetimes&
$0.06$ & $0.1$ & $0  $ & $0.1$ & $0$ & $0$ \\
gluon splitting &
$0.22$ & $0.1$ &$0.1$& $0.1$ & $0.1$ & $0.1$ \\
c fragmentation &
$0.10$ & $0.2$ & $0.1$ & $0.2$ & $0.1$ & $0.1$ \\
light quarks&
$0.07$ & $0.2$ & $0.1$ & $0.1$ & $0$ & $0  $ \\
beam polarisation&
$0$ & $0$ & $0$ & $0$ & $0.5$ & $0.4$ \\
\hline
total &
$0.65$ & $3.1$ & $1.7$ & $3.5$ & $2.0$ & $2.6$ \\
\hline
\end{tabular}
\end{center}
\caption[]{
The dominant error sources for the electroweak parameters from the 14-parameter
fit.
}
\label{tab:hferrbk}
\end{table}

Amongst the non-electroweak observables the B semileptonic branching
fraction ($\Brbl \, = \, 0.1062 \pm 0.0021$) is of special interest. 
The dominant error source on this quantity is the dependence on the 
semileptonic decay models $\bl$, $\cl$ with 
\begin{equation}
\Delta \Brbl (\bl-\rm{modelling})  = 0.0011.
\end{equation}
Extensive studies are made to understand the size of this error.
Amongst the electroweak quantities the quark asymmetries with leptons
depend also on the assumptions on the decay model while the
asymmetries using other methods usually do not. The fit implicitly
requires that the different methods give consistent results. This
effectively constrains the decay model and thus reduces the error from
this source in the fit result for $\Brbl$.

To get a conservative estimate of the modelling error in $\Brbl$ the
fit is repeated removing all asymmetry measurements. The result
of this fit is
\begin{equation}
\Brbl \, = \, 0.1065 \pm 0.0023
\end{equation}
with
\begin{equation}
\Delta \Brbl (\bl-\rm{modelling}) = 0.0014.
\end{equation}


\boldmath
\chapter{The Hadronic Charge Asymmetry $\avQfb$}
\label{sec-QFB}
\unboldmath

\updates{DELPHI and OPAL revert to their published result.  While all
  results are now final, the combination procedure itself is still
  preliminary.}

\noindent
The LEP experiments ALEPH\cite{ALEPHcharge1996},
DELPHI\cite{DELPHIcharge}, L3\cite{ref:ljet} and OPAL\cite{OPALcharge}
provide measurements of the hadronic charge asymmetry based on
the mean difference in jet charges measured in the forward and
backward event hemispheres, $\avQfb$. DELPHI also provides a
related measurement of the total charge asymmetry by making a charge
assignment on an event-by-event basis and performing a likelihood
fit\cite{DELPHIcharge}.  The experimental values quoted for the
average forward-backward charge difference, $\avQfb$, cannot be
directly compared as some of them include detector dependent effects
such as acceptances and efficiencies.  Therefore the effective
electroweak mixing angle, $\swsqeffl$, as defined in
Section~\ref{sec-SW}, is used as a means of combining the experimental
results summarised in Table~\ref{partab}.

\begin{table}[htb]
\begin{center}
\renewcommand{\arraystretch}{1.1}
\begin{tabular}{|ll||c|}
\hline
Experiment & & $\swsqeffl$ \\
\hline
\hline
ALEPH & (90-94), final & $0.2322\pm0.0008\pm0.0011$ \\
DELPHI& (91-91), final & $0.2345\pm0.0030\pm0.0027$ \\
L3    & (91-95), final & $0.2327\pm0.0012\pm0.0013$ \\
OPAL  & (90-91), final & $0.2326\pm0.0012\pm0.0029$ \\
\hline
\hline
LEP Average  &           & $0.2324\pm0.0012$ \\
\hline
\end{tabular}
\caption[]{
  Summary of the determination of $\swsqeffl$ from inclusive hadronic
  charge asymmetries at LEP. For each experiment, the first error is
  statistical and the second systematic. The latter, amounting to
  0.0010 in the average, is dominated by fragmentation and decay
  modelling uncertainties.  }
\label{partab}
\end{center}
\end{table}

The dominant source of systematic error arises from the modelling of
the charge flow in the fragmentation process for each flavour. All
experiments measure the required charge properties for $\Zzero\ra\bb$
events from the data. ALEPH also determines the charm charge
properties from the data. The fragmentation model implemented in the
JETSET Monte Carlo program\cite{JETSET} is used by all experiments as
reference; the one of the HERWIG Monte Carlo program\cite{HERWIG} is
used for comparison. The JETSET fragmentation parameters are varied to
estimate the systematic errors. The central values chosen by the
experiments for these parameters are, however, not the same. 
The smaller of
the two fragmentation errors in any pair of results is treated as
common to both.  The present average of $\swsqeffl$ from $\avQfb$ and
its associated error are not very sensitive to the treatment of common
uncertainties.
The ambiguities due to QCD corrections may cause changes in the
derived value of $\swsqeffl$. These are, however, well below the
fragmentation uncertainties and experimental errors. The effect of
fully correlating the estimated systematic uncertainties from this
source between the experiments has a negligible effect upon the
average and its error.

There is also some correlation between these results and those for
$\Abb$ using jet charges. The dominant source of correlation is again
through uncertainties in the fragmentation and decay models used. The
typical correlation between the derived values of $\swsqeffl$ from
the $\avQfb$ and the $\Abb$ jet charge measurements is estimated
to be about 20\% to 25\%. This leads to only a small change in the
relative weights for the $\Abb$ and $\avQfb$ results when averaging
their $\swsqeffl$ values (Section~\ref{sec-SW}). 
Thus, the correlation between $\avQfb$ and $\Abb$ from
jet charge has little impact on the overall Standard Model fit,
and is neglected at present.

\boldmath
\chapter{Photon-Pair Production at \LEPII}
\label{sec-GG}
\unboldmath

\updates{ This is a new chapter.  LEP results on photon-pair
  production are combined.  These combination results became available
  after the summer conferences and were first presented at Siena, in
  October 2001. }


\section{Introduction}
The reaction $\eeggga$ provides a clean test of QED at LEP energies
and is well suited to detect the presence of non-standard physics.
The differential QED cross-section at the Born level in the
relativistic limit is given by:
\begin{equation}
\xb = \frac{\alpha^2}{s} 
\frac{1+\cos^2\theta}{1 -\cos^2\theta} \; .
\end{equation}
Since the two final state particles are identical the polar angle
$\theta$ is defined such that $\costh > 0$. Various models with 
deviations from this cross-section will be discussed in section \ref{gg:sec:fit}.
Results on the $\ge$2-photon  final state using the high energy data 
collected by the four LEP collaborations are reported by the individual
experiments \cite{gg:ref:LEPGG}.
Here the results of the LEP working group 
dedicated to the combination of the $\eeggga$ measurements
are reported.  Results are given for the averaged total cross-section
and for global fits to the differential cross-sections.

\section{Event Selection}
This channel is very clean and the event selection, which is similar
for all experiments, is based on the presence of at least two
energetic clusters in the electromagnetic calorimeters.  A minimum
energy is required, typically $(E_1+ E_2)/\sqrt{s}$ larger than 0.3 to
0.6, where $E_1$ and $E_2$ are the energies of the two most energetic
photons.  In order to remove $\ee$ events, charged tracks are in
general not allowed except when they can be associated to a photon
conversion in one hemisphere.

The polar angle is defined in order to minimise effects due to initial state radiation
as
\[
\costh =\left.\left| \sin (\frac{\theta_1 - \theta_2}{2}) \right| 
        \right/ \sin (\frac{\theta_1 + \theta_2}{2}) \; ,   \] 
where $\theta_1$ and $\theta_2$ are the polar angles of the two most energetic photons.
The acceptance in polar angle is in the range of 0.90 to 0.96 on 
$|\costh|$, depending on the experiment.

With these criteria, the selection efficiencies are in the range of
$68\%$ to $95\%$ and the residual background (from $\ee$ events and
from $\eetautau$ with $\tau^{\pm} \rightarrow\rm e^{\pm}\nu
\bar{\nu}$) is very small, $0.1\%$ to $1\%$.  Detailed descriptions of
the event selections performed by the four collaborations can be found
in \cite{gg:ref:LEPGG}.

\section{Total cross-section}

\begin{figure}[hbt]
   \begin{center} \mbox{
          \epsfxsize=16.0cm
           \epsffile{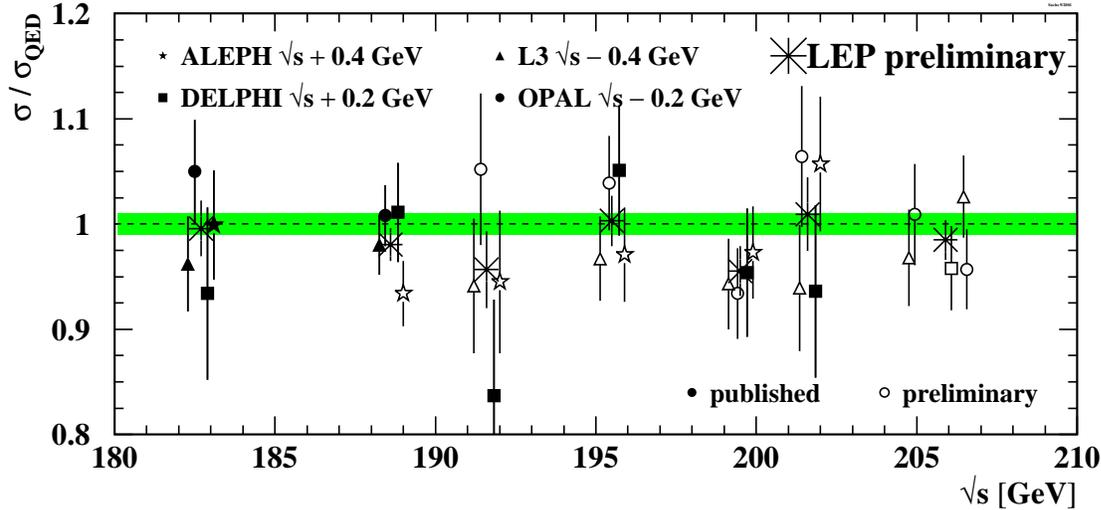}
           } \end{center}
\vspace{-1cm}
\caption{Cross-section ratios 
$r = \sigma_{\rm meas} / \sigma_{\rm QED}$ at different energies.
The measurements of the single experiments are displaced by 
$\pm$ 200 or 400 \MeV\ from the actual energy for clarity. Filled symbols
indicate published results, open symbols stand for preliminary numbers.
The average over the experiments at each energy is shown as a star. 
Measurements between 203 and 209 \GeV\ are averaged to one energy point. 
The theoretical error is not included in the experimental errors 
but is represented as the shaded band.
}
\label{gg:fig:xsn}
\end{figure}

The total cross-sections are combined using a $\chi^2$ minimisation.
Given the different angular acceptances,
only the ratios of the measured cross-sections relative to the QED 
expectation \mbox{$r = \sigma_{\rm meas} / \sigma_{\rm QED}$}
are averaged. Figure \ref{gg:fig:xsn} shows the measured ratios $r_{i,k}$ 
of the experiments $i$ at energies $k$ with their statistical
and systematic errors 
added in quadrature. Systematic errors are uncorrelated between
experiments as the error on the theory is not included in the 
experimental errors. 

Denoting with $\Delta$ the vector of residuals between the measurements
and the expected ratios, three different averages are performed:
\begin{enumerate}
\item per energy $k=1,\ldots,7$: $\Delta_{i,k} = r_{i,k} - x_k$ 
\item per experiment $i=1,\ldots,4$: $\Delta_{i,k} = r_{i,k} - y_i$ 
\item global value:  $\Delta_{i,k} = r_{i,k} - z$ 
\end{enumerate}

The seven fit parameters per energy $x_k$ are shown in Figure 
\ref{gg:fig:xsn} as LEP combined cross-sections. They are correlated
with correlation coefficients ranging from 10\% to 20\%. 
The four fit-parameters per experiment $y_i$ are uncorrelated
between each other, the results are given in Table \ref{gg:tab:xsn}
together with the single global fit parameter $z$.

No significant deviations from the QED expectations are found.
The global ratio is below unity by 1.5$\sigma$ not accounting for the
error on the radiative corrections (1\%) which is of similar size as
the experimental error (1.2\%).

\begin{table}[hbt]
\begin{center}
\begin{tabular}{|l|r@{$\pm$}l|}\hline
Experiment & \multicolumn{2}{c|}{cross-section ratio} \\\hline
ALEPH  & 0.963 & 0.025 \\
DELPHI & 0.974 & 0.032 \\
L3     & 0.982 & 0.021 \\
OPAL   & 1.000 & 0.021 \\ \hline
global & 0.982 & 0.012 \\ \hline
\end{tabular}
\caption[]{Cross-section ratios 
$r = \sigma_{\rm meas} / \sigma_{\rm QED}$ for the four LEP experiments
averaged over all energies and the global average over all experiments
and energies. The error includes the statistical and experimental
systematic error but no error from theory.
}
\label{gg:tab:xsn}
\end{center}
\end{table}

\section{Global fit to the differential cross-sections}
\label{gg:sec:fit}

\begin{table}[bt]
\begin{center}
\begin{tabular}{|l|c|c|c|c|c|} \hline
  & \multicolumn{2}{c|}{data used} & \multicolumn{2}{c|}{sys. error $[ \% ]$}&$\left | \rm{cos} \theta \right |$ \\ 
  & published & preliminary & experimental & theory & \\\hline
ALEPH  &  & 189 -- 202 & 2 & 1& 0.95 \\
DELPHI & 189 -- 202 & 206 & 2.5 & 1 & 0.90 \\
L3     & 183 -- 189 & 192 -- 207 & 2.1 & 1 & 0.96 \\
OPAL   & 183 -- 189 & 192 -- 207 & 1.1 & 1 & 0.90 \\\hline
\end{tabular}
\caption[]{ The data samples used for the global fit to the
  differential cross-sections, the systematic errors, the assumed
  error on the theory and the polar angle acceptance for the LEP
  experiments.}
\label{gg:tab:stat} 
\end{center} 
\end{table}

\begin{figure}[btp]
   \begin{center} 
   \mbox{\epsfxsize=8.0cm\epsffile{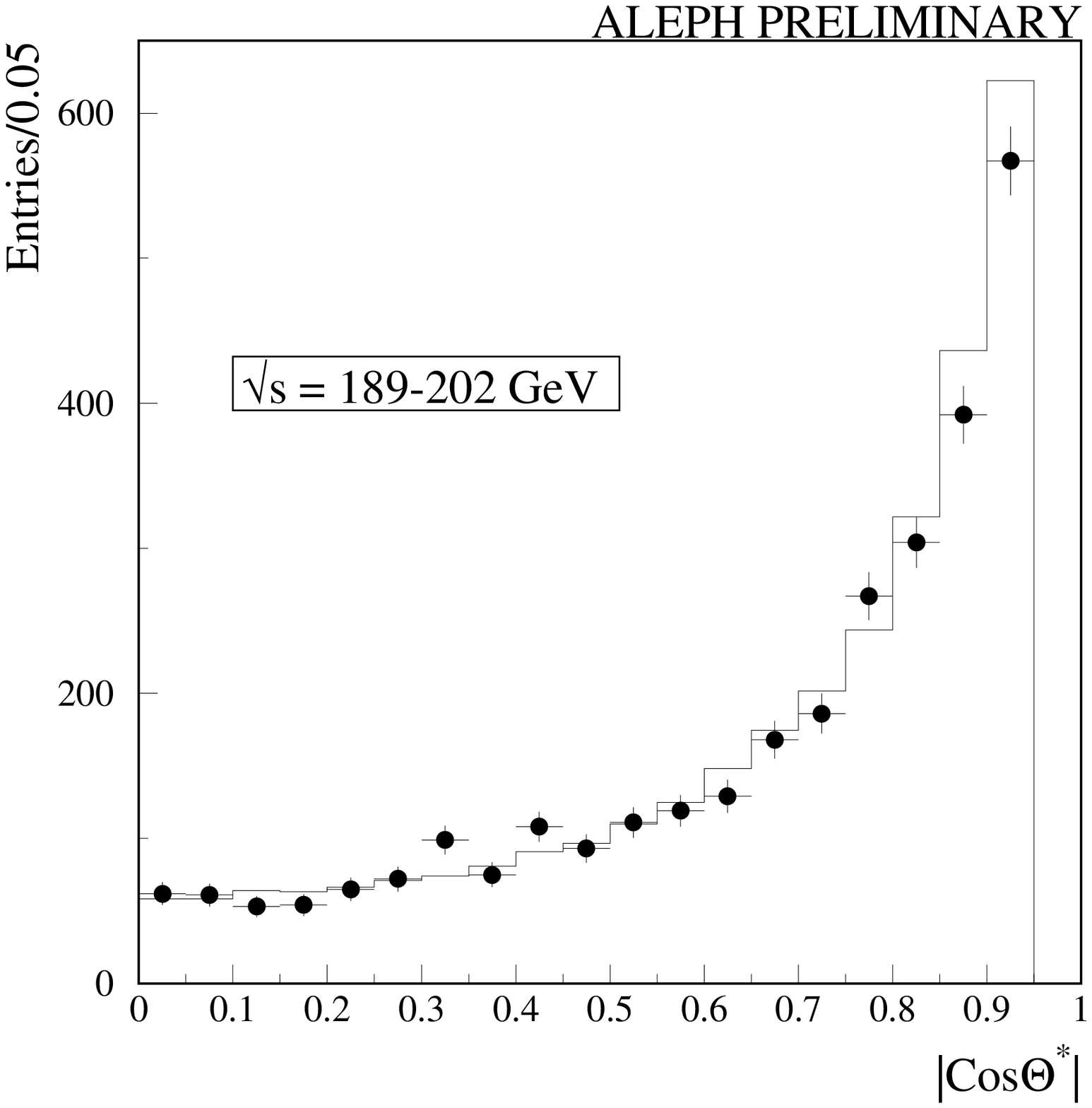}}
   \raisebox{1.5cm}{\epsfxsize=8.0cm\epsffile{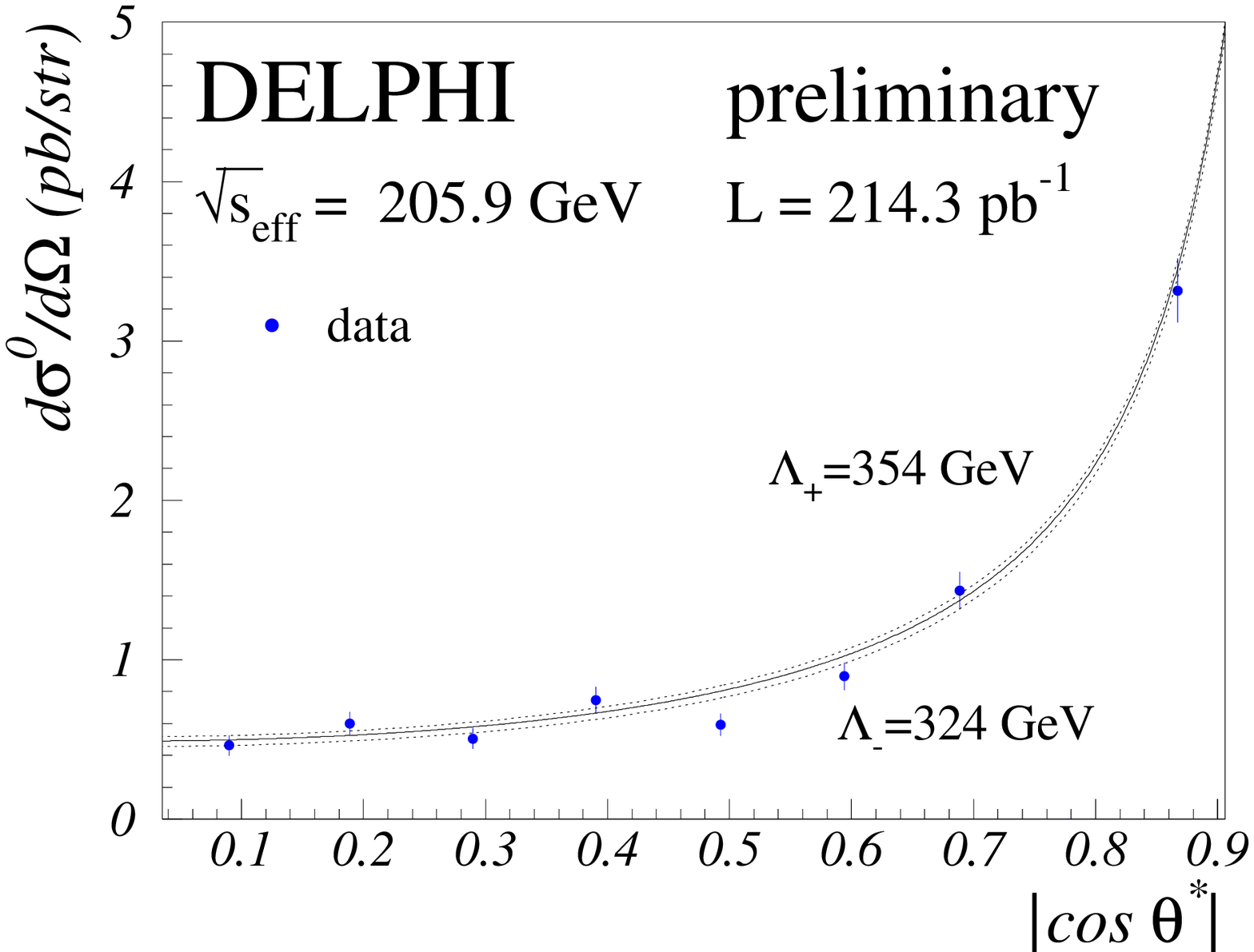}}\\[-2cm]
   \mbox{\epsfxsize=8.0cm\epsffile{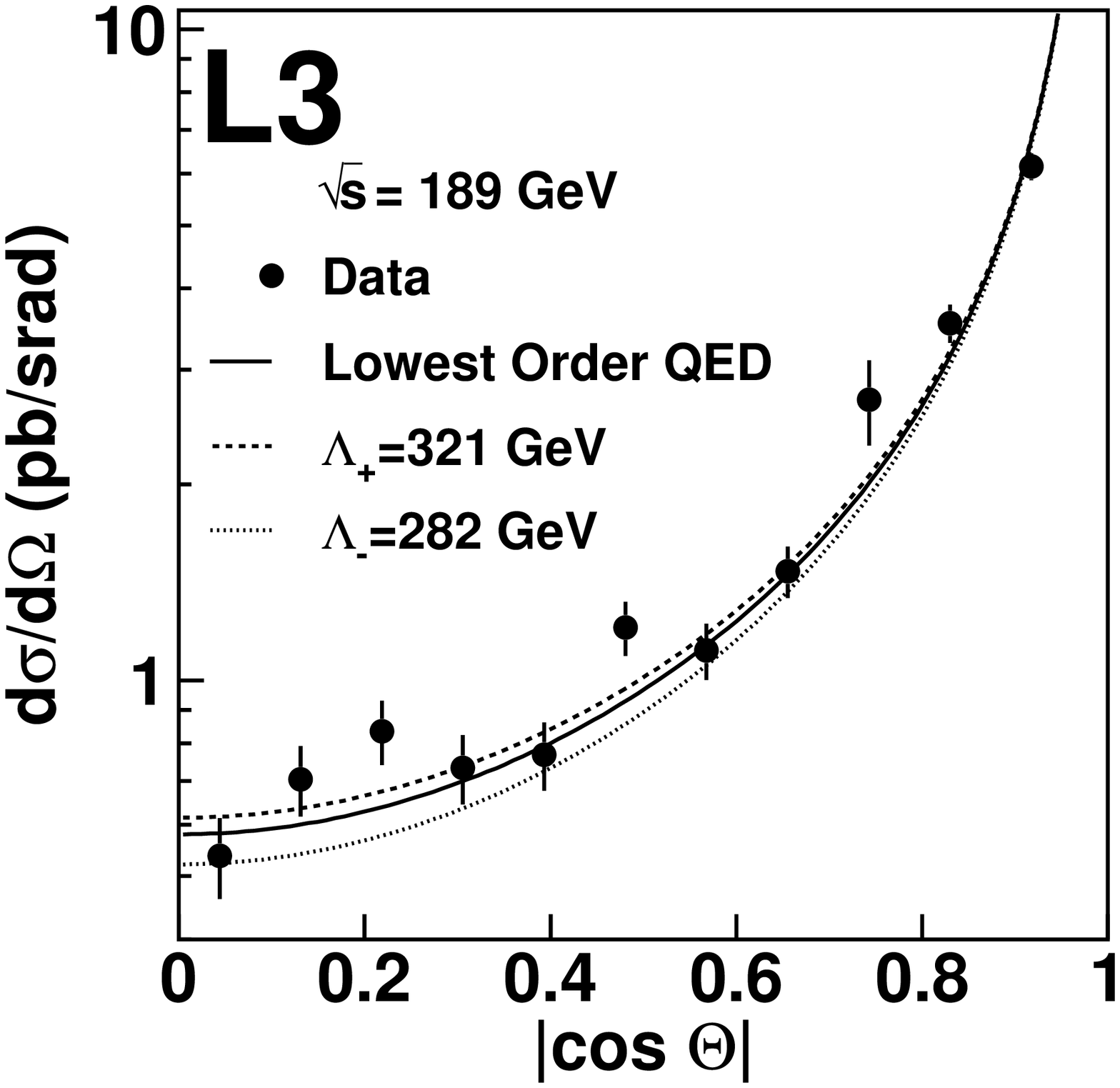}}
   \mbox{\epsfxsize=8.0cm\epsffile{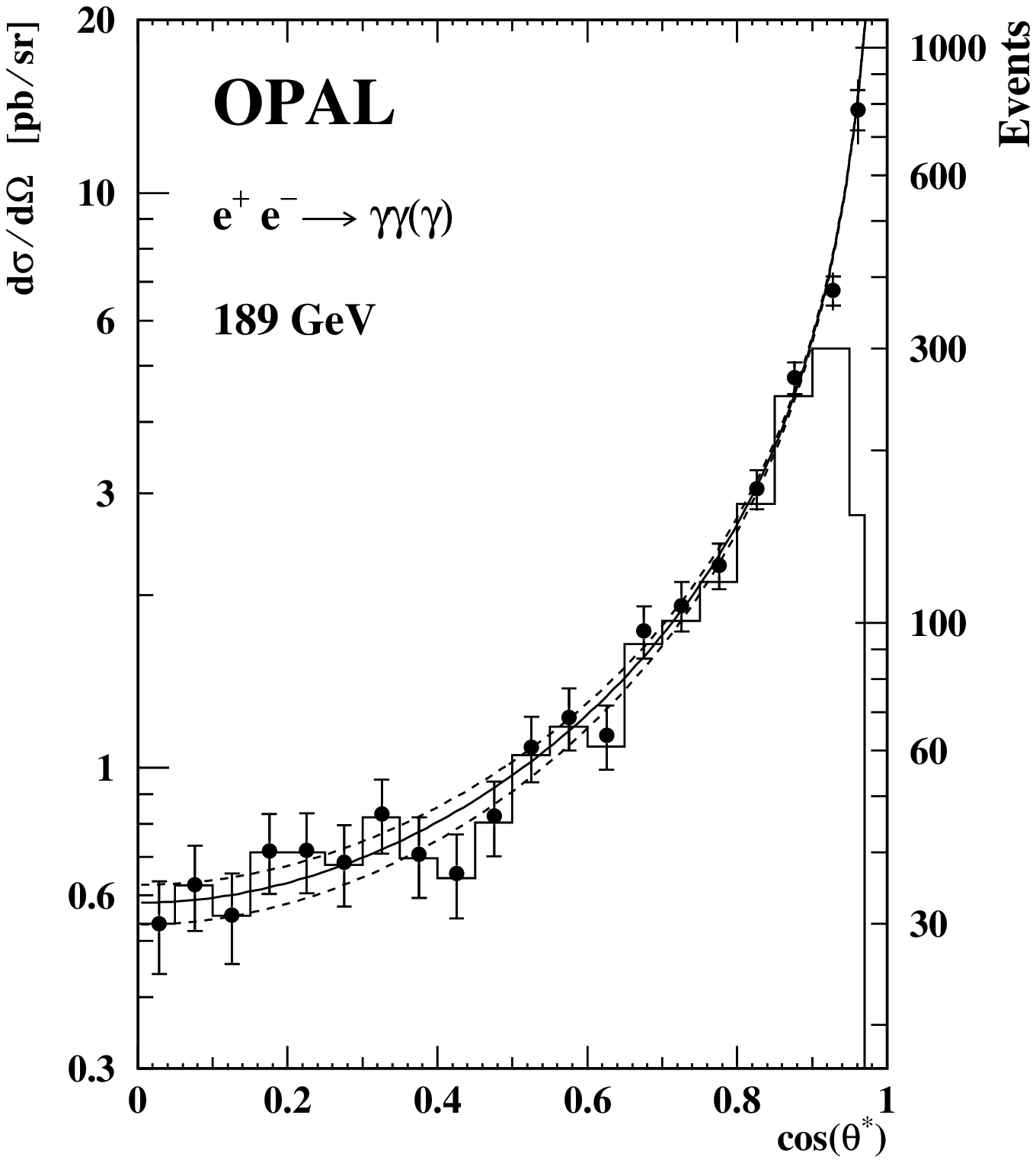}}
   \end{center}
\caption{Examples for angular distributions of the four LEP experiments.
Points are the data and the curves are the QED prediction (solid) and
the individual fit results for $\Lpm$ (dashed). ALEPH shows the
uncorrected number of observed events, the expectation is presented as
histogram. For OPAL the histogram represents the number of observed
events, before efficiency and radiative corrections are applied.
}
\label{gg:fig:ADLO}
\end{figure}

The global fit is based on angular distributions at energies between
183 and 207 \GeV\ from the individual experiments. As an example
angular distributions from each experiment are shown in
Figure~\ref{gg:fig:ADLO}. Combined differential cross-sections are not
available yet, since they need a common binning of the histograms.
All four experiments give preliminary results; DELPHI, L3 and OPAL
include the whole year 2000 data-taking, as shown in Table
\,\ref{gg:tab:stat}.  The systematic errors arise from the luminosity
evaluation (including theory uncertainty on the small-angle Bhabha
cross-section computation), from the selection efficiency and the
background evaluations and from radiative corrections. The last
contribution, owing to the fact that the available $\eeggga$
cross-section calculation is based on $\cal O$$(\alpha^3)$ code,
is assumed to be 
$ 1\%$ and is considered correlated among energies and experiments.

Various model predictions
are fitted to these angular distributions taking into account the
experimental systematic error correlated between energies for each
experiment and the error on the theory.
A binned log likelihood fit is performed
with one free parameter for the model and five fit parameters
used to keep the normalisation free within the systematic errors
of the theory and the four experiments.

The following models of new physics are considered. In some cases they
give rise to identical distortions of the predictions; hence their
parameters can be transformed into each other.

Cut-off parameter $\Lpm$ \cite{gg:ref:drell,gg:ref:low}: 
\begin{equation}
\xl   =  \xb \pm \frac{\alpha^2 s}{2\Lambda_\pm^4}(1+\cos^2{\theta})
\end{equation}

Effective Lagrangian theory \cite{gg:ref:eboli} describing anomalous
$\eeg$ couplings in dimension 6 
\mbox{($\Lambda_6^4 = \frac{2}{\alpha} \Lpm^4$)}
or contact interactions for dimensions 7 and 8
($\Lambda_7 = \Lambda'$; $\Lambda_8^4 = m_{\rm e} \Lambda_7^3$):
\begin{equation}
\xq  =  \xb + \frac{s^2}{32\pi}\frac{1}{\Lambda'{}^6} 
\end{equation}

Low scale gravity in extra dimensions \cite{gg:ref:ad}, where 
$M_s$ is related to the string scale and expected to be of order
${\cal O}(\rm TeV)$:
\begin{equation}
\xg = \xb - \frac{\alpha s}{2\pi} \; \frac{\lambda}{M_s^4}\;(1+\cos^2{\theta})
    + \frac{s^3}{16 \pi^2} \;  \frac{\lambda^2}{M_s^8} \;(1-\cos^4{\theta}) 
    \; , \; \lambda = \pm 1
\end{equation}

Excited electrons \cite{gg:ref:vachon}
with mass $\mestar$ and chiral magnetic
coupling described by the Lagrangian
\begin{equation}
{\cal L} = \frac{1}{2\Lambda} \bar{\ell^\ast} \sigma^{\mu\nu}
\left[ g f \frac{\tau}{2}W_{\mu\nu} + g' f' \frac{Y}{2} B_{\mu\nu}
\right] \ell_L + \mbox{h.c.} \; ,
\end{equation}
where $g$ and $g'$ are the coupling constants of SU(2)$_L$ and U(1)$_Y$,
respectively.
For the two photon final state this leads to the following cross-section:
\begin{equation}
\xe = \xb + \frac{\alpha^2}{4}\frac{f_\gamma^4}{\Lambda^4}\mestar^2 \left[
\frac{p^4}{(p^2-\mestar^2)^2} + \frac{q^4}{(q^2-\mestar^2)^2} +
\frac{\frac{1}{2} s^2 \sin^2\theta}{(p^2-\mestar^2)(q^2-\mestar^2)} \right] 
\; , \end{equation}
with $f_\gamma = -\frac{1}{2}(f+f')$, $p^2=-\frac{s}{2}(1-\costh)$ 
and $q^2=-\frac{s}{2}(1+\costh)$ and $\Lambda=\mestar$.

\section{Fit Results}

\begin{figure}[hbtp]
\vspace*{-1cm}
   \begin{center}\mbox{
          \epsfxsize=15.0cm
           \epsffile{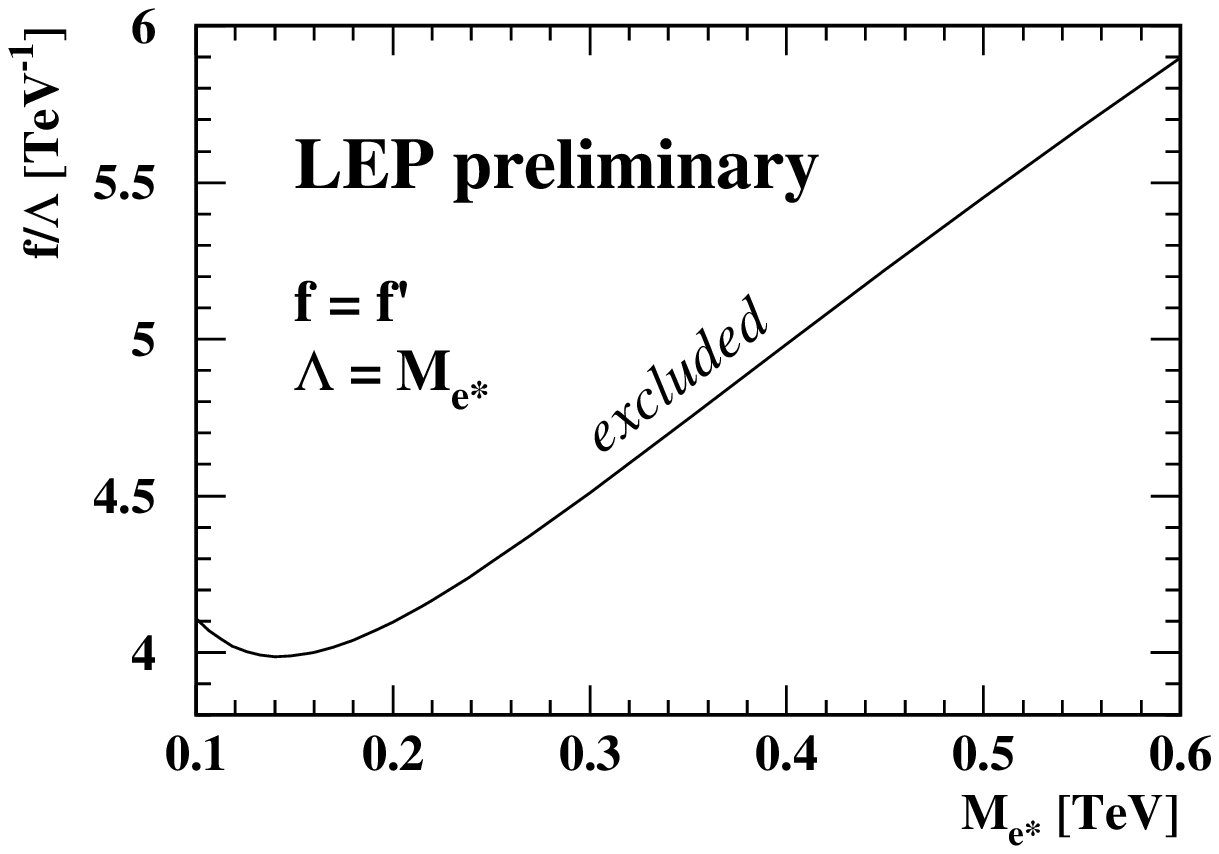}
           } \end{center}
\vspace{-1cm}
\caption{95\% CL limits on $f_\gamma/\Lambda$ as a function of $\mestar$.
In the case of $f=f'$ it follows that $f_\gamma = f$.
It is assumed that $\Lambda=\mestar$.}
\label{gg:fig:estar}
\vspace*{-2cm}
   \begin{center}\mbox{
          \epsfxsize=15.0cm
           \epsffile{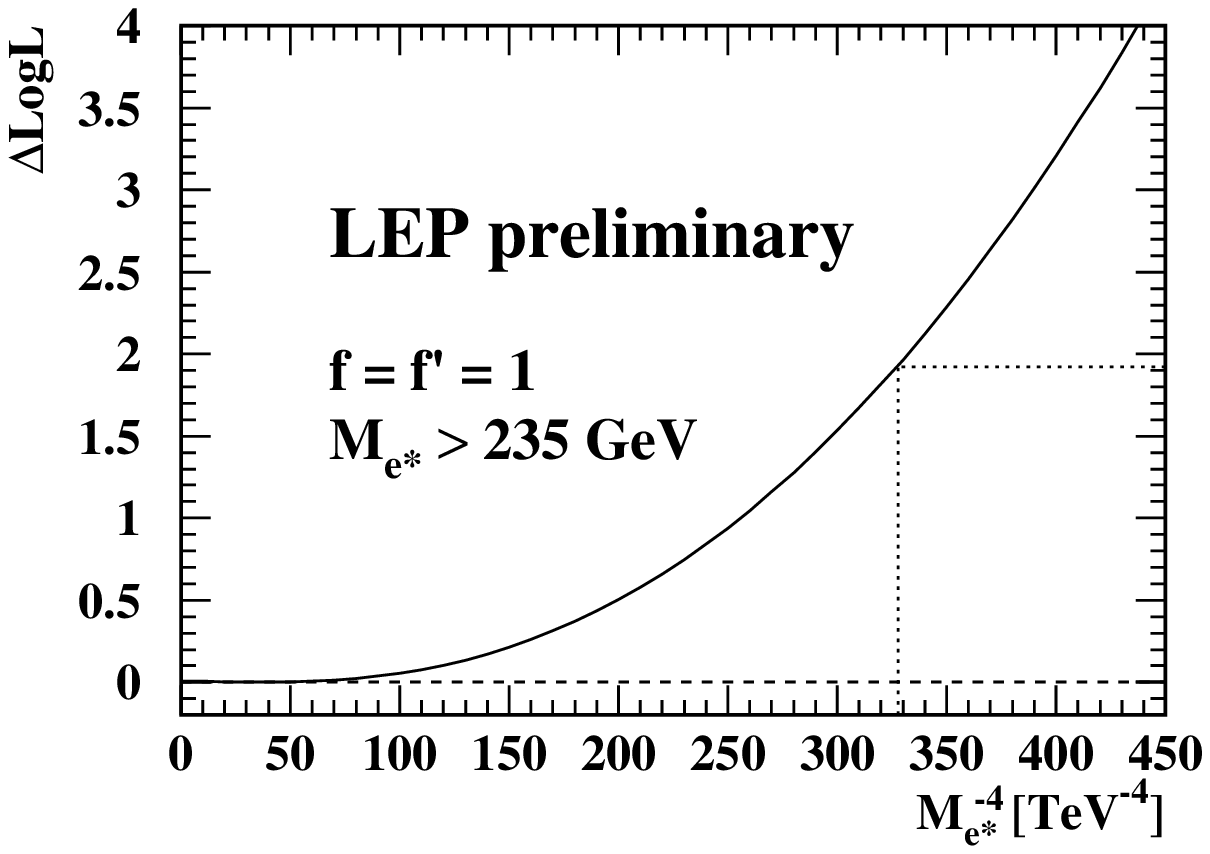}
           } \end{center}
\vspace{-1cm}
\caption{Log likelihood difference 
$\Delta\mbox{LogL} = -\ln{\cal L}+\ln{\cal L}_{\rm max}$
as a function of $\mestar^{-4}$. The coupling is fixed at $f = f' = 1$. 
The value corresponding to $\Delta\mbox{LogL} = 1.92$ is $\mestar$ = 235 \GeV.}
\label{gg:fig:ll}
\end{figure}

Where possible the fit parameters are chosen such that the likelihood
function is approximately Gaussian. The preliminary results of the
fits to the differential cross-sections are given in
Table~\ref{gg:tab:results}.  No significant deviations with respect to
the QED expectations are found (all the parameters are compatible with
zero) and therefore 95$\%$ confidence level limits are obtained by
renormalising the probability distribution of the fit parameter to the
physically allowed region. For limits on $f_\gamma/\Lambda$ a scan
over $\mestar$ is performed and presented in Figure
\ref{gg:fig:estar}. Only for $\mestar$ is the cross-section nonlinear
in the fit parameter.  The obtained negative log likelihood is shown
in Figure \ref{gg:fig:ll} and the limit is determined at 1.92 units
above the minimum.


\begin{table}[htb]
\begin{center}
\renewcommand{\arraystretch}{1.5}
\begin{tabular}{|c|c|r@{ }l|}\hline
Fit parameter & Fit result &
\multicolumn{2}{c|}{95\%\ CL limit [\GeV]}\\  \hline
 &  & $\Lambda_+ >$ &  365 \\
 \raisebox{2.2ex}[-2.2ex]{$\Lpm^{-4}$} &
 \raisebox{2.2ex}[-2.2ex]{$
 \left(4.6{+27.0 \atop -26.5}\right)\cdot 10^{-12}$ \GeV$^{-4}$}
 & $\Lambda_- > $&  379  \\\hline
$\Lambda_7^{-6}$ & $ \left(0.18{+1.95 \atop -1.92}\right)\cdot
                                  10^{-18}$ \GeV$^{-6}$
& $\Lambda_7 > $&  794   \\ \hline
\multicolumn{2}{|c|}{derived from $\Lambda_+$}& $\Lambda_6 > $&  1484   \\
\multicolumn{2}{|c|}{derived from $\Lambda_7$}& $\Lambda_8 > $&  22.5   \\\hline
 &  & $\lambda = +1$: $M_s >$ &  972  \\
 \raisebox{2.2ex}[-2.2ex]{$\lambda/M_s^4$} &
 \raisebox{2.2ex}[-2.2ex]{ $ \left(-0.106{+0.609 \atop -0.615}\right)\cdot
                                         10^{-12}$ \GeV$^{-4} $ }
 & $\lambda = -1$: $M_s >$&  940 \\ \hline
 $f_\gamma^4 (\mestar=200 \rm \GeV)$ & 
  $0.036{+0.414 \atop -0.400 }$ & 
 \multicolumn{2}{c|}{$f_\gamma/\Lambda <  4.1 \mbox{ \TeV}^{-1}$} \\ \hline
\end{tabular}
\caption[]{ The preliminary combined fit parameters 
and the 95$\%$  confidence level limits for the four LEP experiments.}
\label{gg:tab:results} 
\end{center} 
\end{table}

\section{Conclusion}
The LEP collaborations study the $\eeggga$ channel up to the highest
available centre-of-mass energies. The total cross-section results are
combined in terms of the ratios with respect to the QED expectations.
No deviations are found. The differential cross-sections are fit
following different parametrisations from models predicting deviations
from QED. No evidence for deviations is found and therefore combined
95$\%$ confidence level limits are given.

\boldmath
\chapter{Fermion-Pair Production at \LEPII}
\label{sec-FF}
\unboldmath

\updates{ Additional preliminary results based on the data collected
  in the year 2000 are included. } 

\section{Introduction}

Since the start of the $\LEPII$ program LEP has delivered collisions
at centre-of-mass 
energies from $\sim 130$ $\GeV$ to $\sim 209$ $\GeV$. The four LEP experiments
make measurements of the $\eeff$ processes over this range of energies,
and a preliminary combination of these data is discussed in this Chapter.
 
In the years 1995 to 1999 LEP delivered luminosity at a number of 
distinct centre-of-mass energy points. In 2000 most of the luminosity
was delivered close to 2 distinct energies, but there was also
a significant fraction of the luminosity delivered in, more-or-less, a 
continuum of energies. To facilitate the combination of the measurements,
the four LEP experiments all divided the data they collected in 2000 
into two energy bins: from 202.5 to 205.5 $\GeV$; and 205.5 $\GeV$ and above.
The nominal and actual centre-of-mass energies to which the LEP data 
are averaged for each year are given in Table~\ref{ff:tab-ecms}.

\begin{table}[htbp]
 \begin{center}
 \begin{tabular}{|c|c|c|c|}
  \hline
   Year & Nominal Energy & Actual Energy & Luminosity \\
        &     $\GeV$     &    $\GeV$     &  pb$^{-1}$ \\
  \hline
  \hline
   1995 &      130       &    130.2      & $\sim 3  $ \\
        &      136       &    136.2      & $\sim 3  $ \\
  \cline{2-4}
        &  $133^{\ast}$ &     133.2      & $\sim 6  $ \\
  \hline
   1996 &      161       &    161.3      & $\sim 10 $ \\
        &      172       &    172.1      & $\sim 10 $ \\
  \cline{2-4}
        &  $167^{\ast}$ &     166.6      & $\sim 20 $ \\
  \hline
   1997 &      130       &    130.2      & $\sim 2  $ \\
        &      136       &    136.2      & $\sim 2  $ \\
        &      183       &    182.7      & $\sim 50 $ \\
  \hline
   1998 &      189       &    188.6      & $\sim 170$ \\
  \hline
   1999 &      192       &    191.6      & $\sim 30 $ \\
        &      196       &    195.5      & $\sim 80 $ \\
        &      200       &    199.5      & $\sim 80 $ \\
        &      202       &    201.6      & $\sim 40 $ \\
  \hline
   2000 &      205       &    204.9      & $\sim 80 $ \\
        &      207       &    206.7      & $\sim140 $ \\
  \hline
 \end{tabular}
 \end{center}
 \caption{The nominal and actual centre-of-mass energies for data
          collected during $\LEPII$ operation in each year. The approximate
          average luminosity analysed per experiment at each energy is also
          shown. Values marked with 
          a $^{\ast}$ are average energies for 1995 and 1996 used 
          for heavy flavour results. The data taken at nominal energies of
          130 and 136 in 1995 and 1997 are combined by most experiments.}
 \label{ff:tab-ecms}
\end{table}

A number of measurements on the process $\eeff$ exist and are
combined.
The preliminary averages of cross-section and forward-backward asymmetry
measurements are discussed in Section \ref{ff:sec-ave-xsc-afb}.
The results presented in this section update those presented 
in~\cite{bib-EWEP-00,ff:ref:lepff-moriond2001,ff:ref:lepff-osaka,ff:ref:lepff-moriond2000, ff:ref:lepff-tamp}.
Complete results of the combinations are available on the 
web page~\cite{ff:ref:ffbar_web}.
In Section~\ref{ff:sec-dsdc} a preliminary average of the differential
cross-section measurements, $\dsdc$, for the channels 
$\eemumu$ and $\eetautau$ is presented. 
In Section~\ref{ff:sec-hvflv} a preliminary combination of the
heavy flavour results $\Rb$, $\Rc$, $\Abb$ and $\Acc$ from $\LEPII$ is 
presented. In Section~\ref{ff:sec-interp} the combined results are interpreted
in terms of contact interactions and the exchange of $\Zprime$ bosons.
The results are summarised in section~\ref{ff:sec-conc}.

There are significant changes with respect to results presented in
Summer 2000 \cite{bib-EWEP-00,ff:ref:lepff-osaka}:
\begin{itemize}
 \item The method of combining the cross-sections and leptonic 
       forward-backward asymmetries is improved.
 \item The combinations are updated using new data:
  \begin{itemize}
   \item updated preliminary cross-sections and leptonic forward-backward 
         asymmetries for data taken at centre-of-mass energies of 205 and 
         207 $\GeV$, 
   \item new preliminary differential cross-section results for \mumu\ and 
         \tautau\ final states,
   \item new preliminary heavy-flavour results.
  \end{itemize}
 \item The interpretations are updated due to the changes in combined
       LEP results. 

\end{itemize}

\section{Averages for Cross-sections and Asymmetries}
\label{ff:sec-ave-xsc-afb}

In this section the results of the preliminary combination of
cross-sections and asymmetries are given.  The individual experiments'
analyses of cross-sections and forward-backward asymmetries are
discussed in~\cite{ff:ref:expts}.  The preliminary cross-section and
leptonic forward-backward asymmetry results at centre-of-mass energies
of 205 and 207 $\GeV$ are updated with respect
to~\cite{bib-EWEP-00,ff:ref:lepff-osaka}. These are now obtained from
analyses based on the full data set collected in 2000, improving
the precision of the measurements.

Cross-section results are combined for the $\eeqq$, $\eemumu$ and
$\eetautau$ channels, forward-backward asymmetry measurements are
combined for the $\mumu$ and $\tautau$ final states.  At \LEPII\ energies
$\gamma$ radiation is very important, leading in particular to a high
rate for the radiative return to the Z. Events are classified
according to the effective centre of mass energy, $\sqrt{\spr}$,
measured in different ways.  The averages are made for the samples of
events with high $\sqrt{\spr}$, as discussed in the following.

Individual experiments use their own \ff\ signal definitions;
corrections are applied to bring the measurements to two common
signal definitions:

\begin{itemize}
\item {\bf Definition 1:} $\sqrt{\spr}$ is taken to be the mass of the
  $s$-channel propagator, with the $\ff$ signal being defined by the
  cut $\sqrt{\spr/s} > 0.85$. The effects of ISR-FSR photon
  interference is subtracted to render the propagator mass
  unambiguous.

 \item {\bf Definition 2:} For dilepton events, $\sqrt{\spr}$ is taken to be 
 the bare invariant mass of the outgoing difermion pair. For hadronic events,
 it is taken to be the mass of the $s$-channel propagator. In both cases,
 ISR-FSR photon interference is included and the signal is defined by
 the cut $\sqrt{\spr/s} > 0.85$. When calculating the contribution to
 the hadronic cross-section due to ISR-FSR interference, since the propagator 
 mass is ill-defined, it is replaced by the bare $\qq$ mass.
\end{itemize}
\noindent
The measurement corrected to the common signal definition,
$\mathrm{M_{common}}$ is computed from the experimental measurement
$\mathrm{M_{exp}}$,
\begin{displaymath}
{\mathrm{M_{common}}} = {\mathrm{M_{exp}}} + ({\mathrm{P_{common}}} -
                                              {\mathrm{P_{exp}}}),
\end{displaymath}
\noindent
where $\mathrm{P_{exp}}$ is the prediction for the measurement obtained 
for the experiments' signal definition and $\mathrm{P_{common}}$ is the
prediction for the common signal definition. The predictions are computed with
ZFITTER~\cite{ff:ref:ZFITTER}.
The theoretical uncertainties associated with the corrections are obtained 
by comparing ZFITTER, TOPAZ0~v4.4~\cite{ff:ref:TOPAZ0} 
and the Monte Carlo generator KK~v4.02~\cite{ff:ref:KK}. 
The uncertainties are approximately $0.2\%$ for the hadronic cross-sections, $0.7\%$ 
for dilepton cross-sections and 0.003 for the leptonic 
asymmetries~\cite{ff:ref:lepff-tamp}. These uncertainties will be
updated for the final analyses, taking into account the results
of Reference~\citen{ff:ref:lepffwrkshp}. 
These errors are not included in the
combination.
Results are presented extrapolated to full $4\pi$ angular 
acceptance. Events containing additional fermion pairs from radiative 
processes are considered to be signal, providing that the primary 
pair passes the cut on $\sqrt{\spr/s}$ and that the secondary pair
has a mass below 70~$\GeV$.

The average is performed using the Best Linear Unbiased Estimator
(BLUE) technique \cite{common_bib:lyons}, which is based on matrix
algebra and which is equivalent to a $\chi^{2}$ minimisation.
For the first time, all the data, from centre-of-mass energies of 130
to 207 $\GeV$ are averaged together, taking into account correlations
between all $\LEPII$ $\eeff$ measurements.
Previously~\cite{bib-EWEP-00}, the data were treated as three
independent subsamples at (130--189) $\GeV$, (192--202) $\GeV$ and
(205--207) $\GeV$, ignoring correlations between the subsamples.

Particular care is taken to ensure that the correlations between the 
hadronic cross-sections are reasonably estimated. 
As in~\cite{bib-EWEP-00,ff:ref:lepff-osaka} the errors are broken down into 5 categories
\begin{itemize}

\item[1)] The statistical uncertainty plus uncorrelated systematic 
uncertainties, combined in quadrature.

\item[2)] The systematic uncertainty for the final state X which is 
fully correlated between energy points for that experiment.

\item[3)] The systematic uncertainty for experiment Y which is fully
  correlated between different final states for this energy point but
  uncorrelated between energy points.

\item[4)] The systematic uncertainty for the final state X which is 
fully correlated between energy points  and between different experiments.

\item[5)] The systematic uncertainty which is fully correlated between 
energy points and between different experiments for all final states.
\end{itemize}
In previous averages, uncertainties in the hadronic cross-sections 
arising from fragmentation models  and modelling of ISR had been 
treated as uncorrelated between experiments. However, although there are 
some differences between the models used and the methods of evaluating 
the errors, there are significant common elements in the estimation of 
these sources of uncertainty between the experiments. 
For the average reported here, these errors are treated as fully 
correlated between energy points and experiments. 

Table~\ref{ff:tab-xsafbres} gives the averaged cross-sections
and forward-backward asymmetries for all energies for Definition 1.
The differences in the results obtained using Definition 2 are also 
given. 

The $\chi^{2}$ per degree of freedom for the average of the $\LEPII$ $\ff$ 
data is $170/180$. The correlations are rather small, with the largest 
components at any given pair of energies being between the hadronic 
cross-sections. The other off-diagonal terms in the correlation 
matrix are smaller than $10\%$. The correlation matrix between the 
averaged hadronic cross-sections at different centre-of-mass energies 
is given in Table~\ref{ff:tab-hadcorrel}.

Differences in the results with respect to previous combinations at
centre-of-mass energies from
130--202~$\GeV$~\cite{bib-EWEP-00,ff:ref:lepff-osaka,
  ff:ref:lepff-moriond2000} arise mainly from the introduction of
correlations between measurements which were previously taken to be
uncorrelated, and the improved treatment of the correlations
themselves.

\begin{figure}[p]
 \begin{center}
 \mbox{\epsfig{file=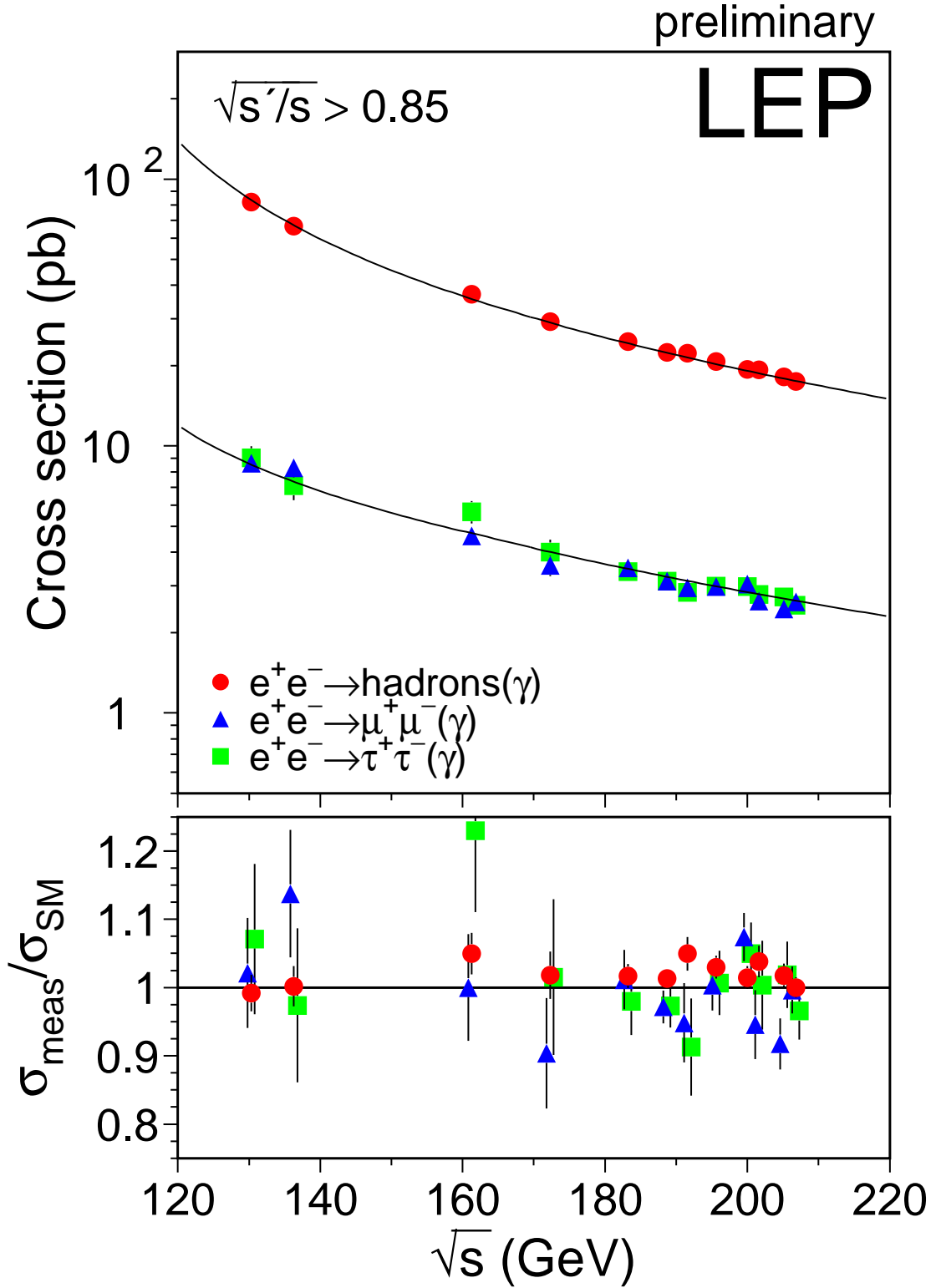,width=15cm}}
 \end{center}
 \caption{Preliminary combined LEP results on the cross-sections for 
          $\qq$, $\mumu$ and $\tautau$ final states, as a function of 
          centre-of-mass energy. The expectations of the SM, 
          computed with ZFITTER~\capcite{ff:ref:ZFITTER}, are shown as curves.
          The lower plot shows the ratio of the data divided by the SM.}
\label{ff:fig-xs_lep}
\end{figure}

\begin{figure}[p]
 \begin{center}
 \mbox{\epsfig{file=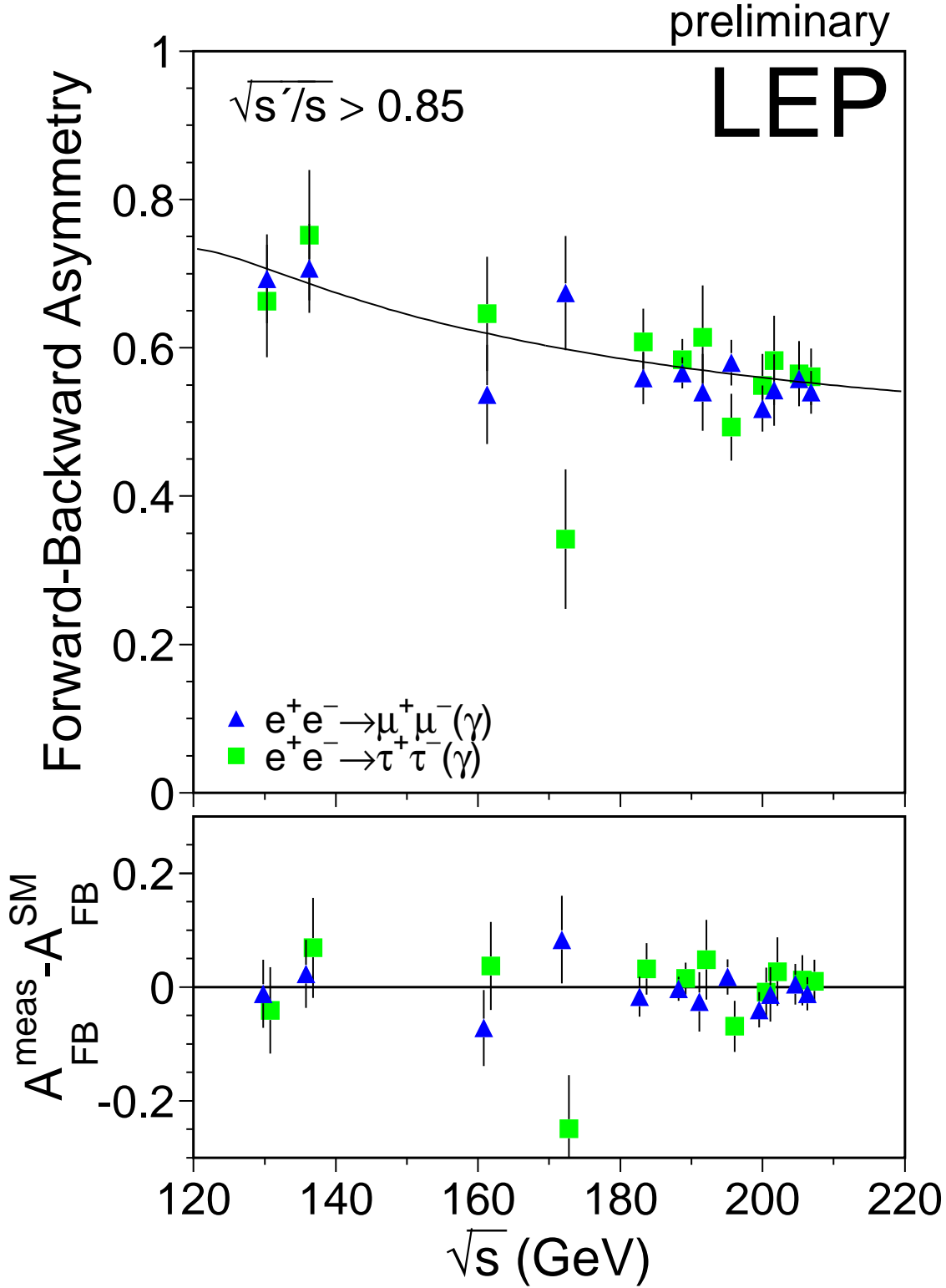,width=15cm}}
 \end{center}
 \caption{Preliminary combined LEP results on the forward-backward 
          asymmetry for $\mumu$  and $\tautau$ final states as a function of 
          centre-of-mass energy. The expectations of the SM 
          computed  with ZFITTER~\capcite{ff:ref:ZFITTER}, are shown as 
          curves. The lower plot shows differences between the data 
          and the SM.}
 \label{ff:fig-afb_lep}
\end{figure}

Figures~\ref{ff:fig-xs_lep} and~\ref{ff:fig-afb_lep} show the LEP 
averaged cross-sections and asymmetries, respectively, as a 
function of the centre-of-mass energy, together with the SM predictions. 
There is good agreement between the SM expectations and the measurements of the
individual experiments and the combined averages.
The measured cross-sections for hadronic final states at most of the energy points 
are somewhat above the SM expectations. Taking into account the correlations
between the data points and also assigning an error of 
$\pm 0.26\%$~\cite{ff:ref:lepffwrkshp} on the absolute SM predictions, 
the difference of the cross-section from the SM expectations averaged over 
all energies is approximately $1.8$ standard deviations.
It is concluded that there is no significant evidence in the results
for physics beyond the SM in the process $\eeff$.

\begin{table}[htbp]
 \begin{center}
 \begin{turn}{90}
 \begin{tabular}{ccc}
 \begin{tabular}{|c|c|r@{$\pm$}r|r|r|}
 \hline
  $\sqrt{s}$ &          &\multicolumn{2}{c|}{Average} &                           &                                 \\ 
    ($\GeV$) & Quantity &\multicolumn{2}{|c|}{value}  &  \multicolumn{1}{|c|}{SM} & \multicolumn{1}{|c|}{$\Delta$}  \\
 \hline\hline
  130 & $\sigma(q\overline{q})$               [pb] & 82.124 &  2.232 & 82.803 & -0.251 \\
      & $\sigma(\mu^{+}\mu^{-})$              [pb] &  8.620 &  0.682 &  8.439 & -0.331 \\
      & $\sigma(\tau^{+}\tau^{-})$            [pb] &  9.036 &  0.930 &  8.435 & -0.108 \\
      & $\mathrm{A_{fb}}(\mu^{+}\mu^{-})$          &  0.693 &  0.060 &  0.705 &  0.012 \\
      & $\mathrm{A_{fb}}(\tau^{+}\tau^{-})$        &  0.663 &  0.076 &  0.704 &  0.012 \\
 \hline
  136 & $\sigma(q\overline{q})$               [pb] & 66.724 &  1.974 & 66.596 & -0.224 \\
      & $\sigma(\mu^{+}\mu^{-})$              [pb] &  8.276 &  0.677 &  7.281 & -0.280 \\
      & $\sigma(\tau^{+}\tau^{-})$            [pb] &  7.086 &  0.820 &  7.279 & -0.091 \\
      & $\mathrm{A_{fb}}(\mu^{+}\mu^{-})$          &  0.707 &  0.060 &  0.684 &  0.013 \\
      & $\mathrm{A_{fb}}(\tau^{+}\tau^{-})$        &  0.752 &  0.088 &  0.683 &  0.014 \\
 \hline
  161 & $\sigma(q\overline{q})$               [pb] & 37.014 &  1.074 & 35.247 & -0.143 \\
      & $\sigma(\mu^{+}\mu^{-})$              [pb] &  4.608 &  0.364 &  4.613 & -0.178 \\
      & $\sigma(\tau^{+}\tau^{-})$            [pb] &  5.673 &  0.545 &  4.613 & -0.061 \\
      & $\mathrm{A_{fb}}(\mu^{+}\mu^{-})$          &  0.537 &  0.067 &  0.609 &  0.017 \\
      & $\mathrm{A_{fb}}(\tau^{+}\tau^{-})$        &  0.646 &  0.077 &  0.609 &  0.016 \\
 \hline
  172 & $\sigma(q\overline{q})$               [pb] & 29.262 &  0.989 & 28.738 & -0.124 \\
      & $\sigma(\mu^{+}\mu^{-})$              [pb] &  3.571 &  0.317 &  3.952 & -0.157 \\
      & $\sigma(\tau^{+}\tau^{-})$            [pb] &  4.013 &  0.450 &  3.951 & -0.054 \\
      & $\mathrm{A_{fb}}(\mu^{+}\mu^{-})$          &  0.674 &  0.077 &  0.591 &  0.018 \\
      & $\mathrm{A_{fb}}(\tau^{+}\tau^{-})$        &  0.342 &  0.094 &  0.591 &  0.017 \\
 \hline
  183 & $\sigma(q\overline{q})$               [pb] & 24.609 &  0.426 & 24.200 & -0.109 \\
      & $\sigma(\mu^{+}\mu^{-})$              [pb] &  3.490 &  0.147 &  3.446 & -0.139 \\
      & $\sigma(\tau^{+}\tau^{-})$            [pb] &  3.375 &  0.174 &  3.446 & -0.050 \\
      & $\mathrm{A_{fb}}(\mu^{+}\mu^{-})$          &  0.559 &  0.035 &  0.576 &  0.018 \\
      & $\mathrm{A_{fb}}(\tau^{+}\tau^{-})$        &  0.608 &  0.045 &  0.576 &  0.018 \\
 \hline
  189 & $\sigma(q\overline{q})$               [pb] & 22.446 &  0.257 & 22.156 & -0.101 \\
      & $\sigma(\mu^{+}\mu^{-})$              [pb] &  3.116 &  0.077 &  3.207 & -0.131 \\
      & $\sigma(\tau^{+}\tau^{-})$            [pb] &  3.121 &  0.099 &  3.207 & -0.048 \\
      & $\mathrm{A_{fb}}(\mu^{+}\mu^{-})$          &  0.566 &  0.021 &  0.569 &  0.019 \\
      & $\mathrm{A_{fb}}(\tau^{+}\tau^{-})$        &  0.584 &  0.028 &  0.569 &  0.018 \\
 \hline
 \end{tabular}
 & &
 \begin{tabular}{|c|c|r@{$\pm$}r|r|r|}
 \hline
  $\sqrt{s}$ &          & \multicolumn{2}{c|}{Average} &                          &                                 \\ 
    ($\GeV$) & Quantity &\multicolumn{2}{|c|}{value}   & \multicolumn{1}{|c|}{SM} & \multicolumn{1}{|c|}{$\Delta$}  \\
 \hline\hline
  192 & $\sigma(q\overline{q})$               [pb] & 22.291 &  0.523 & 21.237 & -0.098 \\
      & $\sigma(\mu^{+}\mu^{-})$              [pb] &  2.943 &  0.175 &  3.097 & -0.127 \\
      & $\sigma(\tau^{+}\tau^{-})$            [pb] &  2.832 &  0.216 &  3.097 & -0.047 \\
      & $\mathrm{A_{fb}}(\mu^{+}\mu^{-})$          &  0.540 &  0.052 &  0.566 &  0.019 \\
      & $\mathrm{A_{fb}}(\tau^{+}\tau^{-})$        &  0.614 &  0.070 &  0.566 &  0.019 \\
 \hline
  196 & $\sigma(q\overline{q})$               [pb] & 20.729 &  0.338 & 20.127 & -0.094 \\
      & $\sigma(\mu^{+}\mu^{-})$              [pb] &  2.967 &  0.106 &  2.962 & -0.123 \\
      & $\sigma(\tau^{+}\tau^{-})$            [pb] &  2.984 &  0.138 &  2.962 & -0.045 \\
      & $\mathrm{A_{fb}}(\mu^{+}\mu^{-})$          &  0.580 &  0.031 &  0.562 &  0.019 \\
      & $\mathrm{A_{fb}}(\tau^{+}\tau^{-})$        &  0.493 &  0.045 &  0.562 &  0.019 \\
 \hline
  200 & $\sigma(q\overline{q})$               [pb] & 19.372 &  0.319 & 19.085 & -0.090 \\
      & $\sigma(\mu^{+}\mu^{-})$              [pb] &  3.040 &  0.104 &  2.834 & -0.118 \\
      & $\sigma(\tau^{+}\tau^{-})$            [pb] &  2.966 &  0.134 &  2.833 & -0.044 \\
      & $\mathrm{A_{fb}}(\mu^{+}\mu^{-})$          &  0.518 &  0.031 &  0.558 &  0.019 \\
      & $\mathrm{A_{fb}}(\tau^{+}\tau^{-})$        &  0.549 &  0.043 &  0.558 &  0.019 \\
 \hline
  202 & $\sigma(q\overline{q})$               [pb] & 19.278 &  0.430 & 18.572 & -0.088 \\
      & $\sigma(\mu^{+}\mu^{-})$              [pb] &  2.621 &  0.139 &  2.770 & -0.116 \\
      & $\sigma(\tau^{+}\tau^{-})$            [pb] &  2.777 &  0.183 &  2.769 & -0.044 \\
      & $\mathrm{A_{fb}}(\mu^{+}\mu^{-})$          &  0.543 &  0.048 &  0.556 &  0.020 \\
      & $\mathrm{A_{fb}}(\tau^{+}\tau^{-})$        &  0.583 &  0.060 &  0.556 &  0.019 \\
 \hline
  205 & $\sigma(q\overline{q})$               [pb] & 18.119 &  0.316 & 17.811 & -0.085 \\
      & $\sigma(\mu^{+}\mu^{-})$              [pb] &  2.449 &  0.100 &  2.674 & -0.112 \\
      & $\sigma(\tau^{+}\tau^{-})$            [pb] &  2.705 &  0.129 &  2.673 & -0.042 \\
      & $\mathrm{A_{fb}}(\mu^{+}\mu^{-})$          &  0.558 &  0.036 &  0.553 &  0.020 \\
      & $\mathrm{A_{fb}}(\tau^{+}\tau^{-})$        &  0.565 &  0.044 &  0.553 &  0.019 \\
 \hline
  207 & $\sigma(q\overline{q})$               [pb] & 17.423 &  0.263 & 17.418 & -0.083 \\
      & $\sigma(\mu^{+}\mu^{-})$              [pb] &  2.613 &  0.088 &  2.623 & -0.111 \\
      & $\sigma(\tau^{+}\tau^{-})$            [pb] &  2.528 &  0.108 &  2.623 & -0.042 \\
      & $\mathrm{A_{fb}}(\mu^{+}\mu^{-})$          &  0.540 &  0.029 &  0.552 &  0.020 \\
      & $\mathrm{A_{fb}}(\tau^{+}\tau^{-})$        &  0.561 &  0.038 &  0.551 &  0.019 \\
 \hline
 \end{tabular}
 \end{tabular}
 \end{turn}
 \end{center}
\caption{Preliminary combined LEP results for $\eeff$. 
 All the results correspond to the signal Definition~1. The Standard Model
 predictions are from ZFITTER~\capcite{ff:ref:ZFITTER}.
 The difference, $\Delta$, in the averages for the measurements for 
 Definition~2 relative to Definition~1 are given in the final column.
 The quoted uncertainties do not include the theoretical 
 uncertainties on the corrections discussed in the text.}
\label{ff:tab-xsafbres}
\end{table}

\begin{table}[htbp]
 \vskip 3cm
 \begin{center}
 \begin{turn}{90}
 \begin{tabular}{|c|c|c|c|c|c|c|c|c|c|c|c|c|}
 \hline
 $\begin{array}[b]{c}\roots \\ \GeV) \end{array}$
       & 130    & 136    & 161    & 172    & 183    & 189    & 192    & 196    & 200    & 202    & 205    & 207    \\
 \hline\hline
 130 &  1.000 &  0.075 &  0.085 &  0.076 &  0.121 &  0.151 &  0.084 &  0.116 &  0.131 &  0.091 &  0.137 &  0.160 \\
 136 &        &  1.000 &  0.079 &  0.071 &  0.112 &  0.140 &  0.078 &  0.107 &  0.121 &  0.084 &  0.127 &  0.148 \\
 161 &        &        &  1.000 &  0.082 &  0.128 &  0.162 &  0.089 &  0.123 &  0.139 &  0.097 &  0.144 &  0.167 \\
 172 &        &        &        &  1.000 &  0.114 &  0.145 &  0.080 &  0.110 &  0.125 &  0.087 &  0.130 &  0.150 \\
 183 &        &        &        &        &  1.000 &  0.237 &  0.130 &  0.179 &  0.203 &  0.139 &  0.208 &  0.242 \\
 189 &        &        &        &        &        &  1.000 &  0.173 &  0.236 &  0.270 &  0.184 &  0.266 &  0.307 \\
 192 &        &        &        &        &        &        &  1.000 &  0.136 &  0.156 &  0.106 &  0.151 &  0.173 \\
 196 &        &        &        &        &        &        &        &  1.000 &  0.212 &  0.145 &  0.207 &  0.238 \\
 200 &        &        &        &        &        &        &        &        &  1.000 &  0.166 &  0.236 &  0.271 \\
 202 &        &        &        &        &        &        &        &        &        &  1.000 &  0.162 &  0.185 \\
 205 &        &        &        &        &        &        &        &        &        &        &  1.000 &  0.282 \\
 207 &        &        &        &        &        &        &        &        &        &        &        &  1.000 \\
 \hline
 \end{tabular}
 \end{turn}
 \end{center}
\caption{The correlation coefficients between averaged hadronic cross-sections
         at different energies.}
\label{ff:tab-hadcorrel}
 \vskip 5cm
\end{table}

\clearpage
 
\section{Averages for Differential Cross-sections}
\label{ff:sec-dsdc}

The LEP experiments measure the differential cross-section, $\dsdc$, 
for the $\eemumu$ and $\eetautau$ channels for samples of events 
with $\sqrt{s'/s}>0.85$. A preliminary combination of these results 
is made using a $\chi^{2}$ fit to the measured differential cross sections,
using the expected error on the differential cross sections, computed from 
the expected cross sections and the expected numbers of events in each 
experiment. Using a Monte Carlo simulation it is shown that this method 
provides a good approximation to the exact likelihood method
based on Poisson statistics~\cite{ff:ref:lepff-osaka}.

The combination included data from 183 to 207 $\GeV$, but not all experiments
provided measurements at all energies. Since~\cite{bib-EWEP-00,ff:ref:lepff-osaka},
new, preliminary, results for centre-of-mass energies of 205 and 207~$\GeV$
are made available by all experiments. In addition, new, preliminary, 
results for $\eemumu$ at energies from 192--202 $\GeV$ from L3 are made 
available. The data used in the combination are summarised in 
Table~\ref{ff:tab:inputs}. 

Each experiments' data are binned in 10 bins of $\cos\theta$ at each 
energy, using their own signal definition. The scattering angle, $\theta$, is
the angle of the negative lepton with respect to the incoming electron 
direction in the lab coordinate system. The outer acceptances of the most 
forward and most backward bins for which the four experiments present 
their data are different. This is accounted for as part of the correction to 
a common signal definition. The ranges in $\cos\theta$ for the measurements of 
the individual experiments and the average are given in 
Table~\ref{ff:tab:acpt}. The signal definition used corresponded 
to Definition 1 of Section~\ref{ff:sec-ave-xsc-afb}.

Correlated small systematic errors between different experiments,
channels and energies, arising from uncertainties on the overall
normalisation are considered in the averaging procedure.

\begin{figure}[p]
 \begin{center}
  \epsfig{file=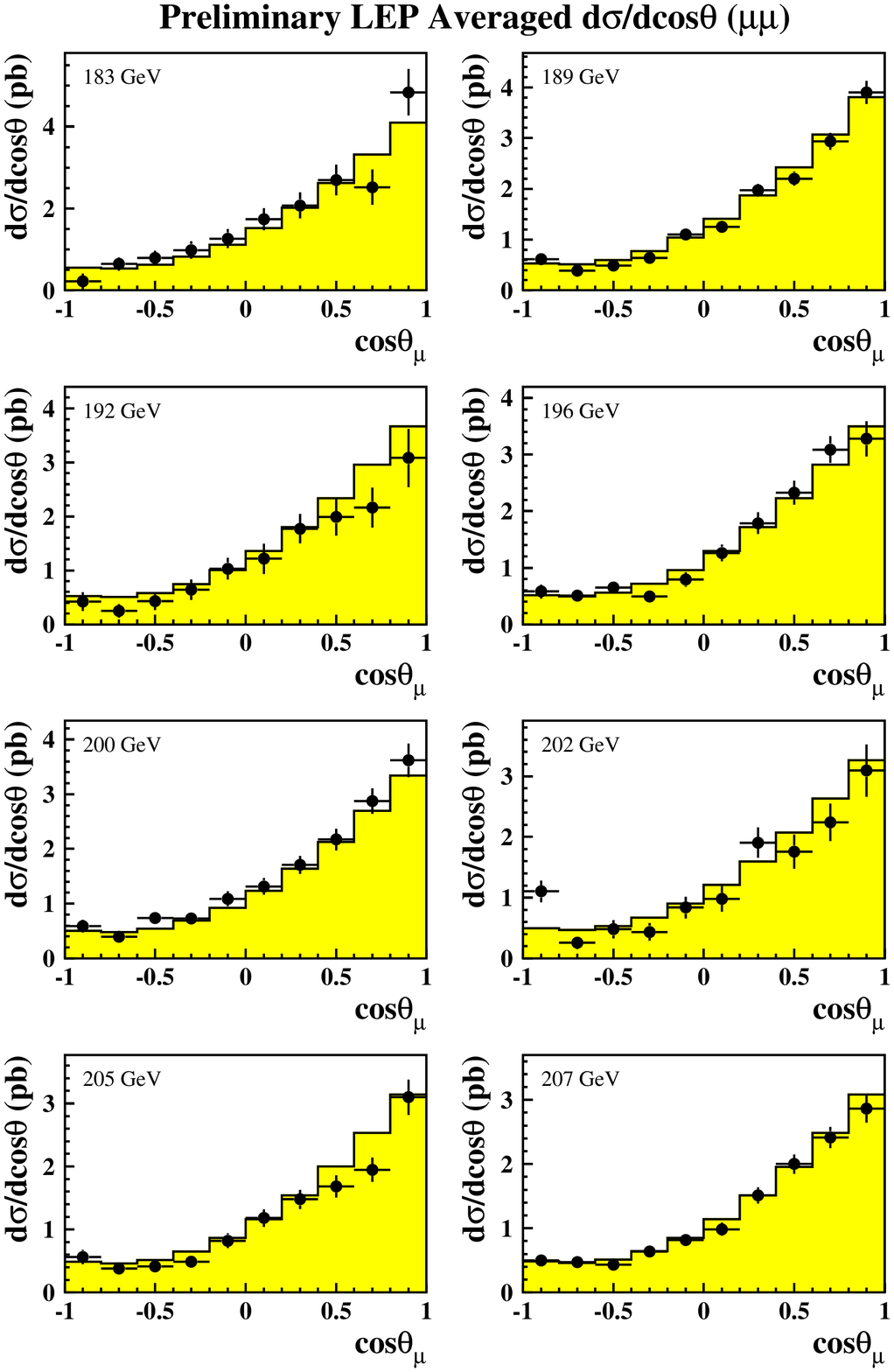,width=0.98\textwidth}
 \end{center}
 \caption{LEP averaged differential cross-sections for $\eemumu$ at
          energies of 183--207 $\GeV$. The SM
          predictions, shown as solid histograms, are computed with
          ZFITTER~\capcite{ff:ref:ZFITTER}.}
 \label{ff:fig:dsdc-res-mm}
 \vskip 2cm 
\end{figure}

\begin{figure}[p]
 \begin{center}
  \epsfig{file=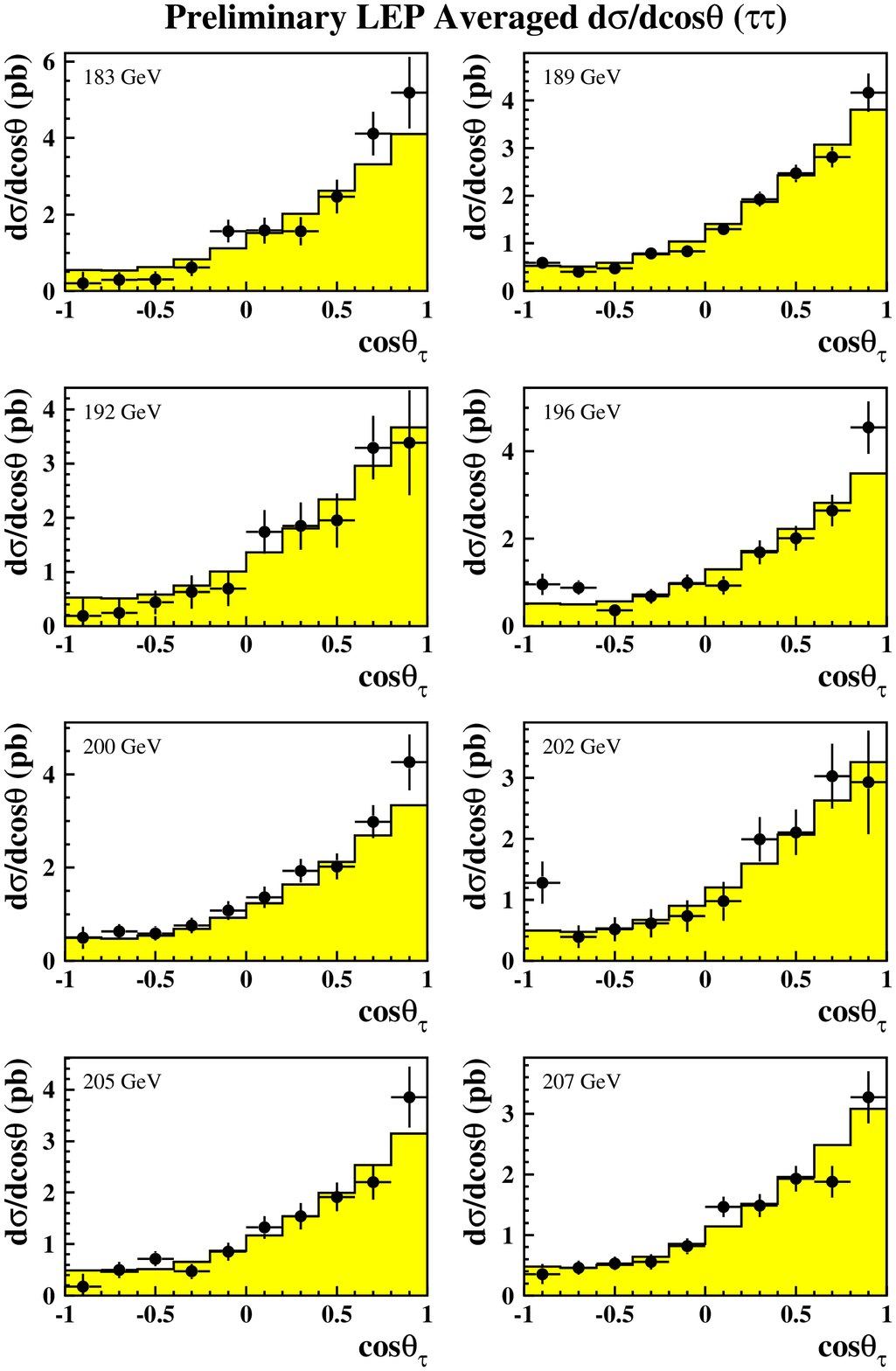,width=0.98\textwidth}
 \end{center}
 \caption{LEP averaged differential cross-sections for $\eetautau$ at 
          energies of 183--207 $\GeV$. The SM
          predictions, shown as solid histograms, are computed with
          ZFITTER~\capcite{ff:ref:ZFITTER}.}
 \label{ff:fig:dsdc-res-tt}
 \vskip 2cm 
\end{figure}

Three separate averages are performed; one for 183 and 189 $\GeV$ data,
one for 192--202 $\GeV$ data and for 205 and 207 $\GeV$ data.
The averages for the 183--189 data set are not updated
with respect to~\cite{bib-EWEP-00,ff:ref:lepff-osaka}. 
The results of the averages are shown in Figures~\ref{ff:fig:dsdc-res-mm} 
and~\ref{ff:fig:dsdc-res-tt}. 

The correlations between bins in the average are less than 
$2\%$ of the total error on the averages in each bin.
The overall agreement between the averaged data and the predictions
is reasonable, with a $\chi^{2}$ of $191$ for $160$ degrees of freedom. 
At 202 $\GeV$ the cross-section in the most backward bin,
$-1.0 < \cos\theta < -0.8$, for both muon and tau final states is 
above the predictions. For the muons the excess in the data corresponds to $3.4$
standard deviations. For the taus the excess is $2.3$ standard deviations,
however, for this measurement the individual experiments are somewhat
inconsistent, having a $\chi^{2}$ with respect to the average of
$10.5$ for $2$ degrees of freedom. The data at 202 $\GeV$ suffer
from rather low delivered luminosity, with fewer than four events
expected in each experiment in each channel in this backward 
$\cos\theta$ bin. The agreement between the data
and the predictions in the same $\cos\theta$ bin is better at 
higher energies. 

\begin{table}[htbp]
 \begin{center}
 \begin{tabular}{|l|cccc|cccc|}
 \hline
                     & \multicolumn{4}{|c|}{$\eemumu$}           
                     & \multicolumn{4}{|c|}{$\eetautau$}         \\
 \cline{2-9}
  $\sqrt{s}$($\GeV$) &      A   &      D   &      L   &      O   
                     &      A   &      D   &      L   &      O   \\
 \hline 
 \hline
  183                & {\sc{-}} & {\sc{F}} & {\sc{-}} & {\sc{F}} 
                     & {\sc{-}} & {\sc{F}} & {\sc{-}} & {\sc{F}} \\
 \hline
  189                & {\sc{P}} & {\sc{F}} & {\sc{F}} & {\sc{F}}  
                     & {\sc{P}} & {\sc{F}} & {\sc{F}} & {\sc{F}} \\
 \hline
  192--202           & {\sc{P}} & {\sc{P}} & {\sc{P}} & {\sc{P}} 
                     & {\sc{P}} & {\sc{P}} & {\sc{-}} & {\sc{P}} \\
 \hline
  205--207           & {\sc{P}} & {\sc{P}} & {\sc{P}} & {\sc{P}} 
                     & {\sc{P}} & {\sc{P}} & {\sc{-}} & {\sc{P}} \\
 \hline
 \end{tabular}
 \end{center}
 \caption{Differential cross-section data provided by the LEP 
          collaborations (ALEPH, DELPHI, L3 and OPAL) for $\eemumu$ and 
          $\eetautau$ combination at different centre-of-mass energies. 
          Data indicated with {\sc{F}} are final, published data. 
          Data marked with {\sc{P}} are preliminary. 
          Data marked with a {\sc{-}} are not available for combination.}
 \label{ff:tab:inputs}
\end{table}

\begin{table}[htbp]
 \begin{center}
 \begin{tabular}{|l|c|c|}
  \hline
  Experiment                   & $\cos\theta_{min}$ & $\cos\theta_{max}$ \\
  \hline
  \hline 
   ALEPH                       &    $-0.95$         &     $0.95$         \\
   DELPHI ($\eemumu$ 183)      &    $-0.94$         &     $0.94$         \\
   DELPHI ($\eemumu$ 189--207) &    $-0.97$         &     $0.97$         \\
   DELPHI ($\eetautau$)        &    $-0.96$         &     $0.96$         \\
   L3                          &    $-0.90$         &     $0.90$         \\
   OPAL                        &    $-1.00$         &     $1.00$         \\
  \hline
  \hline
   Average                     &    $-1.00$         &     $1.00$         \\
  \hline
 \end{tabular}
 \end{center}
 \caption{The acceptances for which experimental data are presented 
          and the acceptance for the LEP average.
          For DELPHI the acceptance is shown for the different channels and 
          for the muons for different centre of mass energies. For all other
          experiments the acceptance is the same for muon and tau-lepton 
          channels and for all energies provided.}
 \label{ff:tab:acpt}
\end{table}


\clearpage

\section{Averages for Heavy Flavour Measurements}
\label{ff:sec-hvflv}

This section presents a preliminary combination of both 
published~\cite{ff:ref:hfpublished}
and preliminary~\cite{ff:ref:hfpreliminary} measurements of the 
ratios\footnote{Unlike at $\LEPI$, $\Rq$ is defined as 
$\mathrm{\frac{\sigma_{q \overline{q}}}{\sigma_{had}}}$.}
$\Rb$ and $\Rc$ and the forward-backward asymmetries, $\Abb$ and 
$\Acc$, from the LEP collaborations at centre-of-mass 
energies in the range of 130 to 207 $\GeV$. The averages are 
updated with respect to~\cite{bib-EWEP-00,ff:ref:lepff-osaka}. New preliminary results 
from DELPHI and L3 at centre-of-mass energies of 205 and 207 $\GeV$, based on 
analyses of the full 2000 data sets, are also included. New, preliminary, results
from ALEPH at lower energies are also combined.
Table~\ref{ff:tab:hfinput} summarises all the inputs that are combined.

A common signal definition is defined for all the measurements, requiring:
\begin{list}{$\bullet$}{\setlength{\itemsep}{0ex}
                        \setlength{\parsep}{0ex}
                        \setlength{\topsep}{0ex}}
 \item{an effective centre-of-mass energy $\sqrt{s^{\prime}} > 0.85 \sqrt{s}$}
 \item{the inclusion of ISR and FSR photon interference contribution and}
 \item{extrapolation to full angular acceptance.}
\end{list}
Systematic errors are divided into three categories: uncorrelated errors, 
errors correlated between the measurements of each experiment, and 
errors common to all experiments.
Full details concerning the combination procedure
can be found in~\cite{ff:ref:hfconfnote}.

\begin{figure}[p]
\begin{center}
\mbox{\epsfig{file=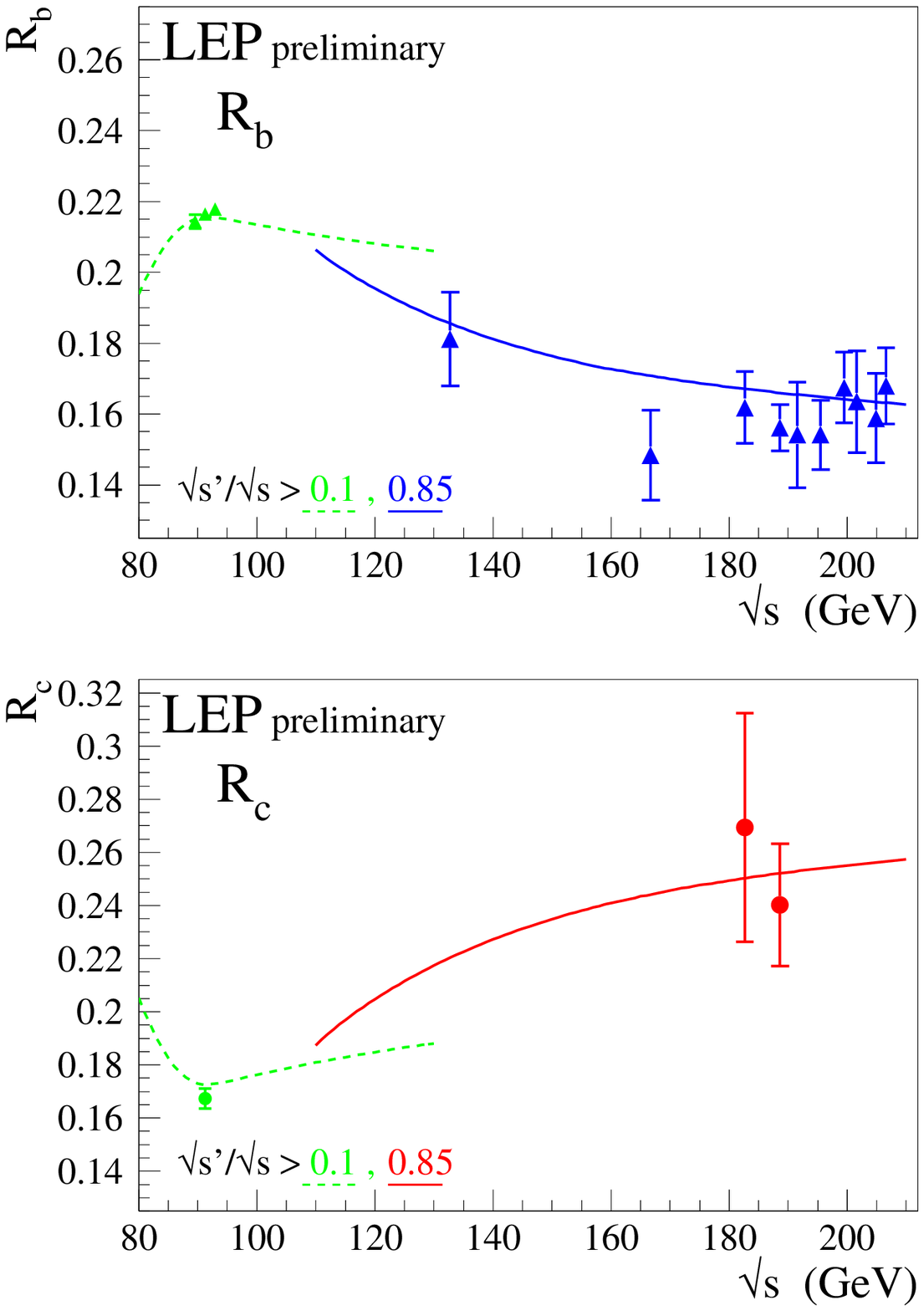,height=20cm}}
\end{center}
\caption{Preliminary combined LEP measurements of $\Rb$ and $\Rc$. Solid lines
         represent the Standard Model prediction for the signal definition and
         dotted lines the inclusive prediction. Both are computed with 
         ZFITTER\capcite{ff:ref:hfzfit}. The $\LEPI$ measurements are  
         taken from \capcite{ff:ref:hflep1-99}.}
\label{ff:fig:hfres1}
\end{figure}

\begin{figure}[p]
\begin{center}
\mbox{\epsfig{file=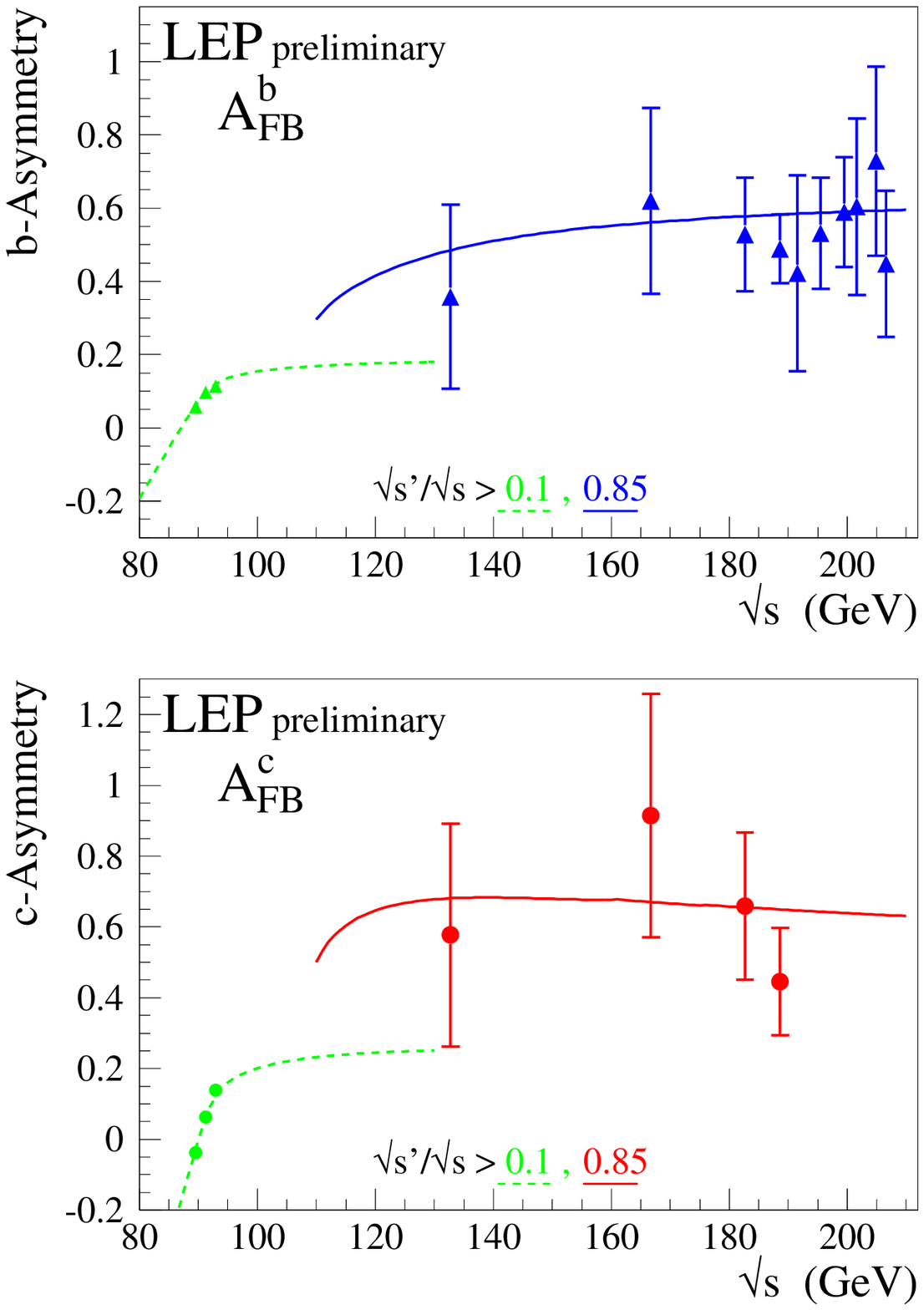,height=20cm}}
\end{center}
\caption{Preliminary combined LEP measurements of the forward--backward  
         asymmetries $\Abb$ and $\Acc$. Solid lines represent the Standard 
         Model prediction for the signal definition and dotted lines the 
         inclusive prediction. Both are computed with  
         ZFITTER~\capcite{ff:ref:hfzfit}. The $\LEPI$ measurements are
         taken from~\capcite{ff:ref:hflep1-99}.}
\label{ff:fig:hfres2} 
\end{figure}

The results of the combination are presented in Table~\ref{ff:tab:hfresults} 
and Figures~\ref{ff:fig:hfres1} and~\ref{ff:fig:hfres2}. 
The results are consistent with the Standard Model predictions of ZFITTER.

Because of the large correlation (-0.36) with $\Rc$ at 183~$\GeV$ and 
189~$\GeV$, the errors on the corresponding measurements of $\Rb$ receive 
an additional contribution which is absent at the other energy points.
For other energies where there is no measurement of $\Rc$, the
Standard Model value of $\Rc$ is used in extracting $\Rb$ (the
error on the Standard Model prediction of $\Rc$ is estimated to be
negligible compared to the other uncertainties on $\Rb$).

A list of the error contributions from the combination at 189~$\GeV$ is shown 
in Table~\ref{ff:tab:hferror}.

\begin{table}[htbp]
\begin{center}
\begin{tabular}{|l|cccc|cccc|cccc|cccc|}
\hline 
 $\sqrt{s}$ ($\GeV$)
            & \multicolumn{4}{|c|}{$\Rb$}
            & \multicolumn{4}{|c|}{$\Rc$}
            & \multicolumn{4}{|c|}{$\Abb$}
            & \multicolumn{4}{|c|}{$\Acc$} \\
\cline{2-17}
            & A & D & L & O & A & D & L & O & A & D & L & O & A & D & L & O \\
\hline\hline
133         & {\sc{F}} & {\sc{F}} & {\sc{F}} & {\sc{F}}
            & {\sc{-}} & {\sc{-}} & {\sc{-}} & {\sc{-}}  
            & {\sc{-}} & {\sc{F}} & {\sc{-}} & {\sc{F}}  
            & {\sc{-}} & {\sc{F}} & {\sc{-}} & {\sc{F}} \\
\hline
167         & {\sc{F}} & {\sc{F}} & {\sc{F}} & {\sc{F}}
            & {\sc{-}} & {\sc{-}} & {\sc{-}} & {\sc{-}} 
            & {\sc{-}} & {\sc{F}} & {\sc{-}} & {\sc{F}} 
            & {\sc{-}} & {\sc{F}} & {\sc{-}} & {\sc{F}}   \\
\hline 
183         & {\sc{F}} & {\sc{P}} & {\sc{F}} & {\sc{F}}
            & {\sc{F}} & {\sc{-}} & {\sc{-}} & {\sc{-}}  
            & {\sc{F}} & {\sc{-}} & {\sc{-}} & {\sc{F}} 
            & {\sc{P}} & {\sc{-}} & {\sc{-}} & {\sc{F}}  \\
\hline 
189         & {\sc{P}} & {\sc{P}} & {\sc{F}} & {\sc{F}}
            & {\sc{P}} & {\sc{-}} & {\sc{-}} & {\sc{-}}  
            & {\sc{P}} & {\sc{P}} & {\sc{F}} & {\sc{F}} 
            & {\sc{P}} & {\sc{-}} & {\sc{-}} & {\sc{F}} \\
\hline 
192 to 202  & {\sc{P}} & {\sc{P}} & {\sc{P}} & {\sc{-}} 
            & {\sc{-}} & {\sc{-}} & {\sc{-}} & {\sc{-}} 
            & {\sc{P}} & {\sc{P}} & {\sc{-}} & {\sc{-}} 
            & {\sc{-}} & {\sc{-}} & {\sc{-}} & {\sc{-}} \\ 
\hline 
205 and 207 & {\sc{-}} & {\sc{P}} & {\sc{P}} & {\sc{-}} 
            & {\sc{-}} & {\sc{-}} & {\sc{-}} & {\sc{-}} 
            & {\sc{-}} & {\sc{P}} & {\sc{-}} & {\sc{-}} 
            & {\sc{-}} & {\sc{-}} & {\sc{-}} & {\sc{-}} \\
\hline
\end{tabular}
\end{center}
\caption{Data provided by the ALEPH, DELPHI, L3, OPAL collaborations 
         for combination at different centre-of-mass energies. 
         Data indicated with {\sc{F}} are final, published data. 
         Data marked with {\sc{P}} are preliminary. Data marked with a 
         {\sc{-}} are not supplied for combination.}
\label{ff:tab:hfinput} 
\end{table}
\begin{table}[htbp]
\begin{center}
\begin{tabular}{|l|c|c|c|c|}
\hline 
$\sqrt{s}$ ($\GeV$) & $\Rb$
                    & $\Rc$ 
                    & $\Abb$
                    & $\Acc$ \\
\hline\hline
133      & 0.1811 $\pm$ 0.0132 & -                 & 0.358 $\pm$ 0.251 &  0.577 $\pm$ 0.314   \\
         & (0.1853)            & -                 & (0.487)           &  (0.681)  \\
\hline
167      & 0.1484 $\pm$ 0.0127 &  -                & 0.620 $\pm$ 0.254 &  0.915 $\pm$ 0.344  \\ 
         & (0.1708)            & -                 & (0.561)           &  (0.671) \\
\hline 
183      & 0.1619 $\pm$ 0.0101 & 0.269 $\pm$ 0.043 & 0.528 $\pm$ 0.155 & 0.658 $\pm$ 0.209 \\
         & (0.1671)            & (0.250)           & (0.578)           &  (0.656) \\
\hline 
189      & 0.1562 $\pm$ 0.0065 & 0.240 $\pm$ 0.023 & 0.488 $\pm$ 0.094 & 0.446 $\pm$ 0.151 \\ 
         & (0.1660)            & (0.252)           &  (0.583)          & (0.649) \\
\hline
192      & 0.1541 $\pm$ 0.0149 &  -                & 0.422 $\pm$ 0.267 &  - \\
         & (0.1655)            &   -               & (0.585)           & - \\ 
\hline 
196      & 0.1542 $\pm$ 0.0098 &  -                & 0.531 $\pm$ 0.151 &  - \\ 
         & (0.1648)            &   -               & (0.587)           & - \\ 
\hline
200      & 0.1675 $\pm$ 0.0100 &  -                & 0.589 $\pm$ 0.150 & - \\
         & (0.1642)            &   -               & (0.590)           & - \\ 
\hline
202      & 0.1635 $\pm$ 0.0143 &  -                & 0.604 $\pm$ 0.241 & - \\
         & (0.1638)            &   -               & (0.593)           & - \\ 
\hline
205      & 0.1588 $\pm$ 0.0126 &  -                & 0.728 $\pm$ 0.258 & - \\
         & (0.1634)            &   -               & (0.594)           & - \\ 
\hline
207      & 0.1680 $\pm$ 0.0108 &  -                & 0.447 $\pm$ 0.200 & - \\
         & (0.1632)            &   -               & (0.593)           & - \\ 
\hline
\end{tabular}
\end{center}
\caption{Results of the global fit, compared to the Standard Model 
         predictions computed with ZFITTER~\capcite{ff:ref:hfzfit}, for the 
         signal definition in parentheses.  The quoted errors  
         are the statistical and systematic errors added in quadrature. 
         Because of the large correlation with $\Rc$ at 183~$\GeV$ and 
         189~$\GeV$, the errors on the corresponding measurements of $\Rb$ 
         receive an additional contribution which is absent at the 
         other energy points.}
\label{ff:tab:hfresults}
\end{table}
\begin{table}[htbp]
\begin{center}
\begin{tabular}{|l|c|c|c|c|}
\hline 
Error list & $\Rb$ (189 $\GeV$) 
           & $\Rc$ (189 $\GeV$) 
           & $\Abb$ (189 $\GeV$) 
           & $\Acc$ (189 $\GeV$)  \\

\hline\hline
statistics    & 0.00606  & 0.0179 & 0.0884 & 0.1229 \\ 
\hline 
internal syst & 0.00232  & 0.0123 & 0.0296 & 0.0481 \\
common syst   & 0.00082  & 0.0078 & 0.0138 & 0.0735 \\
total syst    & 0.00246  & 0.0145 & 0.0327 & 0.0878 \\ 
\hline 
total error   & 0.00654  & 0.0231 & 0.0942 & 0.1510 \\ 
\hline 
\end{tabular}
\end{center}
\caption{Error breakdown at 189 $\GeV$.}
\label{ff:tab:hferror} 
\end{table}

\clearpage

\section{Interpretation}
\label{ff:sec-interp}

The combined cross-sections and asymmetries and results on heavy flavour 
production are interpreted in a variety of models.
The cross-section and asymmetry results are used to place limits 
on the mass of a possible additional heavy neutral boson, $\Zprime$, in
several models. Limits on contact interactions between leptons and on 
contact interaction between electrons and $b$ and $c$ quarks are obtained. 
These results are of particular 
interest since they are inaccessible to ${\mathrm{p\bar{p}}}$ or ep colliders. 
The results update those provided in~\cite{bib-EWEP-00,ff:ref:lepff-osaka}.

\subsection{Models with $\mathbf{\Zprime}$ Bosons}

The combined hadronic and leptonic cross-sections and the leptonic 
forward-backward asymmetries are used to fit the data to models including 
an additional, heavy, neutral boson, $\Zprime$.
The results are updated with respect to those given 
in~\cite{bib-EWEP-00,ff:ref:lepff-osaka} due to the updated cross-section and
leptonic forward-backward asymmetry results.

Fits are made to the mass of a $\Zprime$, $\MZp$, for 4 different models 
referred to as $\chi$, $\psi$, $\eta$ and L-R~\cite{ff:ref:zprime-thry} and 
for the Sequential Standard Model \cite{ff:ref:sqsm}, which proposes the 
existence of a $\Zprime$ with exactly the same coupling to fermions as 
the standard Z. The $\LEPII$ data alone does not significantly constrain
the mixing angle between the Z and $\Zprime$ fields, $\thtzzp$.
However, results from a single experiment in which $\LEPI$ data is used in the 
fit show that the mixing is consistent with zero (see for 
example~\cite{ff:ref:lep1zprime}, giving limits of 30~mrad or less
depending on model).  So for these fits $\thtzzp$ is fixed to 
zero.

No significant evidence is found for the existence of a $\Zprime$ boson
in any of the models. $95\%$ confidence level lower limits on $\MZp$ are 
obtained, by integrating the likelihood function\footnote{To be able to obtain 
confidence limits from the likelihood function it is necessary to convert 
the likelihood to a probability density function; this is done by 
multiplying by a prior probability function. Simply integrating the 
likelihood is equivalent to multiplying by a uniform prior probability 
function.}. The lower limits on the ${\Zprime}$ mass are shown in 
Table~\ref{ff:tab:zprime_mass_lim}.

\begin{table}[htbp]
\begin{center}
\renewcommand{\arraystretch}{1.5}
\begin{tabular}{|r|c|c|c|c|c|}
\hline
 Model                    & $\chi$  & $\psi$ & $\eta$ & L-R  & SSM   \\
\hline \hline
 $\MZplim$ ($\GeV$) & 678     & 463    & 436    &  800 & 1890  \\
\hline
\end{tabular}
\end{center}
\caption{The $95\%$ confidence level lower limits on the $\Zprime$ mass and
         $\chi$, $\psi$, $\eta$, L-R and SSM models.}
\label{ff:tab:zprime_mass_lim}
\end{table}

\subsection{Contact Interactions between Leptons}
\label{ff:sec-cntc}

The averages of cross-sections and forward-backward asymmetries for 
muon-pair and tau-lepton pair final states are used to search for 
contact interactions between leptons. 
The results are updated with respect to those given 
in~\cite{bib-EWEP-00,ff:ref:lepff-osaka} due to the updated cross-section and
leptonic forward-backward asymmetry results.

Following~\cite{ff:ref:ELPthr}, contact interactions are parameterised 
by an effective Lagrangian, $\cal{L}_{\mathrm{eff}}$, which is added to the 
Standard Model Lagrangian and has the form:
\begin{displaymath}
 \mbox{$\cal{L}$}_{\mathrm{eff}} = 
                        \frac{g^{2}}{(1+\delta)\Lambda^{2}} 
                          \sum_{i,j=L,R} \eta_{ij} 
                           \overline{e}_{i} \gamma_{\mu} e_{i}
                            \overline{f}_{j} \gamma^{\mu} f_{j},
\end{displaymath}
where $g^{2}/{4\pi}$ is taken to be 1 by convention, $\delta=1 (0)$ for 
$f=e ~(f \neq e)$, $\eta_{ij}=\pm 1$ or $0$,
$\Lambda$ is the scale of the contact interactions,
$e_{i}$ and $f_{j}$ are left or right-handed spinors. 
By assuming different helicity coupling between the initial 
state and final state currents, a set of different models can be defined
from this Lagrangian~\cite{ff:ref:Kroha}, with either
constructive ($+$) or destructive ($-$) interference between the 
Standard Model process and the contact interactions. The models and 
corresponding choices of $\eta_{ij}$ are given in Table~\ref{ff:tab:cntcdef}.
The models LL, RR, VV, AA, LR, RL, V0, A0 are considered here since 
these models lead to large deviations in the $\eemumu$ and $\eetautau$ 
channels. The total hadronic cross section on its own does not allow
stringent limits to be placed on contact interactions involving
quarks.

For the purpose of fitting contact interaction models to the data, 
a new parameter $\epsilon=1/\Lambda^{2}$ is defined; 
$\epsilon=0$ in the limit that there are no contact interactions. 
This parameter is allowed to take both positive and negative values in 
the fits. 
Theoretical uncertainties on the Standard Model predictions are taken 
from~\cite{ff:ref:lepffwrkshp}, see above.

\begin{figure}[htbp]
 \begin{center}
  \epsfig{file=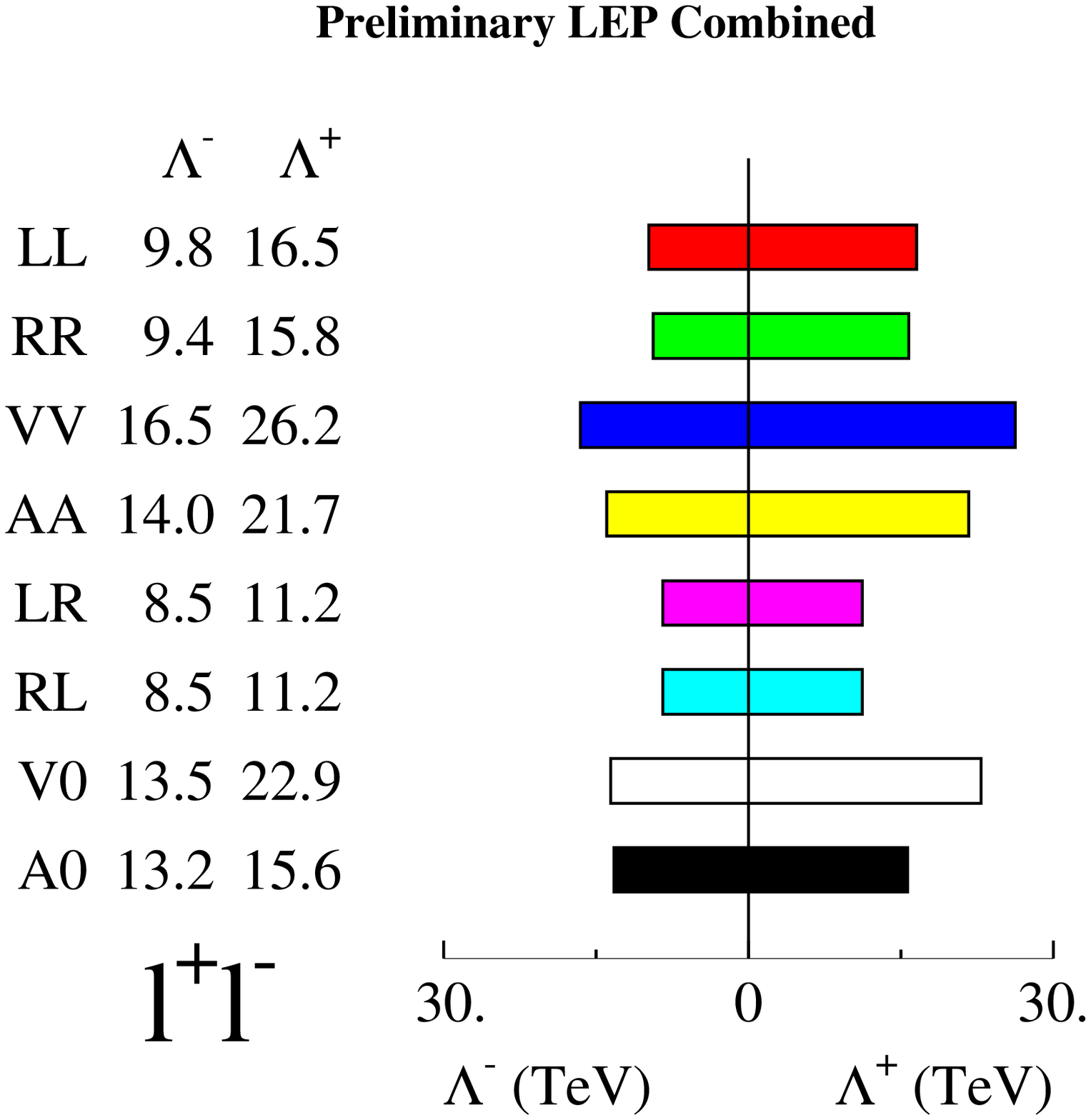,width=0.75\textwidth}
 \end{center}
 \caption{The $95\%$ CL exclusion limits on $\Lambda$ for $\eell$ assuming
          universality in the contact interactions between 
          $\eemumu$ and $\eetautau$.}
 \label{ff:fig:cntc}
\end{figure}

The values of $\epsilon$ extracted for each model are all compatible 
with the Standard Model expectation $\epsilon=0$, at the two standard 
deviation level. These errors on $\epsilon$ are typically a factor of two 
smaller than those obtained from a single LEP experiment with the same data
set. The fitted values of $\epsilon$ are converted into  
$95\%$ confidence level lower limits on $\Lambda$. 
The limits are obtained
by integrating the likelihood function over the physically allowed values, 
$\epsilon \ge 0$ for each $\Lambda^{+}$ limit and $\epsilon \le 0$ for 
$\Lambda^{-}$ limits. The fitted values of $\epsilon$ and the extracted 
limits are shown in Table \ref{ff:tab:cntc-leptons}. 
Figure \ref{ff:fig:cntc} shows the limits obtained on the scale $\Lambda$ for
the different models assuming universality between contact interactions
for $\eemumu$ and $\eetautau$.

\begin{table}[htbp]
 \begin{center}
  \begin{tabular}{|c|c|c|c|c|}
   \hline
   Model      & $\eta_{LL}$ & $\eta_{RR}$ & $\eta_{LR}$ & $\eta_{RL}$ \\
   \hline\hline
   LL$^{\pm}$ &   $\pm 1$   &      0      &      0      &      0      \\
   \hline
   RR$^{\pm}$ &      0      &   $\pm 1$   &      0      &      0      \\
   \hline
   VV$^{\pm}$ &   $\pm 1$   &   $\pm 1$   &   $\pm 1$   &   $\pm 1$   \\
   \hline
   AA$^{\pm}$ &   $\pm 1$   &   $\pm 1$   &   $\mp 1$   &   $\mp 1$   \\
   \hline
   LR$^{\pm}$ &      0      &      0      &   $\pm 1$   &      0      \\
   \hline
   RL$^{\pm}$ &      0      &      0      &      0      &   $\pm 1$   \\
   \hline
   V0$^{\pm}$ &   $\pm 1$   &   $\pm 1$   &      0      &      0      \\
   \hline
   A0$^{\pm}$ &      0      &      0      &  $\pm 1$    &   $\pm 1$   \\
   \hline
  \end{tabular}
 \end{center}
 \caption{Choices of $\eta_{ij}$ for different contact interaction models}
 \label{ff:tab:cntcdef}.
\end{table}

\begin{table}[htbp]
 \begin{center}
 \renewcommand{\arraystretch}{1.1}

  \begin{tabular}{|c|r|c|c|}
   \hline
   \multicolumn{4}{|c|}{\boldmath $\mathrm{e^{+}e^{-}} \rightarrow \mu^{+}\mu^{-}$\unboldmath} \\
   \hline
   Model  & $\epsilon$ ($\TeV^{-2}$)    &  $\Lambda^{-} (\TeV)$ & $\Lambda^{+} (\TeV)$ \\
   \hline
   \hline
   LL & -0.0056$^{+ 0.0042}_{- 0.0037}$ &    8.8 &   14.4 \\
   \hline
   RR & -0.0060$^{+ 0.0051}_{- 0.0046}$ &    8.4 &   13.8 \\
   \hline
   VV & -0.0014$^{+ 0.0016}_{- 0.0012}$ &   15.5 &   22.2 \\
   \hline
   AA & -0.0025$^{+ 0.0018}_{- 0.0023}$ &   12.1 &   20.1 \\
   \hline
   LR &  0.0014$^{+ 0.0043}_{- 0.0074}$ &    7.4 &    9.3 \\
   \hline
   RL &  0.0014$^{+ 0.0043}_{- 0.0074}$ &    7.4 &    9.3 \\
   \hline
   V0 & -0.0036$^{+ 0.0032}_{- 0.0013}$ &   12.2 &   19.9 \\
   \hline
   A0 &  0.0008$^{+ 0.0020}_{- 0.0031}$ &   12.7 &   13.0 \\
   \hline
  \end{tabular}

 \vskip 0.25cm

  \begin{tabular}{|c|r|c|c|}
   \hline
   \multicolumn{4}{|c|}{\boldmath $\mathrm{e^{+}e^{-}} \rightarrow \tau^{+}\tau^{-}$\unboldmath} \\
   \hline
   Model  & $\epsilon$ ($\TeV^{-2}$)    &  $\Lambda^{-} (\TeV)$ & $\Lambda^{+} (\TeV)$ \\
   \hline
   \hline
   LL & -0.0033$^{+ 0.0056}_{- 0.0050}$ &    8.9 &   11.4 \\
   \hline
   RR & -0.0036$^{+ 0.0061}_{- 0.0056}$ &    8.4 &   10.9 \\
   \hline
   VV & -0.0012$^{+ 0.0017}_{- 0.0020}$ &   14.0 &   19.1 \\
   \hline
   AA & -0.0004$^{+ 0.0025}_{- 0.0027}$ &   13.1 &   14.2 \\
   \hline
   LR & -0.0053$^{+ 0.0079}_{- 0.2210}$ &    2.1 &    9.2 \\
   \hline
   RL & -0.0053$^{+ 0.0079}_{- 0.2210}$ &    2.1 &    9.2 \\
   \hline
   V0 & -0.0011$^{+ 0.0023}_{- 0.0033}$ &   12.3 &   15.7 \\
   \hline
   A0 & -0.0028$^{+ 0.0041}_{- 0.0043}$ &    9.3 &   12.9 \\
   \hline
  \end{tabular}

 \vskip 0.25cm

  \begin{tabular}{|c|r|c|c|}
   \hline
   \multicolumn{4}{|c|}{\boldmath $\mathrm{e^{+}e^{-}} \rightarrow \ell^{+}\ell^{-}$\unboldmath} \\
   \hline
   Model  & $\epsilon$ ($\TeV^{-2}$)    &  $\Lambda^{-} (\TeV)$ & $\Lambda^{+} (\TeV)$ \\
   \hline
   \hline
   LL & -0.0042$^{+ 0.0027}_{- 0.0028}$ &    9.8 &   16.5 \\
   \hline
   RR & -0.0046$^{+ 0.0037}_{- 0.0034}$ &    9.4 &   15.8 \\
   \hline
   VV & -0.0014$^{+ 0.0012}_{- 0.0012}$ &   16.5 &   26.2 \\
   \hline
   AA & -0.0018$^{+ 0.0016}_{- 0.0019}$ &   14.0 &   21.7 \\
   \hline
   LR & -0.0023$^{+ 0.0051}_{- 0.0045}$ &    8.5 &   11.2 \\
   \hline
   RL & -0.0023$^{+ 0.0051}_{- 0.0045}$ &    8.5 &   11.2 \\
   \hline
   V0 & -0.0020$^{+ 0.0016}_{- 0.0019}$ &   13.5 &   22.9 \\
   \hline
   A0 & -0.0011$^{+ 0.0025}_{- 0.0023}$ &   13.2 &   15.6 \\
   \hline
 \end{tabular}
 \end{center}
 \caption{Fitted values of $\epsilon$
          and $95\%$ confidence limits on the scale,
          $\Lambda$, for constructive ($+$) and destructive interference
          ($-$) with the Standard Model, for the contact interaction models
          discussed in the text. Results are given for $\eemumu$, $\eetautau$
          and $\eell$, assuming universality in the contact interactions 
          between $\eemumu$ and $\eetautau$.}
 \label{ff:tab:cntc-leptons}
\end{table}

\subsection{Contact Interactions from Heavy Flavour Averages}

Limits on contact interactions between electrons and $b$ and $c$ quarks
are obtained. 
The formalism for describing contact interactions including heavy flavours 
is identical to that described above for leptons.

All heavy flavour $\LEPII$ combined results from 133 to 207 $\GeV$ given in 
Table~\ref{ff:tab:hfresults} are used as inputs. 
For the purpose of fitting contact interaction models to the data, 
$\Rb$ and $\Rc$ are converted to cross sections 
$\sigma_{\bb}$ and $\sigma_{\cc}$ using the averaged ${\qq}$ cross section of 
section \ref{ff:sec-ave-xsc-afb} corresponding to signal Definition 2. 
In the calculation of errors, the correlations between $\Rb$, $\Rc$ and 
$\sigma_{\qq}$ are assumed to be negligible.

\begin{figure}[htbp]
 \begin{center}
  \begin{tabular}{c}
    \epsfig{file=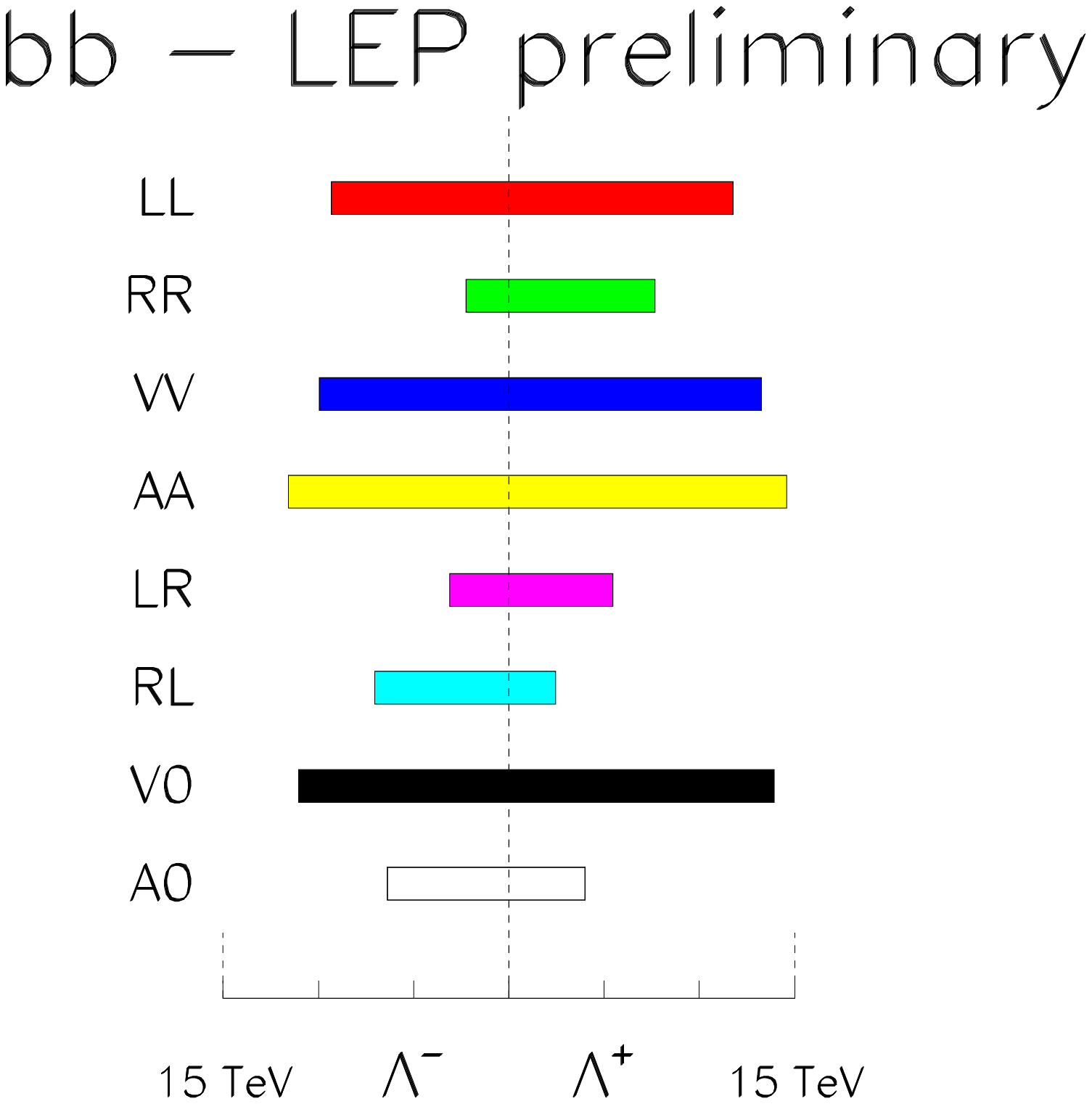,width=0.60\textwidth} \\
    \\
    \epsfig{file=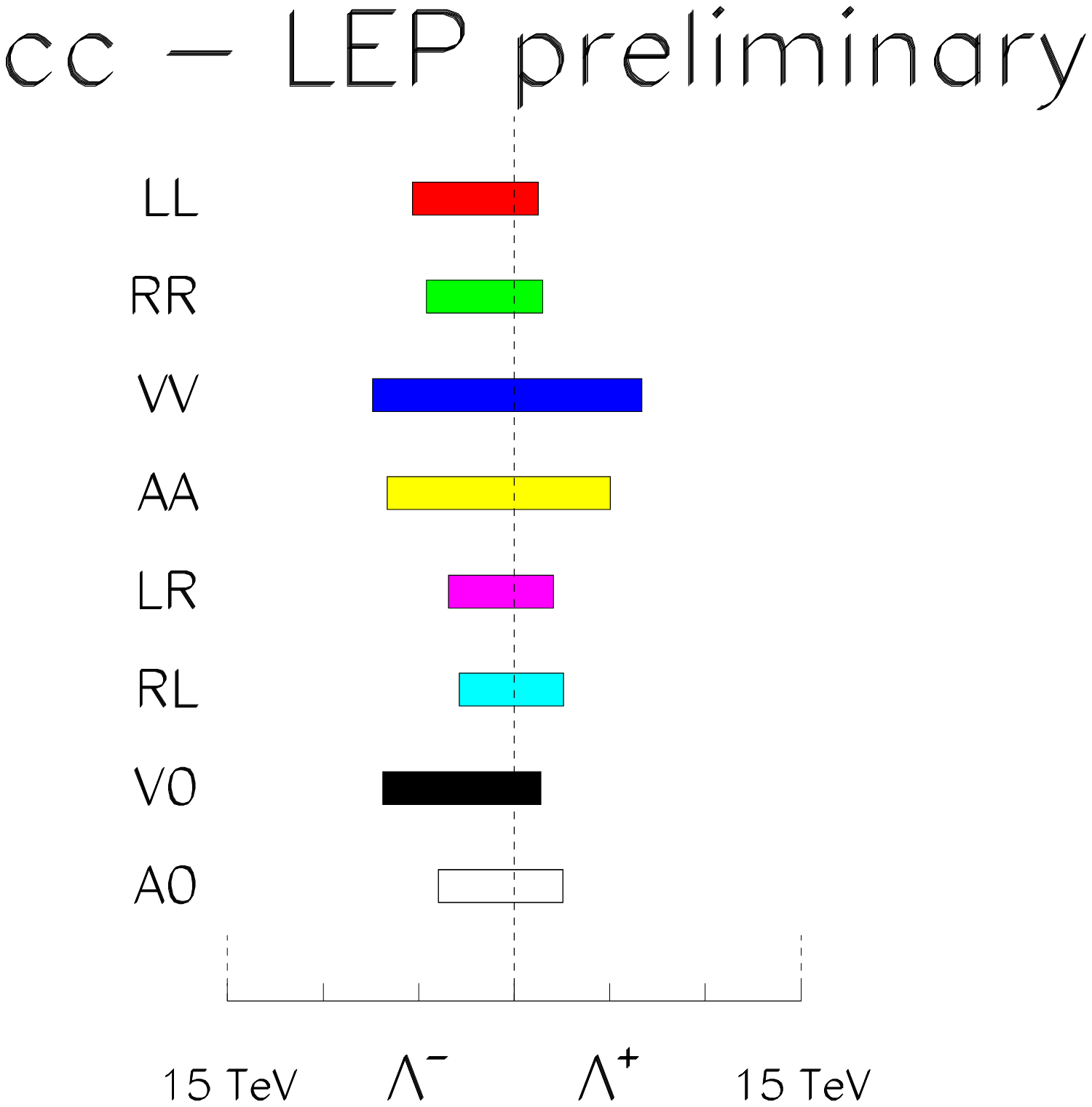,width=0.60\textwidth} \\
  \end{tabular}
 \end{center}
\caption{The $95\%$ CL exclusion limits on the scale of Contact Interactions in $\eebb$
         and $\eecc$ using Heavy Flavour LEP combined results 
         from 133 to 207 $\GeV$.}
 \label{ff:fig:cibc}
\end{figure}

The results are updated with respect to those given 
in~\cite{bib-EWEP-00,ff:ref:lepff-osaka} due to the updated hadronic cross-sections
and heavy flavour results.
No evidence for contact interactions between electrons and $b$ and $c$ 
is found. The fitted values of $\epsilon$ and their 68$\%$ confidence level 
uncertainties together with the 95$\%$ confidence level lower limit 
on ${\mathrm{\Lambda}}$ are shown in Table \ref{ff:tab:cibc}.
Figure \ref{ff:fig:cibc} shows the limits obtained on the scale,
${\mathrm \Lambda}$, of models with different helicity combinations 
involved in the interactions.

\begin{table}[htbp]
 \begin{center}
 \renewcommand{\arraystretch}{1.1}
  \begin{tabular}{|c|r|c|c|}
   \hline
   \multicolumn{4}{|c|}{\boldmath $\mathrm{e^{+}e^{-}} \rightarrow b\overline{b}$\unboldmath} \\
   \hline
   Model  & $\epsilon$ ($\TeV^{-2}$)    &  $\Lambda^{-} (\TeV)$ & $\Lambda^{+} (\TeV)$ \\
   \hline
   \hline
   LL & -0.0030$^{+ 0.0045}_{- 0.0047}$ &    9.3 &   11.8 \\
   \hline
   RR & -0.1755$^{+ 0.1634}_{- 0.0159}$ &    2.2 &    7.7 \\
   \hline
   VV & -0.0029$^{+ 0.0038}_{- 0.0040}$ &   10.0 &   13.3 \\
   \hline
   AA & -0.0018$^{+ 0.0029}_{- 0.0031}$ &   11.6 &   14.6 \\
   \hline
   LR & -0.0491$^{+ 0.0555}_{- 0.0384}$ &    3.1 &    5.5 \\
   \hline
   RL &  0.0065$^{+ 0.1409}_{- 0.0149}$ &    7.0 &    2.5 \\
   \hline
   V0 & -0.0021$^{+ 0.0032}_{- 0.0034}$ &   11.0 &   13.9 \\
   \hline
   A0 &  0.0305$^{+ 0.0203}_{- 0.0348}$ &    6.4 &    4.0 \\
   \hline
  \end{tabular}
\vskip 0.5cm
  \begin{tabular}{|c|r|c|c|}
   \hline
   \multicolumn{4}{|c|}{\boldmath $\mathrm{e^{+}e^{-}} \rightarrow c\overline{c}$\unboldmath} \\
   \hline
   Model  & $\epsilon$ ($\TeV^{-2}$)    &  $\Lambda^{-} (\TeV)$ & $\Lambda^{+} (\TeV)$ \\
   \hline
   \hline
   LL &  0.0146$^{+ 0.5911}_{- 0.0259}$ &    5.3 &    1.3 \\
   \hline
   RR &  0.0492$^{+ 0.3723}_{- 0.0568}$ &    4.6 &    1.5 \\
   \hline
   VV &  0.0008$^{+ 0.0106}_{- 0.0100}$ &    7.4 &    6.7 \\
   \hline
   AA &  0.0081$^{+ 0.0171}_{- 0.0154}$ &    6.6 &    5.0 \\
   \hline
   LR &  0.0913$^{+ 0.1076}_{- 0.1251}$ &    3.5 &    2.1 \\
   \hline
   RL &  0.0145$^{+ 0.0872}_{- 0.0872}$ &    2.9 &    2.6 \\
   \hline
   V0 &  0.0047$^{+ 0.0170}_{- 0.0133}$ &    6.9 &    1.4 \\
   \hline
   A0 &  0.0524$^{+ 0.0736}_{- 0.0780}$ &    4.0 &    2.6 \\
   \hline
  \end{tabular}
 \end{center}
 \caption{Fitted values of $\epsilon$
          and $95\%$ confidence limits on the scale,
          $\Lambda$, for constructive ($+$) and destructive interference
          ($-$) with the Standard Model, for the contact interaction models
          discussed in the text. From
          combined $b \bar b$ and $c \bar c$ results with centre of mass 
          energies from 133 to 207 $\GeV$.}
 \label{ff:tab:cibc}
\end{table}

\section{Summary}
\label{ff:sec-conc}

A preliminary combination of the $\LEPII$ $\eeff$ cross-sections (for hadron, 
muon and tau final states) and
forward-backward asymmetries (for muon and tau final states) 
from LEP running at energies from 130 to 207~$\GeV$ is made. 
The results from the four LEP experiments are in good 
agreement with each other. 

The averages for all energies are shown in Table~\ref{ff:tab-xsafbres}.
Overall the data agree 
with the Standard Model predictions of ZFITTER.
Preliminary differential cross-sections, $\dsdc$, for $\eemumu$ and
$\eetautau$ are combined. Results are shown in 
Figures~\ref{ff:fig:dsdc-res-mm} and~\ref{ff:fig:dsdc-res-tt}.
A preliminary average of results on heavy flavour production at $\LEPII$ 
is also made for measurements of $\Rb$, $\Rc$, $\Abb$
and $\Acc$, using results from LEP centre-of-mass energies
from 130 to 207 $\GeV$. Results are given in Table~\ref{ff:tab:hfresults} and 
shown graphically in Figures~\ref{ff:fig:hfres1} and~\ref{ff:fig:hfres2}.
The results are in good agreement with the predictions of the SM. 

The preliminary averaged cross-section and forward-backward asymmetry
results together with the combined results on heavy flavour production
are interpreted in a variety of models. The $\LEPII$ averaged
cross-sections and lepton asymmetries are used to obtain lower limits
on the mass of a possible $\Zprime$ boson in different models. Limits
range from $436$ to $1890$ $\GeV$ depending on the model. Limits on
the scale of contact interactions between leptons and also between
electrons and $\bb$ and $\cc$ final states are determined.  A
full set of limits are given in Tables~\ref{ff:tab:cntc-leptons}
and~\ref{ff:tab:cibc}.


%

\boldmath
\chapter{W and Four-Fermion Production at \LEPII}
\label{sec-4F}
\unboldmath

\updates{
New preliminary results are presented 
for W-pair, Z-pair and single W production,
based on the full data sample collected in the year 2000 
between 202 and 209 \GeV.
Improved procedures are used for the combination 
of W-pair cross sections and W decay branching fractions.
New averages of the Z-pair and single W cross sections are performed, 
including also preliminary updates below 205 \GeV.
}

\section{Introduction}
\label{4f_sec:introduction}

This Chapter summarises the combination of published and preliminary
results of the four LEP experiments on W-pair, Z-pair and single
W cross sections and on W decay branching fractions, 
prepared for the summer 2001 conferences~\cite{bib-EWEP-00,4f_bib:4f_s01}.
Where available, the published final results 
of the analysis of data collected 
at \CoM\ energies up to 209~\GeV\ 
are used in the combination. 

Most relevant,
with respect to the results presented 
at the summer 2000 conferences~\cite{bib-EWEP-00,4f_bib:4f_s00},
are new measurements of the W-pair, 
Z-pair and single-W cross sections 
at the highest \LEPII\ \CoM\ energies between 202 and 209~\GeV,
using the full data samples collected in the year 2000.
This represents, for energies above 202~\GeV,
an increase in luminosity by more than a factor of two
over the results presented at the summer 2000 conferences
from the year 2000 data available at that time.
Another significant change 
is an improved procedure for the combination
of measured W-pair cross sections 
between 183 and 207~\GeV\
and for the combination of measured W decay branching fractions,
also used to derive the average ratio between 
the measured W-pair cross sections
and the corresponding theoretical predictions from various models.
Finally, new combinations of single-W and Z-pair cross sections are presented 
to take into account the new data available above 200~\GeV,
also including minor changes in the single-W combination procedure
and preliminary updates of Z-pair cross sections between 192 and 202~\GeV.

In the year 2000,
LEP ran at \CoM\  energies larger than 200~\GeV,
up to a maximum of 209~\GeV. 
For the measurements of the W-pair, Z-pair and single-W cross section,
the data collected above 202~\GeV\ is divided~\cite{4f_bib:4f_m01}
in two ranges of \roots, below and above 205.5~\GeV, to enhance the 
sensitivity of the cross-section measurements 
to possible signals of new physics at the highest \epem\ \CoM\ energy. 
The two data sets have mean \CoM\ energies of 204.9 and 206.6~\GeV,
and the respective integrated luminosities used for the analyses
considered in this note are approximately 80 and 130 \pbinv\ per experiment. 

Results from different experiments are combined
in $\chi^2$ minimisations through matrix algebra, 
based on the Best Linear Unbiased Estimate (BLUE) method 
described in Reference~\citen{common_bib:lyons},
and taking into account, when relevant, 
the correlations between the systematic uncertainties, 
which arise mainly from the use of the same \MC\ codes 
to predict the background
and to simulate the hadronisation processes.
The detailed breakdown of the systematic errors for the measurements combined
in this Chapter is described in Appendix~\ref{4f_sec:appendix}.
Experimental results are compared with recent theoretical predictions,
many of which were developed in the framework 
of the \LEPII\ \MC\ workshop~\cite{4f_bib:fourfrep}. 

\section{W-pair production cross section}
\label{4f_sec:WWxsec}

All experiments have published final results 
on the W-pair ({\sc CC03}~\cite{4f_bib:fourfrep}) production cross section for \CoM\ energies 
from~161 to~189~\GeV~\cite{common_bib:adloww161,common_bib:aleww172,
common_bib:delww172,common_bib:ltrww172,common_bib:opaww172,
4f_bib:aleww183,4f_bib:delww183,4f_bib:ltrww183,4f_bib:opaww183,
4f_bib:aleww189,4f_bib:delww189,4f_bib:ltrww189,4f_bib:opaww189}.
The preliminary results contributed by all four collaborations 
at $\roots=192$--202~\GeV\ are unchanged with respect to the summer 2000 
conferences~\cite{4f_bib:aleww1999,4f_bib:delww1999,4f_bib:opaww1999,
4f_bib:opaww2000a,4f_bib:ltrww1999}.
All experiments contribute new preliminary results 
at $\roots=205$--207~\GeV~\cite{4f_bib:aleww2000,4f_bib:delww2000,
4f_bib:ltrww2000,4f_bib:opaww2000b},
based on the analysis of the full data sample collected in the year 2000.
New LEP averages of the measurements 
at the eight \CoM\ energies between 183 and 207~\GeV\ 
are computed for the summer 2001 conferences,
using an improved combination procedure.
In particular, 
the LEP combined cross sections
are now obtained from one global fit to the 32 measurements
performed by the four experiments at each of these eight energies,
taking into account inter-experiment as well as inter-energy correlations,
rather than from eight individual fits at the various energies,
neglecting inter-energy correlations,
as in the case of the previous combination
for the summer 2000~\cite{bib-EWEP-00,4f_bib:4f_s00} conferences.

In the averaging of results at and above $\sqrt{s}=189$~\GeV,
the component of the systematic error from each experiment 
coming from the uncertainty on the 4-jet QCD background
is taken to be fully correlated between experiments.
This is slightly different from the procedure adopted 
for the summer 2000 conferences~\cite{bib-EWEP-00},
where some experiments had also included in the correlated error
the uncertainties due to the modelling of hadronisation 
and final state interactions. 
More importantly, this common error, ranging between 0.04 and 0.12~pb,
is now taken to be also fully correlated between energies.
The remaining sources of systematic errors,
taken as completely uncorrelated between experiments,
are split by each experiment into two categories,
for which 100\% and 0\% correlations 
between different energies are assumed.
The detailed inputs used for the combination 
are given in Appendix~\ref{4f_sec:appendix}.
The measured statistical errors are used for the combination.
After building the full 32$\times$32 covariance matrix for the measurements,
the $\chi^2$ minimisation fit is performed 
as described in Reference~\cite{4f_bib:4f_pdg01}.
More detailed studies on correlated systematic errors are in progress.

The results from each experiment 
for the W-pair production cross section
are shown in Table~\ref{4f_tab:wwxsec}, 
together with the LEP combination at each energy. 
All measurements are defined to represent 
{\sc CC03}~\cite{4f_bib:fourfrep} WW cross sections,
and assume Standard Model values for the W decay branching fractions.
The results for \CoM\ energies between 183 and 207~\GeV,
for which new LEP averages are computed,
supersede the ones presented in~\cite{bib-EWEP-00}:
the effect of the new combination procedure
is to change the LEP combined cross sections 
at these energies by 0.6\% at most, 
generally towards lower values.
The combined LEP cross sections at the eight energies 
are all positively correlated,
with correlations ranging from 9\% to 24\%.
For completeness, 
the measurements at 161~\cite{common_bib:adloww161,4f_bib:lepewwg97}
and 172~\GeV~\cite{common_bib:aleww172,common_bib:delww172,
common_bib:ltrww172,common_bib:opaww172,4f_bib:lepewwg98}
are also listed in the table.
All results from the four experiments
listed in the table are preliminary,
with the exception of those at 161--189~\GeV. 

\renewcommand{\arraystretch}{1.2}
\begin{table}[hbtp]
\begin{center}
\hspace*{-0.5cm}
\begin{tabular}{|c|c|c|c|c|c|r|} 
\hline
\roots & \multicolumn{5}{|c|}{WW cross section (pb)} 
       & \multicolumn{1}{|c|}{$\chi^2/\textrm{d.o.f.}$} \\
\cline{2-6} 
(\GeV)      & \Aleph\                & \Delphi\               &
             \Ltre\                 & \Opal\                 &
             LEP                    &                        \\
\hline
161.3      & $\phz4.23\pm0.75^*$    & 
             $\phz3.67^{\phz+\phz0.99\phz*}_{\phz-\phz0.87}$& 
             $\phz2.89^{\phz+\phz0.82\phz*}_{\phz-\phz0.71}$& 
             $\phz3.62^{\phz+\phz0.94\phz*}_{\phz-\phz0.84}$& 
             $\phz3.69\pm0.45\phs$  & 
             $\left\} \hspace*{2mm} \phz1.3\phz/\phz3 \right.$ \\
172.1      & $11.7\phz\pm1.3\phz^*$ & $11.6\phz\pm1.4\phz^*$ &
             $12.3\phz\pm1.4\phz^*$ & $12.3\phz\pm1.3\phz^*$ &
             $12.0\phz\pm0.7\phz\phs$ & 
             $\left\} \hspace*{2mm} \phz0.22/\phz3 \right.$ \\
182.7      & $15.57\pm0.68^*$       & $15.86\pm0.74^*$       &
             $16.53\pm0.72^*$       & $15.43\pm0.66^*$       &
             $15.79\pm0.36\phs$     & 
             \multirow{8}{20.3mm}{$
               \hspace*{-0.3mm}
               \left\}
                 \begin{array}[h]{rr}
                   &\multirow{8}{8mm}{\hspace*{-4.2mm}27.42/24}\\
                   &\\ &\\ &\\ &\\ &\\ &\\ &\\  
                 \end{array}
               \right.
               $}\\
188.6      & $15.71\pm0.38^*$       & $15.83\pm0.43^*$       &
             $16.24\pm0.43^*$       & $16.30\pm0.38^*$       &
             $16.00\pm0.21\phs$       & \\
191.6      & $17.23\pm0.91\phs$     & $16.90\pm1.02\phs$     &
             $16.39\pm0.93\phs$     & $16.60\pm0.98\phs$     &
             $16.72\pm0.48\phs$     & \\
195.5      & $17.00\pm0.57\phs$     & $17.86\pm0.63\phs$     &
             $16.67\pm0.60\phs$     & $18.59\pm0.74\phs$     &
             $17.43\pm0.32\phs$     & \\
199.5      & $16.98\pm0.56\phs$     & $17.35\pm0.60\phs$     &
             $16.94\pm0.62\phs$     & $16.32\pm0.66\phs$     &
             $16.84\pm0.31\phs$     & \\
201.6      & $16.16\pm0.76\phs$     & $17.67\pm0.84\phs$     &
             $16.95\pm0.88\phs$     & $18.48\pm0.91\phs$     &
             $17.23\pm0.42\phs$     & \\
204.9      & $16.57\pm0.55\phs$     & $17.44\pm0.64\phs$     &
             $17.35\pm0.64\phs$     & $15.97\pm0.64\phs$     &
             $16.71\pm0.31\phs$     & \\
206.6      & $17.32\pm0.45\phs$     & $16.50\pm0.48\phs$     &
             $17.96\pm0.51\phs$     & $17.77\pm0.57\phs$     &
             $17.33\pm0.25\phs$     & \\
\hline
\end{tabular}
\caption{
W-pair production cross section from the four LEP
experiments and combined values at all recorded \CoM\ energies.
All results are preliminary and unpublished,
with the exception of those indicated by~$^*$. 
The measurements between 183 and 207~\GeV\
are combined in one global fit, taking into account 
inter-experiment as well as inter-energy correlations of systematic errors.
The results for the combined LEP W-pair production cross section 
at 161 and 172~\GeV\ are taken 
from~\protect\cite{4f_bib:lepewwg97,4f_bib:lepewwg98} respectively.}
\label{4f_tab:wwxsec}
\end{center}
\vspace*{-4mm}
\end{table}
\renewcommand{\arraystretch}{1.}

\begin{figure}[htbp]
\centering
\epsfig{figure=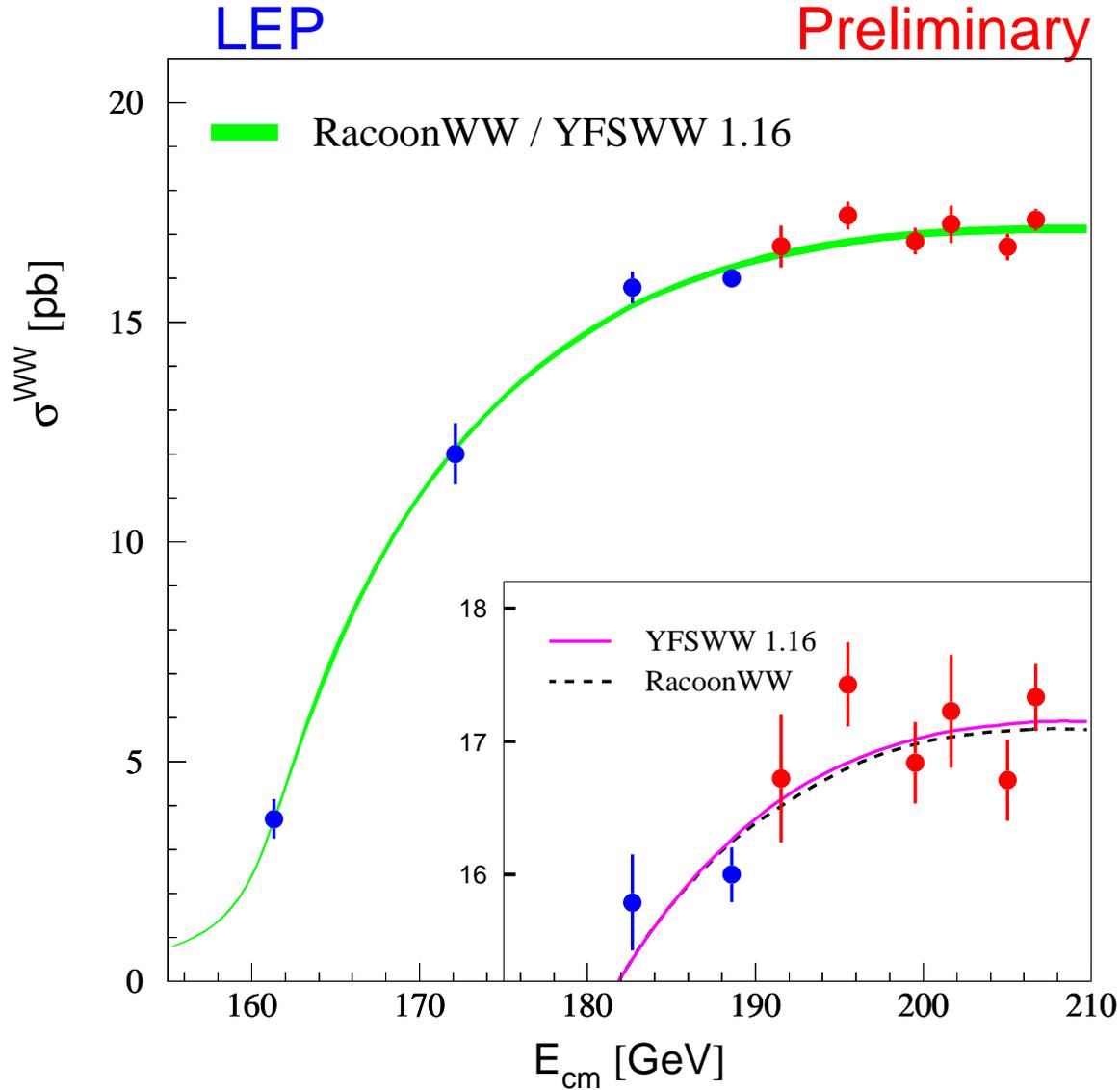,width=1.0\textwidth}
\caption{
  Measurements of the W-pair production cross section,
  compared to the predictions 
  of \RacoonWW~\protect\cite{common_bib:racoonww} 
  and \YFSWW~\protect\cite{common_bib:yfsww}. 
  The shaded area represents the uncertainty on the theoretical predictions,
  estimated to be $\pm$2\% for $\roots\!<\!170$~\GeV\
  and ranging from 0.7 to 0.4\% above 170~\GeV.
}
\label{4f_fig:sww_vs_sqrts}
\end{figure}

Figure~\ref{4f_fig:sww_vs_sqrts} shows the combined LEP W-pair cross section 
measured as a function of the \CoM\ energy.
The combined measurements are compared with the theoretical calculations 
from \YFSWW~\cite{common_bib:yfsww} and \RacoonWW~\cite{common_bib:racoonww} 
between 155 and 215~\GeV\ for $\Mw=80.35$~\GeV.
The two codes have been extensively compared and 
agree at a level better than 0.5\% 
at the \LEPII\ energies~\cite{4f_bib:fourfrep}.
The calculations above 170~\GeV, 
based for the two programs on the so-called leading pole~(LPA) 
or double pole approximations~(DPA)~\cite{4f_bib:dpa}, 
have theoretical uncertainties 
decreasing from 0.7\% at 170~\GeV\
to about 0.4\% at \CoM\ energies larger than 200~\GeV,
while in the threshold region
a larger theoretical uncertainty of 2\% is assigned~\cite{4f_bib:dpaerr}.
This theoretical uncertainty is represented by
the width of the shaded band in Figure~\ref{4f_fig:sww_vs_sqrts}. 
An error of 50 \MeV\ on the W mass would translate 
into additional errors of 0.1\% (3.0\%) 
on the cross-section predictions at 200~\GeV\ (161~\GeV, respectively).
All results, up to the highest \CoM\ energies, 
are in agreement with the two theoretical predictions considered.

\subsection{Ratio of measured and predicted W-pair cross sections}
\label{4f_sec:WWratio}
The agreement between the measured W-pair cross section,
$\sigma_{\mathrm{WW}}^{\mathrm{meas}}$,
and its expectation according to a given theoretical model,
$\sigma_{\mathrm{WW}}^{\mathrm{theo}}$,
can be expressed quantitatively in terms of their ratio
\begin{equation*}
\rww = \frac{\sww^\mathrm{meas}}{\sww^\mathrm{theo}} ,
\end{equation*}
averaged over the measurements performed by the four experiments 
at different energies in the \LEPII\ region.
The above procedure is used to compare the measurements
at the eight energies between 183 and 207~\GeV\ to the predictions of 
\Gentle~\cite{4f_bib:gentle}, \KoralW~\cite{4f_bib:koralw}, 
\YFSWW~\cite{common_bib:yfsww} and \RacoonWW~\cite{common_bib:racoonww}.
The measurements at 161 and 172~\GeV\
are not used in the combination 
because they were performed using data samples of low statistics
and because of the high sensitivity of the cross section 
to the value of the W mass at these energies.

The combination of the ratio $\rww$ is performed
using as input from the four experiments the 32 cross sections
measured at each of the eight energies.
For each model considered,
these are converted into 32 ratios by dividing them
by the corresponding theoretical predictions,
listed in Appendix~\ref{4f_sec:appendix}.
The full 32$\times$32 covariance matrix for the ratios
is built taking into account the same sources
of systematic errors used for the combination 
of the W-pair cross sections at these energies.
The small statistical errors 
on the theoretical predictions at the various energies,
taken as fully correlated for the four experiments
and uncorrelated between different energies,
are also translated into errors 
on the individual measurements of $\rww$.
The theoretical errors on the predictions,
due to the physical and technical precision of the generators used, 
are not propagated to the individual ratios
and are used instead when comparing to
the combined values obtained for $\rww$.
For each of the four models considered,
two fits are performed:
in the first, eight values of $\rww$ at the different energies are extracted,
averaged over the four experiments;
in the second, only one value of $\rww$ is determined,
representing the global agreement of measured and predicted cross sections
over the whole energy range.

The results of the two fits to $\rww$
for each of the four models considered are given
in Table~\ref{4f_tab:wwratio}.
As already qualitatively noted from Figure~\ref{4f_fig:sww_vs_sqrts},
the LEP measurements of the W-pair cross section above threshold
are in very good agreement to the predictions 
of \YFSWW\ and \RacoonWW.
In contrast, the predictions from \Gentle\ and \KoralW\ are
more than 2\% too high with respect to the measurements.
The main differences between these two sets of predictions
come from non-leading $\oa$ electroweak radiative corrections
to the W-pair production process,
which are included 
(in the LPA/DPA approximation~\cite{4f_bib:dpa})
in both \YFSWW\ and \RacoonWW,
but not in \Gentle\ and \KoralW.
Especially interesting is the comparison 
between \KoralW\ and \YFSWW,
as the numerical results provided by the authors for \KoralW\ are
actually those of a downgraded version of \YFSWW,
such that the only differences between the two calculations
are the screening of Coulomb interactions 
according to the prescription of Reference~\citen{4f_bib:screening}
and the inclusion of non-leading $\oa$ 
electroweak radiative corrections to W-pair production
(mainly radiation off W bosons and pure weak corrections).
Of these two effects,
only the latter is found to be relevant 
to the measurement of $\rww$,
while the former has a negligible impact 
on the total W-pair cross section~\cite{4f_bib:melnikov}.

\renewcommand{\arraystretch}{1.2}
\begin{table}[bhtp]
\vspace*{-0mm}
\begin{center}
\hspace*{-0.3cm}
\begin{tabular}{|c|c|c|c|c|} 
\hline
\roots & 
\multicolumn{4}{|c|}{Ratio of measured and expected WW cross sections} \\
\cline{2-5} 
(\GeV)     & $\rww^{\footnotesize\Gentle}$ 
           & $\rww^{\footnotesize\KoralW}$
           & $\rww^{\footnotesize\YFSWW}$  
           & $\rww^{\footnotesize\RacoonWW}$ \\
\hline
182.7      & $1.005\pm0.022$     & $1.011\pm0.023$     &
             $1.028\pm0.023$     & $1.028\pm0.023$     \\
188.6      & $0.961\pm0.013$     & $0.967\pm0.013$     &
             $0.984\pm0.013$     & $0.985\pm0.013$     \\
191.6      & $0.986\pm0.028$     & $0.991\pm0.028$     &
             $1.009\pm0.029$     & $1.012\pm0.029$     \\
195.5      & $1.010\pm0.018$     & $1.015\pm0.018$     &
             $1.035\pm0.019$     & $1.037\pm0.019$     \\
199.5      & $0.964\pm0.018$     & $0.970\pm0.018$     &
             $0.990\pm0.018$     & $0.992\pm0.018$     \\
201.6      & $0.983\pm0.024$     & $0.989\pm0.024$     &
             $1.009\pm0.025$     & $1.012\pm0.025$     \\
204.9      & $0.949\pm0.018$     & $0.955\pm0.018$     &
             $0.976\pm0.018$     & $0.978\pm0.018$     \\
206.6      & $0.984\pm0.014$     & $0.989\pm0.014$     &
             $1.011\pm0.015$     & $1.014\pm0.015$     \\
\hline
$\chi^2$/d.o.f 
           & 27.42/24            & 27.42/24            
           & 27.42/24            & 27.42/24            \\
\hline
\hline
Average    & $0.973\pm0.009$     & $0.979\pm0.009$     &
             $0.998\pm0.009$     & $1.000\pm0.009$     \\
\hline
$\chi^2$/d.o.f 
           & 39.16/31            & 39.20/31            
           & 39.04/31            & 39.14/31            \\
\hline
\end{tabular}
\caption{
Ratios of LEP combined W-pair cross-section measurements
to the expectations according to 
\Gentle~\protect\cite{4f_bib:gentle}, 
\KoralW~\protect\cite{4f_bib:koralw}, 
\YFSWW~\protect\cite{common_bib:yfsww} and 
\RacoonWW~\protect\cite{common_bib:racoonww}.
For each of the four models,
two fits are performed,
one to the LEP combined values 
of $\rww$ at the eight energies between 183 and 207~\GeV,
and another to the LEP combined average of $\rww$ over all energies.
The results of the fits are given in the table
together with the resulting $\chi^2$.
Both fits take into account inter-experiment 
as well as inter-energy correlations of systematic errors.
}
\label{4f_tab:wwratio}
\end{center}
\end{table}
\renewcommand{\arraystretch}{1.}

\begin{figure}[htbp]
\centering
\mbox{
  {\epsfig{figure=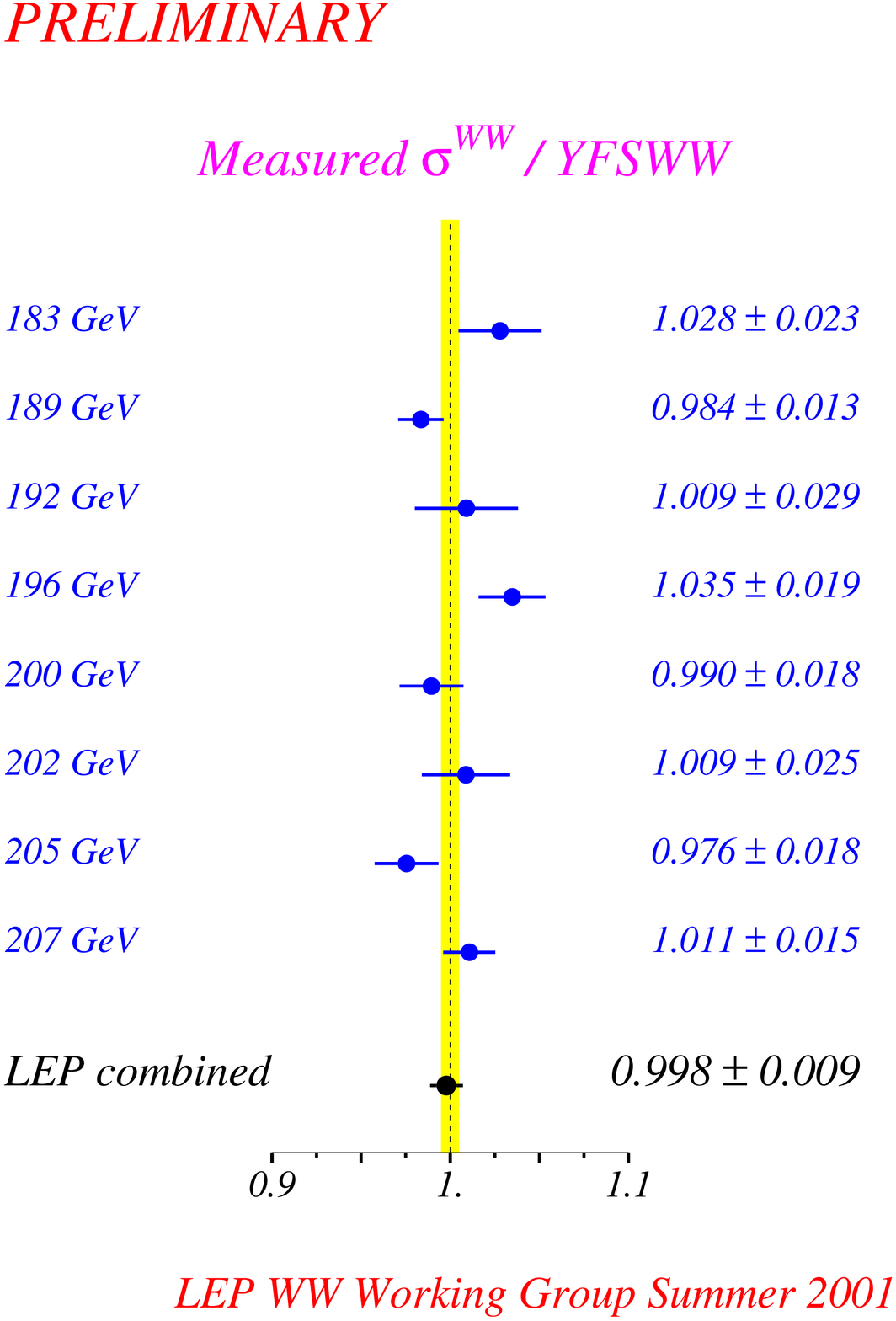,width=0.42\textwidth}}
  \hspace*{0.10\textwidth}
  {\epsfig{figure=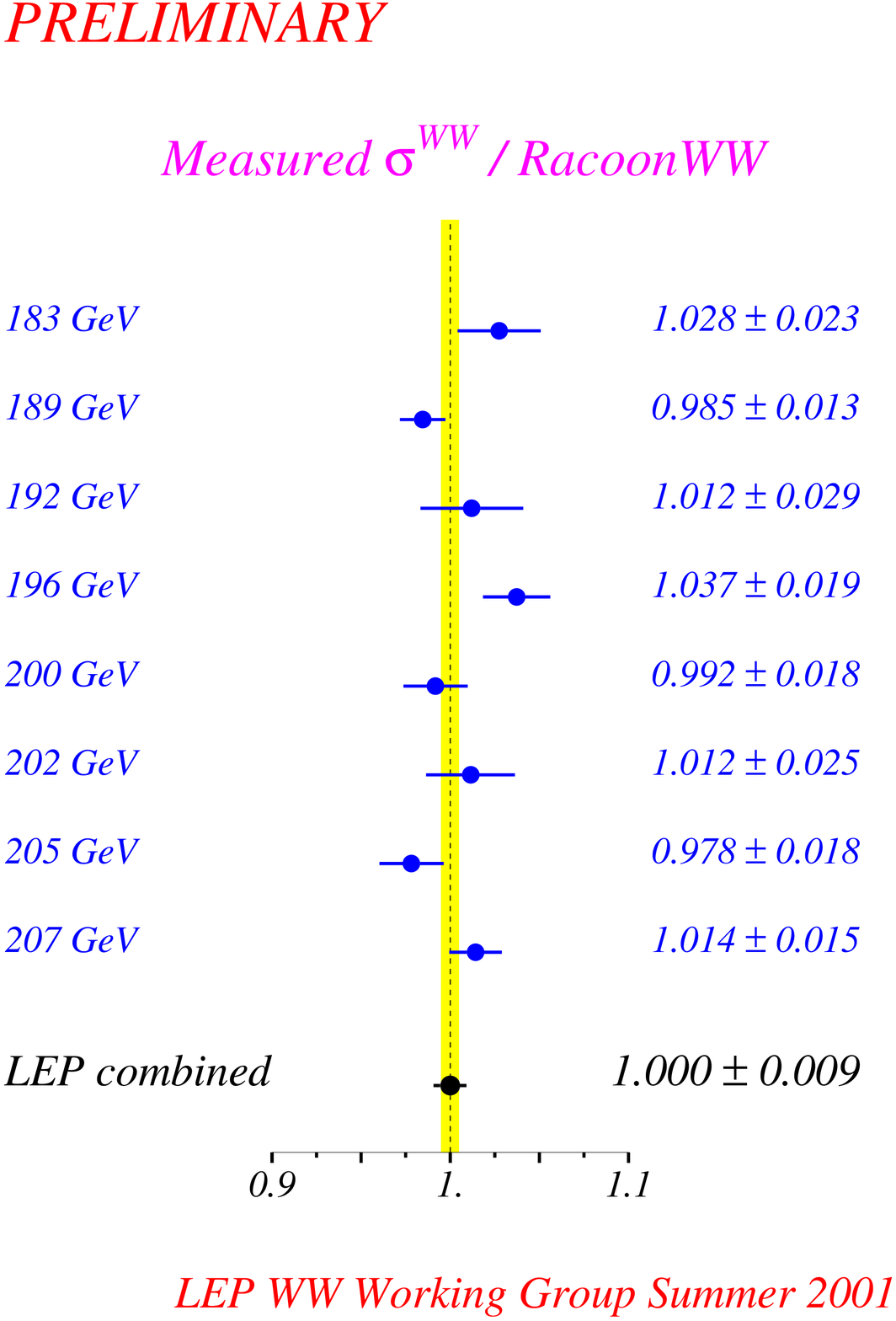,width=0.42\textwidth}}
  }
\vspace*{-0.5truecm}
\caption{
  Ratios of LEP combined W-pair cross-section measurements
  to the expectations according to 
  \YFSWW~\protect\cite{common_bib:yfsww} and 
  \RacoonWW~\protect\cite{common_bib:racoonww}
  The yellow bands represent constant relative errors 
  of 0.5\% on the two 
  cross-section predictions.
}
\label{4f_fig:rww}
\end{figure}

The results of the fits for \YFSWW\ and \RacoonWW are also shown 
in Figure~\ref{4f_fig:rww}, where relative errors of 0.5\% on the 
cross-section predictions are assumed.
For simplicity, 
the energy dependence of the relative error
on the W-pair cross-section predicted by each model
is neglected in Figure~\ref{4f_fig:rww}.

\section{W decay branching fractions}
\label{4f_sec:Wbranching}

\begin{figure}[tbhp]
\centering
\vspace*{-0.5truecm}
\hspace*{-0.2truecm}
\mbox{
  {\epsfig{figure=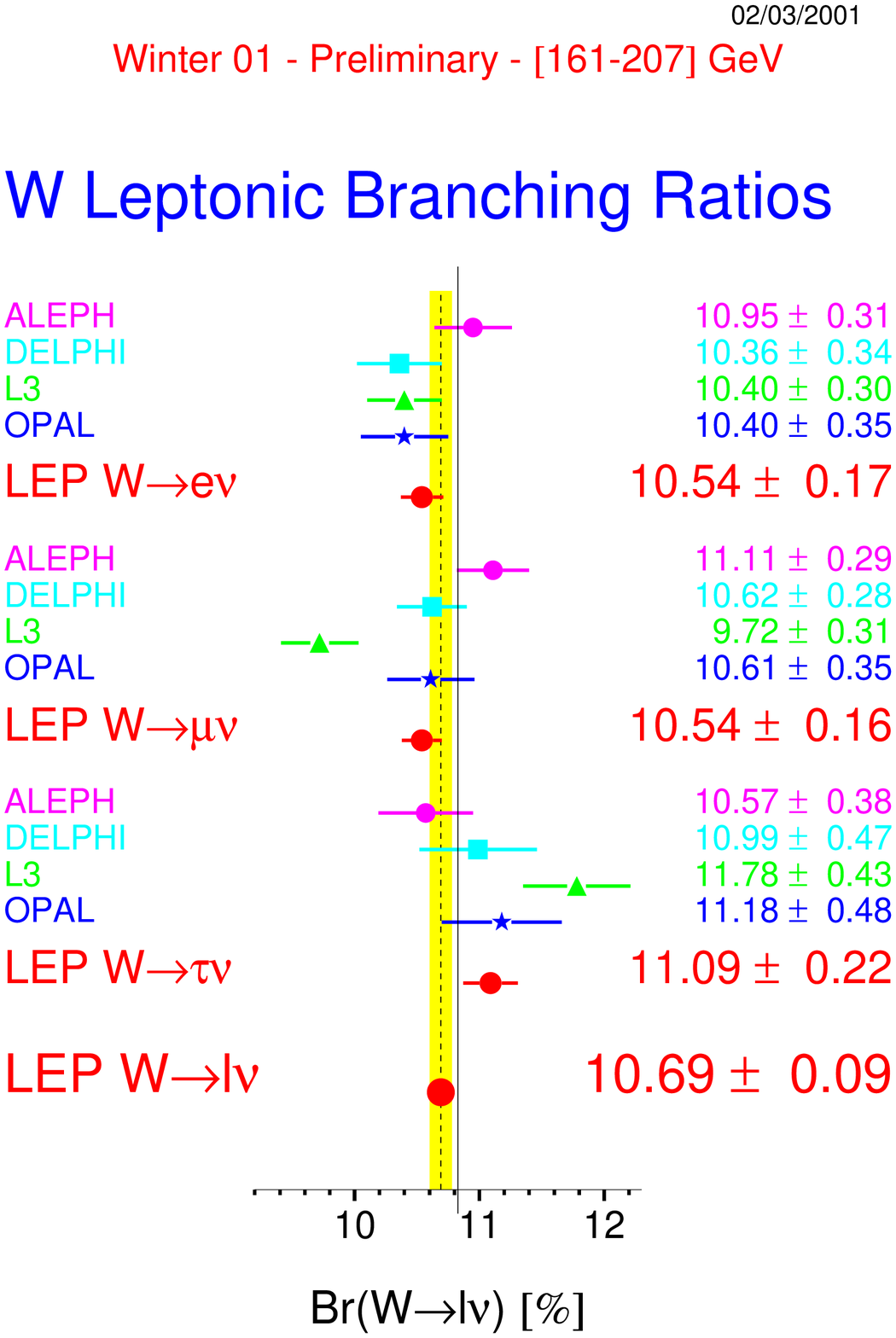,width=0.45\textwidth}  
    \vspace*{-1.0truecm}}
  \hspace*{0.02\textwidth}
  {\epsfig{figure=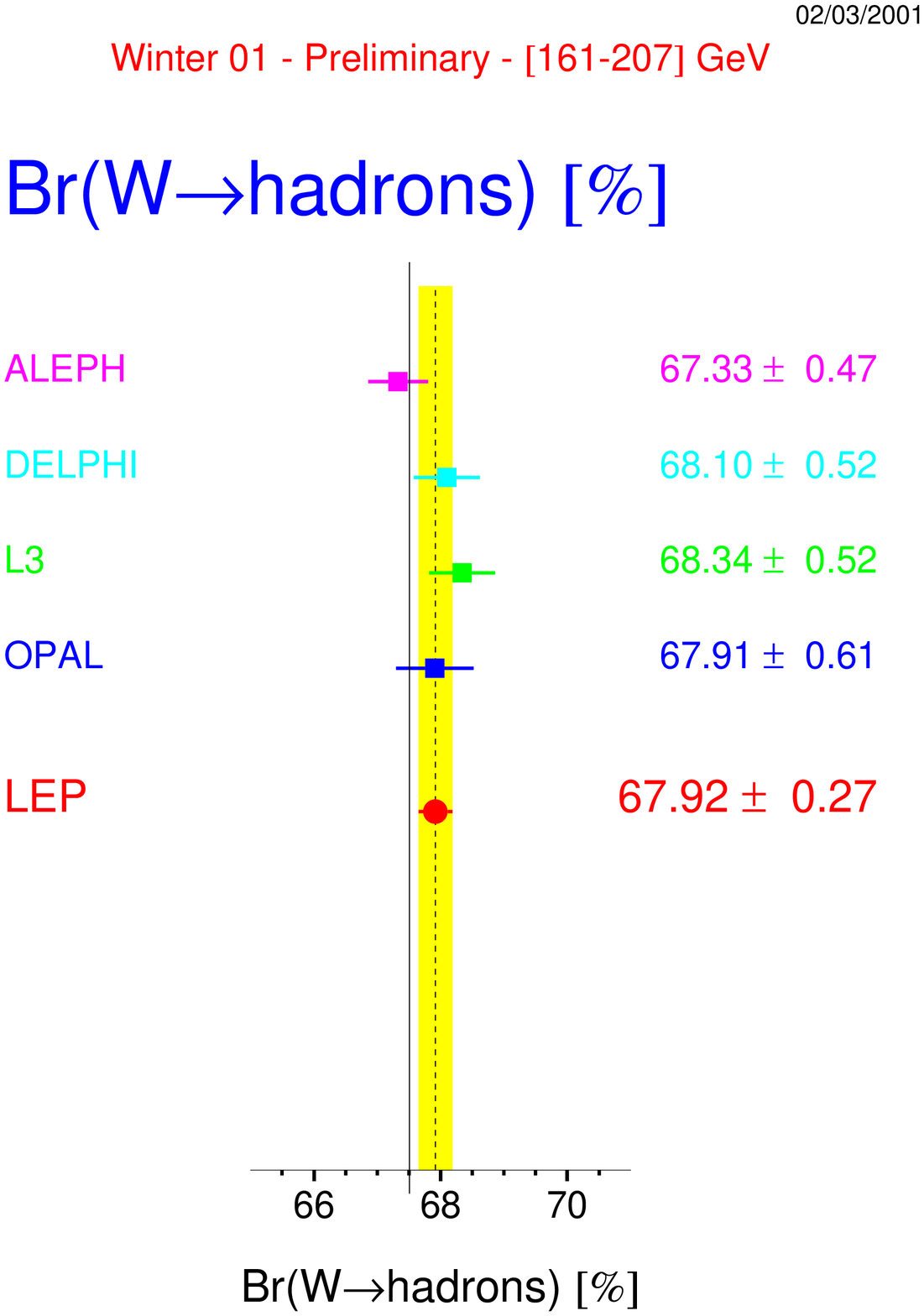,width=0.45\textwidth}
    \vspace*{-1.0truecm}}
  }
\caption{
  Summary of leptonic and hadronic W branching fractions 
  derived from preliminary W-pair production 
  cross-sections measurements up to 207~\GeV\ \CoM\ energy,
  unchanged from winter 2001. The thin vertical line denotes the
  Standard Model expectation.
}
\label{4f_fig:wwbra}
\end{figure}

From the cross sections for the individual 
WW$\rightarrow\mathrm{4f}$ decay channels 
measured by the four experiments at all energies larger than 161~\GeV, 
the W decay branching fractions 
\mbox{$\mathcal{B}(\mathrm{W}\rightarrow\textrm{f}\overline{\textrm{f}}')$}
are determined, with and without the assumption of lepton universality. 
All four experiments update their results 
since the summer 2000 conferences to include 
the full data samples collected in the year 2000 at \CoM\ energies of 205 and
207~\GeV~\cite{4f_bib:aleww2000,4f_bib:delww2000,
4f_bib:ltrww2000,4f_bib:opaww2000b}.
The results from each experiment are given in Table~\ref{4f_tab:wwbra} 
and Figure~\ref{4f_fig:wwbra},
together with the result of the LEP combination. 

\renewcommand{\arraystretch}{1.2}
\begin{table}[bp]
\begin{center}
\begin{tabular}{|c|c|c|c|c|} 
\cline{2-5}
\multicolumn{1}{c|}{$\quad$} & \multicolumn{3}{|c|}{Lepton} & Lepton \\
\multicolumn{1}{c|}{$\quad$} & \multicolumn{3}{|c|}{non--universality} & universality \\
\hline
Experiment 
         & \wwbr(\Wtoenu) & \wwbr(\Wtomnu) & \wwbr(\Wtotnu)  
         & \wwbr({\mbox{$\mathrm{W}\rightarrow\mathrm{hadrons}$}}) \\
         & [\%] & [\%] & [\%] & [\%]  \\
\hline
\Aleph\  & $10.95\pm0.31$ & $11.11\pm0.29$ & $10.57\pm0.38$ & $67.33\pm0.47$ \\
\Delphi\ & $10.36\pm0.34$ & $10.62\pm0.28$ & $10.99\pm0.47$ & $68.10\pm0.52$ \\
\Ltre\   & $10.40\pm0.30$ 
                       & $\phz9.72\pm0.31$ & $11.78\pm0.43$ & $68.34\pm0.52$ \\
\Opal\   & $10.40\pm0.35$ & $10.61\pm0.35$ & $11.18\pm0.48$ & $67.91\pm0.61$ \\
\hline
LEP      & $10.54\pm0.17$ & $10.54\pm0.16$ & $11.09\pm0.22$ & $67.92\pm0.27$ \\
\hline
$\chi^2/\textrm{d.o.f.}$ & \multicolumn{3}{|c|}{14.9/9} & 18.8/11 \\
\hline
\end{tabular}
\caption{
  Summary of leptonic and hadronic
  W branching fractions derived from preliminary W-pair production 
  cross-sections measurements up to 207~\GeV\ \CoM\ energy.
  A common systematic error of \mbox{(0.03--0.06)\%} 
  on the leptonic branching fractions
  is taken into account in the combination.} 
\label{4f_tab:wwbra} 
\end{center}
\vspace*{-0.5truecm}
\end{table}
\renewcommand{\arraystretch}{1.}

The two combinations performed, 
with and without the assumption of lepton universality,
both use as inputs from each of the four experiments 
the three leptonic branching fractions, 
with their systematic and observed statistical errors
and their three by three correlation matrices.
In the fit with lepton universality,
the branching fraction to hadrons is determined from that to leptons
by constraining the sum to unity.
In building the full 12$\times$12 covariance matrix,
it is assumed that 
the 4-jet QCD background components of the systematic error 
are fully correlated between different experiments
both for the same and for different leptonic channels,
as they arise mainly from the uncertainty 
on the WW cross section in the channel 
where both W bosons decay to hadrons.
The combination procedure is consistent with that used
for the combination of the total W-pair cross sections
and outlined in the previous section,
as the same sources of inter-experiment correlations are considered,
while inter-energy correlations of systematic errors
are taken into account internally by each experiment 
when deriving their average branching ratios.
The detailed inputs used for the combination
are given in Appendix~\ref{4f_sec:appendix}.

The results of the fit which does not make use of the lepton universality
assumption show a negative correlation of 21.4\% (18.9\%) between the 
\Wtotnu\  and \Wtoenu\  (\Wtomnu)  branching fractions, while between the
electron and muon decay channels there is a positive correlation of 6.6\%.
The two-by-two comparison of these branching fractions 
constitutes a test of lepton universality
in the decay of on--shell W bosons at the level of 2.9\%:
\begin{eqnarray*}
\centering
\wwbr\mathrm{(\Wtomnu)} \, / \, \wwbr\mathrm{(\Wtoenu)} \,
& = & 1.000 \pm 0.021 \, ,\\
\wwbr\mathrm{(\Wtotnu)} \; / \, \wwbr\mathrm{(\Wtoenu)} \,
& = & 1.052 \pm 0.029 \, ,\\
\wwbr\mathrm{(\Wtotnu)} \, / \, \wwbr\mathrm{(\Wtomnu)} 
& = & 1.052 \pm 0.028 \, .
\end{eqnarray*}
The branching fractions are all consistent with each other within the errors.

Assuming lepton universality,
the measured hadronic branching fraction 
is $[67.92\pm0.17\mathrm{(stat.)}\pm0.21\mathrm{(syst.)}]\%$ 
and the leptonic one 
is $[10.69\pm0.06\mathrm{(stat.)}\pm0.07\mathrm{(syst.)}]\%$.
These results are consistent with their Standard Model expectations,
of 67.51\% and 10.83\% respectively~\cite{4f_bib:yellowreport}.
The systematic error receives equal contributions 
from the correlated and uncorrelated sources.
The high $\chi^2$ of the fit, 
18.8 for 11 degrees of freedom,
is mainly caused by the spread of the \Ltre\ results 
for W decays to muons and taus around the common average.

Within the Standard Model, 
the branching fractions of the W boson depend on the six matrix elements 
$|\mathrm{V}_{\mathrm{qq'}}|$ of the Cabibbo--Kobayashi--Maskawa (CKM) 
quark mixing matrix not involving the top quark. 
In terms of these matrix elements, 
the leptonic branching fraction of the W boson 
$\mathcal{B}(\Wtolnu)$ is given by
\begin{equation*}
  \frac{1}{\mathcal{B}(\Wtolnu)}\quad = \quad 3 
  \Bigg\{ 1 + 
          \bigg[ 1 + \frac{\alpha_s(\mathrm{M}^2_{\mathrm{W}})}{\pi} 
          \bigg] 
          \sum_{\tiny\begin{array}{c}i=(u,c),\\j=(d,s,b)\\\end{array}}
          |\mathrm{V}_{ij}|^2 
  \Bigg\},
\end{equation*} 
where $\alpha_s(\mathrm{M}^2_{\mathrm{W}})$ is the strong coupling
constant. Taking $\alpha_s(\mathrm{M}^2_{\mathrm{W}})=0.121\pm0.002$,
the measured leptonic branching fraction of the W yields
\begin{equation*}
  \sum_{\tiny\begin{array}{c}i=(u,c),\\j=(d,s,b)\\\end{array}}
  |\mathrm{V}_{ij}|^2 
  \quad = \quad
  2.039\,\pm\,0.025\,(\wwbr_{\scriptsize\Wtolnu})\,\pm\,0.001\,(\alpha_s),
\end{equation*} 
where the first error 
is due to the uncertainty on the branching fraction measurement 
and the second to the uncertainty on $\alpha_s$. 
Using the experimental knowledge~\cite{common_bib:pdg2000} of the sum
$|\mathrm{V}_{ud}|^2+|\mathrm{V}_{us}|^2+|\mathrm{V}_{ub}|^2+
 |\mathrm{V}_{cd}|^2+|\mathrm{V}_{cb}|^2=1.0477\pm0.0074$, 
the above result can be interpreted as a measurement of $|\mathrm{V}_{cs}|$ 
which is the least well determined of these matrix elements:
\begin{equation*}
  |\mathrm{V}_{cs}|\quad=\quad0.996\,\pm\,0.013.
\end{equation*}
The error includes 
a $\pm0.0006$ contribution from the uncertainty on $\alpha_s$
and a $\pm0.004$ contribution from the uncertainties 
on the other CKM matrix elements,
the largest of which is that on $|\mathrm{V}_{cd}|$.
These contributions are negligible
in the error on this determination of $|\mathrm{V}_{cs}|$,
which is dominated by the $\pm0.013$ experimental error 
from the measurement of the W branching fractions.

\section{Z-pair production cross section}
\label{4f_sec:ZZxsec}

All experiments have published final results
~\cite{4f_bib:alezz189,4f_bib:delzz189,common_bib:ltrzz183a,4f_bib:ltrzz183b,
common_bib:ltrzz189,common_bib:opazz189} 
on the Z-pair production cross section 
at $\roots=183$ and 189~\GeV,
already presented in~\cite{4f_bib:4f_s00}.
Since the summer 2000 conferences,
\Ltre~\cite{4f_bib:ltrzz1999} has published 
its updated final results between 192 and 202~\GeV,
\Opal~\cite{common_bib:opazz2000new} 
has provided preliminary updates of its previous measurements 
at those energies~\cite{4f_bib:opazz1999},
whereas the corresponding preliminary results from
\Aleph~\cite{4f_bib:alezz1999,4f_bib:alezz2000} and
\Delphi~\cite{4f_bib:delzz1999a,4f_bib:delzz1999b} are unchanged.
All experiments also contribute preliminary results 
at 205 and 207~\GeV~\cite{4f_bib:alezz2000,4f_bib:delzz2000,
common_bib:ltrzz2000new,common_bib:opazz2000new},
based on the analysis of the full data sample collected in the year 2000.

The results of the individual experiments and the LEP averages are
summarised for the different \CoM\ energies in Table~\ref{4f_tab:zzxsec}.
The combination of final results at $\roots=183$ and 189~\GeV\
is the same that was given for the summer 2000 conferences,
while the results above 189~\GeV\ supersede 
those previously presented~\cite{bib-EWEP-00},
and are all preliminary with the exception 
of the \Ltre\ results between 192 and 202~\GeV.

\renewcommand{\arraystretch}{1.2}
\begin{table}[tp]
\begin{center}
\begin{tabular}{|c|c|c|c|c|c|c|} 
\hline
\roots & \multicolumn{5}{|c|}{ZZ cross section (pb)} 
       & $\chi^2/\textrm{d.o.f.}$ \\
\cline{2-6} 
(\GeV) & 
\Aleph\ & 
\Delphi\ &  
\Ltre\ & 
\Opal\ &
LEP &
\\ 
\hline
182.7 & 
$0.11^{\phz+\phz0.16\phz*}_{\phz-\phz0.12}$ & 
$0.38\pm0.18^*$ &
$0.31^{\phz+\phz0.17\phz*}_{\phz-\phz0.15}$ & 
$0.12^{\phz+\phz0.20\phz*}_{\phz-\phz0.18}$ &
$0.23\pm0.08\phs$ & 
2.28/3 \\
188.6 & 
$0.67^{\phz+\phz0.14\phz*}_{\phz-\phz0.13}$ & 
$0.60\pm0.15^*$ &
$0.73^{\phz+\phz0.15\phz*}_{\phz-\phz0.14}$ & 
$0.80^{\phz+\phz0.15\phz*}_{\phz-\phz0.14}$ &
$0.70\pm0.08\phs$ & 
0.97/3 \\
191.6 & 
$0.53^{\phz+\phz0.34}_{\phz-\phz0.27}\phzs$ &
$0.55\pm0.34\phs$ &
$0.29\pm0.22^*$ & 
$1.13^{\phz+\phz0.47}_{\phz-\phz0.41}\phzs$ &
$0.60\pm0.18\phs$ & 
2.88/3 \\
195.5 & 
$0.69^{\phz+\phz0.23}_{\phz-\phz0.20}\phzs$ & 
$1.17\pm0.29\phs$ &
$1.18\pm0.26^*$ & 
$1.19^{\phz+\phz0.28}_{\phz-\phz0.26}\phzs$ &
$1.04\pm0.13\phs$ & 
3.23/3 \\
199.5 & 
$0.70^{\phz+\phz0.22}_{\phz-\phz0.20}\phzs$ & 
$1.08\pm0.26\phs$ &
$1.25\pm0.27^*$ & 
$1.09^{\phz+\phz0.26}_{\phz-\phz0.24}\phzs$ &
$1.01\pm0.13\phs$ & 
2.80/3 \\
201.6 & 
$0.70^{\phz+\phz0.33}_{\phz-\phz0.28}\phzs$ & 
$0.87\pm0.33\phs$ &
$0.95\pm0.39^*$ & 
$0.94^{\phz+\phz0.38}_{\phz-\phz0.33}\phzs$ &
$0.86\pm0.18\phs$ & 
0.32/3 \\
204.9 & 
$1.21^{\phz+\phz0.26}_{\phz-\phz0.23}\phzs$ & 
$1.05\pm0.26\phs$ &
$0.84\pm0.23\phs$ & 
$1.07^{\phz+\phz0.28}_{\phz-\phz0.26}\phzs$ &
$1.03\pm0.13\phs$ & 
1.11/3 \\
206.6 & 
$1.01^{\phz+\phz0.19}_{\phz-\phz0.17}\phzs$ & 
$0.98\pm0.22\phs$ &
$1.20\pm0.21\phs$ & 
$1.07^{\phz+\phz0.22}_{\phz-\phz0.21}\phzs$ &
$1.06\pm0.11\phs$ & 
0.76/3 \\
\hline
\end{tabular}
\caption{
Z-pair production cross section from the four LEP
experiments and combined values 
for the eight energies between 183 and 207~\GeV.
All results are preliminary with the exception of those indicated by $^*$. 
A common systematic error of (0.01--0.07) pb is taken
into account in the averaging procedure.}
\label{4f_tab:zzxsec}
\end{center}
\end{table}
\renewcommand{\arraystretch}{1.}

\begin{figure}[htbp]
\centering
\epsfig{figure=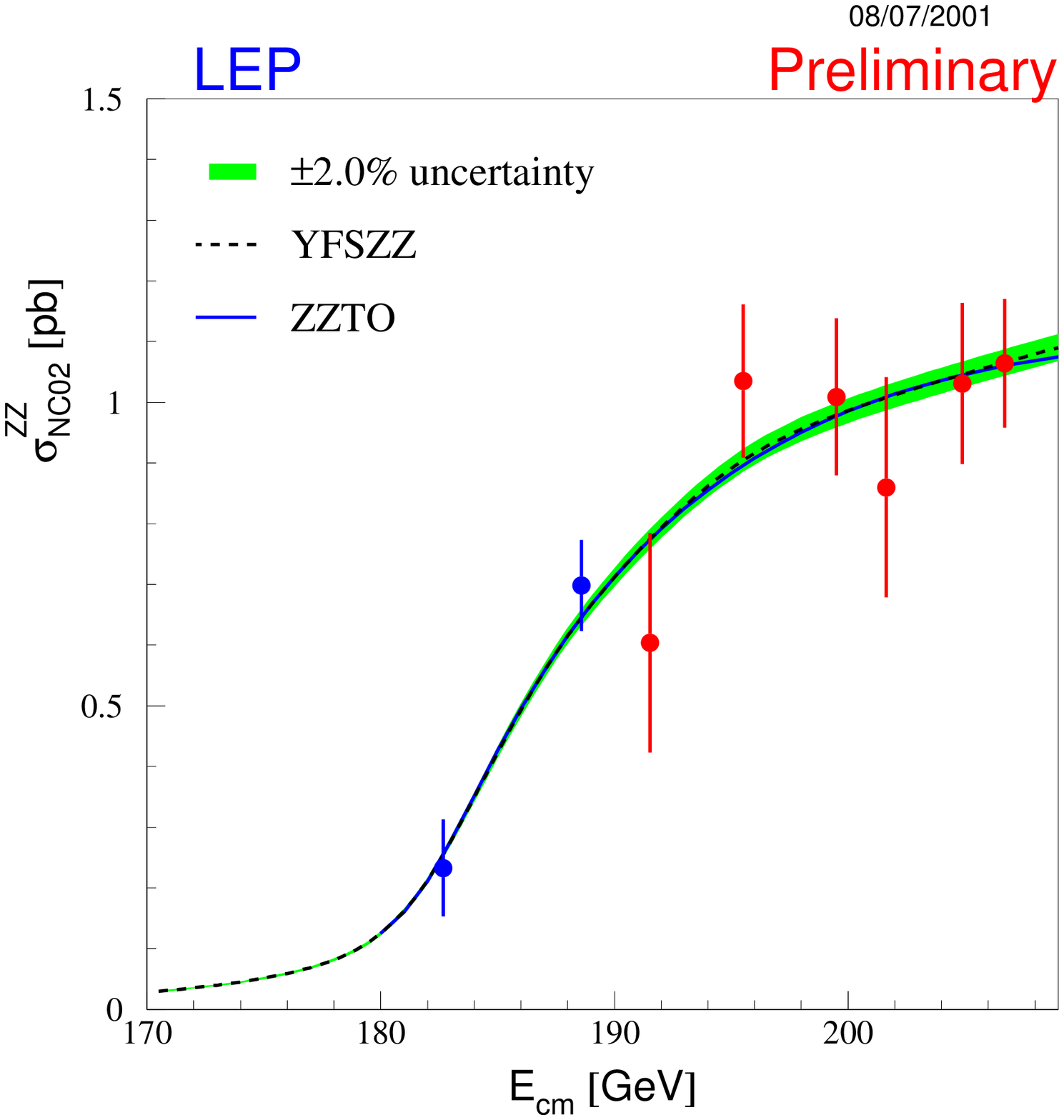,width=1.0\textwidth}
\caption{
  Measurements of the Z-pair production cross section,
  compared to the predictions 
  of \YFSZZ~\protect\cite{4f_bib:yfszz} and \ZZTO~\protect\cite{4f_bib:zzto}. 
  The shaded area represents the $\pm2$\% uncertainty 
  on the predictions.
}
\label{4f_fig:szz_vs_sqrts}
\end{figure} 

All numerical results presented in this Section
are defined to represent {\sc NC02}~\cite{4f_bib:fourfrep} ZZ cross sections.
The combination of results is performed 
using the symmetrized expected statistical error of each analysis, 
to avoid biases due to the limited number of events selected. 
As in the combination performed 
for the summer 2000 conferences~\cite{4f_bib:4f_s00},
the component of the systematic errors 
that is considered as correlated between experiments includes
the uncertainty on the backgrounds
from q$\mathrm{\bar{q}}$, WW, Zee and We$\nu$ processes
and the uncertainty on the {\it b} quark modelling.
Summing these contributions together,
the common error ranges between 0.01 and 0.07 pb
for the various experiments,
as described in Appendix~\ref{4f_sec:appendix}.

The measurements are shown in Figure~\ref{4f_fig:szz_vs_sqrts} 
as a function of the LEP \CoM\ energy,
where they are compared to the YFSZZ~\cite{4f_bib:yfszz} and
ZZTO~\cite{4f_bib:zzto} predictions.
Both these calculations have an estimated 
uncertainty of $\pm2\%$~\cite{4f_bib:fourfrep}. 
The data do not show any significant deviation 
from the theoretical expectations.
 
\section{Single-W production cross section}
\label{4f_sec:wenxsec}

Since the summer 2000~\cite{4f_bib:4f_s00} conferences,
only \Aleph~\cite{4f_bib:alesw2000} and \Delphi~\cite{4f_bib:delsw2000} 
present new measurements of the single-W cross section,
from the analysis of the full data sample collected 
in the year 2000 at 205 and 207~\GeV,
while \Ltre\ has published unchanged its results 
at 189~\GeV~\cite{4f_bib:ltrsw189}, 
already presented as preliminary in the summer 2000.
None of the other results previously presented 
by the four experiments are updated:
these include the results published by 
\Aleph~\cite{4f_bib:alesw183}
and \Ltre~\cite{4f_bib:ltrswdef,4f_bib:ltrsw183} at 183~\GeV,
and the preliminary measurements by 
\Aleph\ at 189--202~\cite{4f_bib:alesw189},
\Delphi\ at 189--202~\cite{4f_bib:delsw189a,4f_bib:delsw189b},
\Ltre\ at 192--202~\cite{4f_bib:ltrsw1999}
and \Opal\ at 189~\GeV~\cite{4f_bib:opasw189}.

A new combination of LEP results 
for the summer 2001 conferences is performed
not only at 205--207~\GeV to include the new results by \Aleph\ and \Delphi,
but also at all energies between 183 and 202~\GeV.
This is done to include the \Delphi\ results
for the total single-W cross section
at these energies~\cite{4f_bib:delsw189b} 
defined according to the common LEP prescription 
of Reference~\citen{4f_bib:swmor00},
which accounts for all decays of the W boson, including those to taus.
In contrast, for the previous combination performed
for the winter 2000 conferences~\cite{4f_bib:swmor00},
previous \Delphi\ results~\cite{4f_bib:delsw189a} had been used,
accounting only for decays of the W boson to hadrons, electrons or muons.
In the new average for the summer 2001 conferences,
results are combined assuming
uncorrelated systematic errors between experiments
and consistently using expected statistical errors for all measurements,
given the limited statistical precision 
of the single-W cross-section measurements.
This also differs slightly from the procedure previously used
for the winter 2000 conferences,
where expected statistical errors had only been used
for a few measurements on very limited data samples,
reverting to measured statistical errors elsewhere.

The measurements of the hadronic and total single-W cross sections by
the four LEP experiments between 183 and 207~\GeV are listed in
Tables~\ref{4f_tab:swxsechad} and~\ref{4f_tab:swxsectot}, together
with the corresponding LEP combined values.  All numerical results
presented in this Section represent single-W cross sections according
to the common LEP definition given in~\cite{4f_bib:swmor00}.  Single-W
production is considered as the complete $t$-channel subset of Feynman
diagrams contributing to e$\nu_\mathrm{e}$f$\bar{\mathrm{f}}'$ final
states, with additional cuts on kinematic variables to exclude the
regions of phase space dominated by multiperipheral diagrams, where
the cross-section calculation is affected by large uncertainties.  The
kinematic cuts used in the signal definitions are:
\mbox{$m_{\qq}>45$~\GeV} for the $\enu\qq$ final states,
\mbox{$E_\ell>20$~\GeV} for the $\enu\lnu$ final states with $\ell=\mu$
or $\tau$, and finally \mbox{$|\cos\theta_\mathrm{e^-}|>0.95$},
\mbox{$|\cos\theta_\mathrm{e^+}|<0.95$} and
\mbox{$E_\mathrm{e^+}>20$~\GeV} (or the charge conjugate cuts) for the
$\enu\enu$ final states.  The measurements performed on the small
amount of data below 183~\GeV, by \Ltre\ at
130--172~\GeV~\cite{4f_bib:ltrsw172,4f_bib:ltrsw183} and \Aleph\ at
161--172~\GeV~\cite{4f_bib:alesw183}, are not converted into the
single-W common LEP definition and are absent from the tables and the
following plot.

\begin{figure}[p]
\centering
\epsfig{figure=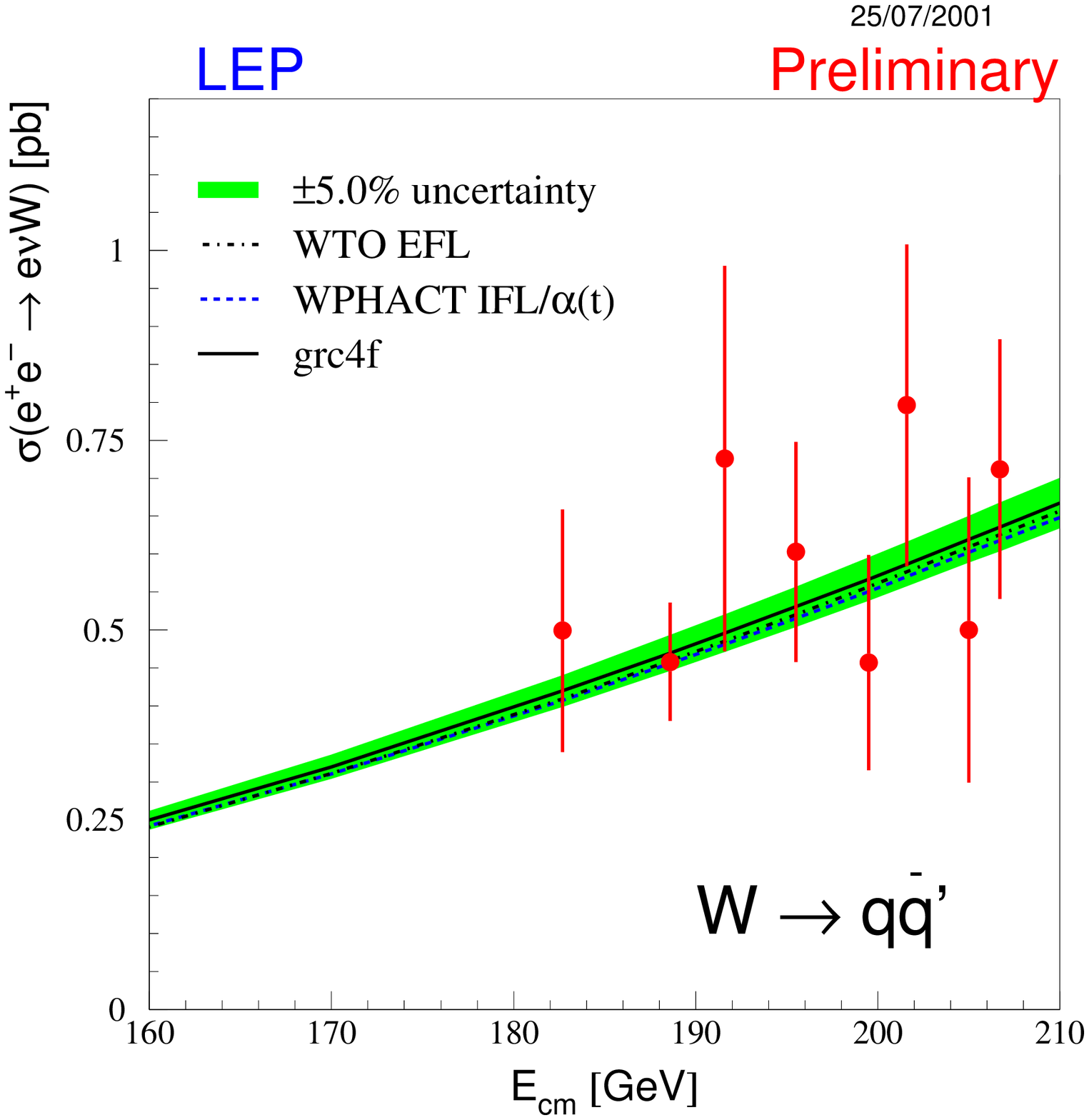,width=0.66\textwidth}\\
\vskip -0.5cm
\epsfig{figure=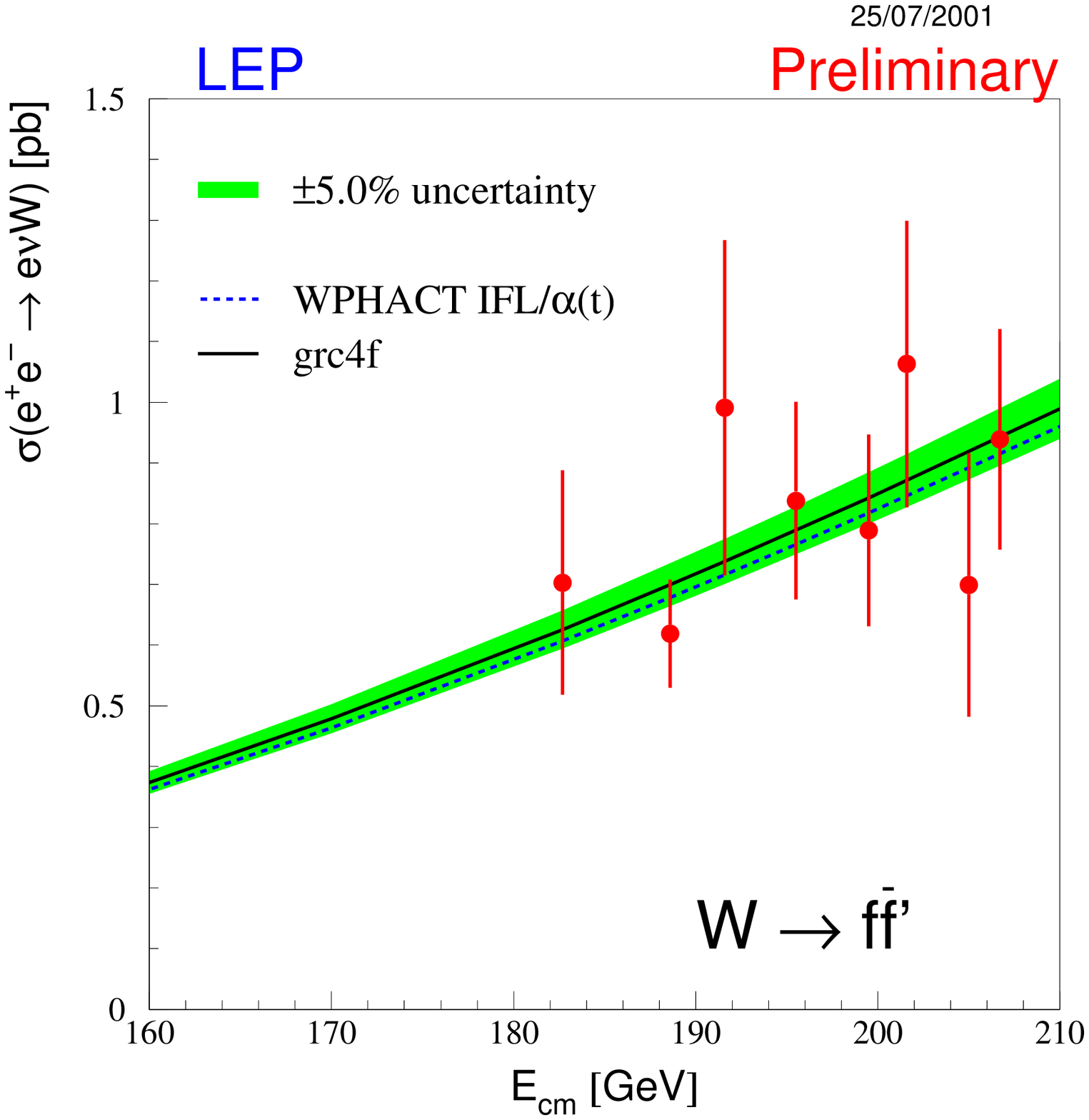,width=0.66\textwidth}
\vskip -0.5cm
\caption{
  Measurements of the single-W production cross section.  Top:
  hadronic decay channel of the W boson; bottom: total production
  cross section.  Also shown are the predictions of
  \WTO~\protect\cite{4f_bib:wto} (hadronic decay channel only),
  \WPHACT~\protect\cite{4f_bib:wphact} and
  \Grace~\protect\cite{4f_bib:grace}.  The shaded area represents the
  $\pm5$\% uncertainty on the predictions.  }
\label{4f_fig:swen}
\end{figure}

The LEP measurements of the single-W cross section are shown,
as a function of the LEP \CoM\ energy, 
in Figure~\ref{4f_fig:swen} for the hadronic decays
and  for all decays of the W boson.
In the two figures, 
the measurements are compared with the expected values 
from \WTO~\cite{4f_bib:wto}, \WPHACT~\cite{4f_bib:wphact} 
and \Grace~\cite{4f_bib:grace}.
As discussed more in detail 
in~\cite{bib-EWEP-00} and~\cite{4f_bib:fourfrep},
the theoretical predictions are scaled upward 
to correct for the implementation of QED radiative corrections 
at the wrong momentum transfer scale s.
The full correction factor of~4\%,
derived~\cite{4f_bib:fourfrep} by the comparison 
to the theoretical predictions from \SWAP~\cite{4f_bib:swap},
is conservatively taken as a systematic error.
This uncertainty dominates the~$\pm$5\% theoretical error 
currently assigned to these 
predictions~\cite{bib-EWEP-00,4f_bib:fourfrep},
represented by the shaded area 
in Figure~\ref{4f_fig:swen}.
All results, up to the highest \CoM\ energies, 
are in agreement with the theoretical predictions.

\renewcommand{\arraystretch}{1.2}
\begin{table}[p]
\begin{center}
\hspace*{-0.0cm}
\begin{tabular}{|c|c|c|c|c|c|c|} 
\hline
\roots & \multicolumn{5}{|c|}{Single-W hadronic cross section (pb)} 
       & $\chi^2/\textrm{d.o.f.}$ \\
\cline{2-6} 
(\GeV) & \Aleph\ & \Delphi\ & \Ltre\ & \Opal\ & LEP & \\
\hline
182.7 & 
$0.40\pm0.24^*$ & 
--- &
$0.58^{\phz+\phz0.23\phz*}_{\phz-\phz0.20}$ & 
--- &
$0.50\pm0.16\phs$ &
0.31/1 \\
188.6 & 
$0.31\pm0.14\phs$ & 
$0.44^{\phz+\phz0.28}_{\phz-\phz0.25}\phs$ &
$0.52^{\phz+\phz0.14\phz*}_{\phz-\phz0.13}$ & 
$0.53^{\phz+\phz0.14}_{\phz-\phz0.13}\phs$ &
$0.46\pm0.08\phs$ &
1.47/3 \\
191.6 & 
$0.94\pm0.44\phs$ & 
$0.01^{\phz+\phz0.19}_{\phz-\phz0.07}\phs$ &
$0.85^{\phz+\phz0.45\phz}_{\phz-\phz0.37}\phs$ & 
--- &
$0.73\pm0.25\phs$ &
1.94/2 \\
195.5 & 
$0.45\pm0.23\phs$ & 
$0.78^{\phz+\phz0.38}_{\phz-\phz0.34}\phs$ &
$0.66^{\phz+\phz0.25\phz}_{\phz-\phz0.23}\phs$ & 
--- &
$0.60\pm0.15\phs$ &
0.77/2 \\
199.5 & 
$0.82\pm0.26\phs$ & 
$0.16^{\phz+\phz0.29}_{\phz-\phz0.17}\phs$ &
$0.34^{\phz+\phz0.23\phz}_{\phz-\phz0.20}\phs$ & 
--- &
$0.46\pm0.14\phs$ &
3.60/2 \\
201.6 & 
$0.68\pm0.35\phs$ & 
$0.55^{\phz+\phz0.47}_{\phz-\phz0.40}\phs$ &
$1.09^{\phz+\phz0.42\phz}_{\phz-\phz0.37}\phs$ & 
--- &
$0.80\pm0.21\phs$ &
1.13/2 \\
204.9 & 
$0.50\pm0.25\phs$ & 
$0.50^{\phz+\phz0.35}_{\phz-\phz0.31}\phs$ & 
--- &
--- &
$0.50\pm0.20\phs$ &
0.00/1 \\
206.6 & 
$0.95\pm0.24\phs$ & 
$0.37^{\phz+\phz0.24}_{\phz-\phz0.21}\phs$ & 
--- &
--- &
$0.71\pm0.17\phs$ &
2.77/1 \\
\hline
\end{tabular}
\end{center}
\vspace*{-0.3cm}
\caption{
  Single-W production cross section from the four LEP
  experiments and combined values 
  for the eight energies between 183 and 207~\GeV,
  in the hadronic decay channel of the W boson.
  All results are preliminary with the exception of those indicated by $^*$.}
\label{4f_tab:swxsechad}
\end{table}
\renewcommand{\arraystretch}{1.}

\renewcommand{\arraystretch}{1.2}
\begin{table}[p]
\vspace*{-0.3cm}
\begin{center}
\hspace*{-0.0cm}
\begin{tabular}{|c|c|c|c|c|c|c|} 
\hline
\roots & \multicolumn{5}{|c|}{Single-W total cross section (pb)} 
       & $\chi^2/\textrm{d.o.f.}$ \\
\cline{2-6} 
(\GeV) & \Aleph\ & \Delphi\ & \Ltre\ & \Opal\ & LEP & \\
\hline
182.7 & 
$0.61\pm0.27^*$ & 
--- &
$0.80^{\phz+\phz0.28\phz*}_{\phz-\phz0.25}$ & 
--- &
$0.70\pm0.19\phs$ &
0.26/1 \\
188.6 & 
$0.45\pm0.15\phs$ & 
$0.75^{\phz+\phz0.30}_{\phz-\phz0.26}\phs$ &
$0.69^{\phz+\phz0.16\phz*}_{\phz-\phz0.15}$ &
$0.67^{\phz+\phz0.17}_{\phz-\phz0.15}\phs$ &
$0.62\pm0.09\phs$ &
1.60/3 \\
191.6 & 
$1.31\pm0.48\phs$ & 
$0.17^{\phz+\phz0.34}_{\phz-\phz0.18}\phs$ &
$1.06^{\phz+\phz0.49}_{\phz-\phz0.42}\phs$ & 
--- &
$0.99\pm0.28\phs$ &
2.38/2 \\
195.5 & 
$0.65\pm0.25\phs$ & 
$0.94^{\phz+\phz0.41}_{\phz-\phz0.36}\phs$ &
$0.98^{\phz+\phz0.28}_{\phz-\phz0.27}\phs$ & 
--- &
$0.84\pm0.16\phs$ &
0.92/2 \\
199.5 & 
$0.99\pm0.27\phs$ & 
$0.51^{\phz+\phz0.33}_{\phz-\phz0.32}\phs$ &
$0.79^{\phz+\phz0.27}_{\phz-\phz0.24}\phs$ & 
--- &
$0.79\pm0.16\phs$ &
1.40/2 \\
201.6 & 
$0.75\pm0.36\phs$ & 
$1.15^{\phz+\phz0.55}_{\phz-\phz0.46}\phs$ &
$1.38^{\phz+\phz0.47}_{\phz-\phz0.42}\phs$ & 
--- &
$1.06\pm0.24\phs$ &
1.38/2 \\
204.9 & 
$0.78\pm0.27\phs$ & 
$0.56^{\phz+\phz0.36}_{\phz-\phz0.32}\phs$ & 
--- & 
--- &
$0.70\pm0.22\phs$ &
0.24/1 \\
206.6 & 
$1.19\pm0.25\phs$ & 
$0.58^{\phz+\phz0.26}_{\phz-\phz0.23}\phs$ & 
--- & 
--- &
$0.94\pm0.18\phs$ &
2.71/1 \\
\hline
\end{tabular}
\end{center}
\vspace*{-0.3cm}
\caption{
  Single-W total production cross section from the four LEP
  experiments and combined values 
  for the eight energies between 183 and 207~\GeV.
  All results are preliminary with the exception of those indicated by $^*$.}
\label{4f_tab:swxsectot}
\vspace*{-0.3cm}
\end{table}
\renewcommand{\arraystretch}{1.}

\boldmath
\chapter{Electroweak Gauge Boson Self Couplings}
\label{sec-GC}
\unboldmath

\updates{ Additional preliminary results based on the data collected
  in the year 2000 are included.  No results on charged TGCs are
  included as the effects of newly calculated radiative corrections on
  the couplings results derived from W-pair production are still under
  investigation. }

\section{Introduction}
\label{sec:gc_introduction}
 

The measurement of  gauge boson couplings and the
search for possible anomalous contributions due to the effects of new
physics beyond the Standard Model are among the principal physics
aims at \LEPII~\cite{gc_bib:LEP2YR}.
Combined preliminary measurements of the neutral triple gauge boson
couplings and quartic gauge couplings are presented here. The results
for the neutral couplings already include the full data set for all but
the OPAL results from Z$\gamma$-production. For the
quartic gauge couplings the whole data set is analysed so far only by
L3, and by ALEPH for the $\nngg$ -channel.

The W-pair production process, $\mathrm{e^+e^-\rightarrow\WW}$,
involves the charged triple gauge boson vertices between the $\WW$ and
the Z or the photon.  During \LEPII\ operation, about 10,000 W-pair
events are collected by each experiment.  Single W ($\enw$) and single
photon ($\nng$) production at LEP are also sensitive to the $\WWg$
vertex.  Results from these channels have been combined for previous
summer conferences.

For the charged TGCs, new Monte Carlo calculations
(RacoonWW~\cite{common_bib:racoonww} and
YFSWW~\cite{common_bib:yfsww}) including $O(\alpha_{em})$
corrections to the WW production process have recently become
available. They have the potential to largely affect the measurements
of the charged TGCs in W-pair production. Their implications are still
under investigation. Preliminary results including these
$O(\alpha_{em})$ corrections are so far available only from
ALEPH~\cite{gc_bib:ALEPH-cTGC-Oalpha}. Therefore, as for the winter
conferences this year, no new combinations are made for these
measurements.

At centre-of-mass energies exceeding twice the Z boson mass, pair
production of Z bosons is kinematically allowed. Here, one searches
for the possible existence of triple vertices involving only neutral
electroweak gauge bosons. Such vertices could also contribute to
Z$\gamma$ production.  In contrast to triple gauge boson vertices with
two charged gauge bosons, purely neutral gauge boson vertices do not
occur in the Standard Model of electroweak interactions.

Within the Standard Model, quartic electroweak gauge boson vertices
with at least two charged gauge bosons do exist. In $\ee$ collisions at
\LEPII\ centre-of-mass energies, the $\WWZg$ and $\WWgg$ vertices
contribute to $\WWg$ and $\nngg$ production in $s$-channel and
$t$-channel, respectively.  The effect of the Standard Model quartic
electroweak vertices is below the sensitivity of \LEPII.  Thus only
anomalous quartic vertices are searched for in the production of
$\WWg$, $\nngg$ and also $\Zgg$ final states. No results from the $\Zgg$
final state analysis are included in the combinations due to current
investigations of differences in the description of the anomalous contributions
to this vertex~\cite{Montagna:2001ej}.

\subsection{Neutral Triple Gauge Boson Couplings}
\label{sec:gc_nTGCs}

There are two classes of Lorentz invariant structures associated with
neutral TGC vertices which preserve $U(1)_{em}$ and Bose symmetry, as
described in~\cite{Hagiwara:1987vm,Gounaris:2000tb}.

The first class refers to anomalous Z$\gamma\gamma^*$ and $\rm Z\gamma
\rm Z^*$ couplings which are accessible at LEP in the process
$\mathrm{e^{+} e^{-}} \rightarrow {\rm Z} \gamma$. The parametrisation
contains eight couplings: $h_i^{V}$ with $i=1,...,4$ and $V=\gamma$,Z.
The superscript $\gamma$ refers to Z$\gamma\gamma^*$ couplings and
superscript Z refers to $\rm Z\gamma \rm Z^*$ couplings.  The photon
and the Z boson in the final state are considered as on-shell
particles, while the third boson at the vertex, the $s$-channel
internal propagator, is off shell.  The couplings $h_{1}^{V}$ and
$h_{2}^{V}$ are CP-odd while $h_{3}^{V}$ and $h_{4}^{V}$ are CP-even.

The second class refers to anomalous $\rm{ZZ}\gamma^*$ and
$\rm{ZZZ}^*$ couplings which are accessible at \LEPII\ in the process
$\mathrm{e^{+} e^{-}} \rightarrow$ ZZ.  This anomalous vertex is
parametrised in terms of four couplings: $f_{i}^{V}$ with $i=4,5$ and
$V=\gamma$,Z.  The superscript $\gamma$ refers to ZZ$\gamma^*$
couplings and the superscript Z refers to $\rm{ZZZ}^*$ couplings,
respectively.  Both Z bosons in the final state are assumed to be
on-shell, while the third boson at the triple vertex, the $s$-channel
internal propagator, is off-shell.
The couplings $f_{4}^{V}$ are CP-odd whereas $f_{5}^{V}$ are CP-even.

Note that the $h_i^{V}$ and $f_{i}^{V}$ couplings are independent of
each other.  They are assumed to be  real and they vanish at tree level 
in the Standard Model.

\subsection{Quartic Gauge Boson Couplings}
\label{sec:gc_QGCs}

Anomalous contributions to electroweak quartic vertices are treated in
the framework of
References~\cite{Belanger:1992qi,Stirling:1999ek,Stirling:1999xa}.
Considered are the three lowest-dimensional operators leading to
quartic vertices not causing anomalous TGCs.  According to a more
recent description of the QGCs~\cite{Belanger:1999aw}, anomalous
contributions to the $\WWgg$ and $\ZZgg$ vertex are treated
separately, although their structure is the same.  The corresponding
couplings are parametrised by $\azvl$ and $\acvl$, where $\Lambda$
represents the energy scale of new physics and V=W,Z for the
respective $\WWgg$ and $\ZZgg$ vertices.  An anomalous contribution to
the $\WWZg$ vertex is parametrised by $\anl$.  The couplings $\azvl$
and $\acvl$ conserve C and P, while the coupling $\anl$ is
CP-violating.  The production of $\WWg$ depends on all three $\azwl$,
$\acwl$, and $\anl$ couplings.  The production of $\nngg$ and $\Zgg$
depend only on $\azvl$ and $\acvl$ (for V=W,Z or Z respectively), as
they do not involve the $\WWZg$ vertex.  The coupling parameters are
assumed to be real and they vanish at tree level in the Standard
Model. At present there are differences between the Monte Carlo
descriptions of \cite{Stirling:1999ek} and \cite{Montagna:2001ej} of the
quartic gauge coupling vertex, especially in the $\Zgg$-final state.
This issue is still under investigation as stated in
\cite{Montagna:2001ej} and currently effort is going on to repeat the
measurement using the latter description. No new results are available
using this framework so far and therefore no new combinations are
presented for the $\ZZgg$ couplings.  The analyses of the $\nngg$
final state do not include possible contributions from the $\ZZgg$
vertex and hence the presented measurements here assume a vanishing
$\ZZgg$ vertex measuring only \azwl, \acwl\ and \anl\ accordingly.

\section{Measurements}
\label{sec:gc_data}

The combined results presented here are obtained from updated neutral
electroweak gauge boson coupling measurements and quartic gauge
coupling measurements as discussed above. The individual references
should be consulted for details about the data samples used.

The $h$-coupling analyses of ALEPH, DELPHI and L3 use the data
collected at \LEPII\ up to centre-of-mass energies of 209~\GeV. The
OPAL measurements so far use the data at 189~\GeV.  The results of the
$f$-couplings are now obtained from the whole data set above the
ZZ-production threshold by all of the experiments.  The experiments
already pre-combine different processes and final states for each of
the couplings.  For the neutral TGCs, the analyses use measurements of
the total cross sections of Z$\gamma$ and ZZ production and the
differential distributions: the $h_i^V$
couplings~\cite{gc_bib:ALEPH-nTGC,gc_bib:DELPHI-nTGC,gc_bib:L3-hTGC,
  gc_bib:OPAL-hTGC} and the $f_i^V$
couplings~\cite{gc_bib:ALEPH-nTGC,gc_bib:DELPHI-nTGC,gc_bib:L3-fTGC,
  gc_bib:OPAL-fTGC} are determined.

For QGCs, the combined results are based on measurements from $\WWg$
and $\nngg$ production.  In addition to the total cross section, the
photon energy is used as a sensitive variable in the $\WWg$ channel.
The analyses in the $\nngg$ channel generally restrict to low recoil
masses where contributions from the Standard Model and a possible
$\ZZgg$ vertex are small.  The QGCs $\azvl$, $\acvl$ and
$\anl$~\cite{gc_bib:ALEPH-QGC,gc_bib:L3-QGC,gc_bib:OPAL-QGC} are
determined, where the whole data set is analysed by L3 and
ALEPH, while OPAL uses the data at 189 \GeV.

\section{Combination Procedure}
\label{sec:gc_combination}

The combination procedure is identical  to the previous LEP
combination of electroweak gauge boson couplings~\cite{gc_bib:moriond01}.

Each experiment provides the negative log likelihood, $\LL$, as a
function of the coupling parameters (one or two) to be
combined.  The single-parameter analyses are performed fixing 
all other parameters to their Standard Model values.  The
two-parameter analyses are performed setting the remaining
parameters to their Standard Model values. 


The $\LL$ functions from each experiment include statistical as well
as those systematic uncertainties which are considered as
uncorrelated between experiments.  For both single- and
multi-parameter combinations, the individual $\LL$ functions are
added.  It is necessary to use the $\LL$ functions directly in the
combination, since in some cases they are not parabolic, and hence it is not
possible to combine the results correctly by simply taking weighted
averages of the measurements.

The main contributions to the systematic uncertainties that are
uncorrelated between experiments arise from detector effects,
background in the selected signal samples, limited Monte Carlo
statistics and the fitting method.  Their importance varies for each
experiment and the individual references should be consulted for
details.

The systematic uncertainties arising from the theoretical cross
section prediction in Z$\gamma$-production ($\simeq 1\%$ in the
$\qq\gamma$- and $\simeq 2\%$ in the $\nng$ channel) are treated as
correlated.
For ZZ production, the uncertainty on the theoretical cross section
prediction is small compared to the statistical accuracy and therefore
is neglected.  Smaller sources of correlated systematic uncertainties,
such as those arising from the LEP beam energy, are for simplicity
treated as uncorrelated.

The correlated systematic uncertainties in the $h$-coupling analyses
are taken into account by scaling the combined log-likelihood
functions by the squared ratio of the sum of statistical and
uncorrelated systematic uncertainty over the total uncertainty
including all correlated uncertainties.  For the general case of
non-Gaussian probability density functions, this treatment of the
correlated errors is only an approximation; it also neglects
correlations in the systematic uncertainties between the parameters in
multi-parameter analyses.

The one standard deviation uncertainties (68\% confidence level) are
obtained by taking the coupling values for which $\Delta\LL=+0.5$
above the minimum.  The 95\% confidence level (C.L.)  limits are given
by the coupling values for which $\Delta\LL=+1.92$ above the minimum.
These cut-off values are used for obtaining the results of both
single- and multi-parameter analyses reported here.  Note that in the
case of the neutral TGCs, double minima structures appear in the
negative log-likelihood curves.  For multi-parameter analyses, the two
dimensional 68\%~C.L.  contour curves for any pair of couplings are
obtained by requiring $\Delta\LL=+1.15$, while for the 95\% C.L.
contour curves $\Delta\LL=+3.0$ is required.

\section{Results}

We present results from the four LEP experiments on the various
electroweak gauge boson couplings, and their combination.  The results
quoted for each individual experiment are calculated using the method
described in Section~\ref{sec:gc_combination}.  Thus they may differ
slightly from those reported in the individual references. In
particular for the $h$-coupling result from OPAL and DELPHI, a slightly
modified estimate of the systematic uncertainty due to the theoretical
cross section prediction is responsible for slightly different limits
compared to the published results.  Furthermore, for the QGC, L3
integrates the likelihood in order to determine the 95\%CL, whereas
here it is read off the $\LL$-curve at $\Delta\LL=1.92$ as for
Gaussian shaped likelihood functions.

\subsection{Neutral Triple Gauge Boson Couplings in Z\boldmath$\gamma$ 
Production}

The individual analyses and results of the experiments for the $h$-couplings 
are described in~\cite{gc_bib:ALEPH-nTGC,gc_bib:DELPHI-nTGC,
gc_bib:L3-hTGC,gc_bib:OPAL-hTGC}.

\subsubsection*{Single-Parameter Analyses}
\begin{figure}[p]
\begin{center}
\includegraphics[width=\linewidth]{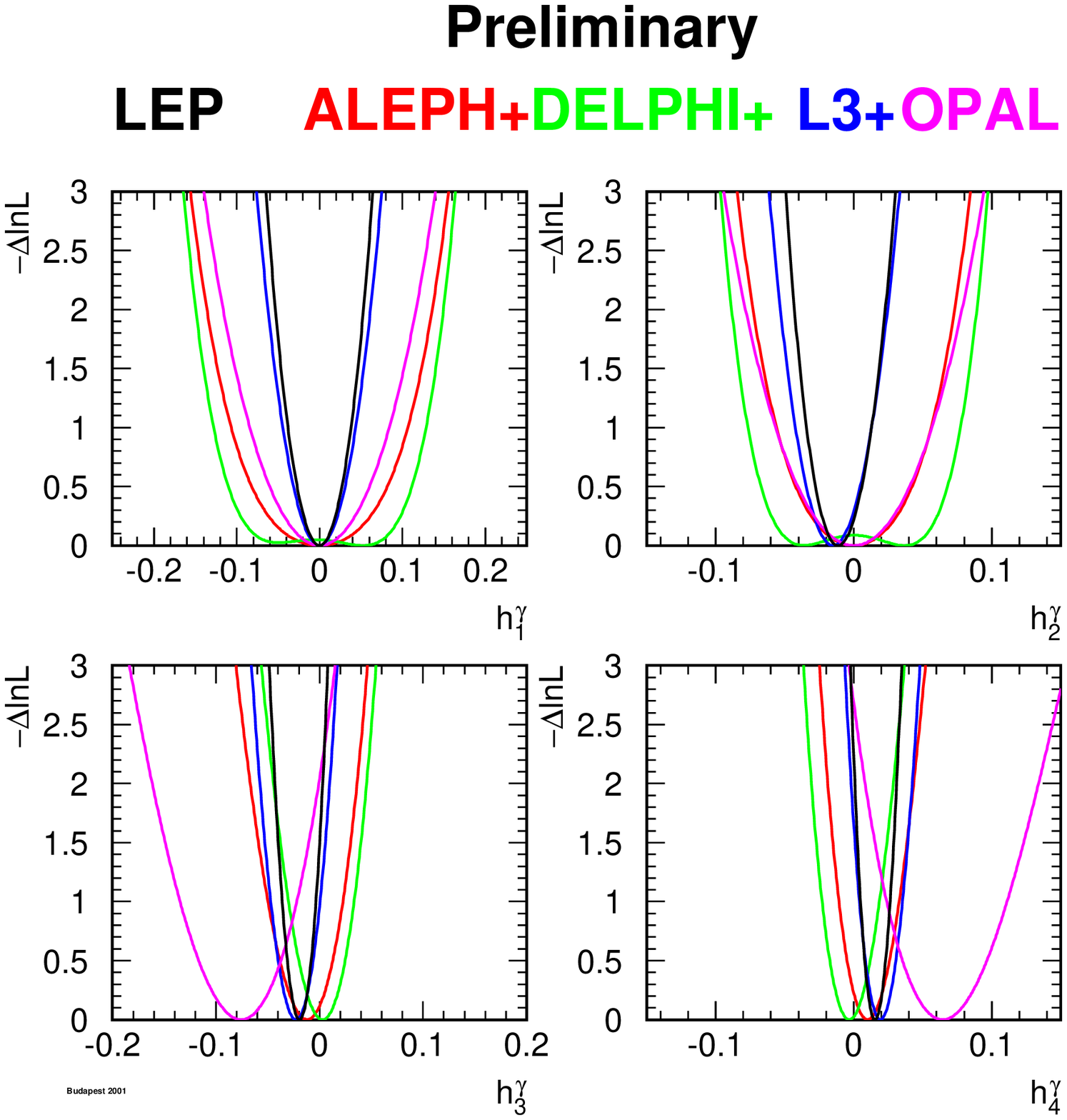}
\caption[]{
  The $\LL$ curves of the four experiments, and the LEP combined curve
  for the four neutral TGCs $h_i^\gamma,~i=1,2,3,4$. In each case, the
  minimal value is subtracted.  }
\label{fig:gc_hgTGC-1}
\end{center}
\end{figure}

\begin{figure}[p]
\begin{center}
\includegraphics[width=\linewidth]{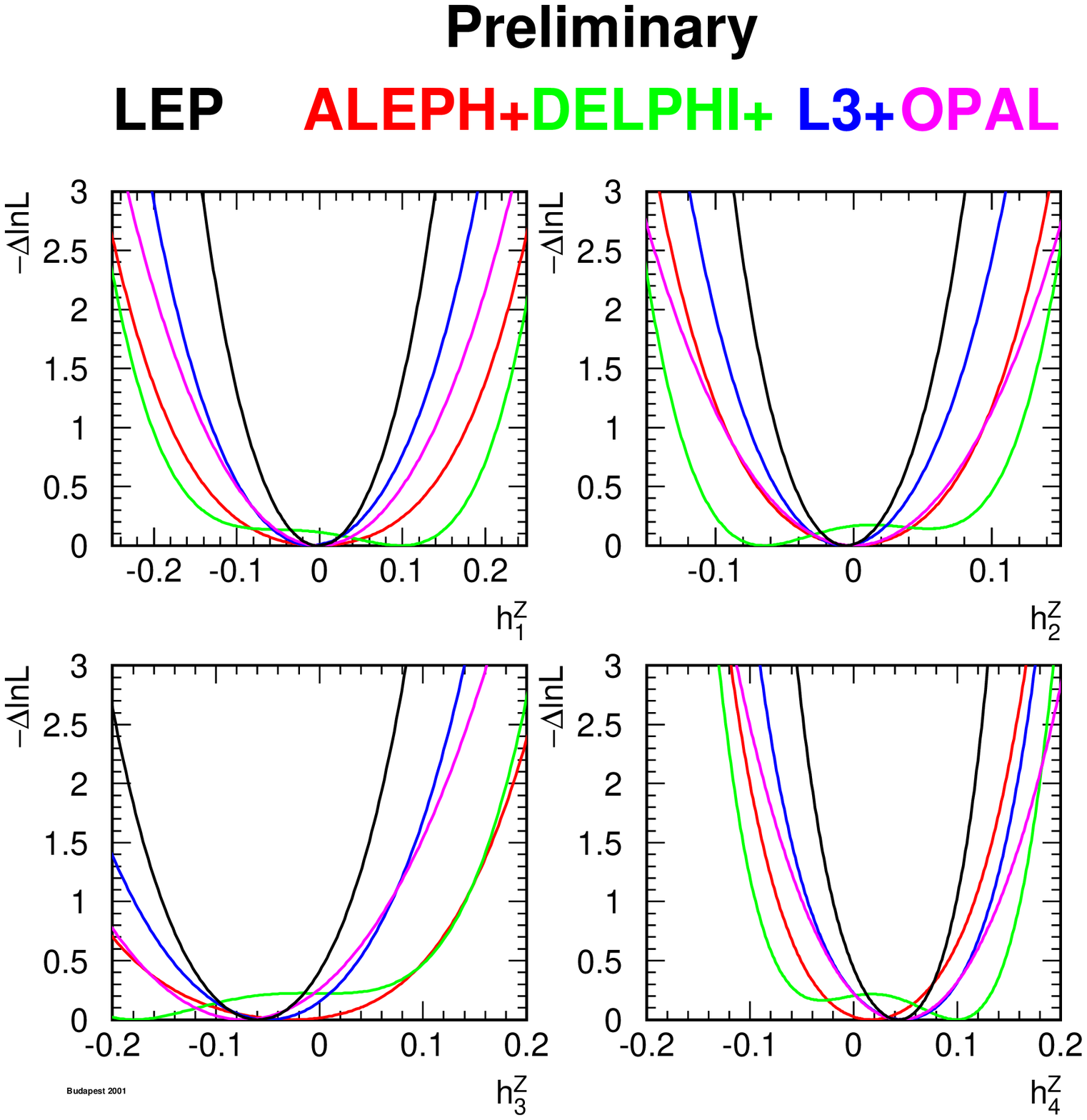}
\caption[]{
  The $\LL$ curves of the four experiments, and the LEP combined curve
  for the four neutral TGCs $h_i^Z,~i=1,2,3,4$.  In each case, the
  minimal value is subtracted.  }
\label{fig:gc_hzTGC-1}
\end{center}
\end{figure}

The results for each experiment are shown in
Table~\ref{tab:gc_hTGC-1-ADLO}, where the errors include both
statistical and systematic uncertainties.  The individual $\LL$ curves
and their sum are shown in Figures~\ref{fig:gc_hgTGC-1}
and~\ref{fig:gc_hzTGC-1}.  The results of the combination are given in
Table~\ref{tab:gc_hTGC-1-LEP}.  From
Figures~\ref{fig:gc_hgTGC-1} and \ref{fig:gc_hzTGC-1} it is clear that the
sensitivity of the L3 analysis~\cite{gc_bib:L3-hTGC} is the highest
amongst the LEP experiments. This is partially due to the use of a larger
phase space region, which increases the statistics by about a factor two,
and partially due to added information from using an Optimal Observable
technique.

\begin{table}[htbp]
\begin{center}
\renewcommand{\arraystretch}{1.3}
\begin{tabular}{|l||r|r|r|r|} 
\hline
Parameter  & ALEPH & DELPHI  &  L3  & OPAL \\
\hline
\hline
$h_1^\gamma$ & [$-0.14,~~+0.14$]  & [$-0.15,~~+0.15$]   & [$-0.06,~~+0.06$]   & [$-0.13,~~+0.13$] \\
\hline                             
$h_2^\gamma$ & [$-0.07,~~+0.07$]  & [$-0.09,~~+0.09$]   & [$-0.053,~~+0.024$] & [$-0.089,~~+0.089$] \\
\hline                             
$h_3^\gamma$ & [$-0.069,~~+0.037$]& [$-0.047,~~+0.047$] & [$-0.062,~~-0.014$] & [$-0.16,~~+0.00$] \\
\hline                             
$h_4^\gamma$ & [$-0.020,~~+0.045$]& [$-0.032,~~+0.030$] & [$-0.004,~~+0.045$] & [$+0.01,~~+0.13$] \\
\hline                             
$h_1^Z$      & [$-0.23,~~+0.23$]  & [$-0.24,~~+0.25$]   & [$-0.17,~~+0.16$]   & [$-0.22,~~+0.22$] \\
\hline                             
$h_2^Z$      & [$-0.12,~~+0.12$]  & [$-0.14,~~+0.14$]   & [$-0.10,~~+0.09$]   & [$-0.15,~~+0.15$] \\
\hline                             
$h_3^Z$      & [$-0.28,~~+0.19$]  & [$-0.32,~~+0.18$]   & [$-0.23,~~+0.11$]   & [$-0.29,~~+0.14$] \\
\hline                             
$h_4^Z$      & [$-0.10,~~+0.15$]  & [$-0.12,~~+0.18$]   & [$-0.08,~~+0.16$]   & [$-0.09,~~+0.19$] \\
\hline
\end{tabular}
\caption[]{The 95\% C.L. intervals ($\Delta\LL=1.92$) measured by
  the ALEPH, DELPHI, L3 and OPAL.  In each case the parameter listed is varied
  while the remaining ones are fixed to their Standard Model values.
  Both statistical and systematic uncertainties are included.  }
\label{tab:gc_hTGC-1-ADLO}
\end{center}
\end{table}

\begin{table}[htbp]
\begin{center}
\renewcommand{\arraystretch}{1.3}
\begin{tabular}{|l||c|} 
\hline
Parameter     & 95\% C.L.      \\
\hline
\hline
$h_1^\gamma$  & [$-0.056,~~+0.055$]  \\ 
\hline
$h_2^\gamma$  & [$-0.045,~~+0.025$]  \\ 
\hline
$h_3^\gamma$  & [$-0.049,~~-0.008$]  \\ 
\hline
$h_4^\gamma$  & [$-0.002,~~+0.034$]  \\ 
\hline
$h_1^Z$       & [$-0.13,~~+0.13$]  \\ 
\hline
$h_2^Z$       & [$-0.078,~~+0.071$]  \\ 
\hline
$h_3^Z$       & [$-0.20,~~+0.07$]  \\ 
\hline
$h_4^Z$       & [$-0.05,~~+0.12$]  \\ 
\hline
\end{tabular}
\caption[]{ The 95\% C.L. intervals ($\Delta\LL=1.92$) obtained
  combining the results from the four experiments.  In each case the
  parameter listed is varied while the remaining ones are fixed to
  their Standard Model values.  Both statistical and systematic
  uncertainties are included.  }
 \label{tab:gc_hTGC-1-LEP}
\end{center}
\end{table}

\subsubsection*{Two-Parameter Analyses}

\begin{figure}[p]
\begin{center}
\includegraphics[width=0.49\linewidth]{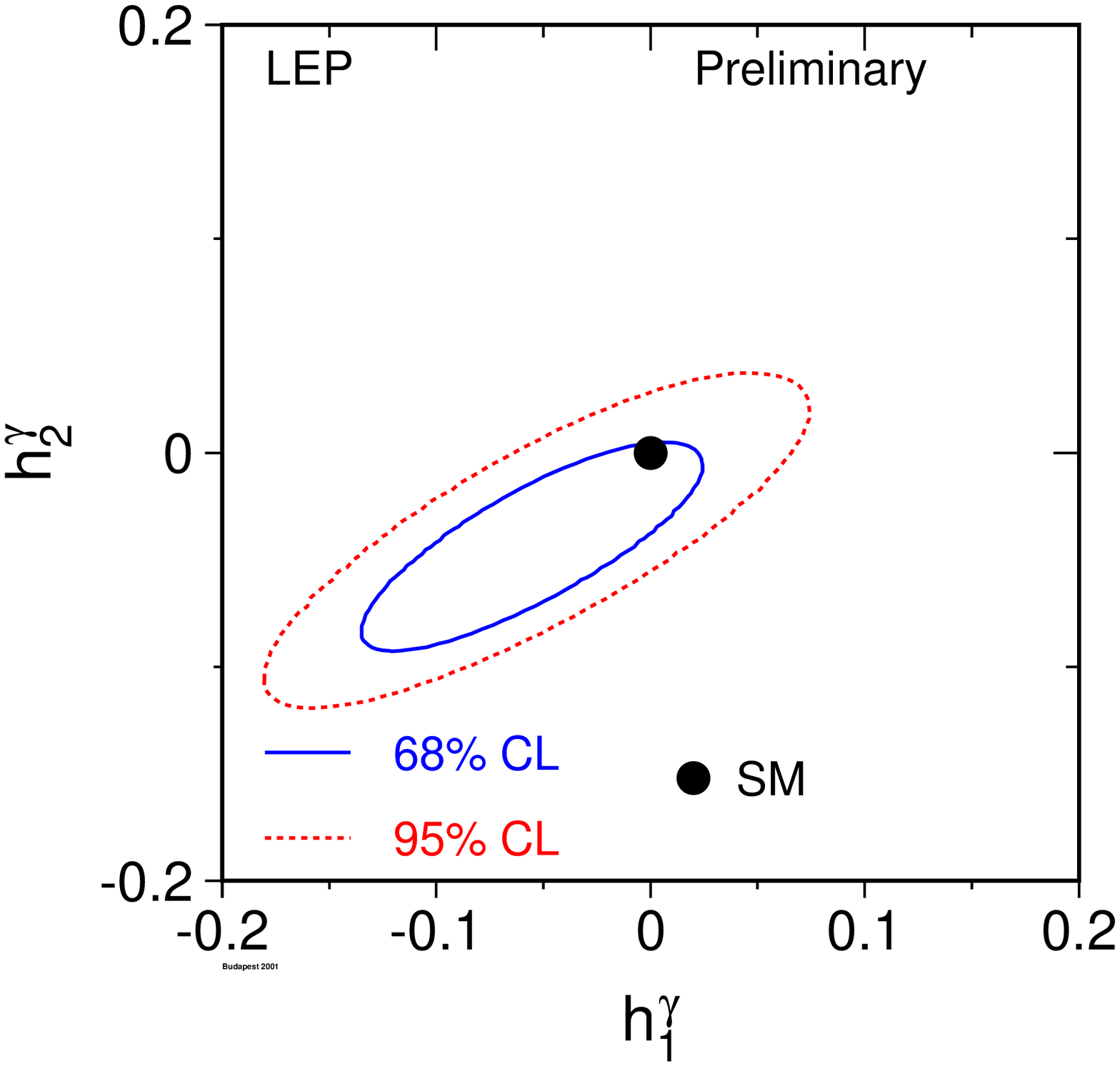}\hfill
\includegraphics[width=0.49\linewidth]{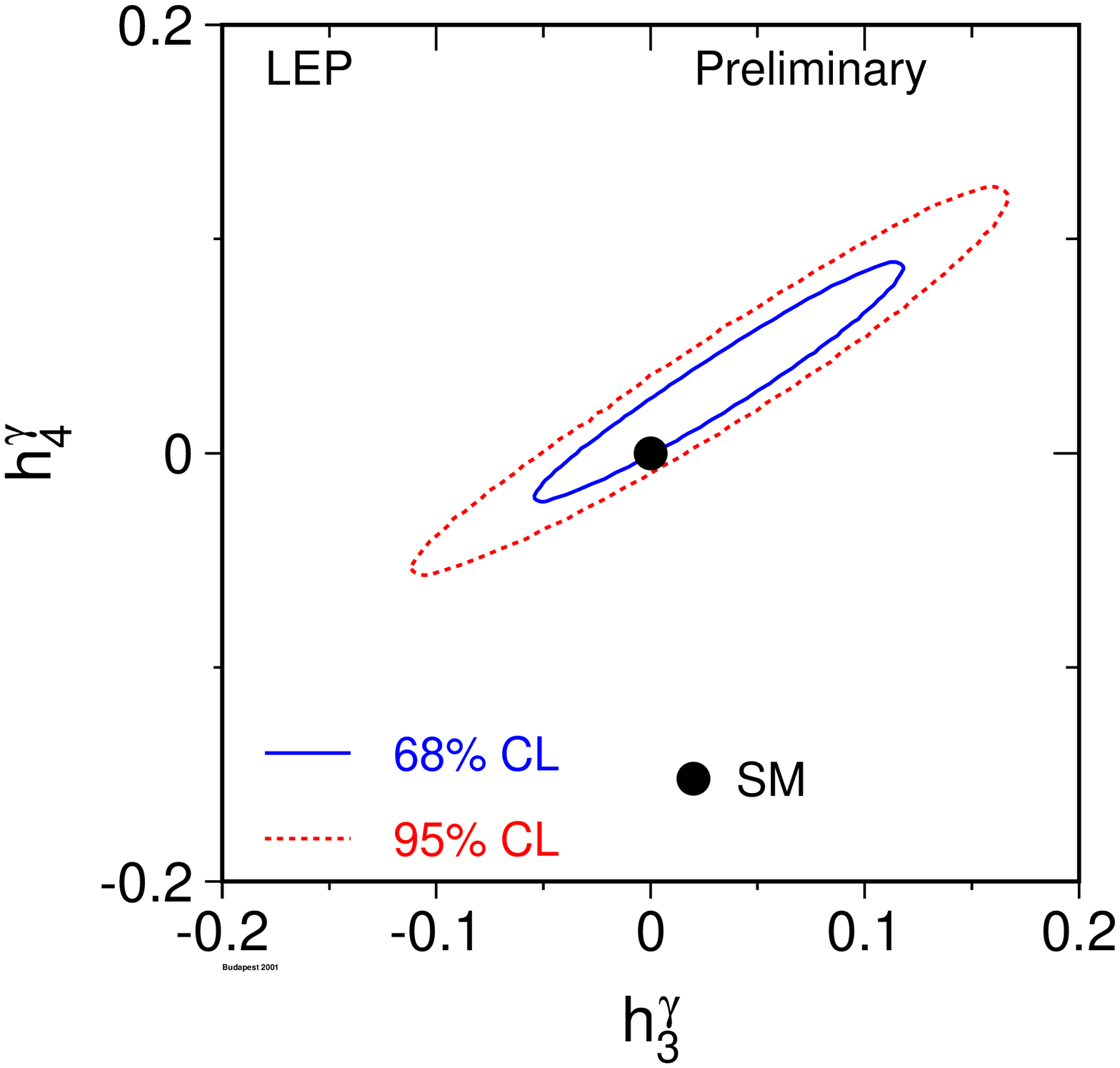}\\
\includegraphics[width=0.49\linewidth]{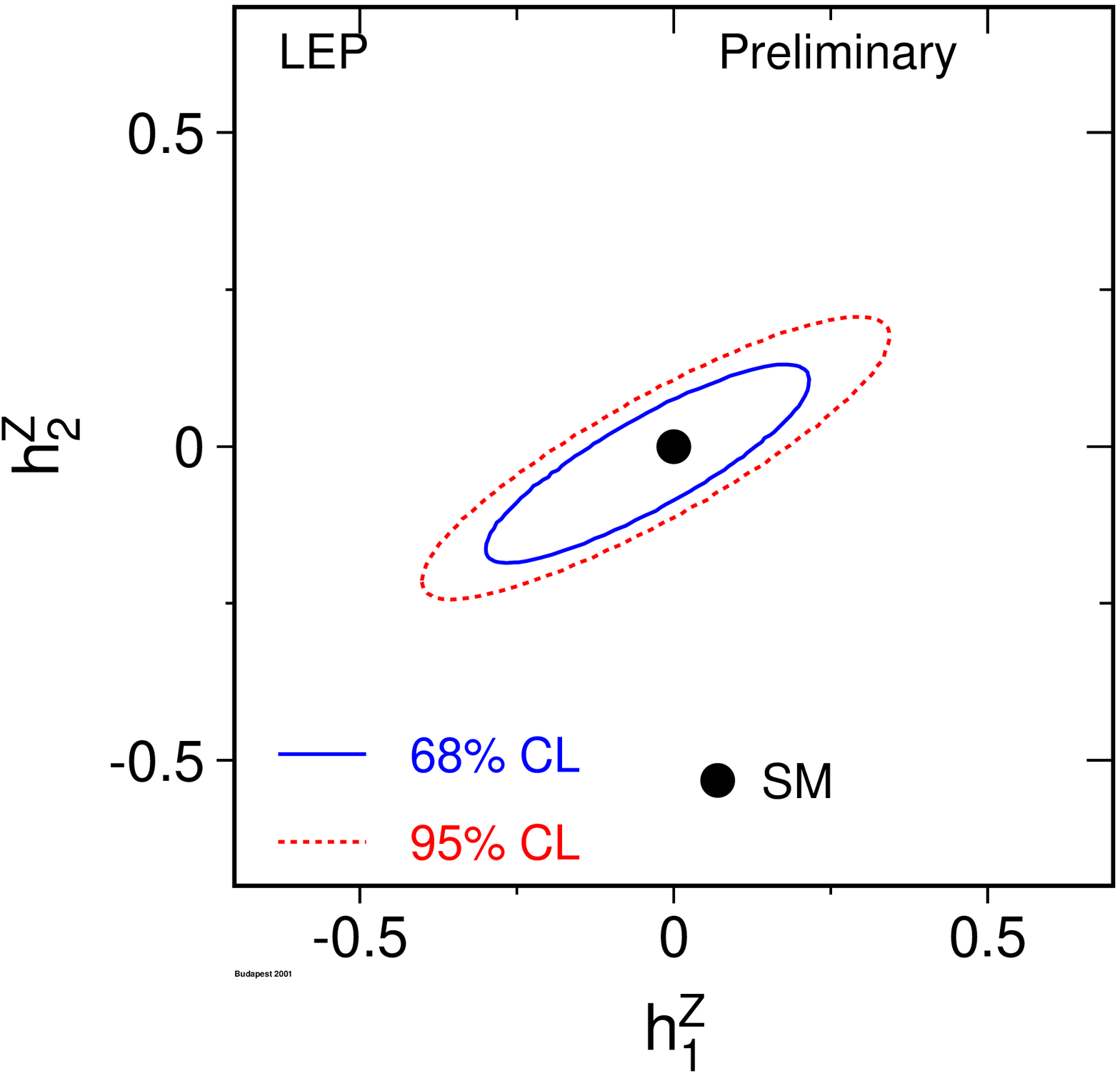}\hfill
\includegraphics[width=0.49\linewidth]{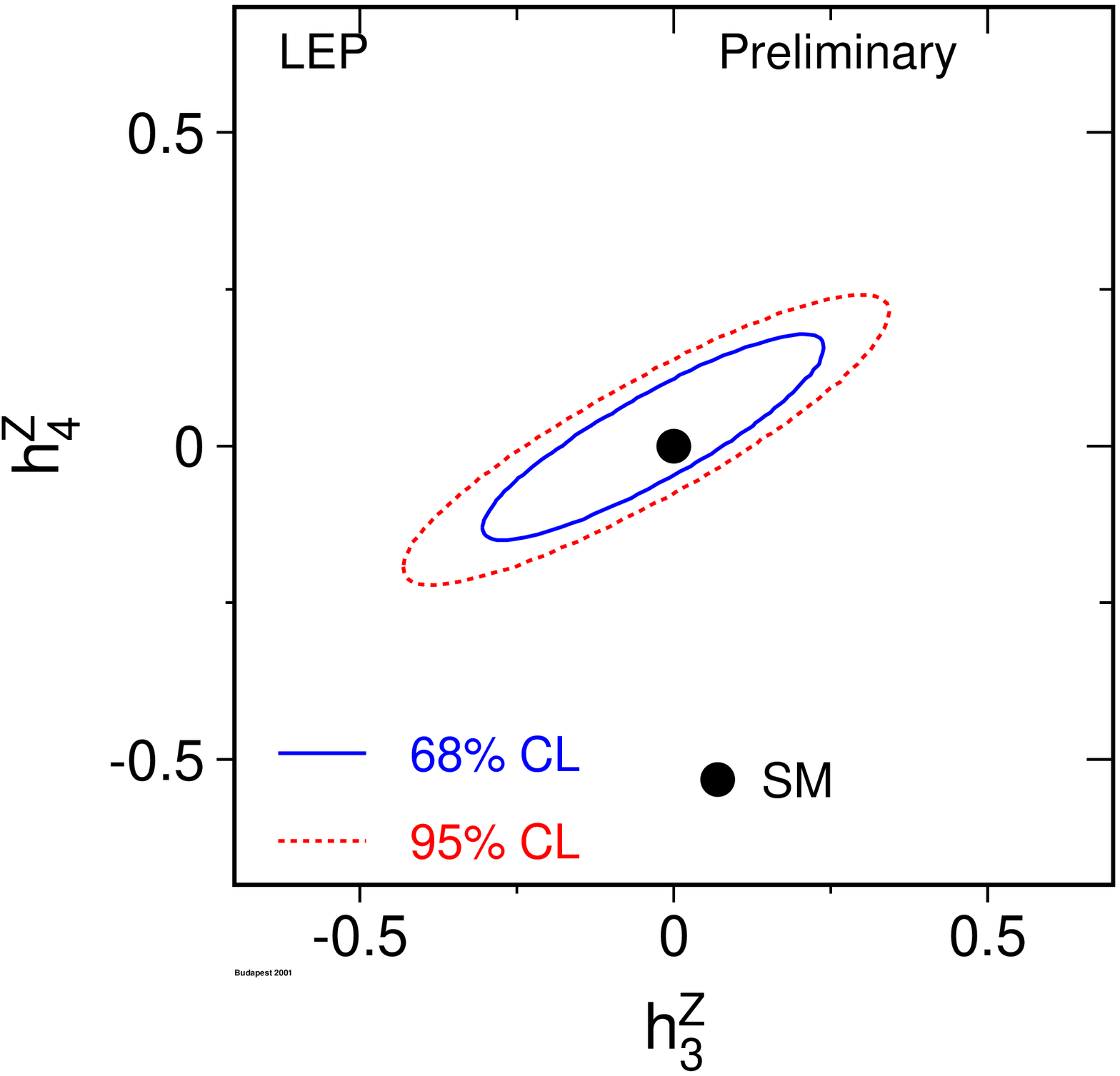}
\caption[]{
  Contour curves of 68\% C.L. and 95\% C.L. in the planes
  $(h_1^\gamma,h_2^\gamma)$, $(h_3^\gamma,h_4^\gamma)$,
  $(h_1^Z,h_2^Z)$ and $(h_3^Z,h_4^Z)$ showing the LEP combined result.
  }
\label{fig:gc_hTGC-2}
\end{center}
\end{figure}

The results for each experiment are shown in
Table~\ref{tab:gc_hTGC-2-ADLO}, where the errors include both
statistical and systematic uncertainties.  The 68\% C.L. and 95\% C.L.
contour curves resulting from the combinations of the two-dimensional
likelihood curves are shown in Figure~\ref{fig:gc_hTGC-2}.  The LEP
average values are given in Table~\ref{tab:gc_hTGC-2-LEP}.

\begin{table}[p]
\begin{center}
\renewcommand{\arraystretch}{1.3}
\begin{tabular}{|l||r|r|r|} 
\hline
Parameter  & ALEPH & DELPHI  &  L3  \\
\hline
\hline
$h_1^\gamma$ & [$-0.32,~~+0.32$] & [$-0.28,~~+0.28$] & [$-0.17,~~+0.04$] \\
$h_2^\gamma$ & [$-0.18,~~+0.18$] & [$-0.17,~~+0.18$] & [$-0.12,~~+0.02$] \\
\hline                                               
$h_3^\gamma$ & [$-0.17,~~+0.38$] & [$-0.48,~~+0.20$] & [$-0.09,~~+0.13$] \\
$h_4^\gamma$ & [$-0.08,~~+0.29$] & [$-0.08,~~+0.15$] & [$-0.04,~~+0.11$] \\
\hline                                               
$h_1^Z$      & [$-0.54,~~+0.54$] & [$-0.45,~~+0.46$] & [$-0.48,~~+0.33$] \\
$h_2^Z$      & [$-0.29,~~+0.30$] & [$-0.29,~~+0.29$] & [$-0.30,~~+0.22$] \\
\hline
$h_3^Z$      & [$-0.58,~~+0.52$] & [$-0.57,~~+0.38$] & [$-0.43,~~+0.39$] \\
$h_4^Z$      & [$-0.29,~~+0.31$] & [$-0.31,~~+0.28$] & [$-0.23,~~+0.28$] \\
\hline
\end{tabular}
\caption[]{The 95\% C.L. intervals ($\Delta\LL=1.92$) measured  by
  ALEPH, DELPHI and L3.  In each case the two parameters listed are varied
  while the remaining ones are fixed to their Standard Model values.
  Both statistical and systematic uncertainties are included.  }
\label{tab:gc_hTGC-2-ADLO}
\end{center}
\end{table}

\begin{table}[p]
\begin{center}
\renewcommand{\arraystretch}{1.3}
\begin{tabular}{|l||c|rr|} 
\hline
Parameter  & 95\% C.L. & \multicolumn{2}{|c|}{Correlations} \\
\hline
\hline
$h_1^\gamma$  & [$-0.16,~~+0.05$]    & $ 1.00$ & $+0.79$ \\ 
$h_2^\gamma$  & [$-0.11,~~+0.02$]    & $+0.79$ & $ 1.00$ \\ 
\hline
$h_3^\gamma$  & [$-0.08,~~+0.14$]    & $ 1.00$ & $+0.97$ \\ 
$h_4^\gamma$  & [$-0.04,~~+0.11$]    & $+0.97$ & $ 1.00$ \\ 
\hline
$h_1^Z$       & [$-0.35,~~+0.28$]    & $ 1.00$ & $+0.77$ \\ 
$h_2^Z$       & [$-0.21,~~+0.17$]    & $+0.77$ & $ 1.00$ \\ 
\hline
$h_3^Z$       & [$-0.37,~~+0.29$]    & $ 1.00$ & $+0.76$ \\ 
$h_4^Z$       & [$-0.19,~~+0.21$]    & $+0.76$ & $ 1.00$ \\ 
\hline
\end{tabular}
\caption[]{ The 95\% C.L. intervals ($\Delta\LL=1.92$) obtained
  combining the results from ALEPH, DELPHI and L3.  In each case the two
  parameters listed are varied while the remaining ones are fixed to
  their Standard Model values.  Both statistical and systematic
  uncertainties are included.  Since the shape of the log-likelihood
  is not parabolic, there is some ambiguity in the definition of the
  correlation coefficients and the values quoted here are approximate.
  }
 \label{tab:gc_hTGC-2-LEP}
\end{center}
\end{table}

\clearpage

\subsection{Neutral Triple Gauge Boson Couplings in ZZ Production}

The individual analyses and results of the experiments for the $f$-couplings 
are described in~\cite{gc_bib:ALEPH-nTGC,gc_bib:DELPHI-nTGC,gc_bib:L3-fTGC,gc_bib:OPAL-fTGC}.

\subsubsection*{Single-Parameter Analyses}

\begin{figure}[p]
\begin{center}
\includegraphics[width=\linewidth]{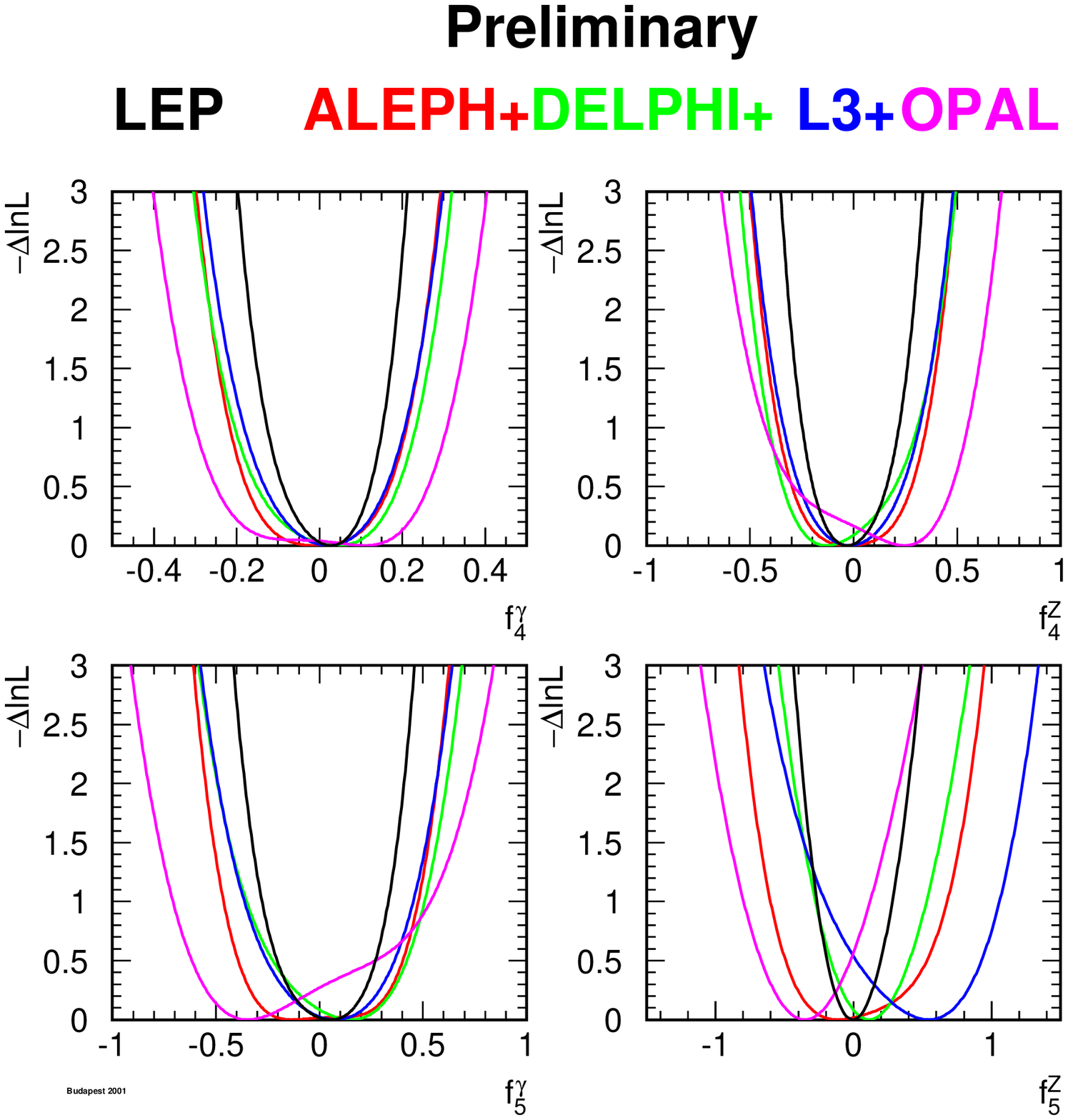}
\caption[]{
  The $\LL$ curves of the four experiments, and the LEP combined curve
  for the four neutral TGCs $f_i^V,~V=\gamma,Z,~i=4,5$.  In each case,
  the minimal value is subtracted.  }
\label{fig:gc_fTGC-1}
\end{center}
\end{figure}

The results for each experiment are shown in
Table~\ref{tab:gc_fTGC-1-ADLO}, where the errors include both statistical
and systematic uncertainties.  The individual $\LL$ curves and their sum are
shown in Figure~\ref{fig:gc_fTGC-1}.  The results of the combination are
given in Table~\ref{tab:gc_fTGC-1-LEP}.

\begin{table}[p]
\begin{center}
\renewcommand{\arraystretch}{1.3}
\begin{tabular}{|l||r|r|r|r|} 
\hline
Parameter  & ALEPH & DELPHI  &  L3   & OPAL  \\
\hline
\hline
$f_4^\gamma$ & [$-0.26,~~+0.26$] & [$-0.26,~~+0.28$] & [$-0.24,~~+0.26$] & [$-0.36,~~+0.36$] \\ 
\hline
$f_4^Z$      & [$-0.44,~~+0.43$] & [$-0.49,~~+0.42$] & [$-0.43,~~+0.41$] & [$-0.55,~~+0.64$] \\ 
\hline
$f_5^\gamma$ & [$-0.54,~~+0.56$] & [$-0.48,~~+0.61$] & [$-0.48,~~+0.56$] & [$-0.82,~~+0.72$] \\ 
\hline
$f_5^Z$      & [$-0.73,~~+0.83$] & [$-0.42,~~+0.69$] & [$-0.46,~~+1.2$] & [$-0.96,~~+0.31$] \\ 
\hline
\end{tabular}
\caption[]{The 95\% C.L. intervals ($\Delta\LL=1.92$) measured by
  ALEPH, DELPHI, L3 and OPAL.  In each case the parameter listed is varied
  while the remaining ones are fixed to their Standard Model values.
  Both statistical and systematic uncertainties are included.  }
\label{tab:gc_fTGC-1-ADLO}
\end{center}
\end{table}

\begin{table}[p]
\begin{center}
\renewcommand{\arraystretch}{1.3}
\begin{tabular}{|l||c|} 
\hline
Parameter     & 95\% C.L.     \\
\hline
\hline
$f_4^\gamma$  & [$-0.17,~~+0.19$]  \\ 
\hline
$f_4^Z$       & [$-0.31,~~+0.28$]  \\ 
\hline
$f_5^\gamma$  & [$-0.36,~~+0.40$]  \\ 
\hline
$f_5^Z$       & [$-0.36,~~+0.39$]  \\ 
\hline
\end{tabular}
\caption[]{ The 95\% C.L. intervals ($\Delta\LL=1.92$) obtained
  combining the results from all four experiments.  In each case the
  parameter listed is varied while the remaining ones are fixed to
  their Standard Model values.  Both statistical and systematic
  uncertainties are included.  }
 \label{tab:gc_fTGC-1-LEP}
\end{center}
\end{table}

\subsubsection*{Two-Parameter Analyses}

\begin{figure}[p]
\begin{center}
\includegraphics[width=0.55\linewidth]{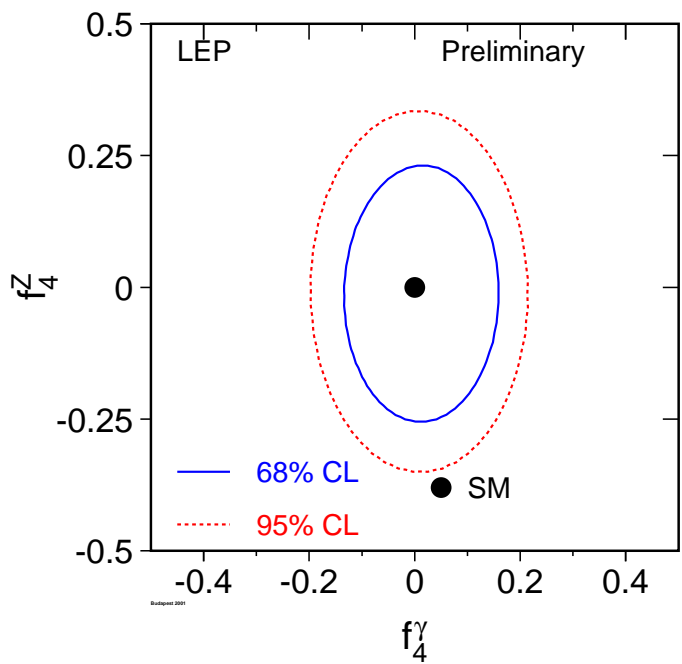}\\
\includegraphics[width=0.55\linewidth]{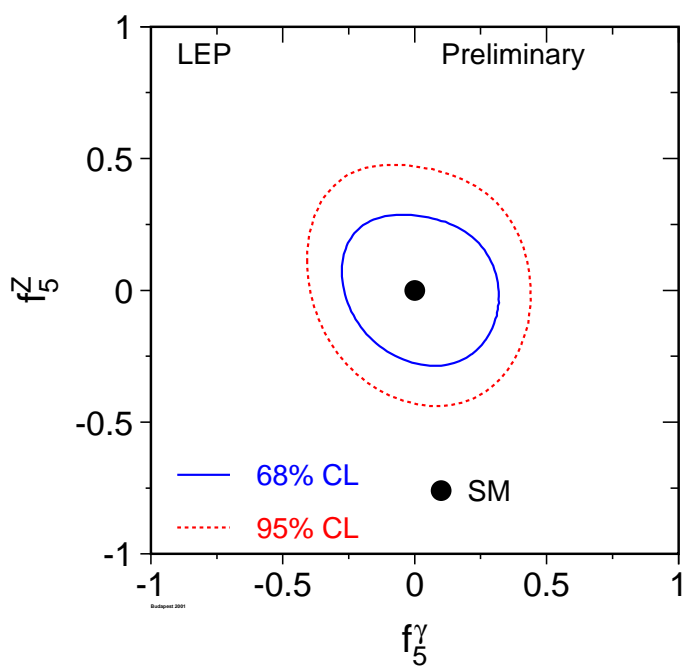}
\caption[]{
  Contour curves of 68\% C.L. and 95\% C.L. in the plane
  $(f_4^\gamma,f_4^Z)$ and $(f_5^\gamma,f_5^Z)$showing the LEP
  combined result.  }
\label{fig:gc_fTGC-2}
\end{center}
\end{figure}

The results from each experiment are shown in
Table~\ref{tab:gc_fTGC-2-ADLO}, where the errors include both
statistical and systematic uncertainties.  The 68\% C.L. and 95\% C.L.
contour curves resulting from the combinations of the two-dimensional
likelihood curves are shown in Figure~\ref{fig:gc_fTGC-2}.  The LEP
average values are given in Table~\ref{tab:gc_fTGC-2-LEP}.

\begin{table}[p]
\begin{center}
\renewcommand{\arraystretch}{1.3}
\begin{tabular}{|l||r|r|r|r|} 
\hline
Parameter  & ALEPH & DELPHI  &  L3   & OPAL  \\
\hline
\hline
$f_4^\gamma$ & [$-0.26,~~+0.26$] & [$-0.26,~~+0.28$]& [$-0.24,~~+0.26$] & [$-0.36,~~+0.36$] \\ 
$f_4^Z$      & [$-0.44,~~+0.43$] & [$-0.49,~~+0.42$]& [$-0.43,~~+0.41$] & [$-0.54,~~+0.63$] \\ 
\hline                                              
$f_5^\gamma$ & [$-0.52,~~+0.53$] & [$-0.52,~~+0.61$]& [$-0.48,~~+0.56$] & [$-0.77,~~+0.73$] \\ 
$f_5^Z$      & [$-0.77,~~+0.86$] & [$-0.44,~~+0.69$]& [$-0.46,~~+1.2$] & [$-0.96,~~+0.44$] \\ 
\hline
\end{tabular}
\caption[]{The 95\% C.L. intervals ($\Delta\LL=1.92$) measured by
  ALEPH, DELPHI, L3 and OPAL.  In each case the two parameters listed are
  varied while the remaining ones are fixed to their Standard Model
  values.  Both statistical and systematic uncertainties are included.
  }
\label{tab:gc_fTGC-2-ADLO}
\end{center}
\end{table}

\begin{table}[p]
\begin{center}
\renewcommand{\arraystretch}{1.3}
\begin{tabular}{|l||c|rr|} 
\hline
Parameter     & 95\% C.L. & \multicolumn{2}{|c|}{Correlations} \\
\hline
\hline
$f_4^\gamma$  &[$-0.17,~~+0.19$] & $ 1.00$ & $+0.10$\\ 
$f_4^Z$       &[$-0.30,~~+0.28$] & $+0.10$ & $ 1.00$\\ 
\hline
$f_5^\gamma$  &[$-0.34,~~+0.38$] & $ 1.00$ & $-0.18$\\ 
$f_5^Z$       &[$-0.36,~~+0.38$]   & $-0.18$ & $ 1.00$\\ 
\hline
\end{tabular}
\caption[]{ The 95\% C.L. intervals ($\Delta\LL=1.92$) obtained
  combining the results from all four experiments.  In each case the
  two parameters listed are varied while the remaining ones are fixed
  to their Standard Model values.  Both statistical and systematic
  uncertainties are included. Since the shape of the log-likelihood is
  not parabolic, there is some ambiguity in the definition of the
  correlation coefficients and the values quoted here are approximate.
  }
 \label{tab:gc_fTGC-2-LEP}
\end{center}
\end{table}


\subsection{Quartic Gauge Boson Couplings}

The individual analyses and results of the experiments for the quartic gauge
couplings are described in~\cite{gc_bib:ALEPH-QGC,gc_bib:L3-QGC,
gc_bib:OPAL-QGC}.

\begin{figure}[p]
\begin{center}
\includegraphics[width=\linewidth]{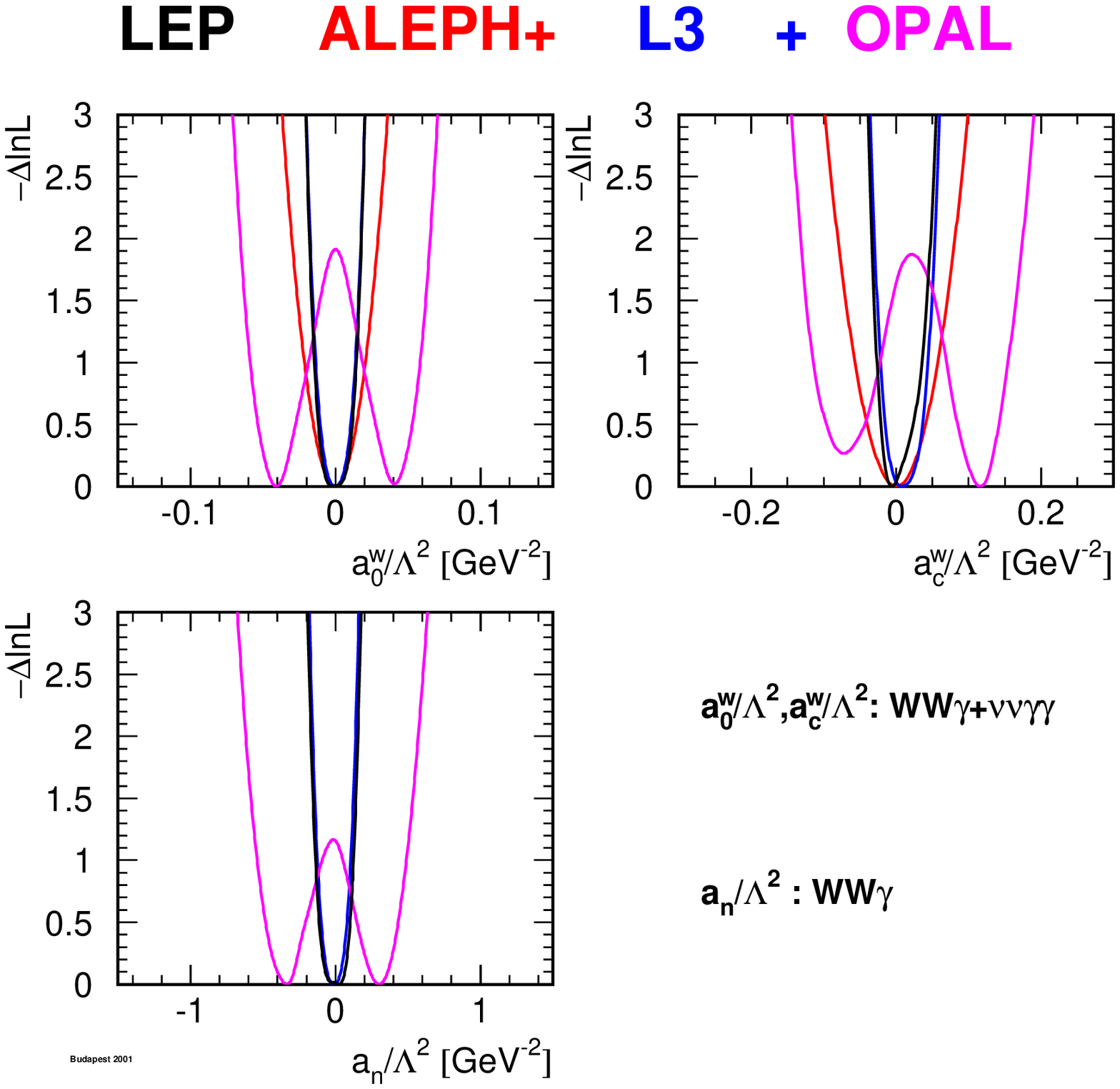}
\caption[]{
  The $\LL$ curves of ALEPH, L3 and OPAL, and the LEP combined curve
  for the QGCs $\azwl$, $\acwl$ and $\anl$.  In each
  case, the minimal value is subtracted.  }
\label{fig:gc_wQGC-1}
\end{center}
\end{figure}

The results for each experiment are shown in
Table~\ref{tab:gc_QGC-1-ADLO}, where the uncertainties include both statistical
and systematic effects.  The individual $\LL$ curves and their sum are
shown in Figures~\ref{fig:gc_wQGC-1}.  The results
of the combination are given in Table~\ref{tab:gc_QGC-1-LEP}.

\begin{table}[p]
\begin{center}
\renewcommand{\arraystretch}{1.3}
\begin{tabular}{|l||r|r|r|} 
\hline
Parameter $[\GeV^{-2}]$     & ALEPH   &  L3   & OPAL  \\
\hline
\hline
\azwl    & [$-0.029,~~+0.029$] & [$-0.017,~~+0.017$] & [$-0.065,~~+0.065$] \\ 
\hline
\acwl    & [$-0.079,~~+0.080$] &  [$-0.03,~~+0.05$]  &  [$-0.13,~~+0.17$] \\ 
\hline
\anl     &              ---    &  [$-0.15,~~+0.14$]  &  [$-0.61,~~+0.57$] \\  
\hline
\end{tabular}
\caption[]{The 95\% C.L. intervals ($\Delta\LL=1.92$) measured by
  ALEPH, L3 and OPAL.  In each case the parameter listed is varied
  while the remaining ones are fixed to their Standard Model values.
  Both statistical and systematic uncertainties are included.}
\label{tab:gc_QGC-1-ADLO}
\end{center}
\end{table}

\begin{table}[p]
\begin{center}
\renewcommand{\arraystretch}{1.3}
\begin{tabular}{|l||c|} 
\hline
Parameter $[\GeV^{-2}]$ & 95\% C.L.      \\
\hline
\hline
$\azwl$     & [$-0.018,~~+0.018$]  \\ 
\hline
$\acwl$     & [$-0.033,~~+0.047$]  \\ 
\hline
$\anl$      & [$-0.17,~~+0.15$]  \\ 
\hline
\end{tabular}
\caption[]{ The 95\% C.L. intervals ($\Delta\LL=1.92$) obtained
  combining the results from ALEPH, L3 and OPAL.  In each case the
  parameter listed is varied while the remaining ones are fixed to
  their Standard Model values.  Both statistical and systematic
  uncertainties are included.}
 \label{tab:gc_QGC-1-LEP}
\end{center}
\end{table}

\section*{Conclusions}

No significant deviation
from the Standard Model prediction is seen for any of the electroweak
gauge boson couplings studied. 

%







\clearpage

\boldmath
\chapter{W-Boson Mass and Width at \LEPII}
\label{sec-MW}
\unboldmath

\updates{ Additional preliminary results based on the data collected
  in the year 2000 are included. }

\section{W Mass Measurements}

The W boson mass results presented in this Chapter are obtained from
data recorded over a range of centre-of-mass energies,
$\roots=161-209$~\GeV, during the 1996-2000 operation of the LEP
collider. The results reported by the ALEPH, DELPHI and L3
collaborations include an analysis of the year 2000 data, and have an
integrated luminosity per experiment of about $700$~\ipb. The OPAL
collaboration has analysed the data up to and including 1999 and has
an integrated luminosity of approximately $450$~\ipb.

The results on the W mass and width quoted below correspond to a 
definition based on a Breit-Wigner denominator with an $s$-dependent width,
$|(s-\Mw^2) + i s \Gw/\Mw|$. 

Since 1996 the LEP $\epem$ collider has been operating above the
threshold for $\WW$ pair production. Initially, 10~\ipb\ of data
were recorded close to the $\WW$ pair production threshold. At this
energy the $\WW$ cross section is sensitive to the W boson mass, $\Mw$.
Table \ref{mw:tab:wmass_threshold} summarises the W mass results from the four 
LEP collaborations based 
on these data~\cite{common_bib:adloww161}.
\begin{table}[htbp]
 \begin{center}
  \begin{tabular}{|r|c|}\hline
     \multicolumn{2}{|c|}{THRESHOLD ANALYSIS~\cite{common_bib:adloww161}} \\
Experiment &   \Mw(threshold)/\GeVm     \\ \hline
   ALEPH    & $80.14\pm0.35$  \\ 
   DELPHI   & $80.40\pm0.45$  \\
   L3       & $80.80^{+0.48}_{-0.42}$  \\
   OPAL     & $80.40^{+0.46}_{-0.43}$  \\ \hline
\end{tabular}
 \caption{W mass measurements from the $\WW$ threshold cross section 
          at $\roots=161$~\GeV. The errors
          include statistical and systematic contributions.}
 \label{mw:tab:wmass_threshold}
\end{center}
\end{table} 

Subsequently LEP has operated at energies significantly above the
$\WW$ threshold, where the $\epem\rightarrow\WW$ cross section has
little sensitivity to $\Mw$. For these higher energy data $\Mw$ is
measured through the direct reconstruction of the W boson's invariant
mass from the observed jets and leptons.  Table
\ref{mw:tab:wmass_experiments} summarises the W mass results presented
individually by the four LEP experiments using the direct
reconstruction method.  The combined values of $\Mw$ from each
collaboration take into account the correlated systematic
uncertainties between the decay channels and between the different
years of data taking. In addition to the combined numbers, each
experiment presents mass measurements from $\WWqqln$ and $\WWqqqq$
channels separately.  The DELPHI and OPAL collaborations provide
results from independent fits to the data in the $\qqln$ and $\qqqq$
decay channels separately and hence account for correlations between
years but do not include correlations between the two channels. The
$\qqln$ and $\qqqq$ results quoted by the ALEPH and L3 collaborations
are obtained from a simultaneous fit to all data which, in addition to
other correlations, takes into account the correlated systematic
uncertainties between the two channels. The L3 result is unchanged
when determined through separate fits.  The large variation in the
systematic uncertainties in the $\WWqqqq$ channel are caused by
differing estimates of the possible effects of Colour Reconnection
(CR) and Bose-Einstein Correlations (BEC); this is discussed below.
The systematic errors in the $\WWqqln$ channel are dominated by
uncertainties from hadronisation, with estimates ranging from
15 to 30~$\MeVm$.

\begin{table}[htbp]
 \begin{center}
  \begin{tabular}{|r|c|c||c|}\hline
    \multicolumn{1}{|c|}{ } & \multicolumn{3}{c|}{DIRECT RECONSTRUCTION } \\
           & \WWqqln         & \WWqqqq         & Combined        \\   
Experiment & \Mw/\GeVm        & \Mw/\GeVm        & \Mw/\GeVm        \\ \hline
     ALEPH \cite{mw:bib:A-mw183,mw:bib:A-mw189,mw:bib:A-mw20x}
           & $80.456 \pm 0.051 \pm 0.032$ & $80.507 \pm 0.054 \pm 0.045$
& $80.477 \pm 0.038 \pm 0.032$ \\ 
     DELPHI \cite{common_bib:delww172,mw:bib:D-mw183,mw:bib:D-mw189,mw:bib:D-mw20X}  
           & $80.414 \pm 0.074 \pm 0.048$ & $80.384 \pm 0.053 \pm 0.065$
& $80.399 \pm 0.045 \pm 0.049$ \\ 
     L3 \cite{common_bib:ltrww172,mw:bib:L-mw183,mw:bib:L-mw189,mw:bib:L-mw19X,mw:bib:L-mw20X}      
           & $80.314 \pm 0.074 \pm 0.045$ & $80.478 \pm 0.063 \pm 0.069$
& $80.389 \pm 0.048 \pm 0.051$ \\ 
     OPAL\cite{common_bib:opaww172,mw:bib:O-mw183,mw:bib:O-mw189,mw:bib:O-mw19X,mw:bib:O-mwlvlv}
           & $80.516 \pm 0.067 \pm 0.030$ & $80.408 \pm 0.066 \pm 0.100$
& $80.491 \pm 0.053 \pm 0.038$ \\ \hline
\end{tabular}
 \caption{Preliminary W mass measurements from direct reconstruction
         ($\roots=172-209$~\GeV). The first error is statistical and the second systematic. Results are given for the
         semi-leptonic, fully-hadronic channels and the combined value.
           The $\WWqqln$ results from the 
         ALEPH and OPAL collaborations include mass information from 
         the $\WWlnln$ channel.
       }
 \label{mw:tab:wmass_experiments}
\end{center}
\end{table}

\section{Combination Procedure}
 
A combined LEP W mass measurement is obtained from the results
of the four experiments. In order to perform a reliable combination of
the measurements, a more detailed input than that given in
Table~\ref{mw:tab:wmass_experiments} is required.  Each experiment
provided a W mass measurement for both the $\WWqqln$ and $\WWqqqq$
channels for each of the data taking years (1996-2000) that it had
analysed. In addition to the four threshold measurements a total of 36
direct reconstruction measurements are supplied: ALEPH and DELPHI
provided 10 measurements (1996-2000), L3 gave 8 measurements
(1996-2000) having already combined the 1996 and 1997 results and OPAL
provided 8 measurements (1996-1999). The $\WWlnln$ channel is also
analysed by the ALEPH(1997-2000) and OPAL(1997-2000)
collaborations; the lower precision results obtained from this channel
are combined by the experiments with their $\WWqqln$ channel mass
determinations.

Subdividing the results by data-taking years enables a proper
treatment of the correlated systematic uncertainty from the LEP beam
energy and other dependences on the centre-of-mass energy or
data-taking period.  A detailed breakdown of the sources of systematic
uncertainty are provided for each result and the correlations
specified. The inter-year, inter-channel and inter-experiment
correlations are included in the combination. The main sources of
correlated systematic errors are: colour reconnection, Bose-Einstein
correlations, hadronisation, the LEP beam energy, and uncertainties
from initial and final state radiation. The full correlation matrix
for the LEP beam energy is employed\cite{mw:bib:energy}.
The combination is performed and the evaluation of the components of
the total error assessed using the Best Linear Unbiased Estimate
(BLUE) technique, see Reference~\citen{common_bib:lyons}.

\begin{figure}[tbp]
\begin{center}
 \mbox{\epsfxsize=12cm\epsffile{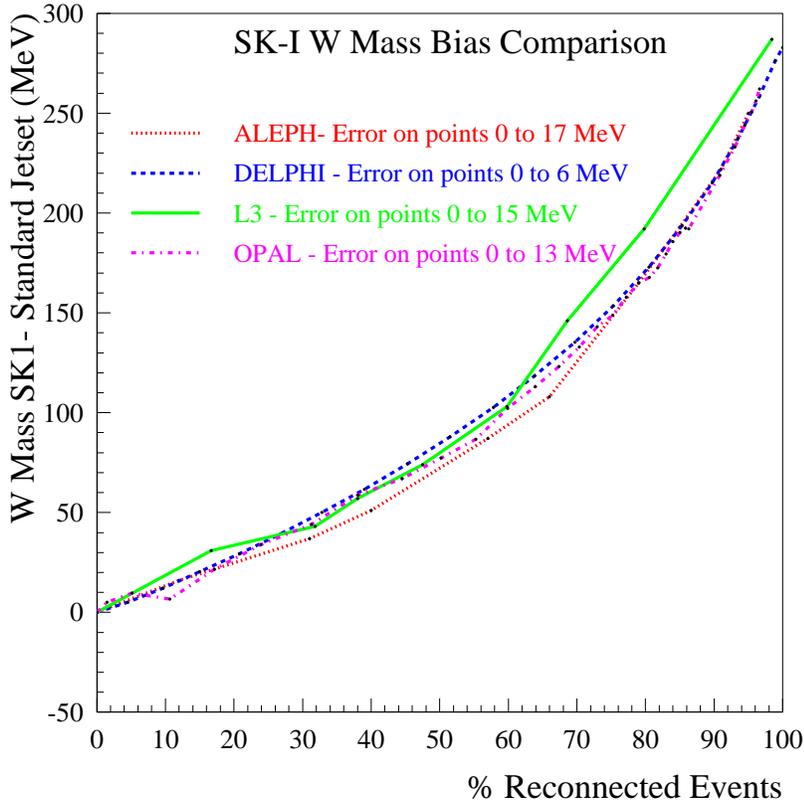}}
\caption{W mass bias obtained in the SK-I model of colour reconnection
  relative to a simulation without colour reconnection as a function
  of the fraction of events reconnected, at a centre of
  mass energy of 189 \GeV\ and for the fully-hadronic decay channel.  The
  analyses of the four LEP experiments show similar sensitivity to
  this effect. The points connected by the lines have correlated
  uncertainties increasing to the right in the range indicated.}
 \label{mw:fig:sk1}
\end{center}
\end{figure}

The four LEP collaborations gave different estimates of the systematic
errors arising from final state interactions: these varied from
30-66~$\MeVm$ for colour reconnection and from 20-67~$\MeVm$ for
Bose-Einstein correlations. This range of estimates could be due to
different experimental sensitivities to these effects or,
alternatively, simply a reflection of the different phenomenological
models used to assess the uncertainties. This question is
resolved by comparing the results of the experiments when analysing
simulation samples with and without CR effects in the SK-I model
\cite{mw:bib:ski}. Studies of these samples demonstrate that the four
experiments are equally sensitive to colour reconnection effects, {\em
  i.e.} when looking at the same CR model similar biases are seen by
all experiments. This is shown in Figure \ref{mw:fig:sk1} as a
function of the fraction of reconnected events, a reconnection
fraction of 30$\%$ of events is typically assumed by the experiments
for the assessment of systematic uncertainties.

\label{mw:sec:cr}

For this reason a common value of the CR systematic uncertainty is
used in the combination. For Bose-Einstein Correlations, no similar
test is made of the respective experimental sensitivities.
However, in the absence of evidence that the experiments have
different sensitivities to the effect, a common value of the
systematic uncertainty from BEC is assumed. In the combination a
common colour reconnection error of 40~\MeVm\ and a common
Bose-Einstein systematic uncertainty of 25~\MeVm\ are used. These
values are chosen as representative averages of the estimates of the
different LEP experiments, resulting in the same final error on $\Mw$
as obtained when using the BEC and CR estimates of the experiments.
Applying this procedure changes the value of $\Mw$ from the fit by 7
\MeVm.

\section{LEP Combined W Boson Mass }

The combined W mass from direct reconstruction is
\begin{eqnarray*}
        \Mw(\mathrm{direct}) = 80.450\pm0.026(\mathrm{stat.})\pm0.030(\mathrm{syst.})~\GeVm,
\end{eqnarray*}
with a $\chi^2$/d.o.f. of 31.1/35, corresponding to a $\chi^2$ probability 
of 66\%.
The weight of the fully-hadronic channel in the combined fit is
0.27. This reduced weight is a consequence of the relatively large
size of the current estimates of the systematic errors from 
CR and BEC. Table \ref{mw:tab:errors} gives a breakdown of the contribution
to the total error of the various sources of systematic errors. The largest
contribution to the systematic error comes from hadronisation uncertainties,
which are conservatively treated as correlated between the two channels, between experiments and between years. In the absence of systematic effects the current LEP statistical precision on $\Mw$ would be $22$~$\MeVm$: the statistical error contribution in the LEP combination is larger than this (26~$\MeVm$) due to the significantly reduced weight of the fully-hadronic channel.
\begin{table}[tbp]
 \begin{center}
  \begin{tabular}{|l|r|r||r|}\hline
       Source  &  \multicolumn{3}{|c|}{Systematic Error on \Mw\ ($\MeVm$)}  \\  
                             &  \qqln & \qqqq  & Combined  \\ \hline   
 ISR/FSR                     &  8 &  9 &  8 \\
 Hadronisation               & 19 & 17 & 17 \\
 Detector Systematics        & 12 &  8 & 10 \\
 LEP Beam Energy             & 17 & 17 & 17 \\
 Colour Reconnection         & $-$& 40 & 11 \\
 Bose-Einstein Correlations  & $-$& 25 &  7 \\
 Other                       &  4 &  4 &  3 \\ \hline
 Total Systematic            & 29 & 54 & 30 \\ \hline
 Statistical                 & 33 & 30 & 26 \\ \hline\hline
 Total                       & 44 & 62 & 40 \\ \hline
 & & & \\
 Statistical in absence of Systematics  & 32 & 29 & 22 \\ \hline

\end{tabular}
 \caption{Error decomposition for the combined LEP W mass results. 
          Detector systematics include uncertainties
          in the jet and lepton energy scales and resolution. The `Other'
          category refers to errors, all of which are uncorrelated
          between experiments, arising from: simulation statistics,
          background estimation, four-fermion treatment, fitting method 
          and event selection. The error decomposition 
          in the $\qqln$ and $\qqqq$
          channels refers to the independent fits to the results from 
          the two channels separately.}
 \label{mw:tab:errors}
\end{center}
\end{table}

In addition to the above results, the W boson mass is measured at
LEP from the 10~\ipb\ per experiment of
data recorded at threshold for W pair production:
\begin{eqnarray*}
      {\Mw(\mathrm{threshold}) = 
  80.40\pm0.20(\mathrm{stat.})\pm
          0.07(\mathrm{syst.})\pm0.03(\mathrm{E_{beam}})~\GeVm}.
\end{eqnarray*}
When the threshold measurements are combined with the much more precise results obtained from direct reconstruction one achieves a W mass measurement of 
\begin{eqnarray*}
           \Mw = 80.450\pm0.026(\mathrm{stat.})\pm0.030(\mathrm{syst.}) \GeVm.
\end{eqnarray*}
The LEP beam energy uncertainty is the only correlated systematic error source 
between the threshold and direct reconstruction measurements. 
The threshold measurements have a weight of only $0.02$ in the combined fit.
This LEP combined result is compared with the  results (threshold and direct reconstruction combined) of the four LEP experiments in Figure \ref{mw:fig:mwgw}.

\section{Consistency Checks}

The difference between the combined W boson mass measurements
obtained from the fully-hadronic and semi-leptonic channels,
$\Delta\Mw(\qqqq-\qqln)$, is determined:
\begin{eqnarray*}
 \Delta\Mw(\qqqq-\qqln) =  +9\pm44~\MeVm.    
\end{eqnarray*}

A significant non-zero value for $\Delta\Mw$ could indicate that FSI
effects are biasing the value of \Mw\ determined from \WWqqqq\ events.
Since $\Delta\Mw$ is primarily of interest as a check of the possible
effects of final state interactions, the errors from CR and BEC are
set to zero in its determination. The result is obtained from a fit
where the imposed correlations are the same as those for the results
given in the previous sections. This result is almost unchanged if the
systematic part of the error on $\Mw$ from hadronisation effects is
considered as uncorrelated between channels, although the uncertainty
increases by 16\%.
The study of the mass difference and the equivalent analysis for the W
width are not used to place limits on colour reconnection, for example
using the study of the W mass bias in the SK-I colour reconnection
model reported in Section \ref{mw:sec:cr}.  This is because only one
model is analysed there, and, taken in isolation, the results are not
sufficiently precise.

\begin{figure}[hbt]
\begin{center}
 \mbox{\epsfxsize=8.25cm\epsffile{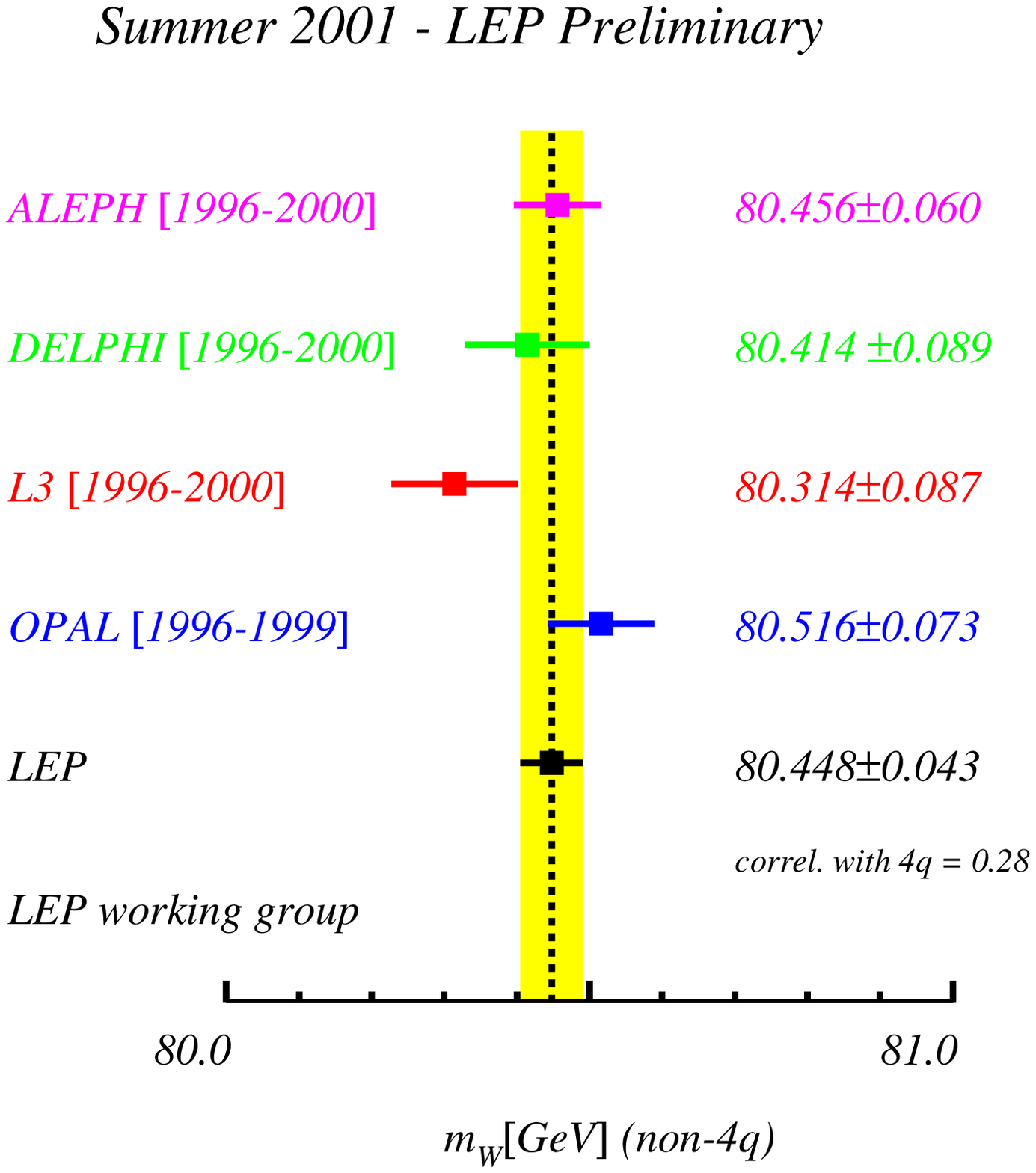} \newline
       \epsfxsize=8.25cm\epsffile{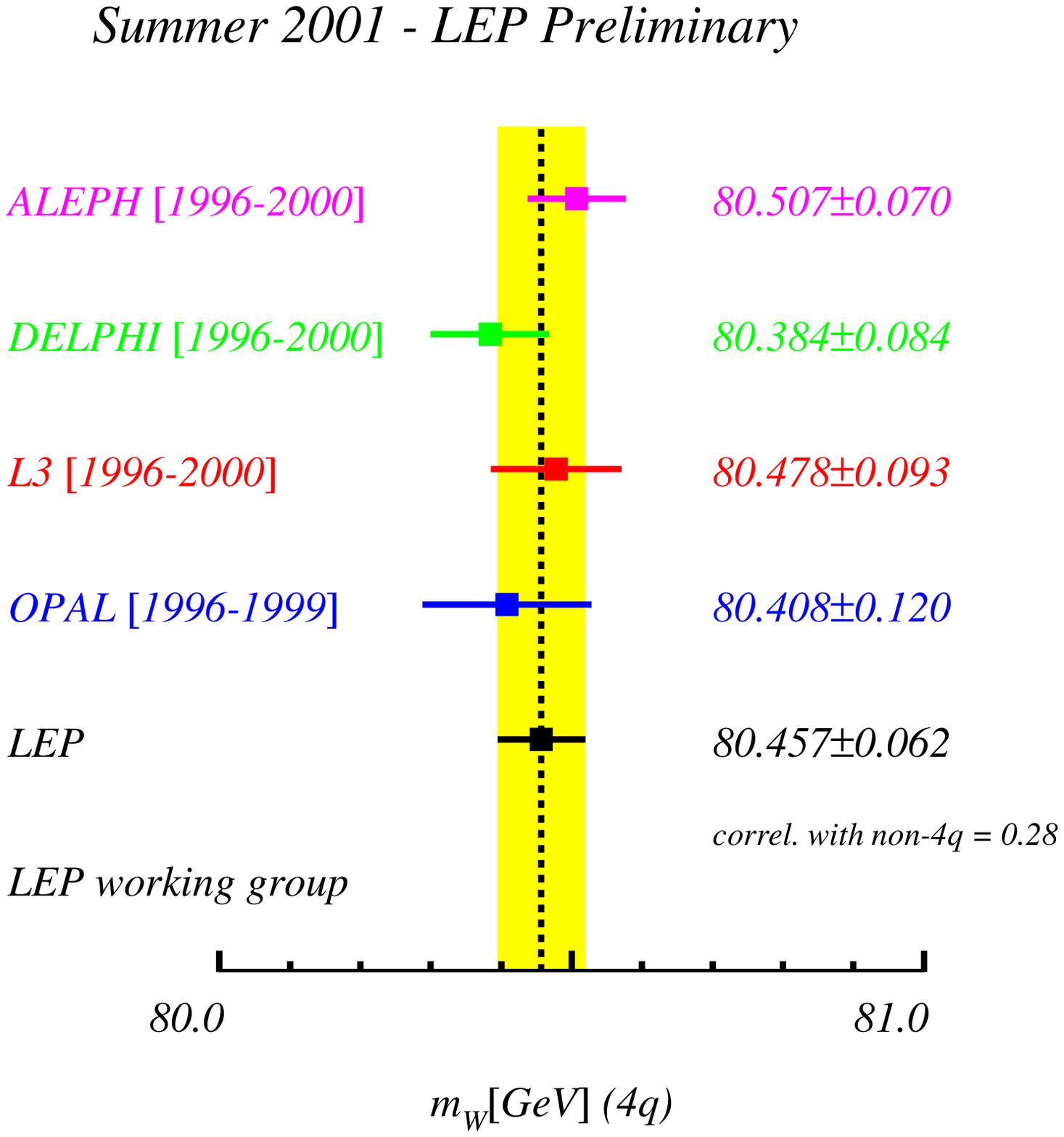}} 
\vspace*{-0.5cm}
\caption{\label{mw:fig-qqlnqqqq} 
          The W mass measurements
          from the $\WWqqln$ (left) and $\WWqqqq$ (right) channels 
          obtained by the four LEP collaborations compared to the 
          combined value. The combined values take into account 
          correlations between experiments, years and the two channels.
          In the LEP combination of 
          the $\qqqq$ results common values (see text) for the CR and BEC
          errors are used. 
          The ALEPH and L3 $\qqln$ and $\qqqq$ results are
          correlated since they are obtained from a fit to both channels 
          taking into account inter-channel correlations. }
 \end{center}
\end{figure}

The masses from the two channels obtained from this fit with the BEC and CR errors now included are:
\begin{eqnarray*}
\Mw(\WWqqln) = 80.448\pm0.033(\mathrm{stat.})\pm0.028(\mathrm{syst.})~\GeVm,\\
\Mw(\WWqqqq) = 80.457\pm0.030(\mathrm{stat.})\pm0.054(\mathrm{syst.})~\GeVm.  
\end{eqnarray*}
These two results are correlated and have a correlation coefficient of 
0.28. The value of $\chi^2$/d.o.f is 31.1/34, corresponding to a
$\chi^2$ probability of 62$\%$.  
These results and the correlation between them
can be used to combine the two measurements or to form the mass 
difference. The LEP combined results from the two channels 
are compared with those quoted by the individual experiments in 
Figure \ref{mw:fig-qqlnqqqq}. 

Experimentally, separate $\Mw$ measurements are obtained from the
$\WWqqln$ and $\WWqqqq$ channels for each of the years of data. 
The combination using only the $\qqlv$ measurements yields:
\begin{eqnarray*}
 \Mwindep(\WWqqln) = 80.448\pm0.033(\mathrm{stat.})\pm0.029(\mathrm{syst.})~\GeVm. 
\end{eqnarray*}
The  systematic error is dominated by
hadronisation uncertainties  ($\pm19$~$\MeVm$) and the 
uncertainty in the LEP beam energy  ($\pm17$~$\MeVm$).
The combination using only the $\qqqq$ measurements gives:
\begin{eqnarray*}
 \Mwindep(\WWqqqq) = 80.447\pm0.030(\mathrm{stat.})\pm0.054(\mathrm{syst.})~\GeVm.  
\end{eqnarray*}
where the dominant contributions to the systematic error arise from 
BEC/CR ($\pm47$~$\MeVm$), hadronisation ($\pm17$~$\MeVm$)
and from the uncertainty in the LEP beam energy ($\pm17$~$\MeVm$).

\section{LEP Combined W Boson Width}

\begin{figure}[hbt]
\begin{center}
 \mbox{\epsfxsize=8.25cm\epsffile{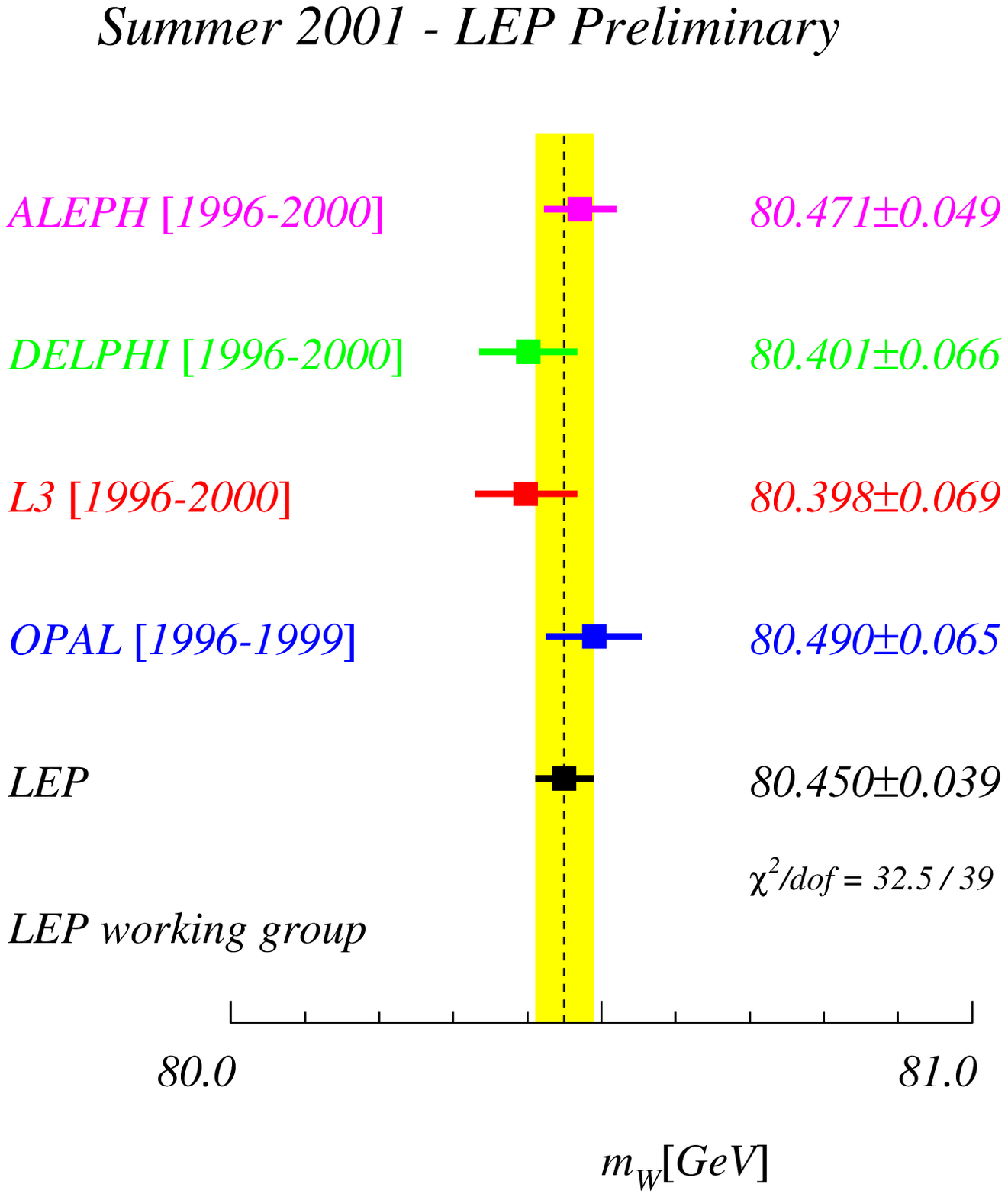}
       \epsfxsize=8.25cm\epsffile{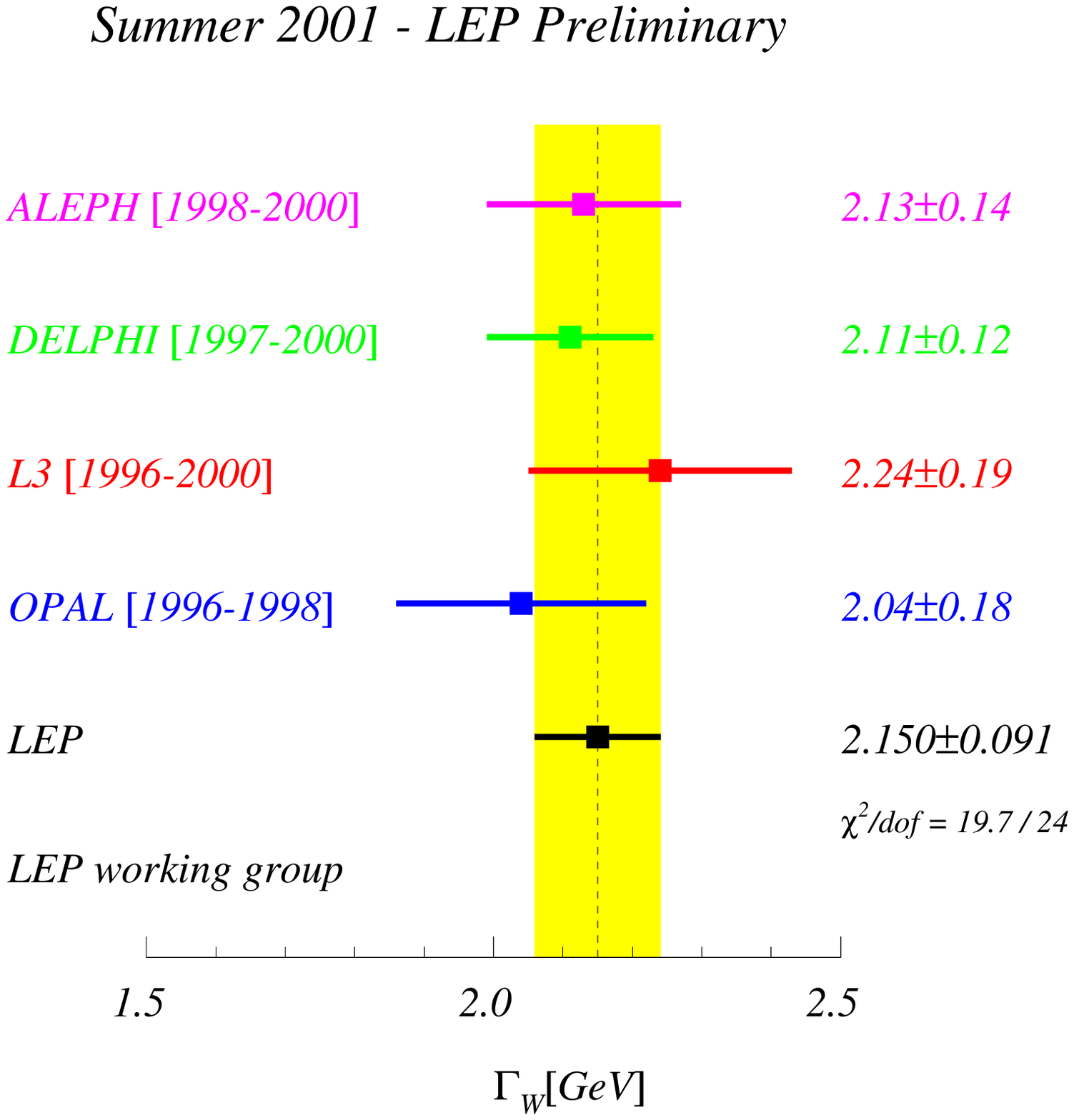}} 
\vskip-0.5cm
\caption{\label{mw:fig:mwgw} 
          The combined results for the measurements of the
          W mass (left) and W width (right) compared to the results  
          obtained by the four LEP collaborations. The combined
          values take into account correlations between experiments
          and years and hence, in general, do not give the same central 
          value as a simple average. In the LEP combination of 
          the $\qqqq$ results common values (see text) for the CR and BEC
          errors are used. The individual and combined $\Mw$ results 
          include the measurements from
          the threshold cross section.}
 \end{center}
\end{figure}

The method of direct reconstruction is also well suited to the
direct measurement of the width of the W boson. The results of the four 
LEP experiments are shown in Table \ref{mw:tab:wwidth_experiments}
and in Figure \ref{mw:fig:mwgw}.
\begin{table}[htbp]
 \begin{center}
  \begin{tabular}{|c|c|}\hline
  Experiment & \Gw\ (\GeVm)        \\ \hline
   ALEPH    & $2.13\pm0.11\pm0.09$ \\ 
   DELPHI   & $2.11\pm0.10\pm0.07$ \\
   L3       & $2.24\pm0.11\pm0.15$ \\
   OPAL     & $2.04\pm0.16\pm0.09$ \\ \hline
\end{tabular}
 \caption{Preliminary W width measurements ($\roots=172-209$~\GeV) 
         from the individual experiments. The first error is statistical
         and the second systematic.}
 \label{mw:tab:wwidth_experiments}
\end{center}
\end{table}

Each experiment provided a W width measurement for both $\WWqqln$ and
$\WWqqqq$ channels for each of the data taking years (1996-2000) that
it has analysed. A total of 25 measurements are supplied: ALEPH
provided 3 $\WWqqqq$ results (1998-2000) and two $\WWqqln$ results
(1998-1999), DELPHI 8 measurements (1997-2000), L3 8 measurements
(1996-2000) having already combined the 1996 and 1997 results and OPAL
provided 4 measurements (1996-1998) where for the first two years the
$\WWqqln$ and $\WWqqqq$ results are already combined.

A common colour reconnection error of 65 $\MeVm$ and a common
Bose-Einstein correlation error of 35 $\MeVm$ are used in the
combination.  This procedure resulted in the same error on \Gw\ as
obtained using the BEC/CR errors supplied by the experiments. The
change in the value of the width is only 2 \MeVm. 

A simultaneous fit to the results of the four LEP collaborations is
performed in the same way as for the $\Mw$ measurement. Correlated
systematic uncertainties are taken into account and the combination gives: 
\begin{eqnarray*}
      \Gw = 2.150\pm0.068(\mathrm{stat.})\pm0.060
                                     (\mathrm{syst.})~\GeVm,
\end{eqnarray*}
with a $\chi^2$/d.o.f. of 19.7/24, corresponding to a
$\chi^2$ probability of 71$\%$.


\section{Summary}

The results of the four LEP experiments on the mass and width of the W
boson are combined taking into account correlated systematic
uncertainties, giving:
\begin{eqnarray*}
       \Mw & = & 80.450\pm0.039~\GeVm, \\
       \Gw & = &  2.150\pm0.091~\GeVm.
\end{eqnarray*}

%

\boldmath
\chapter{Effective Couplings of the Neutral Weak Current}
\label{sec-eff}
\unboldmath

\updates{Effective vector and axial-vector coupling constants are
  also determined for the heavy quark flavours.}

\boldmath
\section{The Coupling Parameters $\cAf$}
\label{sec-AF}
\unboldmath

The coupling parameters $\cAf$ are defined in terms of the effective
vector and axial-vector neutral current couplings of fermions
(Equation~(\ref{eqn-cAf})).  The LEP measurements of the
forward-backward asymmetries of charged leptons (Chapter~\ref{sec-LS})
and b and c quarks (Chapter~\ref{sec-HF}) determine the products
$\Afbzf=\frac{3}{4}\cAe\cAf$ (Equation~(\ref{eqn-apol})). The LEP
measurements of the $\tau$ polarisation (Chapter~\ref{sec-TP}),
$\ptau(\cos\theta)$, determine $\cAt$ and $\cAe$ separately
(Equation~(\ref{eqn-taupol})).  Owing to polarised beams at SLC, SLD
measures the coupling parameters directly with the left-right and
forward-backward left-right asymmetries (Chapters~\ref{sec-ALR}
and~\ref{sec-HF}).

Table~\ref{tab-AF-L} shows the results for the leptonic coupling
parameter $\cAl$ from the LEP and SLD measurements, assuming lepton
universality.

\begin{table}[tbp]
\begin{center}
\renewcommand{\arraystretch}{1.15}
\begin{tabular}{|c||c|c|r|}
\hline
                   & $\cAl$            & Cumulative Average & $\chi^2$/d.o.f.\\
\hline
\hline
$\Afbzl$           & $0.1512\pm0.0042$ &                    &                \\
$\ptau $           & $0.1465\pm0.0033$ & $0.1482\pm 0.0026$ &  0.8/1         \\
\hline
$\cAl$ (SLD)       & $0.1513\pm0.0021$ & $0.1501\pm 0.0016$ &  1.6/2          \\
\hline
\end{tabular}
\end{center}
\caption[]{
  Determination of the leptonic coupling parameter $\cAl$ assuming
  lepton universality. The second column lists the $\cAl$ values
  derived from the quantities listed in the first column. The third
  column contains the cumulative averages of the $\cAl$ results up to
  and including this line.
  The $\chi^2$ per degree of freedom for the cumulative
  averages is given in the last column.  }
\label{tab-AF-L}
\end{table}
\begin{table}[tbp]
\begin{center}
\renewcommand{\arraystretch}{1.15}
\begin{tabular}{|c||c|c|c|c|}
\hline
         &    LEP                  & SLD  & LEP+SLD  & \mcc{Standard} \\
         & ($\cAl=0.1482\pm0.0026$)&      & ($\cAl=0.1501\pm0.0016$) & \mcc{Model fit}\\
\hline
\hline
$\cAb$   & $0.891\pm0.022$    & $0.922 \pm 0.020$  & $0.899\pm0.013$ & 0.935 \\
$\cAc$   & $0.615\pm0.033$    & $0.670 \pm 0.026$  & $0.645\pm0.020$ & 0.668 \\
\hline
\end{tabular}
\end{center}
\caption[]{
  Determination of the quark coupling parameters $\cAb$ and $\cAc$
  from LEP data alone (using the LEP average for $\cAl$), from SLD
  data alone, and from LEP+SLD data (using the LEP+SLD average for
  $\cAl$) assuming lepton universality.  }
\label{tab-AF-Q}
\end{table}

Using the measurements of $\cAl$ one can extract $\cAb$ and $\cAc$
from the LEP measurements of the b and c quark asymmetries.  The SLD
measurements of the left-right forward-backward asymmetries for b and
c quarks are direct determinations of $\cAb$ and $\cAc$.
Table~\ref{tab-AF-Q} shows the results on the quark coupling
parameters $\cAb$ and $\cAc$ derived from LEP measurements
(Equations~\ref{eqn-hf4}) and SLD measurements separately, and from
the combination of LEP+SLD measurements (Equation~\ref{eqn-hf6}).

\begin{figure}[htbp]
  \begin{center}
    \leavevmode
   \mbox{
    \includegraphics[width=0.49\textwidth]{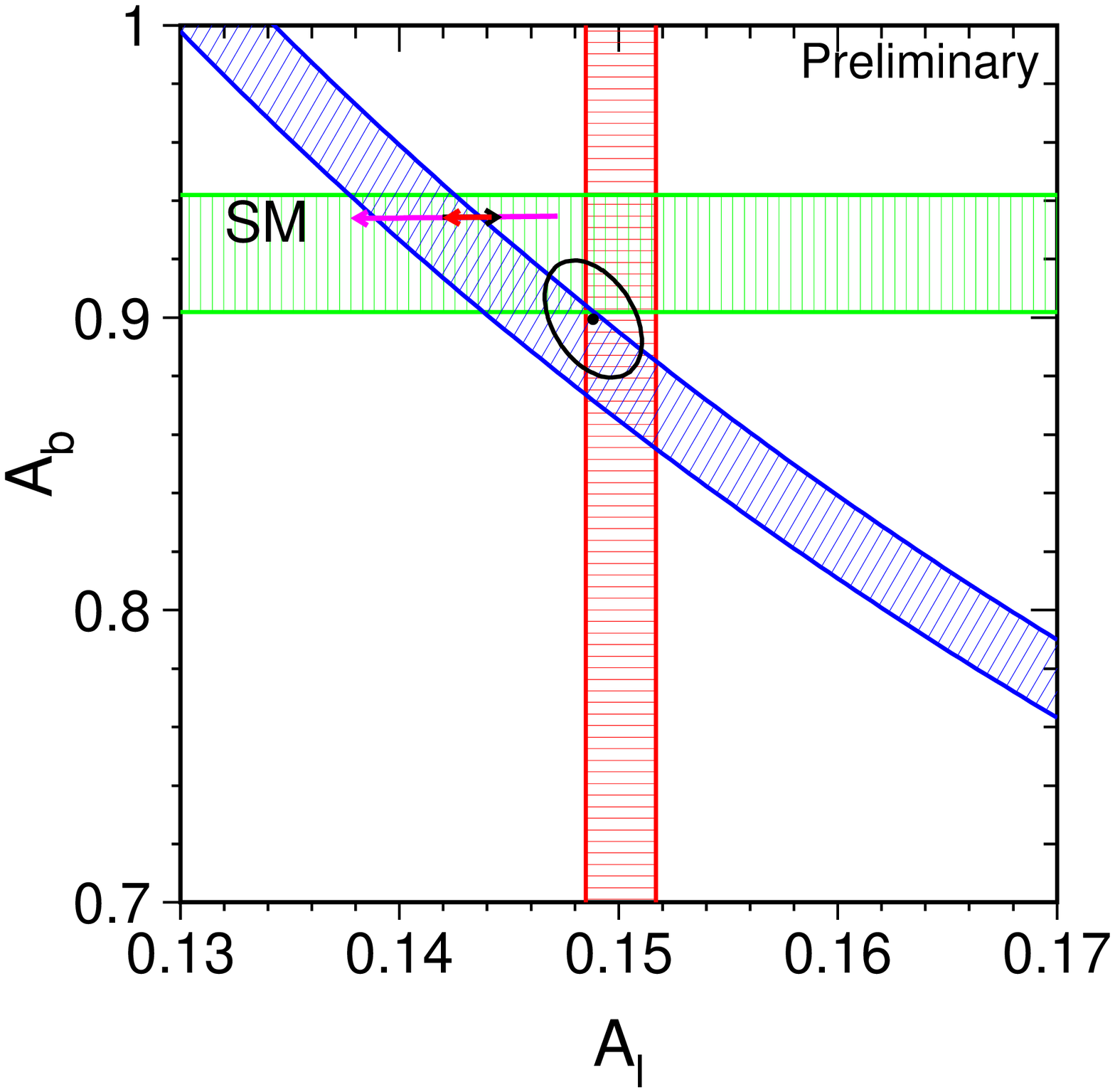}
    \includegraphics[width=0.49\textwidth]{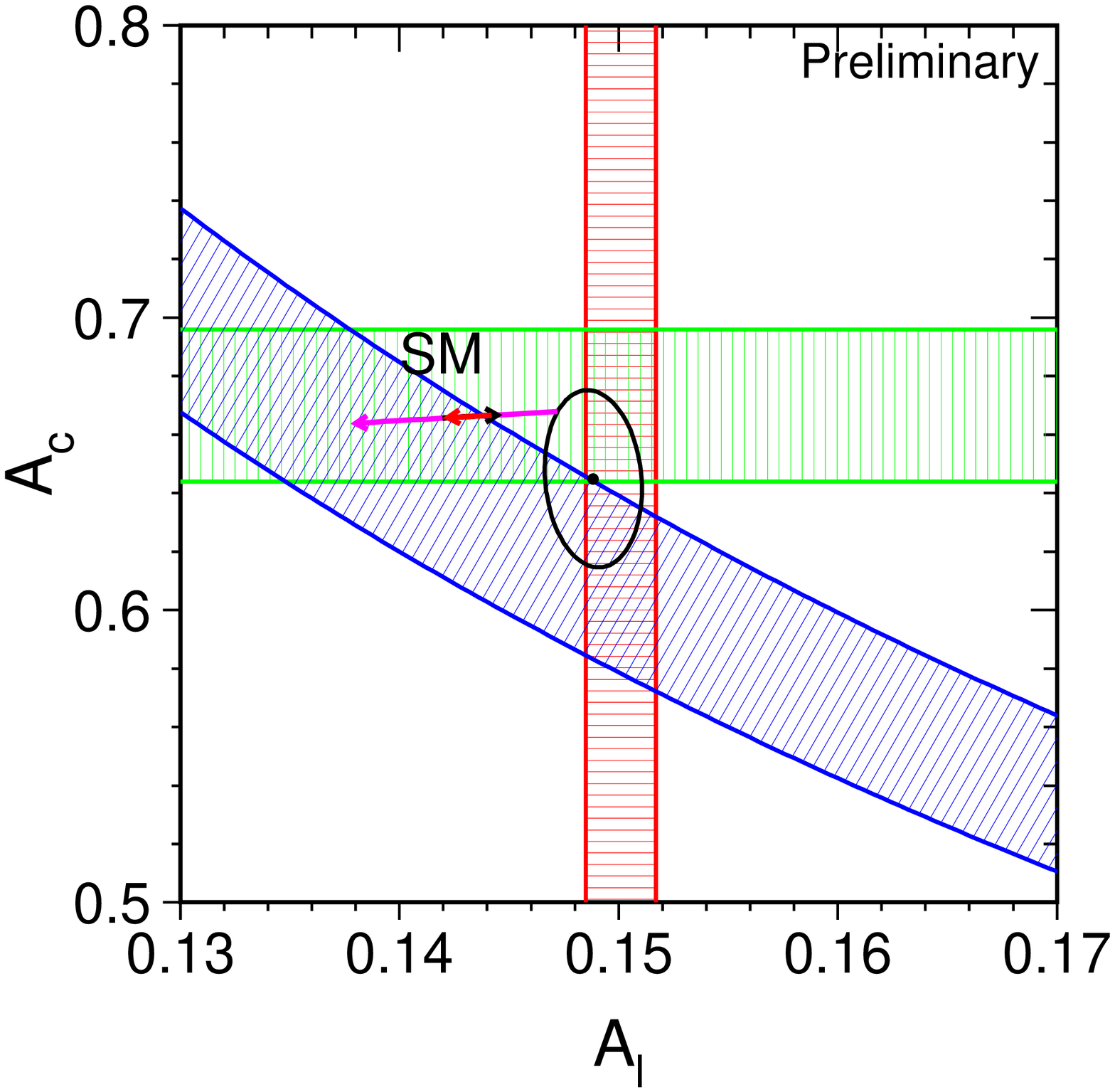}
    }
    \caption{The measurements of the combined LEP+SLD $\cAl$ (vertical
      band), SLD $\cAb$,$\cAc$ (horizontal bands) and LEP
      $\Afbzb$,$\Afbzc$ (diagonal bands), compared to the Standard
      Model expectations (arrows).  The arrow pointing to the left
      shows the variation in the $\SM$ prediction for $\MH$ in the
      range $300^{+700}_{-186}$ \GeV, and the arrow pointing to the
      right for $\Mt$ in the range $174.3 \pm 5.1$ \GeV.  Varying the
      hadronic vacuum polarisation by $\dalhad=0.02761\pm0.00036$
      yields an additional uncertainty on the Standard-Model
      prediction, oriented in direction of the Higgs-boson arrow and
      size corresponding to the top-quark arrow.  Also shown is the
      68\% confidence level contour for the two asymmetry parameters
      resulting from the joint analyses.  Although the $\Afbzb$
      measurements prefer a high Higgs mass, the Standard Model fit to
      the full set of measurements prefers a low Higgs mass, for
      example because of the influence of $\cAl$.  }
    \label{fig-ae_ab}
  \end{center}
\end{figure}

The LEP extracted values of $\cAb$ and $\cAc$ are in agreement with
the SLD measurements, but somewhat lower than the Standard Model
predictions (0.935 and 0.668, respectively, essentially independent of
$\Mt$ and $\MH$).  The combination of LEP and SLD of $\cAb$ is 2.8
sigma below the Standard Model, while $\cAc$ agrees at the 1.2 sigma
level.  This is mainly because the $\cAb$ value, deduced from the
measured $\Afbzb$ and the combined $\cAl$, is significantly lower than
both the Standard Model and the direct measurement of $\cAb$, this can
also be seen in Figure~\ref{fig-ae_ab}.

\boldmath
\section{The Effective Vector and Axial-Vector Coupling Constants}
\label{sec-GAGV}
\unboldmath

The partial widths of the $\Zzero$ into leptons and the lepton
forward-backward asymmetries (Section~\ref{sec-LS}), the $\tau$
polarisation and the $\tau$ polarisation asymmetry
(Section~\ref{sec-TP}) are combined to determine the effective
vector and axial-vector couplings for $\rm e$, $\mu$ and $\tau$. The
asymmetries (Equations~(\ref{eqn-apol}) and~(\ref{eqn-taupol}))
determine the ratio $\gvl/\gal$ (Equation~(\ref{eqn-cAf})),
while the leptonic partial widths determine the sum of the squares of
the couplings:
\begin{eqnarray}
\label{eqn-Gll}
\Gll & = &
{{\GF\MZ^3}\over{6\pi\sqrt 2}}
(\gvl^{2}+\gal^{2})(1+\delta^{QED}_\ell)\,,
\end{eqnarray}
where $\delta^{QED}_\ell=3q^2_\ell\alpha(\MZ^2)/(4\pi)$, with $q_\ell$
denoting the electric charge of the lepton, accounts for final state
photonic corrections. Corrections due to lepton masses, neglected in
Equation~\ref{eqn-Gll}, are taken into account for the results
presented below.

\begin{figure}[htbp]
\begin{center}
  \mbox{\includegraphics[width=0.9\linewidth]{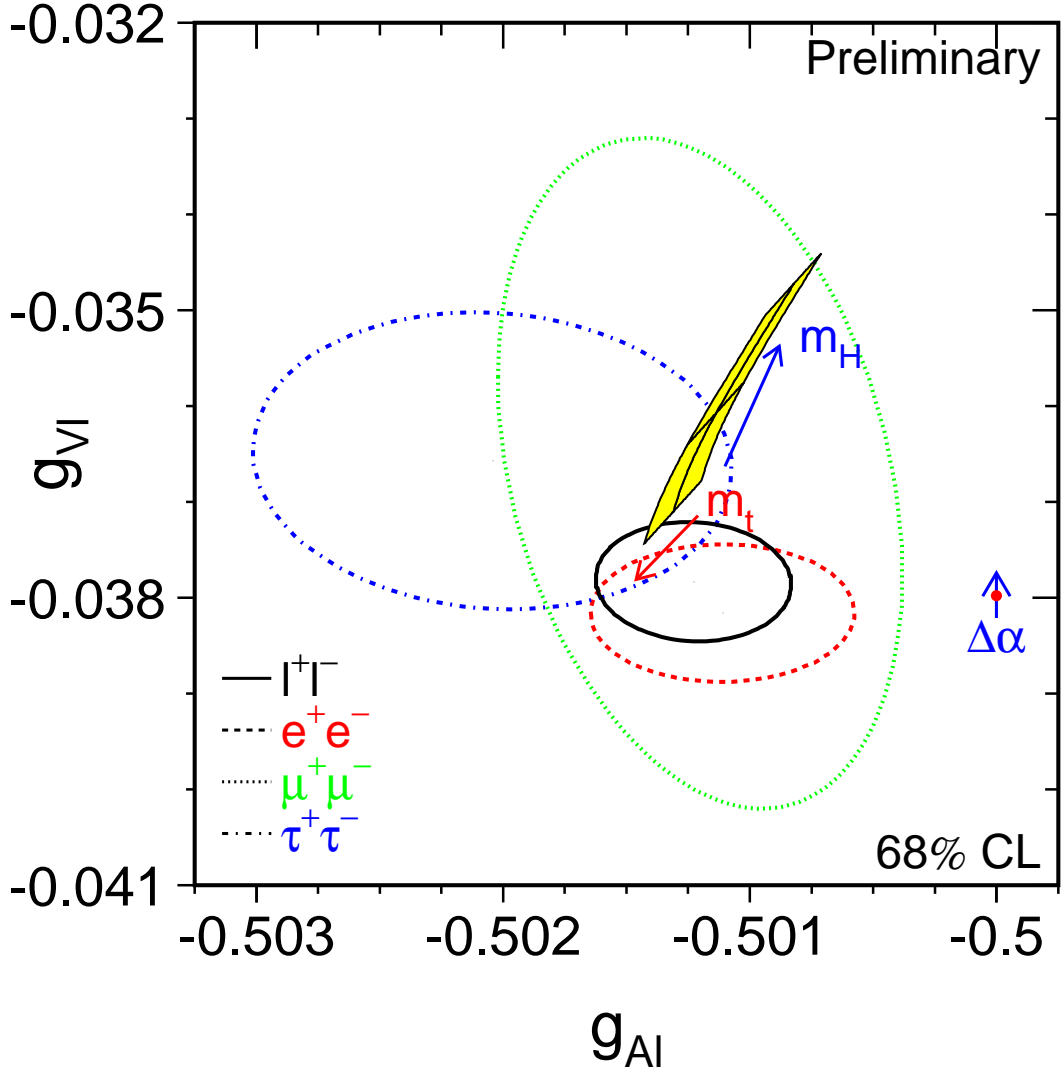}}
\end{center}
\caption[]{
  Contours of 68\% probability in the ($\gvl$,$\gal$) plane from LEP
  and SLD measurements.  The solid contour results from a fit to the
  LEP and SLD results assuming lepton universality. The shaded region
  corresponds to the Standard Model prediction for $\Mt = 174.3 \pm
  5.1$~\GeV{} and $\MH=300^{+700}_{-186}~\GeV$. The arrows point in
  the direction of increasing values of $\Mt$ and $\MH$.  Varying the
  hadronic vacuum polarisation by $\dalhad=0.02761\pm0.00036$ yields
  an additional uncertainty on the Standard-Model prediction indicated
  by the corresponding arrow.}
\label{fig-gagv}
\end{figure}

The averaged results for the effective lepton couplings are given in
Table~\ref{tab-coup} for both the LEP data alone as well as for the
LEP and SLD measurements.  Figure~\ref{fig-gagv} shows the 68\%
probability contours in the $\gal$-$\gvl$ plane for the individual
lepton species. The signs of $\gal$ and $\gvl$ are based on the
convention $\gae < 0$. With this convention the signs of the couplings
of all charged leptons follow from LEP data alone.  The measured
ratios of the $\rm e$, $\mu$ and $\tau$ couplings provide a test of
lepton universality and are shown in Table~\ref{tab-coup}.  All values
are consistent with lepton universality.  The combined results
assuming universality are also given in the table and are shown as a
solid contour in Figure~\ref{fig-gagv}.

The neutrino couplings to the $\Zzero$ can be derived from the
measured value of the invisible width of the $\Zzero$, $\Ginv$ (see
Table~\ref{tab-widths}), attributing it exclusively to the decay into
three identical neutrino generations ($\Ginv=3\Gnn$) and assuming
$\gan\equiv\gvn\equiv\gn$.  The relative sign of $\gn$ is chosen to be
in agreement with neutrino scattering data\cite{ref:CHARMIIgn},
resulting in $\gn = +0.50068\pm 0.00075$.

In addition, the couplings analysis is extended to include also the
heavy-flavour measurements as presented in Section~\ref{sec-HFSUM}.
Assuming neutral-current lepton universality, the effective coupling
constants are determined jointly for leptons as well as for b and c
quarks. QCD corrections, modifying Equation~\ref{eqn-Gll}, are taken
from the Standard Model, as is also done to obtain the quark pole
asymmetries, see Section~\ref{sec:asycorrections}.

\begin{figure}[htbp]
\begin{center}
  \mbox{\includegraphics[width=0.6\linewidth]{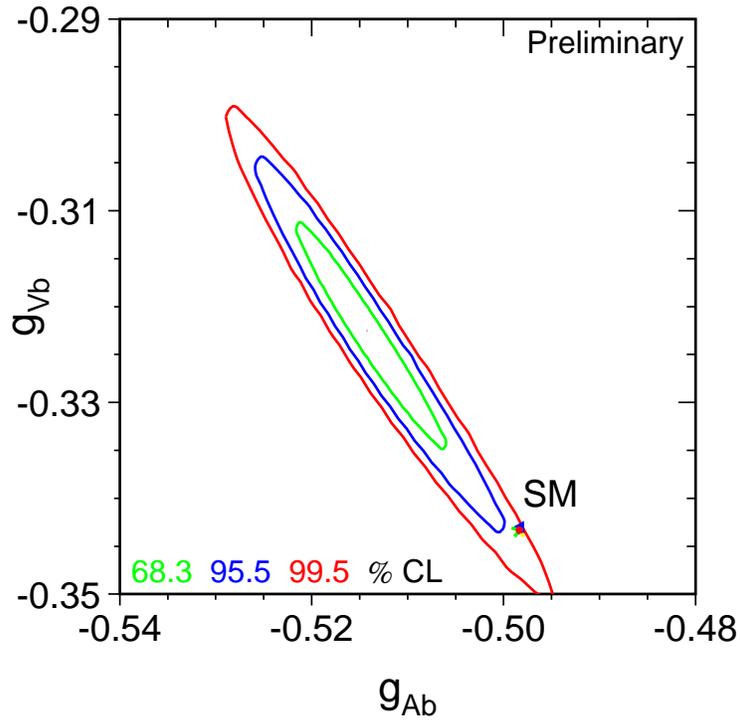}}\\
  \mbox{\includegraphics[width=0.6\linewidth]{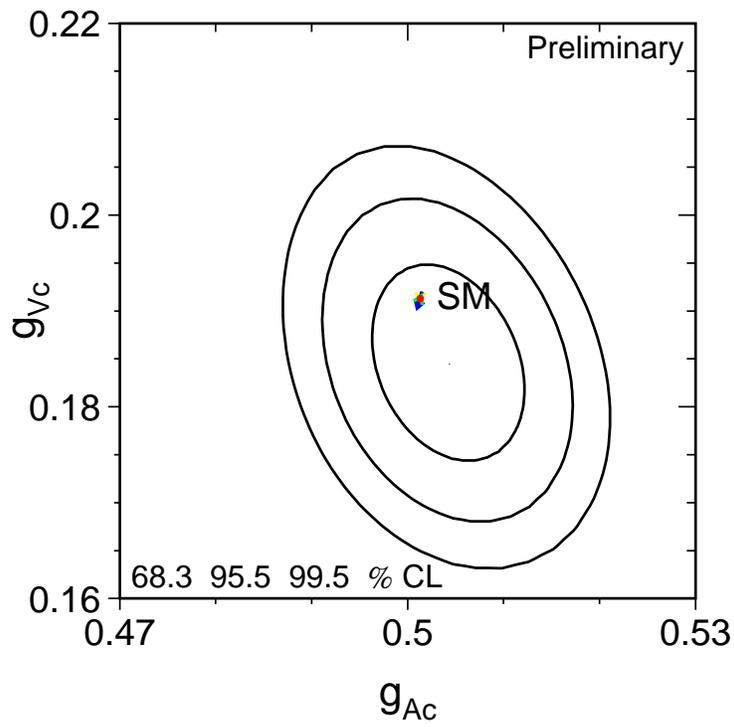}}
\end{center}
\caption[]{
  Contours of 68.3, 95.5 and 99.5\% probability in the ($\gvq$,$\gaq$)
  plane from LEP and SLD measurements for b and c quarks and assuming
  lepton universality. The dot corresponds to the Standard Model
  prediction for $\Mt = 174.3 \pm 5.1$~\GeV{},
  $\MH=300^{+700}_{-186}~\GeV$ and $\dalhad=0.02761\pm0.00036$. }
\label{fig-gaqgvq}
\end{figure}

The results are also reported in Table~\ref{tab-coup} and shown in
Figure~\ref{fig-gaqgvq}.  The deviation of the b-quark couplings from
the Standard-Model expectation is mainly caused by the combined value
of $\cAb$ being low as discussed in Section~\ref{sec-AF} and shown in
Figure~\ref{fig-ae_ab}.

\begin{table}[htbp]
\renewcommand{\arraystretch}{1.15}
\begin{center}
\begin{tabular}{|l||c|c|}
\hline
&\multicolumn{2}{|c|}{Without Lepton Universality:} \\
 & LEP & LEP+SLD\\
\hline
\hline
$\gae$    & $-0.50112 \pm 0.00035$ & $-0.50111 \pm 0.00035$ \\
$\gamu$   & $-0.50115 \pm 0.00056$ & $-0.50120 \pm 0.00054$ \\
$\gatau$  & $-0.50204 \pm 0.00064$ & $-0.50204 \pm 0.00064$ \\
$\gve$    & $-0.0378  \pm 0.0011 $ & $-0.03816 \pm 0.00047$ \\
$\gvmu$   & $-0.0376  \pm 0.0031 $ & $-0.0367  \pm 0.0023 $ \\
$\gvtau$  & $-0.0368  \pm 0.0011 $ & $-0.0366  \pm 0.0010 $ \\
\hline
\hline
&\multicolumn{2}{|c|}{Ratios of couplings:} \\
 & LEP & LEP+SLD\\
\hline
\hline
$\gamu/\gae$ & $1.0001\pm0.0014$  &$1.0002\pm0.0014$\\
$\gatau/\gae$& $1.0018\pm0.0015$  &$1.0019\pm0.0015$\\
$\gvmu/\gve$ & $0.995\pm0.096$    &$0.962\pm0.063$\\
$\gvtau/\gve$& $0.973\pm0.041$    &$0.958\pm0.029$\\
\hline
\hline
&\multicolumn{2}{|c|}{With Lepton Universality:   } \\
 & LEP & LEP+SLD\\
\hline
\hline
$\gal$    & $-0.50126 \pm 0.00026$& $-0.50123 \pm 0.00026$\\
$\gvl$    & $-0.03736 \pm 0.00066$& $-0.03783 \pm 0.00041$\\
\hline
$\gn$     & $+0.50068 \pm 0.00075$& $+0.50068 \pm 0.00075$\\
\hline
\hline
&\multicolumn{2}{|c|}{With Lepton Universality   } \\
&\multicolumn{2}{|c|}{and Heavy Flavour Results: } \\
 & LEP & LEP+SLD\\
\hline
\hline
$\gal$    & $-0.50126 \pm 0.00026$ & $-0.50125 \pm 0.00026$ \\
$\gab$    & $-0.5179  \pm 0.0078 $ & $-0.5146  \pm 0.0051 $ \\
$\gac$    & $+0.5032  \pm 0.0079 $ & $+0.5043  \pm 0.0052 $ \\
$\gvl$    & $-0.03736 \pm 0.00066$ & $-0.03751 \pm 0.00037$ \\
$\gvb$    & $-0.317   \pm 0.012  $ & $-0.3221  \pm 0.0077 $ \\
$\gvc$    & $+0.173   \pm 0.011  $ & $+0.1843  \pm 0.0067 $ \\
\hline
\end{tabular}
\end{center}
\caption[]{%
  Results for the effective vector and axial-vector couplings derived
  from the LEP data and the combined LEP and SLD data without and with 
  the assumption of lepton universality. Note that the results, in
  particular for b quarks, are highly correlated.}
\label{tab-coup}
\end{table}

\clearpage

\boldmath
\section{The Leptonic Effective Electroweak Mixing Angle $\swsqeffl$}
\label{sec-SW}
\unboldmath

The asymmetry measurements from LEP 
and SLD can be combined into a single
parameter, the effective electroweak mixing angle, $\swsqeffl$,
defined as:
\begin{eqnarray}
\label{eqn-sw}
\swsqeffl & \equiv &
\frac{1}{4}\left(1-\frac{\gvl}{\gal}\right)\,,
\end{eqnarray}
without making strong model-specific assumptions.

\begin{figure}[p]
  \begin{center}
    \leavevmode
    \includegraphics[width=0.9\linewidth]{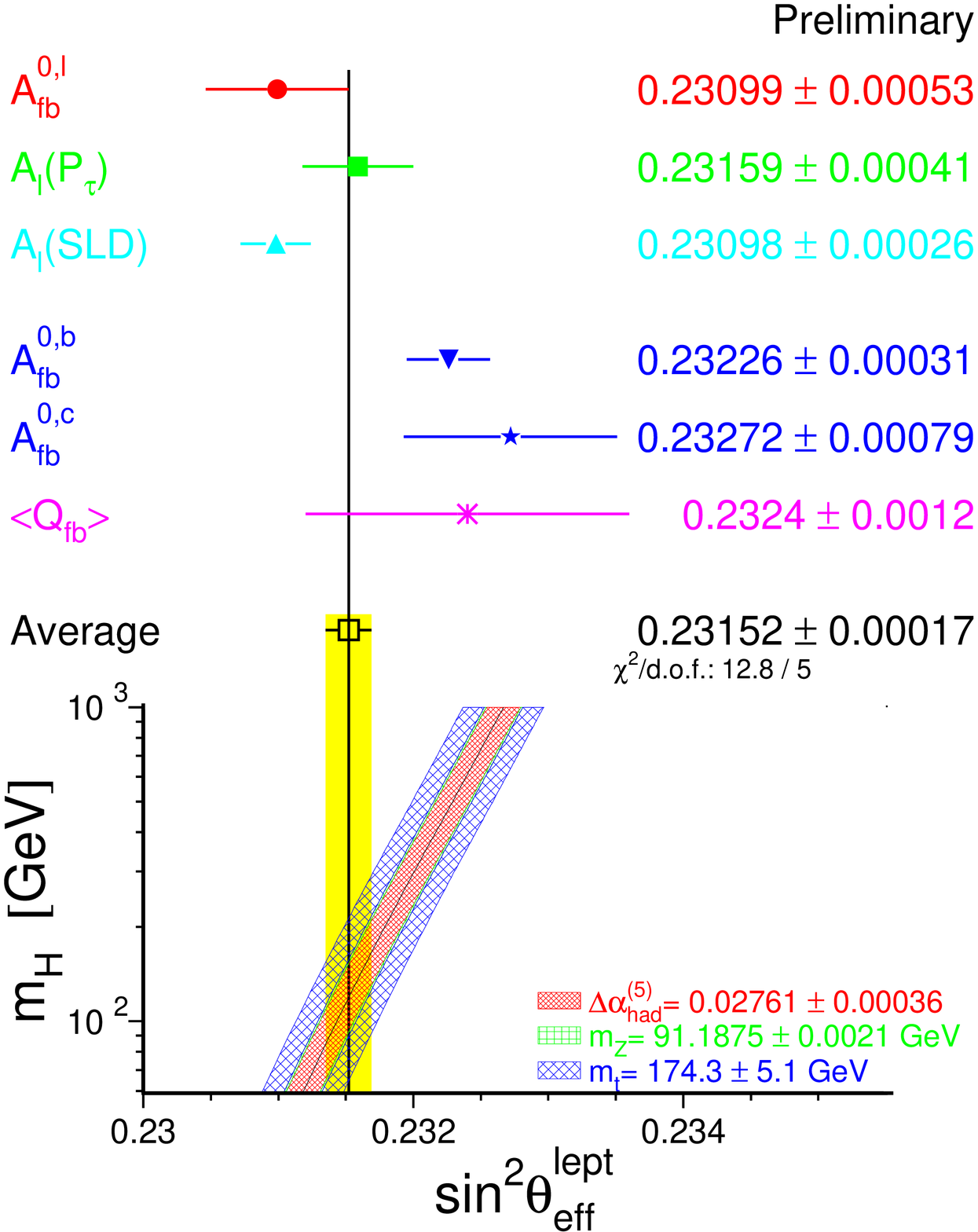}
  \end{center}
  \caption[]{
    Comparison of several determinations of $\swsqeffl$ from 
    asymmetries.  In the average, the small correlation between
    $\Afbzb$ and $\Afbzc$ is included.
    Also shown is the prediction of the Standard Model
    as a function of $\MH$.  The width of the Standard Model band is
    due to the uncertainties in
    $\Delta\alpha_{\mathrm{had}}^{(5)}(\MZ^2)$ (see Chapter~\ref{sec-MSM}),
    $\MZ$ and $\Mt$.
    The total width of the band is the linear sum of these effects.}
  \label{fig-swsq}
\end{figure}

For a combined average of $\swsqeffl$ from $\Afbzl$, $\cAt$ and $\cAe$
only the assumption of lepton universality, already inherent in the
definition of $\swsqeffl$, is needed.  Also the value derived from the
measurements of $\cAl$ from SLD is given.  We also include the
hadronic forward-backward asymmetries, assuming the difference between
$\swsqefff$ for quarks and leptons to be given by the Standard Model.
This is justified within the Standard Model as the hadronic
asymmetries $\Afbzb$ and $\Afbzc$ have a reduced sensitivity to the
small non-universal corrections specific to the quark vertex.  The
results of these determinations of $\swsqeffl$ and their combination
are shown in Table~\ref{tab-swsq} and in Figure~\ref{fig-swsq}.  The
combinations based on the leptonic results plus $\cAl$(SLD) and on the
hadronic forward-backward asymmetries differ by 3.3 standard
deviations, mainly caused by the two most precise measurements of
$\swsqeffl$, $\cAl$ (SLD) dominated by $\ALRz$, and $\Afbzb$ (LEP).
This is the same effect as discussed already in sections~\ref{sec-AF}
and~\ref{sec-GAGV} and shown in Figures~\ref{fig-ae_ab}
and~\ref{fig-gaqgvq}: the deviation in $\cAb$ as extracted from
$\Afbzb$ discussed above is reflected in the value of $\swsqeffl$
extracted from $\Afbzb$ in this analysis.

\begin{table}[htbp]
\renewcommand{\arraystretch}{1.25}
\begin{center}
\begin{tabular}{|l||c|c|c|c|}
\hline
     & $\swsqeffl$&\mco{Average by Group}&Cumulative &       \\
     &            &\mco{of Observations} &Average    &$\chi^2$/d.o.f.\\
\hline
\hline
$\Afbzl$         & $0.23099\pm 0.00053$ &&& \\
$\cAl~(\ptau)$   & $0.23159\pm 0.00041$ &$0.23137\pm0.00033$
                                 &                   &     \\
\hline
$\cAl$ (SLD)     &$0.23098\pm0.00026$ &                   
                                 &$0.23113\pm0.00021$&0.8/1\\
\hline
$\Afbzb$  & $0.23226\pm 0.00031$ &&& \\
$\Afbzc$  & $0.23272\pm 0.00079$ &&& \\
$\avQfb$  & $0.2324 \pm 0.0012 $ &$0.23230\pm0.00029$
                                 &$0.23152\pm0.00017$&12.8/5\\
\hline
\end{tabular}\end{center}
\caption[]{
  Determinations of $\swsqeffl$ from asymmetries.  
  The second
  column lists the $\swsqeffl$ values derived from the quantities
  listed in the first column. The third column contains the averages
  of these numbers by groups of observations, where the groups are
  separated by the horizontal lines. The fourth column shows the
  cumulative averages. The $\chi^2$ per degree of freedom for the
  cumulative averages is also given. The averages are performed
  including the small correlation between $\Afbzb$ and 
  $\Afbzc$. }
\label{tab-swsq}
\end{table}

\boldmath
\chapter{Constraints on the Standard Model}
\label{sec-MSM}
\unboldmath

\updates{A new determination of the hadronic vacuum polarisation is
  used. For the first time, the atomic parity violation parameter
  measured in cesium is included in the analysis.}

\section{Introduction}

The precise electroweak measurements performed at LEP and SLC and
elsewhere can be used to check the validity of the Standard Model and,
within its framework, to infer valuable information about its
fundamental parameters. The accuracy of the measurements makes them
sensitive to the mass of the top quark $\Mt$, and to the mass of the
Higgs boson $\MH$ through loop corrections. While the leading $\Mt$
dependence is quadratic, the leading $\MH$ dependence is logarithmic.
Therefore, the inferred constraints on $\MH$ are much weaker than
those on $\Mt$. 

\section{Measurements}

The LEP and SLD measurements used are summarised in
Table~\ref{tab-SMIN} together with the results of the Standard Model
fit.  Also shown are the results of measurements of $\MW$ from
UA2\cite{bib-UA2MW}, CDF\cite{bib-CDFMW1,bib-CDFMW2}, and
D\O\cite{bib-D0MW}\footnote{See Reference~\citen{bib-MWAVE-00} for a
combination of these $\MW$ measurements.}, measurements of the top
quark mass by CDF\cite{bib-topCDF} and D\O\cite{bib-topD0}\footnote{
See Reference~\citen{bib-Tevatop} for a combination of these $\Mt$
measurements.}, measurements of the neutrino-nucleon neutral to
charged current ratios from CCFR\cite{bib-CCFRnn} and
NuTeV\cite{bib-NuTeV}, and measurements of atomic parity violation in
cesium\cite{QWCs:exp:1, QWCs:exp:2} with the numerical result taken
from\cite{QWCs:theo:2000,QWCs:theo:2001}.
Although the combined preliminary\footnote{The final NuTeV
  result~\cite{bib-NuTeV-final} is not used in this report as it was
  published only after the 2001 summer conferences.} $\nu{\cal N}$ result is
quoted in terms of $\swsq=1-\MW^2/\MZ^2$, radiative corrections result
in small $\Mt$ and $\MH$ dependences\footnote{The formula used is
  $\delta\sin^2\theta_W = -0.00142 \frac{\Mt^2 -
    (175\GeV)^2}{(100\GeV)^2} + 0.00048 \ln(\frac{\MH}{150\GeV}).$ See
  Reference \citen{bib-NuTeV} for details.}  that are included in the
fit.
An additional input parameter, not shown in the table, is the Fermi
constant $G_F$, determined from the $\mu$ lifetime, $G_F = 1.16637(1)
\cdot 10^{-5} \GeV^{-2}$\cite{bib-Gmu}.  The relative error of $G_F$
is comparable to that of $\MZ$; both errors have negligible effects on
the fit results.

\begin{table}[p]
\begin{center}
\renewcommand{\arraystretch}{1.10}
\begin{tabular}{|ll||r|r|r|r|}
\hline
 && \mcc{Measurement with}  &\mcc{Systematic} & \mcc{Standard} & \mcc{Pull} \\
 && \mcc{Total Error}       &\mcc{Error}      & \mcc{Model fit}&            \\
\hline
\hline
&&&&& \\[-3mm]
& $\Delta\alpha^{(5)}_{\mathrm{had}}(\MZ^2)$\cite{bib-BP01}
                & $0.02761 \pm 0.00036$ & 0.00035 &0.02774& $-0.3$ \\
&&&&& \\[-3mm]
\hline
a) & \underline{LEP}     &&&& \\
   & line-shape and      &&&& \\
   & lepton asymmetries: &&&& \\
&$\MZ$ [\GeV{}] & $91.1875\pm0.0021\pz$
                & ${}^{(a)}$0.0017$\pz$ &91.1874$\pz$ & $ 0.0$ \\
&$\GZ$ [\GeV{}] & $2.4952 \pm0.0023\pz$
                & ${}^{(a)}$0.0012$\pz$ & 2.4963$\pz$ & $-0.5$ \\
&$\shad$ [nb]   & $41.540 \pm0.037\pzz$ 
                & ${}^{(b)}$0.028$\pzz$ &41.481$\pzz$ & $ 1.6$ \\
&$\RZ$          & $20.767 \pm0.025\pzz$ 
                & ${}^{(b)}$0.007$\pzz$ &20.739$\pzz$ & $ 1.1$ \\
&$\Afbzl$       & $0.0171 \pm0.0010\pz$ 
                & ${}^{(b)}$0.0003\pz & 0.0165\pz     & $ 0.7$ \\
&+ correlation matrix Table~\ref{tab-zparavg} &&&& \\
&                                             &&&& \\[-3mm]
&$\tau$ polarisation:                         &&&& \\
&$\cAl~(\ptau)$ & $0.1465\pm 0.0033\pz$ 
                & 0.0016$\pz$ & 0.1483$\pz$ & $-0.5$ \\
                      &                       &&&& \\[-3mm]
&$\qq$ charge asymmetry:                      &&&& \\
&$\swsqeffl$
($\avQfb$)      & $0.2324\pm0.0012\pz$ 
                & 0.0010$\pz$ & 0.2314$\pz$ & $ 0.9$ \\
&                                             &&&& \\[-3mm]
&$\MW$ [\GeV{}] & $80.450 \pm 0.039 \pzz$
                & 0.030$\pzz$ &80.398$\pzz$ & $ 1.3$ \\
                                              &&&&& \\[-3mm]
\hline
b) & \underline{SLD}\cite{ref:sld-s99} &&&& \\
&$\cAl$ (SLD)   & $0.1513\pm 0.0021\pz$ 
                & 0.0010$\pz$ & 0.1483$\pz$ & $ 1.5$ \\
&&&&& \\[-3mm]
\hline
c) & \underline{LEP and SLD Heavy Flavour} &&&& \\
&$\Rbz{}$        & $0.21646\pm0.00065$  
                 & 0.00053     & 0.215743    & $ 1.1$ \\
&$\Rcz{}$        & $0.1719\pm0.0031\pz$
                 & 0.0022$\pz$ & 0.1723$\pz$ & $-0.1$ \\
&$\Afbzb{}$      & $0.0990\pm0.0017\pz$
                 & 0.0009$\pz$ & 0.1039$\pz$ & $-2.9$ \\
&$\Afbzc{}$      & $0.0685\pm0.0034\pz$
                 & 0.0017$\pz$ & 0.0743$\pz$ & $-1.7$ \\
&$\cAb$          & $0.922\pm 0.020\pzz$
                 & 0.016$\pzz$ & 0.935$\pzz$ & $-0.6$ \\
&$\cAc$          & $0.670\pm 0.026\pzz$
                 & 0.016$\pzz$ & 0.668$\pzz$ & $ 0.1$ \\
&+ correlation matrix Table~\ref{tab:14parcor} &&&& \\
&                                              &&&& \\[-3mm]
\hline
d) & \underline{$\pp$ and $\nu$N} &&&& \\
&$\MW$ [\GeV{}] ($\pp$\cite{bib-MWAVE-00})
& $80.454 \pm 0.060\pzz$ & 0.050$\pzz$   & 80.398$\pzz$ & $ 0.9$ \\
&$\swsq$ ($\nu{\cal N}$\cite{bib-CCFRnn,bib-NuTeV})
& $0.2255\pm0.0021\pz$   & 0.0010$\pz$   & 0.2226$\pz$  & $ 1.2$ \\
&$\Mt$ [\GeV{}] ($\pp$\cite{bib-Tevatop})
& $174.3\pm 5.1\pzz\pzz$ & 4.0$\pzz\pzz$ & 175.8$\pzz\pzz$ & $-0.3$ \\
&$\QWCs$~\cite{QWCs:theo:2001}
& $-72.5\pm 0.7\pzz\pzz$ & 0.6$\pzz\pzz$ & $-72.9\pzz\pzz$ & $ 0.6$ \\
\hline
\end{tabular}\end{center}
\caption[]{
  Summary of measurements included in the combined analysis of
  Standard Model parameters. Section~a) summarises LEP averages,
  Section~b) SLD results ($\swsqeffl$ includes $\ALR$ and the
  polarised lepton asymmetries), Section~c) the LEP and SLD heavy
  flavour results and Section~d) electroweak measurements from $\pp$
  colliders and $\nu$N scattering.  The total errors in column 2
  include the systematic errors listed in column 3.  Although the systematic
  errors include both correlated and uncorrelated sources, the determination 
  of the systematic part of each error is approximate.  The $\SM$
  results in column~4 and the pulls (difference between measurement
  and fit in units of the total measurement error) in column~5 are
  derived from the Standard Model fit including all data
  (Table~\ref{tab-BIGFIT}, column~5) with the Higgs mass treated as a
  free parameter.\\
  $^{(a)}$\small{The systematic errors on $\MZ$ and $\GZ$ contain the
    errors arising from the uncertainties in the LEP energy only.}\\
  $^{(b)}$\small{Only common systematic errors are indicated.}\\
}
\label{tab-SMIN}
\end{table}

\section{Theoretical and Parametric Uncertainties}

Detailed studies of the theoretical uncertainties in the Standard
Model predictions due to missing higher-order electroweak corrections
and their interplay with QCD corrections are carried out in the
working group on `Precision calculations for the $\Zzero$
resonance'\cite{bib-PCLI}, and more recently in~\cite{bib-PCP99}.
Theoretical uncertainties are evaluated by comparing different but,
within our present knowledge, equivalent treatments of aspects such as
resummation techniques, momentum transfer scales for vertex
corrections and factorisation schemes.  The effects of these
theoretical uncertainties are reduced by the inclusion of
higher-order corrections\cite{bib-twoloop,bib-QCDEW} in the
electroweak libraries\cite{bib-SMNEW}.  

The recently calculated complete fermionic two-loop corrections on
$\MW$~\cite{FHWW-f2l-MW} are currently only used in the determination
of the theoretical uncertainty.  Their effect on $\MW$ is small
compared to the current experimental uncertainty on $\MW$, however,
the naive propagation of this new $\MW$ to
$\swsqeffl=\kappa(1-\MW^2/\MZ^2)$, keeping the electroweak form-factor
$\kappa$ unmodified, shows a more visible effect as $\swsqeffl$ is
measured very precisely.  Thus the corresponding calculations for
$\swsqeffl$ (or $\kappa$) and for the partial Z widths are urgently
needed; in particular since partial cancellations of these new
corrections in the product $\kappa(1-\MW^2/\MZ^2)=\swsqeffl$ are
expected~\cite{BGPWprivate}.

The use of the new QCD corrections\cite{bib-QCDEW} increases the value
of $\alfmz$ by 0.001, as expected.  The effects of missing
higher-order QCD corrections on $\alfmz$ covers missing higher-order
electroweak corrections and uncertainties in the interplay of
electroweak and QCD corrections and is estimated to be at least
0.002~\cite{bib-SMALFAS}.  A discussion of theoretical uncertainties
in the determination of $\alfas$ can be found in
References~\citen{bib-PCLI} and~\citen{bib-SMALFAS}.  The
determination of the size of remaining theoretical uncertainties is
under continued study.  

The theoretical errors discussed above are not included in the results
presented in Table~\ref{tab-BIGFIT}.  At present the impact of
theoretical uncertainties on the determination of $\SM$ parameters
from the precise electroweak measurements is small compared to the
error due to the uncertainty in the value of $\alpha(\MZ^2)$, which is
included in the results.

The uncertainty in $\alpha(\MZ^2)$ arises from the contribution of
light quarks to the photon vacuum polarisation
($\Delta\alpha_{\mathrm{had}}^{(5)}(\MZ^2)$):
\begin{equation}
\alpha(\MZ^2) = \frac{\alpha(0)}%
   {1 - \Delta\alpha_\ell(\MZ^2) -
   \Delta\alpha_{\mathrm{had}}^{(5)}(\MZ^2) -
   \Delta\alpha_{\mathrm{top}}(\MZ^2)} \,,
\end{equation}
where $\alpha(0)=1/137.036$.  The top contribution, $-0.00007(1)$,
depends on the mass of the top quark, and is therefore determined
inside the electroweak libraries\cite{bib-SMNEW}.  The leptonic
contribution is calculated to third order\cite{bib-alphalept} to be
$0.03150$, with negligible uncertainty.

For the hadronic contribution, we no longer use the value $0.02804 \pm
0.00065$\cite{bib-JEG2}, but rather the new evaluation
$0.02761\pm0.0036$~\cite{bib-BP01} which takes into account the
recently published results on electron-positron annihilations into
hadrons at low centre-of-mass energies by the BES
collaboration~\cite{BES_01}.  This reduced uncertainty still causes an
error of 0.00013 on the $\SM$ prediction of $\swsqeffl$, and errors of
0.2~\GeV{} and 0.1 on the fitted values of $\Mt$ and $\log(\MH)$, all
included in the results presented below.  The effect on the $\SM$
prediction for $\Gll$ is negligible.  The $\alfmz$ values for the
$\SM$ fits presented in this Section are stable against a variation of
$\alpha(\MZ^2)$ in the interval quoted.  

There are also several evaluations of
$\Delta\alpha^{(5)}_{\mathrm{had}}(\MZ^2)$%
\cite{bib-Swartz,bib-Zeppe,bib-Alemany,bib-Davier,bib-alphaKuhn,bib-Erler,bib-ADMartin,bib-jeger99}
which are more theory-driven.  One of the most recent of these
(Reference \citen{bib-ADMartin}) also includes the new results from
BES, yielding $0.02738\pm0.00020$.  To show the effects of the
uncertainty of $\alpha(\MZ^2)$, we also use this evaluation of the
hadronic vacuum polarisation.  Note that all these evaluations obtain
values for $\Delta\alpha^{(5)}_{\mathrm{had}}(\MZ^2)$ consistently
lower than - but still in agreement with - the old value of $0.02804
\pm 0.00065$.

\section{Selected Results}

Figure~\ref{fig-gllsef} shows a comparison of the leptonic partial
width from LEP (Table~\ref{tab-widths}) and the effective electroweak
mixing angle from asymmetries measured at LEP and SLD
(Table~\ref{tab-swsq}), with the Standard Model. Good agreement with
the $\SM$ prediction is observed.  The point with the arrow shows the
prediction if among the electroweak radiative corrections only the
photon vacuum polarisation is included, which shows an that
LEP+SLD data are sensitive to non-trivial electroweak corrections.
Note that the error due to the uncertainty on $\alpha(\MZ^2)$ (shown
as the length of the arrow) is not much smaller than the experimental
error on $\swsqeffl$ from LEP and SLD.  This underlines the continued
importance of a precise measurement of
$\sigma(\mathrm{e^+e^-\rightarrow hadrons})$ at low centre-of-mass
energies.

Of the measurements given in Table~\ref{tab-SMIN}, $\RZ$ is one of the
most sensitive to QCD corrections.  For $\MZ=91.1875$~\GeV{}, and
imposing $\Mt=174.3\pm5.1$~\GeV{} as a constraint,
$\alfas=0.1224\pm0.0038$ is obtained.  Alternatively, $\sll$ (see
Table~\ref{tab-widths}) which has higher sensitivity to QCD
corrections and less dependence on $\MH$ yields:
$\alfas=0.1180\pm0.0030$.  Typical errors arising from the variation
of $\MH$ between $100~\GeV$ and $200~\GeV$ are of the order of
$0.001$, somewhat smaller for $\sll$.  These results on $\alfas$, as
well as those reported in the next section, are in very good agreement
with recently determined world averages ($\alfmz=0.118 \pm
0.002$\cite{common_bib:pdg2000}, or $\alfmz=0.1178 \pm 0.0033$ based
solely on NNLO QCD results excluding the LEP lineshape results and
accounting for correlated errors\cite{Siggi-Bethke-alpha-s}).

\section{Standard Model Analyses}

In the following, several different Standard Model fits to the data
reported in Table~\ref{tab-BIGFIT} are discussed.  The $\chi^2$
minimisation is performed with the program MINUIT~\cite{MINUIT}, and
the predictions are calculated with TOPAZ0~\cite{ref:TOPAZ0} and
ZFITTER~\cite{ref:ZFITTER}.  The large $\chi^2$/d.o.f.{} for all of
these fits is caused by the same effect as discussed in the previous
chapter, namely the large dispersion in the values of the leptonic
effective electroweak mixing angle measured through the various
asymmetries.  For the analyses presented here, this dispersion is
interpreted as a fluctuation in one or more of the input measurements,
and thus we neither modify nor exclude any of them.

To test the agreement between the LEP data and the Standard Model, a
fit to the data (including the $\LEPII$ $\MW$ determination) leaving
the top quark mass and the Higgs mass as free parameters is performed.
The result is shown in Table~\ref{tab-BIGFIT}, column~1.  This fit
shows that the LEP data predicts the top mass in good agreement with
the direct measurements. In addition, the data prefer an intermediate
Higgs boson mass, albeit with very large errors.  The strongly
asymmetric errors on $\MH$ are due to the fact that to first order,
the radiative corrections in the Standard Model are proportional to
$\log(\MH)$.

\begin{figure}[htbp]
\begin{center}
  \leavevmode \includegraphics[width=0.9\linewidth]{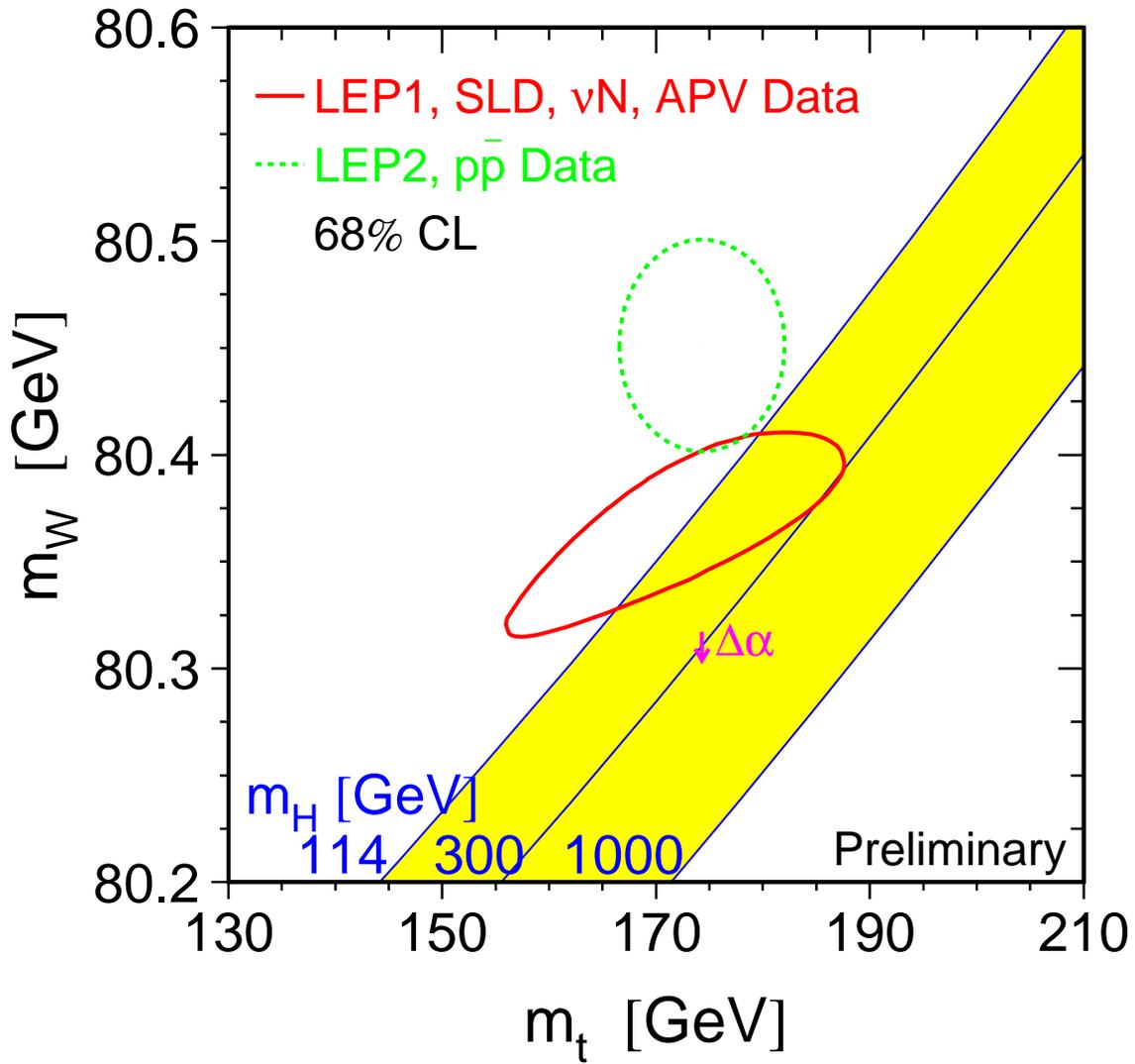}
\caption[]{
  The comparison of the indirect measurements of $\MW$ and $\Mt$
  ($\LEPI$+SLD+$\nu$N+APV data) (solid contour) and the direct
  measurements ($\pp$ colliders and $\LEPII$ data) (dashed contour).  In both
  cases the 68\% CL contours are plotted.  Also shown is the Standard
  Model relationship for the masses as a function of the Higgs mass.}
\label{fig:mtmW}
\end{center}
\end{figure}

The data can also be used within the Standard Model to determine the
top quark and W masses indirectly, which can be compared to the direct
measurements performed at the $\pp$ colliders and \LEPII.  In the
second fit, all the results in Table~\ref{tab-SMIN}, except the
$\LEPII$ and $\pp$ colliders $\MW$ and $\Mt$ results are used.  The
results are shown in column~2 of Table~\ref{tab-BIGFIT}.  The indirect
measurements of $\MW$ and $\Mt$ from this data sample are shown in
Figure~\ref{fig:mtmW}, compared with the direct measurements. Also
shown are the Standard Model predictions for Higgs masses between 114
and 1000~\GeV.  As can be seen in the figure, the indirect and direct
measurements of $\MW$ and $\Mt$ are in good agreement, and both sets
prefer a low value of the Higgs mass.

\begin{figure}[htbp]
\begin{center}
  \mbox{\includegraphics[width=0.9\linewidth]{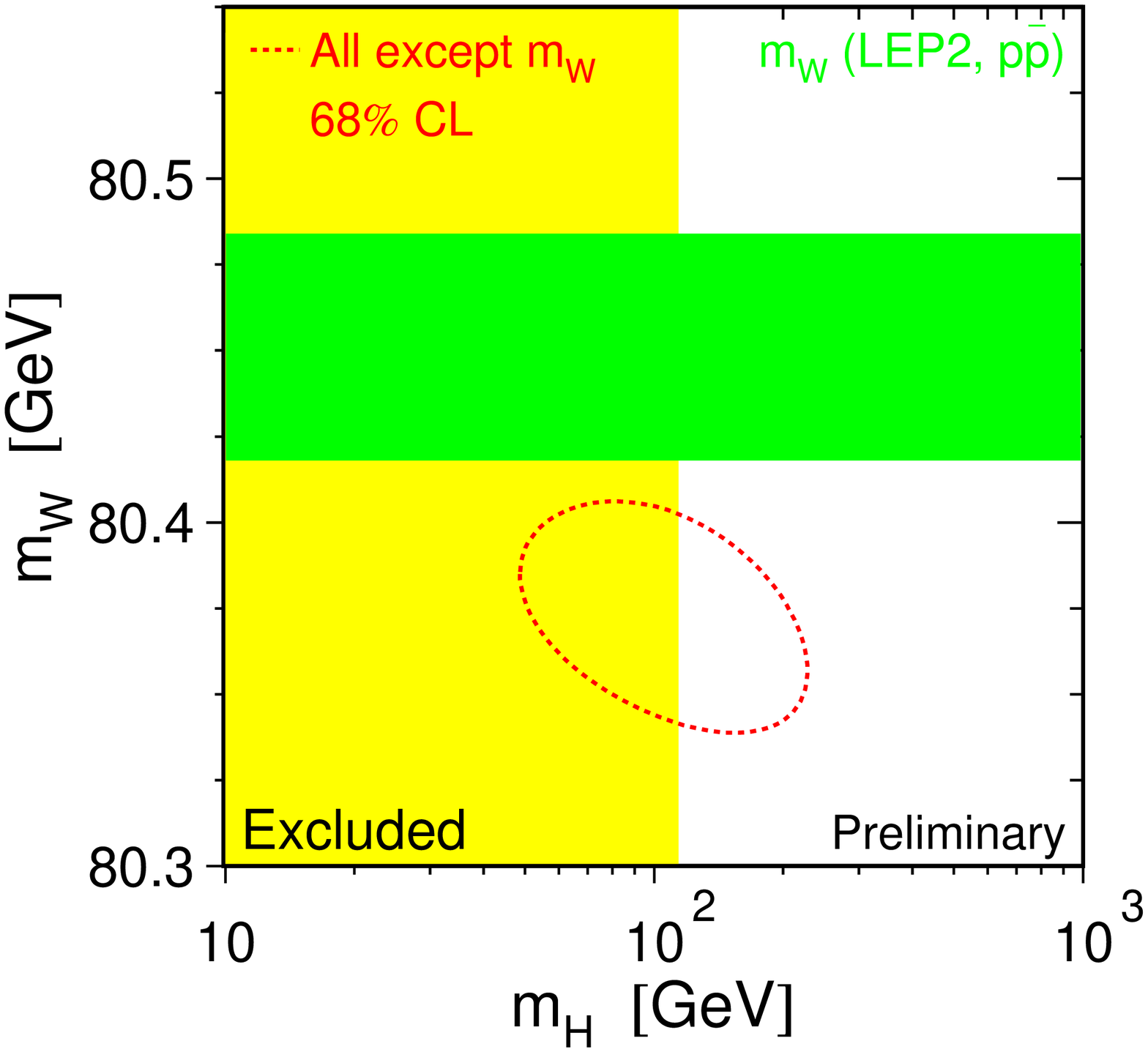}}
\end{center}
\vspace*{-0.6cm}
\caption[]{
  The 68\% confidence level contour in $\Mt$ and $\MW$ for the fit to
  all data except the direct measurement of $\MW$, indicated by the
  shaded horizontal band of $\pm1$ sigma width.  The vertical band
  shows the 95\% CL exclusion limit on $\MH$ from the direct search.
  }
\label{fig-mhmw}
\end{figure}

\begin{figure}[htbp]
\begin{center}
  \mbox{\includegraphics[width=0.9\linewidth]{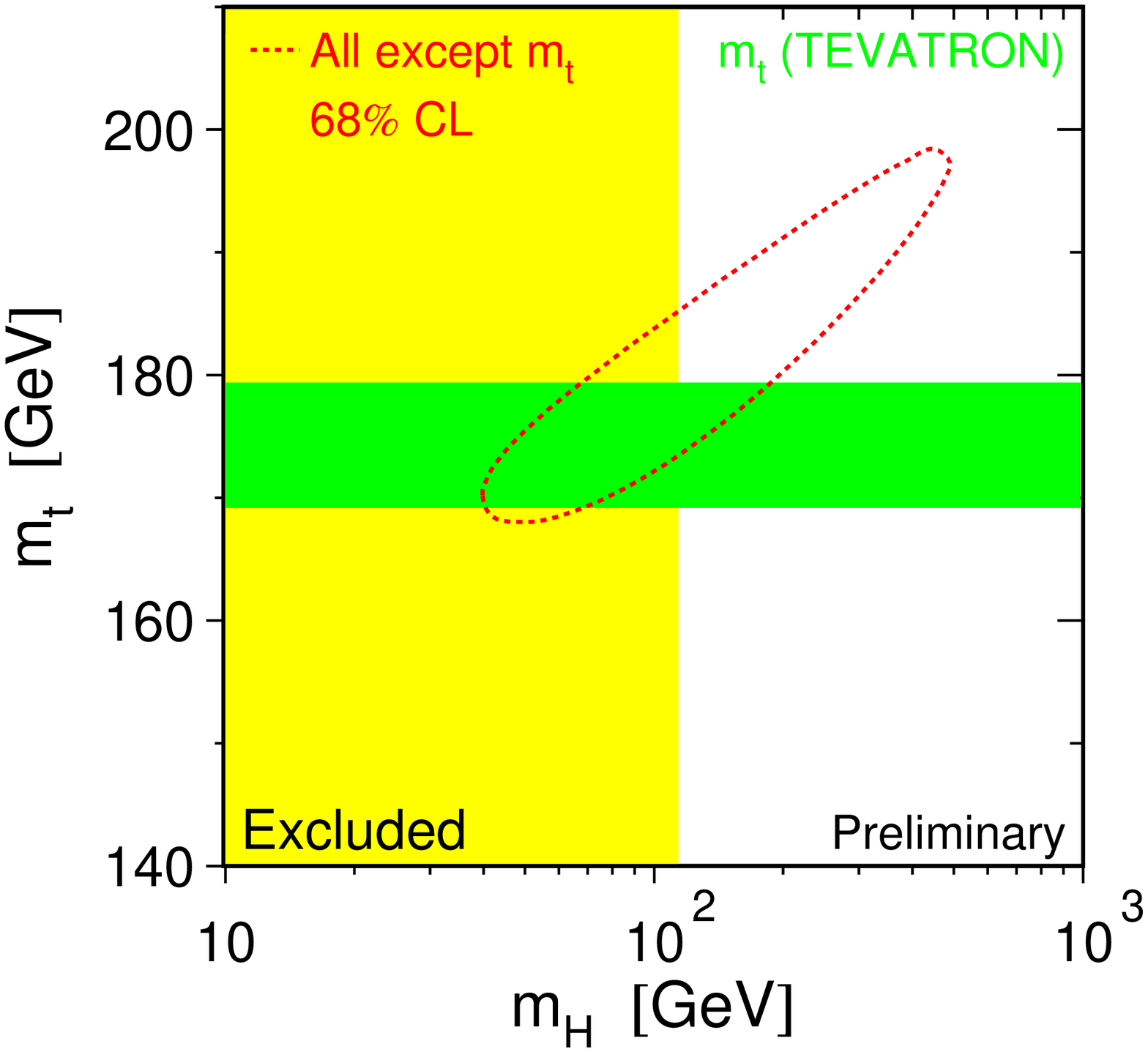}}
\end{center}
\vspace*{-0.6cm}
\caption[]{
  The 68\% confidence level contour in $\Mt$ and $\MH$ for the fit to
  all data except the direct measurement of $\Mt$, indicated by the
  shaded horizontal band of $\pm1$ sigma width.  The vertical band
  shows the 95\% CL exclusion limit on $\MH$ from the direct search.
  }
\label{fig-mhmt}
\end{figure}

For the third fit, the direct $\Mt$ measurement is used to obtain the
best indirect determination of $\MW$.  The result is shown in column~3
of Table~\ref{tab-BIGFIT} and in Figure~\ref{fig-mhmw}.  Also here,
the indirect determination of W boson mass $80.373\pm0.023$ \GeV\ is
in agreement with the combination of direct measurements from \LEPII\
and $\pp$ colliders~\cite{bib-MWAVE-00} of $\MW= 80.451\pm0.033$ \GeV.
For the next fit, (column~4 of Table~\ref{tab-BIGFIT} and
Figure~\ref{fig-mhmt}), the direct $\MW$ measurements from LEP and
$\pp$ colliders are included to obtain $\Mt= 181^{+11}_{-9}$ \GeV, in
very good agreement with the direct measurement of $\Mt = 174.3\pm5.1$
\GeV. Compared to the second fit, the error on $\log\MH$ increases due
to effects from higher-order terms.

\begin{figure}[htbp]
\begin{center}
  \mbox{\includegraphics[width=0.9\linewidth]{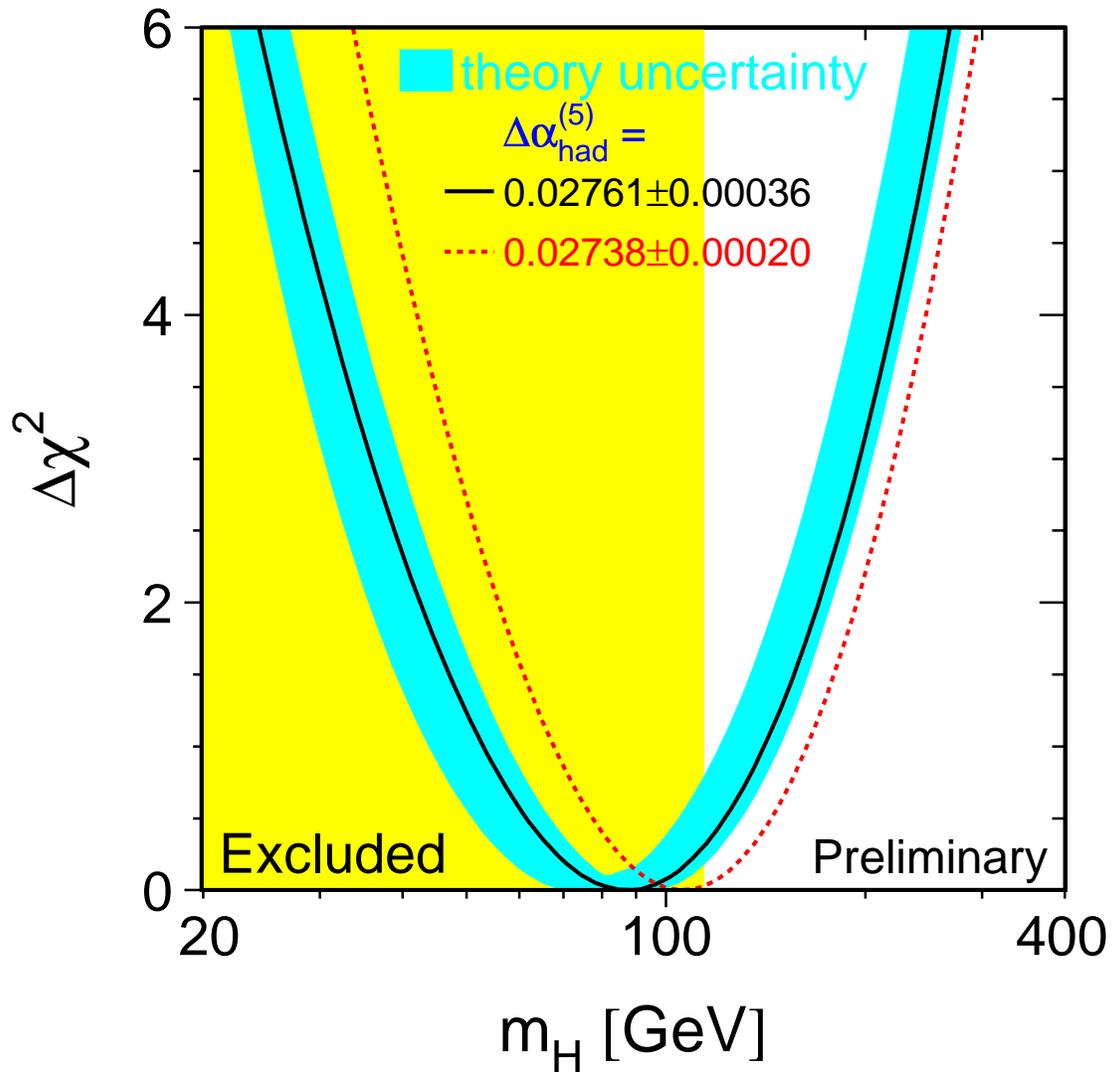}}
\end{center}
\vspace*{-0.6cm}
\caption[]{%
  $\Delta\chi^{2}=\chi^2-\chi^2_{min}$ {\it vs.} $\MH$ curve.  The
  line is the result of the fit using all data (last column of
  Table~\protect\ref{tab-BIGFIT}); the band represents an estimate of
  the theoretical error due to missing higher order corrections.  The
  vertical band shows the 95\% CL exclusion limit on $\MH$ from the
  direct search.  The dashed curve is the result obtained using the
  evaluation of $\Delta\alpha^{(5)}_{\mathrm{had}}(\MZ^2)$ from
  Reference~\citen{bib-ADMartin}. }
\label{fig-chiex}
\end{figure}

Finally, the best constraints on $\MH$ are obtained when all data are
used in the fit.  The results of this fit are shown in column~5 of
Table~\ref{tab-BIGFIT} and Figure~\ref{fig-chiex}.  In
Figure~\ref{fig-chiex} the observed value of $\Delta\chi^2 \equiv
\chi^2 - \chi^2_{\mathrm{min}}$ as a function of $\MH$ is plotted for
the fit including all data.  The solid curve is the result using
ZFITTER, and corresponds to the last column of Table~\ref{tab-BIGFIT}.
The shaded band represents the uncertainty due to uncalculated
higher-order corrections, as estimated by ZFITTER and TOPAZ0.
Compared to previous analyses, its width is enlarged towards lower
Higgs-boson masses due to the effects of the complete fermionic
two-loop calculation of $\MW$ discussed above.
The 95\% confidence level upper limit on $\MH$ (taking the band into
account) is 196 \GeV.  The lower limit on $\MH$ of approximately 114
\GeV{} obtained from direct searches\cite{ref:HiggsOsaka} is not used
in the determination of this limit.  Also shown is the result (dashed
curve) obtained when using $\Delta\alpha^{(5)}_{\mathrm{had}}(\MZ^2)$
of Reference \citen{bib-ADMartin}.  That fit results in
$\log(\MH/\GeV) = 2.03\pm0.19$ corresponding to $\MH= 106^{+57}_{-38}$
\GeV~ and an upper limit on $\MH$ of approximately 222 \GeV{} at 95\%
confidence level.

\begin{figure}[p]
\begin{center}
  \mbox{\includegraphics[width=0.9\linewidth]{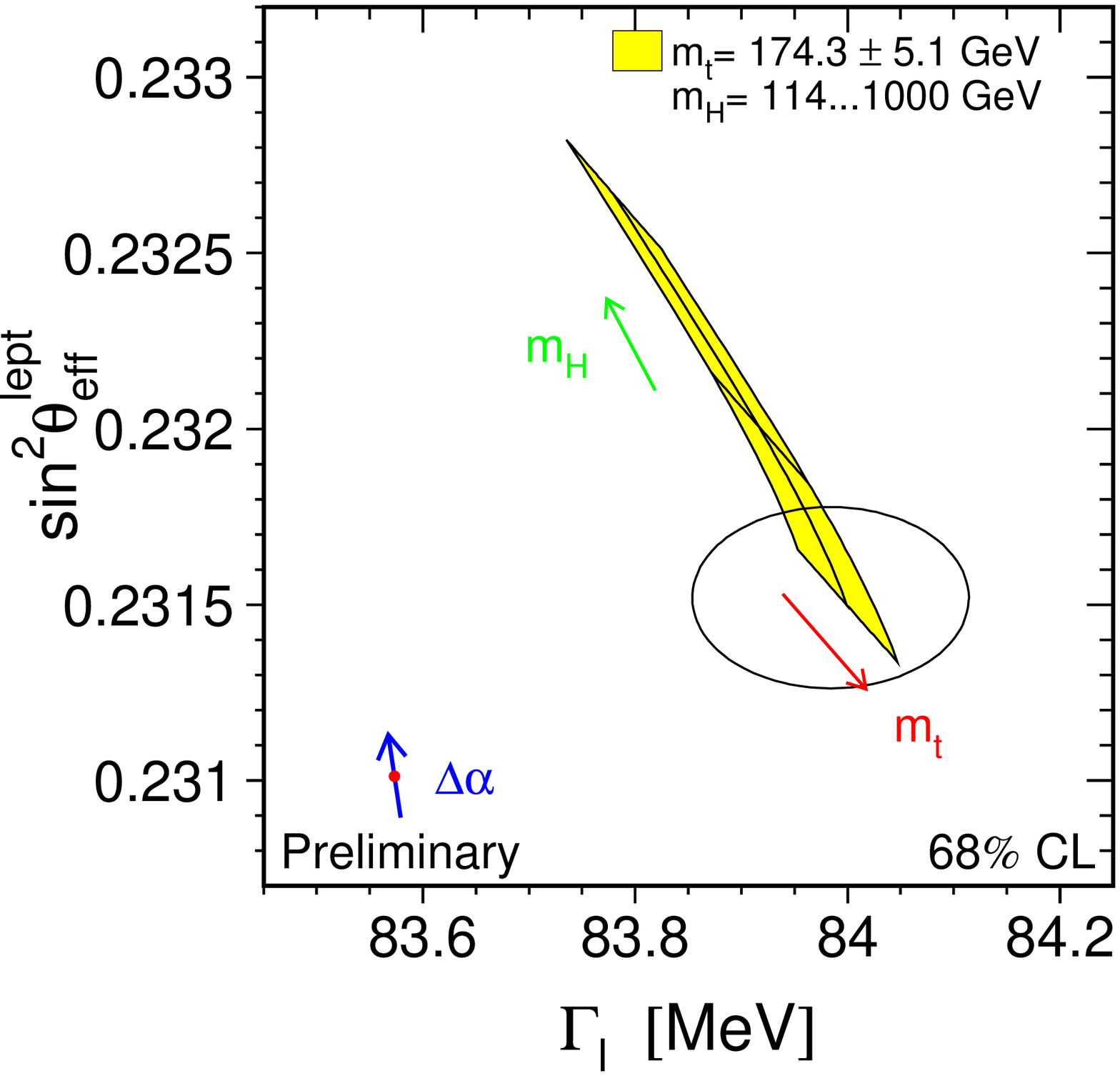}}
\end{center}
\caption[]{%
  $\LEPI$+SLD measurements of $\swsqeffl$ (Table~\ref{tab-swsq}) and
  $\Gll$ (Table~\ref{tab-widths}) and the Standard Model prediction.
  The point shows the predictions if among the electroweak radiative
  corrections only the photon vacuum polarisation is included. The
  corresponding arrow shows variation of this prediction if
  $\alpha(\MZ^2)$ is changed by one standard deviation. This variation
  gives an additional uncertainty to the Standard Model prediction
  shown in the figure.  }
\label{fig-gllsef}
\end{figure}

\begin{figure}[p]
\vspace*{-2.0cm}
\begin{center}
  \mbox{\includegraphics[height=21cm]{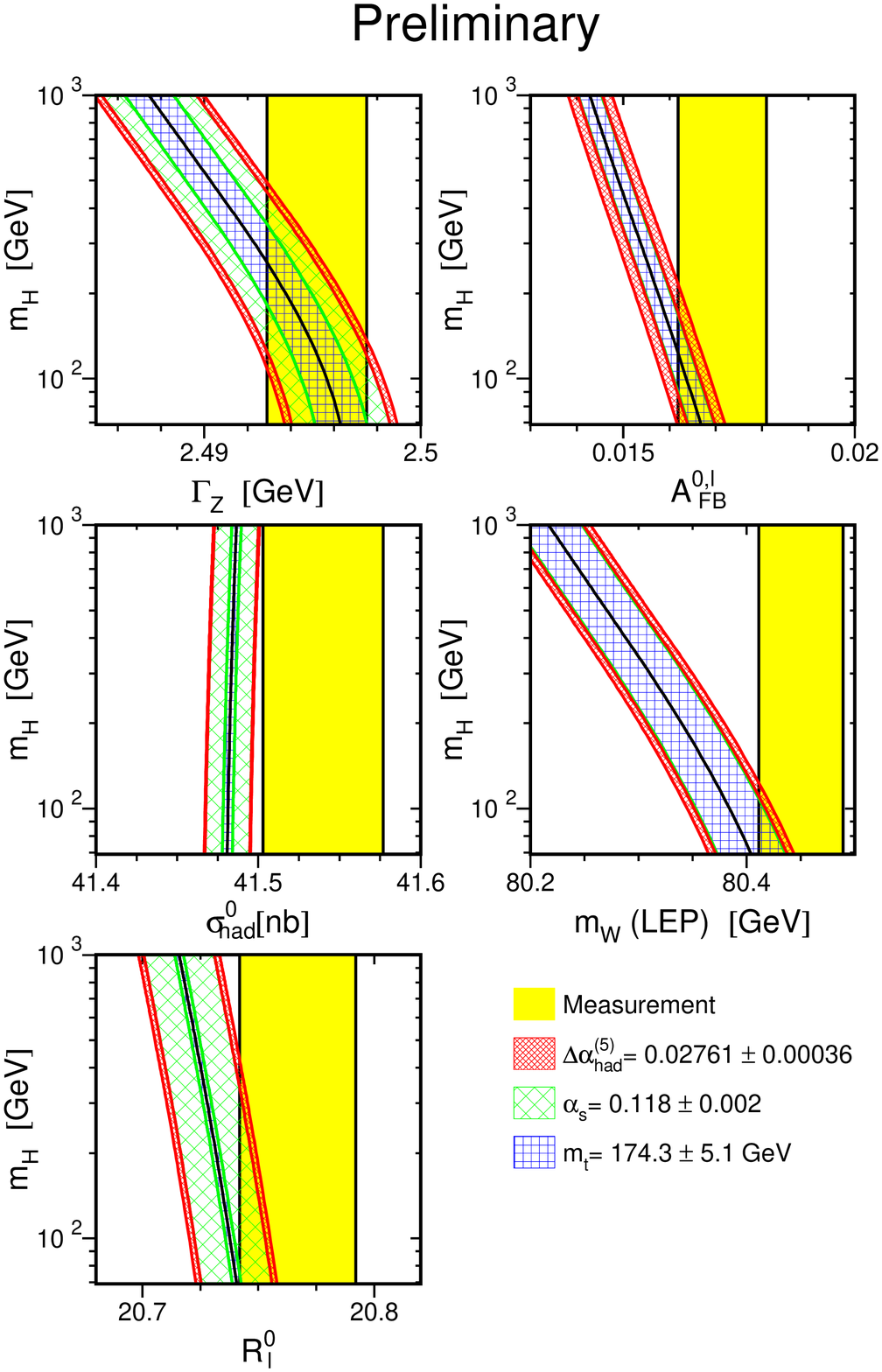}}
\end{center}
\vspace*{-0.6cm}
\caption[]{%
  Comparison of $\LEPI$ measurements with the Standard Model
  prediction as a function of $\MH$.  The measurement with its error
  is shown as the vertical band.  The width of the Standard Model band
  is due to the uncertainties in
  $\Delta\alpha^{(5)}_{\mathrm{had}}(\MZ^2)$, $\alfmz$ and $\Mt$.  The
  total width of the band is the linear sum of these effects.  }
\label{fig-higgs1}
\end{figure}

\begin{figure}[p]
\vspace*{-2.0cm}
\begin{center}
  \mbox{\includegraphics[height=21cm]{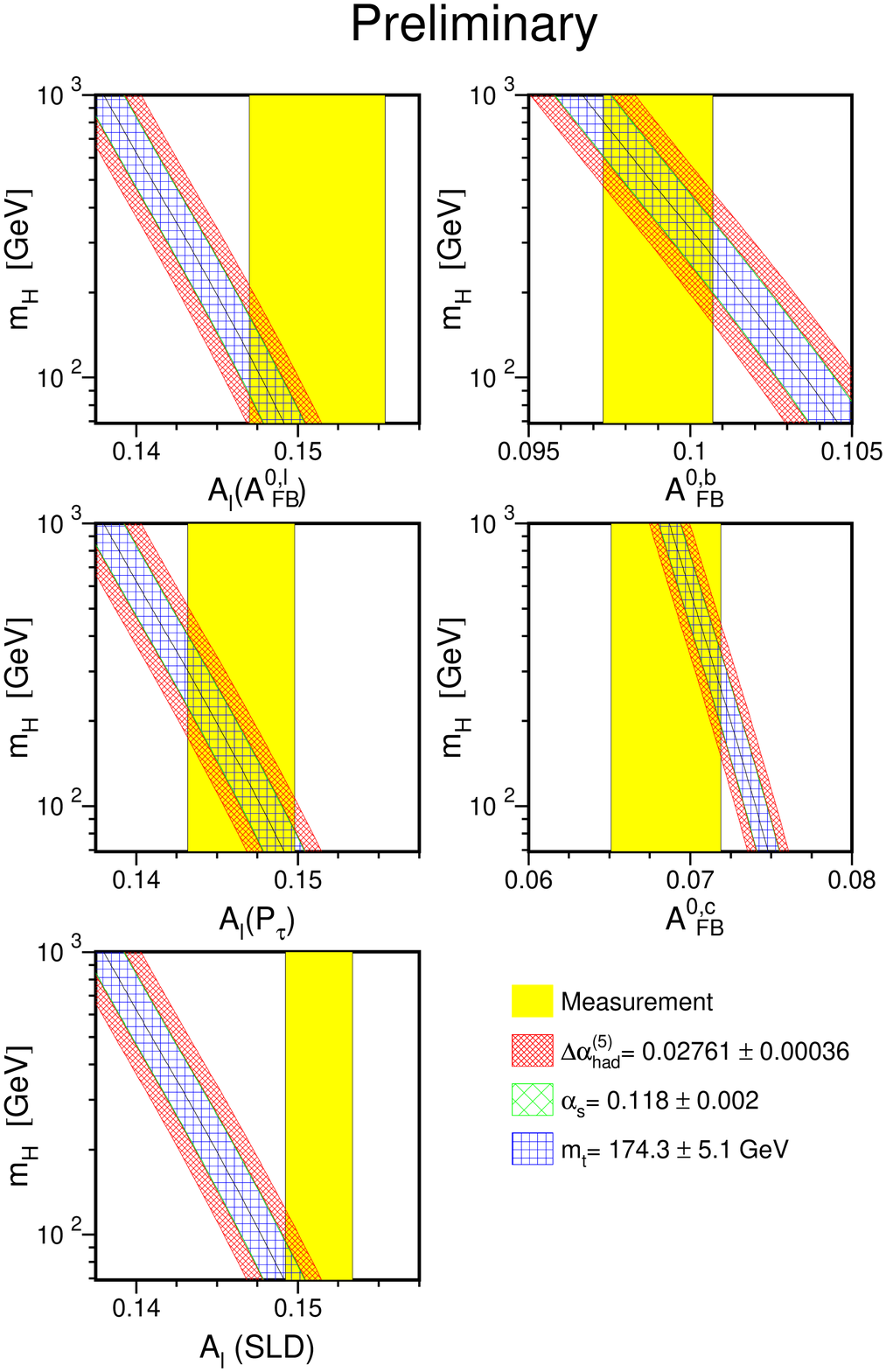}}
\end{center}
\vspace*{-0.6cm}
\caption[]{%
  Comparison of $\LEPI$ measurements with the Standard Model
  prediction as a function of $\MH$.  The measurement with its error
  is shown as the vertical band.  The width of the Standard Model band
  is due to the uncertainties in
  $\Delta\alpha^{(5)}_{\mathrm{had}}(\MZ^2)$, $\alfmz$ and $\Mt$.  The
  total width of the band is the linear sum of these effects.  Also
  shown is the comparison of the SLD measurement of $\cAl$, dominated
  by $\ALRz$, with the Standard Model. }
\label{fig-higgs2}
\end{figure}

\begin{figure}[p]
\vspace*{-2.0cm}
\begin{center}
  \mbox{\includegraphics[height=21cm]{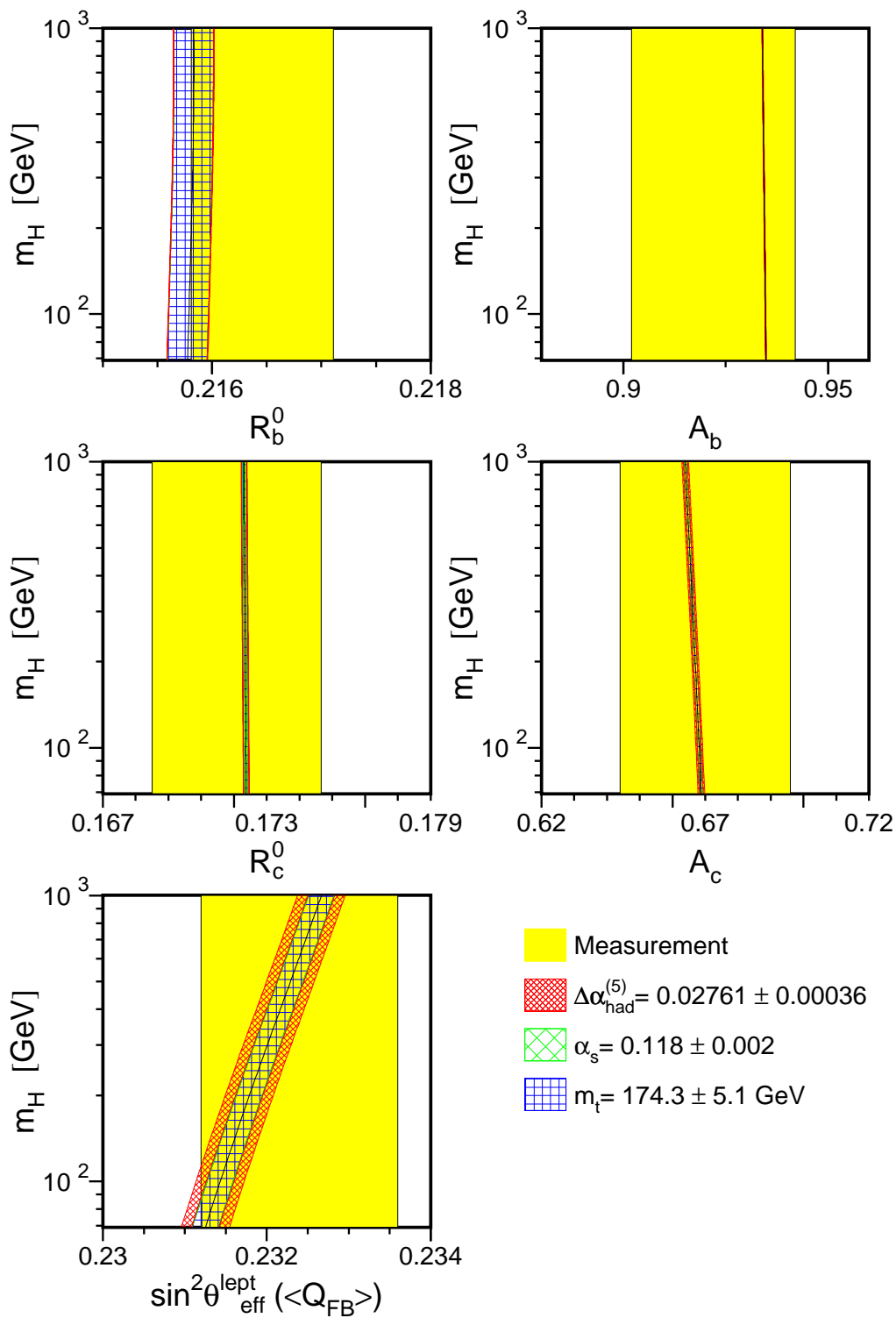}}
\end{center}
\vspace*{-0.6cm}
\caption[]{%
  Comparison of $\LEPI$ and SLD measurements with the Standard Model
  prediction as a function of $\MH$.  The measurement with its error
  is shown as the vertical band.  The width of the Standard Model band
  is due to the uncertainties in
  $\Delta\alpha^{(5)}_{\mathrm{had}}(\MZ^2)$, $\alfmz$ and $\Mt$.  The
  total width of the band is the linear sum of these effects.  }
\label{fig-higgs3}
\end{figure}

In Figures~\ref{fig-higgs1} to~\ref{fig-higgs3} the sensitivity of the
LEP and SLD measurements to the Higgs mass is shown.  As can be seen,
the most sensitive measurements are the asymmetries, \ie, $\swsqeffl$.
A reduced uncertainty for the value of $\alpha(\MZ^2)$ would therefore
result in an improved constraint on $\log\MH$ and thus $\MH$, as
already shown in Figures~\ref{fig-gllsef} and \ref{fig-chiex}.

\vfill

\begin{table}[htbp]
\renewcommand{\arraystretch}{1.5}
  \begin{center}
    \leavevmode
    \begin{tabular}{|c||c|c|c|c|c|}
\hline
  &    - 1 -         &     - 2 -          &      - 3 -           &    - 4 -             &    - 5 -    \\
  & LEP including    & all data except    & all data except      & all data except      &
  all data \\[-3mm]
  & $\LEPII$ $\MW$   & $\MW$ and $\Mt$    &           $\MW$      &           $\Mt$      &           \\
\hline
\hline
$\Mt$\hfill[\GeV] 
& $186^{+13}_{-11}$      & $169^{+12}_{-9}$ & $173.3^{+4.7}_{-4.6}$ 
& $181^{+11}_{- 9}$      & $175.8^{+4.4}_{-4.3}$   \\
$\MH$\hfill[\GeV] 
& $260^{+404}_{-155}$    & $81^{+109}_{-40}$ & $108^{+70}_{-44}$   
& $126^{+182}_{-69}$     & $88^{+53}_{-35}$\\
$\log(\MH/\GeV)$  
& $2.42^{+0.41}_{-0.39}$ & $1.91^{+0.37}_{-0.29} $ &  $2.03^{+0.22}_{-0.23}$ 
& $2.10^{+0.39}_{-0.34}$ & $1.94^{+0.21}_{-0.22}$\\
$\alfmz$          
& $0.1201\pm 0.0030$     & $0.1187\pm 0.0027$ & $0.1189\pm0.0027$   
& $0.1186\pm 0.0028$     & $0.1183\pm 0.0027$ \\
\hline
$\chi^2$/d.o.f.{} 
& $15.5/8$               & $18.9/12$          & $19.1/13$           
& $22.6/14$              & $22.9/15$          \\
\hline
\hline
$\swsqeffl$ & $\pz0.23162$& $\pz0.23150$& $\pz0.23151$
            & $\pz0.23139$& $\pz0.23136$ \\[-1mm]
            & $\pm0.00018$& $\pm0.00016$& $\pm0.00016$
            & $\pm0.00015$& $\pm0.00014$ \\
$\swsq$     & $\pz0.22282$& $\pz0.22333$& $\pz0.22313$
            & $\pz0.22248$& $\pz0.22263$     \\[-1mm]
            & $\pm0.00051$& $\pm0.00063$& $\pm0.00045$
            & $\pm0.00045$& $\pm0.00036$     \\
$\MW$\hfill[\GeV]
& $80.389\pm0.026$ & $80.363\pm0.032$   & $80.373\pm0.023$   
& $80.406\pm0.023$ & $80.398\pm0.019$        \\
\hline
    \end{tabular}
    \caption[]{
      Results of the fits to: (1) LEP data alone, (2) all data except
      the direct determinations of $\Mt$ and $\MW$ ($\pp$ collider and
      $\LEPII$), (3) all data except direct $\MW$ determinations, (4)
      all data except direct $\Mt$ determinations, and (5) all data.
      As the sensitivity to $\MH$ is logarithmic, both $\MH$ as well
      as $\log(\MH/\GeV)$ are quoted.  The bottom part of the table
      lists derived results for $\swsqeffl$, $\swsq$ and $\MW$.  See
      text for a discussion of theoretical errors not included in the
      errors above.  }
    \label{tab-BIGFIT}
  \end{center}
\renewcommand{\arraystretch}{1.0}
\end{table}
\vfill

\clearpage

\boldmath
\chapter{Conclusions}
\label{sec-Conc}
\unboldmath

The combination of the many precise electroweak results yields
stringent constraints on the Standard Model.  In addition, the results
are sensitive to the Higgs mass.  Most measurements agree well with
the predictions.  The spread in values of the various determinations
of the effective electroweak mixing angle is larger than expected.
Within the Standard Model analysis, this seems to be caused by the
measurement of the forward-backward asymmetry in b-quark production,
showing the largest deviation w.r.t. the Standard-Model expectation.

The experiments wish to stress that this report reflects a preliminary
status at the time of the 2001 summer conferences.  A definitive
statement on these results must wait for publication by each
collaboration.

\boldmath
\section*{Prospects for the Future}
\label{sec-Future}
\unboldmath

Most of the measurements from data taken at or near the Z resonance,
both at LEP as well as at SLC, that are presented in this report are
either final or are being finalised.  The main improvements will
therefore take place in the high energy data, with more than
700~pb$^{-1}$ per experiment.  The measurements of $\MW$ are likely to
reach a precision not too far from the uncertainty on the prediction
obtained via the radiative corrections of the \Zzero{} data, providing
a further important test of the Standard Model.  In the measurement of
the triple and quartic electroweak gauge boson self couplings, the
analysis of the complete $\LEPII$ statistics, together with the
increased sensitivity at higher beam energies, will lead to an
improvement in the current precision.
\section*{Acknowledgements}

We would like to thank the CERN accelerator divisions for the
efficient operation of the LEP accelerator, the precise information on
the absolute energy scale and their close cooperation with the four
experiments.  The SLD collaboration would like to thank the SLAC
accelerator department for the efficient operation of the SLC
accelerator.  We would also like to thank members of the CDF, D\O{}
and NuTeV Collaborations for making preliminary results available to
us in advance of the conferences and for useful discussions concerning
their combination.  Finally, the results of the section on Standard
Model constraints would not be possible without the close
collaboration of many theorists.

\clearpage

\begin{appendix}

\chapter{The Measurements used in the Heavy-Flavour Averages}
\label{app-HF-tab}

In the following 20 tables the results used in the combination are listed.
In each case an indication of the dataset used and the type of analysis is
given.
Preliminary results are indicated by the symbol ``\dag''. 
The values of centre-of-mass energy are given where
relevant.  In each table, the result used as input to the average
procedure is given followed by the statistical error, the correlated
and uncorrelated systematic errors, the total systematic error, and
any dependence on other electroweak parameters.  In the case of the
asymmetries, the measurement moved to a common energy (89.55 \GeV{},
91.26 \GeV{} and 92.94 \GeV{}, respectively, for peak$-$2, peak and
peak+2 results) is quoted as {\it corrected\/} asymmetry.

Contributions to the correlated systematic error quoted here are from
any sources of error shared with one or more other results from
different experiments in the same table, and the uncorrelated errors
from the remaining sources. In the case of \cAc{} and \cAb{} from SLD
the quoted correlated systematic error has contributions from any
source shared with one or more other measurements from LEP experiments.
Constants such as $a(x)$ denote the dependence on the assumed value of
$x^{\rm{used}}$, which is also given.
 \begin{table}[htb]
 \begin{center}
 \begin{tabular}{|l||c|c|c|c|c|}
 \hline
           & \mca{1}{ALEPH} & \mca{1}{DELPHI} & \mca{1}{L3} & \mca{1}{OPAL} & \mca{1}{SLD} \\
 \hline
           &92-95 &92-95 &94-95 &92-95 &93-98\dag\\
           &\tmcite{ref:alife} &\tmcite{ref:drb} &\tmcite{ref:lrbmixed} &\tmcite{ref:omixed} &\tmcite{ref:SLD_R_B}\\
 \hline\hline
 \Rbz      &    0.2158 &   0.2163 &   0.2173 &   0.2174 &   0.2164\\
 \hline
 Statistical          &    0.0009 &   0.0007 &   0.0015 &   0.0011 &   0.0009\\
 \hline
 Uncorrelated         &    0.0007 &   0.0004 &   0.0015 &   0.0009 &   0.0006\\
 Correlated           &    0.0006 &   0.0004 &   0.0018 &   0.0008 &   0.0005\\
 \hline
 Total Systematic     &    0.0009 &   0.0005 &   0.0023 &   0.0012 &   0.0007\\
 \hline
 \hline
 $a( \Rc      )$ &   -0.0033 &  -0.0041 &  -0.0376 &  -0.0122 &  -0.0057\\
 $\Rc     ^{\mathrm{used}}$ &    0.1720 &   0.1720 &   0.1734 &   0.1720 &   0.1710\\
 \hline
 $a( \Brcl    )$ &           &          &  -0.0133 &  -0.0067 &         \\
 $\Brcl   ^{\mathrm{used}}$ &           &          &     9.80 &     9.80 &         \\
 \hline
 $a( \fDp     )$ &   -0.0010 &  -0.0010 &  -0.0086 &  -0.0029 &  -0.0008\\
 $\fDp    ^{\mathrm{used}}$ &    0.2330 &   0.2330 &   0.2330 &   0.2380 &   0.2370\\
 \hline
 $a( \fDs     )$ &   -0.0001 &   0.0001 &  -0.0005 &  -0.0001 &  -0.0003\\
 $\fDs    ^{\mathrm{used}}$ &    0.1020 &   0.1030 &   0.1030 &   0.1020 &   0.1140\\
 \hline
 $a( \fLc     )$ &    0.0002 &   0.0003 &   0.0008 &   0.0003 &  -0.0003\\
 $\fLc    ^{\mathrm{used}}$ &    0.0650 &   0.0630 &   0.0630 &   0.0650 &   0.0730\\
 \hline
 \end{tabular}
 \caption{The measurements of 
 \Rbz     .
 All measurements use a lifetime tag enhanced by other features like
 invariant mass cuts or high $p_T$ leptons.
 }
 \label{tab:Rbinp}
 \end{center}
 \end{table}
 \begin{table}[htb]
 \begin{center}
 \begin{tabular}{|l||c|c|c|c|c|c|c|c| }
 \hline
           & \mca{3}{ALEPH} & \mca{2}{DELPHI} & \mca{2}{OPAL} & \mca{1}{SLD} \\
 \hline
           &91-95   &91-95 &92-95 &92-95 & 92-95 &91-94 & 90-95 &93-97\dag\\
           &c-count &D meson &lepton &c-count &D meson& c-count & D meson &vtx-mass\\
           &\tmcite{ref:arcc} &\tmcite{ref:arcd} &\tmcite{ref:arcd} &\tmcite{ref:drcc} 
           &\tmcite{ref:drcd,ref:drcc} &\tmcite{ref:orcc} &\tmcite{ref:orcd}&\tmcite{ref:SLD_R_C}\\
 \hline\hline
 \Rcz               &    0.1735 &   0.1682 &  0.1670 & 0.1693& 0.1610 & 0.1642 & 0.1760 &  0.1738\\
 \hline                                    
 Statistical       &    0.0051 &   0.0082 &  0.0062 & 0.0050 & 0.0104 & 0.0122 & 0.0095  &  0.0031\\
 \hline                                    
 Uncorrelated      &    0.0057 &   0.0077 &  0.0059 & 0.0050 & 0.0064 & 0.0126 & 0.0102 & 0.0019\\
 Correlated        &    0.0094 &   0.0028 &  0.0009 & 0.0077 & 0.0060 & 0.0099 & 0.0062 & 0.0008\\
 \hline                                    
 Total Systematic  &    0.0110 &   0.0082 &  0.0059 & 0.0092 & 0.0088 & 0.0161 & 0.0120& 0.0021\\
 \hline
 \hline
 $a( \Rb      )$ & &   -0.0050 &          & & & & & -0.0433\\
 $\Rb     ^{\mathrm{used}}$ & &    0.2159 & & &   & &      &   0.2166\\
 \hline
 $a( \Brcl    )$ & &           &  -0.1646 & & & & &       \\
 $\Brcl   ^{\mathrm{used}}$ & &           &     9.80 && & & &         \\
 \hline
 \end{tabular}
 \end{center}
 \caption{The measurements of 
 $\Rcz$. 
``c-count'' denotes the determination of \Rcz{} from the sum of production rates
of weakly decaying charmed hadrons. ``D meson'' denotes any single/double tag
analysis using exclusive and/or inclusive  D meson reconstruction.
}
 \label{tab:Rcinp}
 \end{table}
 \begin{table}[htb]
 \begin{center}
 \begin{sideways}
 \begin{minipage}[b]{\textheight}
 \begin{center}
 \begin{tabular}{|l||c|c|c|c|c|c|c|c|c|c|c|}
 \hline
           & \mca{4}{ALEPH} & \mca{3}{DELPHI} & \mca{1}{L3} & \mca{3}{OPAL} \\
 \hline
           &90-95 &90-95 &90-95 &91-95 &93-95\dag &92-95 &92-95 &90-95 &91-95 &90-95\dag &90-95\\
           &lepton &lepton &lepton &multi &lepton &$D$-meson &jet charge &lepton &jet charge &lepton &$D$-meson\\
           &\tmcite{ref:alasy} &\tmcite{ref:alasy} &\tmcite{ref:alasy} &\tmcite{ref:ajet} &\tmcite{ref:dlasy} &\tmcite{ref:ddasy} &\tmcite{ref:djasy} &\tmcite{ref:llasy} &\tmcite{ref:ojet} &\tmcite{ref:olasy} &\tmcite{ref:odsac}\\
 \hline\hline
 \roots\ (\GeV)       &  88.380   & 89.380   & 90.210   & 89.470   & 89.433   & 89.434   & 89.550   & 89.500   & 89.440   & 89.490   & 89.490  \\
 \hline
 \Abl     &     -3.53 &     5.47 &     9.10 &     4.36 &     5.90 &     5.64 &     6.80 &     6.15 &     4.10 &     3.56 &    -9.20\\
 \hline
 \hline\hline
 \Abl     Corrected &   \mca{3}{5.87}
&     4.55 &     6.18 &     5.92 &     6.80 &     6.27 &     4.36 &     3.70 &    -9.06\\
 \hline
 Statistical          &    \mca{3}{1.90}
 &     1.19 &     2.20 &     7.59 &     1.80 &     2.93 &     2.10 &     1.73 &    10.80\\
 \hline
 Uncorrelated         &   \mca{3}{0.39}   
 &     0.05 &     0.08 &     0.91 &     0.12 &     0.37 &     0.25 &     0.16 &     2.51\\
 Correlated           &    \mca{3}{0.70} 
 &     0.01 &     0.08 &     0.08 &     0.01 &     0.19 &     0.02 &     0.04 &     1.41\\
 \hline
 Total Systematic     &     \mca{3}{0.80} 
 &     0.05 &     0.12 &     0.91 &     0.13 &     0.41 &     0.25 &     0.16 &     2.87\\
 \hline
 \hline
 $a( \Rb      )$ &     \mca{3}{-0.3069}
 &  -9.5 &  -1.1543 &          &  -0.1962 &  -1.4467 &  -0.7300 &  -0.1000 &         \\
 $\Rb     ^{\mathrm{used}}$ &   \mca{3}{0.2192} 
 &   0.2150 &   0.2164 &          &   0.2158 &   0.2170 &   0.2150 &   0.2155 &         \\
 \hline
 $a( \Rc      )$ &   \mca{3}{0.0362} 
&   0.3100 &   1.0444 &          &   0.3200 &   0.3612 &   0.0700 &   0.1000 &         \\
 $\Rc     ^{\mathrm{used}}$ &  \mca{3}{0.1710}
 &   0.1725 &   0.1671 &          &   0.1720 &   0.1734 &   0.1730 &   0.1720 &         \\
 \hline
 $a( \Acl     )$ &   \mca{3}{-0.2244}
 &  -0.2955 &          &          &          &  -0.1000 &  -0.3156 &          &         \\
 $\Acl    ^{\mathrm{used}}$ &      \mca{3}{-2.34}
 &    -2.87 &          &          &          &    -2.50 &    -2.81 &          &         \\
 \hline
 $a( \Brbl    )$ &    \mca{3}{-0.2486}
 &          &  -1.0154 &          &          &  -1.0290 &          &   0.3406 &         \\
 $\Brbl   ^{\mathrm{used}}$ &        \mca{3}{11.34}
 &          &    10.56 &          &          &    10.50 &          &    10.90 &         \\
 \hline
 $a( \Brbclp  )$ &    \mca{3}{-0.1074}
&          &  -0.1424 &          &          &  -0.1440 &          &  -0.5298 &         \\
 $\Brbclp ^{\mathrm{used}}$ &     \mca{3}{7.86}  
 &          &     8.07 &          &          &     8.00 &          &     8.30 &         \\
 \hline
 $a( \Brcl    )$ &   \mca{3}{-0.0474} 
 &          &   0.7224 &          &          &   0.5096 &          &   0.1960 &         \\
 $\Brcl   ^{\mathrm{used}}$ &        \mca{3}{9.80}
 &          &     9.90 &          &          &     9.80 &          &     9.80 &         \\
 \hline
 $a( \chiM    )$ &    \mca{3}{5.259}
&          &   1.3054 &          &          &          &          &          &         \\
 $\chiM   ^{\mathrm{used}}$ &  \mca{3}{ 0.12460}
 &          &  0.11770 &          &          &          &          &          &         \\
 \hline
 $a( \fDp     )$ &   \mca{3}{}      
     &          &          &   0.5083 &   0.0949 &          &          &          &         \\
 $\fDp    ^{\mathrm{used}}$ &   \mca{3}{}       
          &          &          &   0.2210 &   0.2330 &          &          &          &         \\
 \hline
 $a( \fDs     )$ &           \mca{3}{}
         &          &          &   0.1742 &   0.0035 &          &          &          &         \\
 $\fDs    ^{\mathrm{used}}$ &          \mca{3}{}
           &          &          &   0.1120 &   0.1020 &          &          &          &         \\
 \hline
 $a( \fLc     )$ &       \mca{3}{}
   &          &          &  -0.0191 &  -0.0225 &          &          &          &         \\
 $\fLc    ^{\mathrm{used}}$ &     \mca{3}{} 
      &          &          &   0.0840 &   0.0630 &          &          &          &         \\
 \hline
 $a( \PcDst   )$ &       \mca{3}{} 
  &  -0.1100 &          &          &          &          &          &          &         \\
 $\PcDst  ^{\mathrm{used}}$ &    \mca{3}{}
  &   0.1830 &          &          &          &          &          &          &         \\
 \hline
 \end{tabular}
 \end{center}
 \caption{The measurements of 
 \Abl    . 
  All numbers are given in \%.
}
 \label{tab:Ablinp}
 \end{minipage}
 \end{sideways}
 \end{center}
 \end{table}
 \begin{table}[htb]
 \begin{center}
 \begin{tabular}{|l||c|c|c|c|c|}
 \hline
           & \mca{1}{ALEPH} & \mca{2}{DELPHI} & \mca{2}{OPAL} \\
 \hline
           &91-95 &93-95\dag &92-95 &90-95\dag &90-95\\
           &$D$-meson &lepton &$D$-meson &lepton &$D$-meson\\
           &\tmcite{ref:adsac} &\tmcite{ref:dlasy} &\tmcite{ref:ddasy} &\tmcite{ref:olasy} &\tmcite{ref:odsac}\\
 \hline\hline
 \roots\ (\GeV)       &  89.370   & 89.433   & 89.434   & 89.490   & 89.490  \\
 \hline
 \Acl     &     -1.10 &     1.12 &    -5.02 &    -6.91 &     3.90\\
 \hline
 \hline\hline
 \Acl     Corrected &     -0.02 &     1.82 &    -4.32 &    -6.55 &     4.26\\
 \hline
 Statistical          &      4.30 &     3.60 &     3.69 &     2.44 &     5.10\\
 \hline
 Uncorrelated         &      1.00 &     0.53 &     0.40 &     0.38 &     0.80\\
 Correlated           &      0.09 &     0.16 &     0.09 &     0.23 &     0.30\\
 \hline
 Total Systematic     &      1.00 &     0.55 &     0.41 &     0.44 &     0.86\\
 \hline
 \hline
 $a( \Rb      )$ &           &  -0.2886 &          &  -3.4000 &         \\
 $\Rb     ^{\mathrm{used}}$ &           &   0.2164 &          &   0.2155 &         \\
 \hline
 $a( \Rc      )$ &           &   1.0096 &          &   3.2000 &         \\
 $\Rc     ^{\mathrm{used}}$ &           &   0.1671 &          &   0.1720 &         \\
 \hline
 $a( \Abl     )$ &   -1.3365 &          &          &          &         \\
 $\Abl    ^{\mathrm{used}}$ &      6.13 &          &          &          &         \\
 \hline
 $a( \Brbl    )$ &           &  -1.0966 &          &  -1.7031 &         \\
 $\Brbl   ^{\mathrm{used}}$ &           &    10.56 &          &    10.90 &         \\
 \hline
 $a( \Brbclp  )$ &           &   1.1156 &          &  -1.4128 &         \\
 $\Brbclp ^{\mathrm{used}}$ &           &     8.07 &          &     8.30 &         \\
 \hline
 $a( \Brcl    )$ &           &   1.0703 &          &   3.3320 &         \\
 $\Brcl   ^{\mathrm{used}}$ &           &     9.90 &          &     9.80 &         \\
 \hline
 $a( \chiM    )$ &           &  -0.0856 &          &          &         \\
 $\chiM   ^{\mathrm{used}}$ &           &  0.11770 &          &          &         \\
 \hline
 $a( \fDp     )$ &           &          &  -0.3868 &          &         \\
 $\fDp    ^{\mathrm{used}}$ &           &          &   0.2210 &          &         \\
 \hline
 $a( \fDs     )$ &           &          &  -0.1742 &          &         \\
 $\fDs    ^{\mathrm{used}}$ &           &          &   0.1120 &          &         \\
 \hline
 $a( \fLc     )$ &           &          &  -0.0878 &          &         \\
 $\fLc    ^{\mathrm{used}}$ &           &          &   0.0840 &          &         \\
 \hline
 \end{tabular}
 \end{center}
 \caption{The measurements of 
 \Acl    .  All numbers are given in \%.
 }
 \label{tab:Aclinp}
 \end{table}
 \begin{table}[htb]
 \begin{center}
 \begin{sideways}
 \begin{minipage}[b]{\textheight}
 \begin{center}
 \begin{tabular}{|l||c|c|c|c|c|c|c|c|c|c|c|c|}
 \hline
           & \mca{2}{ALEPH} & \mca{5}{DELPHI} & \mca{2}{L3} & \mca{3}{OPAL} \\
 \hline
           &91-95\dag &91-95 &91-92 &93-95\dag &92-95 &92-95 &92-95\dag&91-95 &90-95 &91-95 &90-95\dag &90-95\\
           &lepton &multi &lepton &lepton &$D$-meson &jet charge &multi&jet charge &lepton &multi&lepton &$D$-meson\\
           &\tmcite{ref:alasy} &\tmcite{ref:ajet} &\tmcite{ref:dlasy} &\tmcite{ref:dlasy} &\tmcite{ref:ddasy} 
&\tmcite{ref:djasy}&\tmcite{ref:dnnasy} &\tmcite{ref:ljet} &\tmcite{ref:llasy} &\tmcite{ref:ojet} &\tmcite{ref:olasy} &\tmcite{ref:odsac}\\
 \hline\hline
 \roots\ (\GeV)       &  91.210   & 91.230   & 91.270   & 91.223   & 91.235   & 91.260   & 91.260 & 91.240   & 91.260   & 91.210   & 91.240   & 91.240  \\
 \hline
 \Abp     &      9.71 &   10.00 &    10.89 &     9.86 &     7.58 &     9.83 &  9.72 &   9.31 &     9.85 &    10.06 &     9.14 &     9.00\\
 \hline
 \hline\hline
 \Abp     Corrected &      9.81 &   10.06 &    10.87 &     9.93 &     7.63 &     9.83 & 9.72 &     9.35 &     9.85 &    10.15 &     9.18 &     9.04\\
 \hline
 Statistical          &      0.40 &     0.27 &     1.30 &     0.64 &     1.97 &     0.47 &  0.35&   1.01 &     0.67 &     0.52 &     0.44 &     2.70\\
 \hline
 Uncorrelated         &      0.16 &     0.11 &     0.33 &     0.15 &     0.76 &     0.13 &  0.21 &    0.51 &     0.27 &     0.41 &     0.14 &     2.14\\
 Correlated           &      0.12 &     0.02 &     0.27 &     0.14 &     0.10 &     0.06 &  0.05 &   0.21 &     0.14 &     0.20 &     0.15 &     0.45\\
 \hline
 Total Systematic     &      0.20 &     0.11 &     0.43 &     0.20 &     0.77 &     0.15 &  0.22 &   0.55 &     0.31 &     0.46 &     0.20 &     2.19\\
 \hline
 \hline
 $a( \Rb      )$ &   -0.9545 &  -9.5 &  -2.8933 &  -2.0201 &          &  -0.1962 & 0.0637 & -9.1622 &  -2.1700 &  -7.6300 &  -0.7000 &         \\
 $\Rb     ^{\mathrm{used}}$ &    0.2172 &   0.2158 &   0.2170 &   0.2164 &          &   0.2158 & 0.2164 &  0.2170 &   0.2170 &   0.2150 &   0.2155 &       \\
 \hline
 $a( \Rc      )$ &    0.6450 &   0.3100 &   1.0993 &   1.1488 &          &   0.8400 & 0.0595 &  1.0831 &   1.3005 &   0.4600 &   0.6000 &         \\
 $\Rc     ^{\mathrm{used}}$ &    0.1720 &   0.1715 &   0.1710 &   0.1671 &          &   0.1720 & 0.1731 &  0.1733 &   0.1734 &   0.1730 &   0.1720 &       \\
 \hline
 $a( \Acp     )$ &           &   0.6849 &          &          &          &      &0.2756    &   1.1603 &   0.9262 &   0.6870 &          &         \\
 $\Acp    ^{\mathrm{used}}$ &           &     6.66 &          &          &          &   &6.89        &     6.91 &     7.41 &     6.19 &          &         \\
 \hline
 $a( \Brbl    )$ &   -1.8480 &          &  -3.8824 &  -2.0308 &          &          &   &       &  -2.0160 &          &  -0.3406 &         \\
 $\Brbl   ^{\mathrm{used}}$ &     10.78 &          &    11.00 &    10.56 &          &   &       &          &    10.50 &          &    10.90 &         \\
 \hline
 $a( \Brbclp  )$ &    0.4233 &          &   0.4740 &  -0.3798 &          &          &   &       &  -0.1280 &          &  -0.3532 &         \\
 $\Brbclp ^{\mathrm{used}}$ &      8.14 &          &     7.90 &     8.07 &          &   &       &          &     8.00 &          &     8.30 &         \\
 \hline
 $a( \Brcl    )$ &    0.5096 &          &   0.7840 &   1.0703 &          &          &    &      &   1.5288 &          &   0.5880 &         \\
 $\Brcl   ^{\mathrm{used}}$ &      9.80 &          &     9.80 &     9.90 &          &    &      &          &     9.80 &          &     9.80 &         \\
 \hline
 $a( \chiM    )$ &    2.9904 &          &   3.4467 &   1.6692 &          &          &    &      &          &          &          &         \\
 $\chiM   ^{\mathrm{used}}$ &   0.12460 &          &  0.12100 &  0.11770 &          &     &     &          &          &          &          &         \\
 \hline
 $a( \fDp     )$ &           &          &          &          &   0.0442 &   0.2761 & -0.0175 &         &          &          &          &         \\
 $\fDp    ^{\mathrm{used}}$ &           &          &   &      &   0.2210 &   0.2330 &  0.2330 &         &          &          &          &         \\
 \hline
 $a( \fDs     )$ &           &          &          &          &  -0.0788 &   0.0106 & -0.0260 &        &          &          &          &         \\
 $\fDs    ^{\mathrm{used}}$ &           &          &          &          &   0.1120 &   0.1020 & 0.1300 &          &          &          &          &         \\
 \hline
 $a( \fLc     )$ &           &          &          &          &  -0.0115 &  -0.0495 &  0.0221 &         &          &          &          &         \\
 $\fLc    ^{\mathrm{used}}$ &           &          &          &          &   0.0840 &   0.0630 &  0.0960 &        &          &          &          &         \\
 \hline
{\tiny $a( \PcDst   )$} &           &  -0.2500 &          &          &          &          &          &          &          &          &          &         \\
{\tiny $\PcDst  ^{\mathrm{used}}$} &           &   0.1830 &          &          &          &          &          &          &          &          &          &         \\
 \hline
 \end{tabular}
 \end{center}
 \caption{The measurements of 
 \Abp    .   All numbers are given in \%.
}
 \label{tab:Abpinp}
 \end{minipage}
 \end{sideways}
 \end{center}
 \end{table}
 \begin{table}[htb]
 \begin{center}
 \begin{tabular}{|l||c|c|c|c|c|c|c|c|}
 \hline
           & \mca{2}{ALEPH} & \mca{3}{DELPHI} & \mca{1}{L3} & \mca{2}{OPAL} \\
 \hline
           &91-95\dag &91-95 &91-92 &93-95\dag &92-95 &90-95 &90-95\dag &90-95\\
           &lepton &$D$-meson &lepton &lepton &$D$-meson &lepton &lepton &$D$-meson\\
           &\tmcite{ref:alasy} &\tmcite{ref:adsac} &\tmcite{ref:dlasy} &\tmcite{ref:dlasy} &\tmcite{ref:ddasy} &\tmcite{ref:llasy} &\tmcite{ref:olasy} &\tmcite{ref:odsac}\\
 \hline\hline
 \roots\ (\GeV)       &  91.210   & 91.220   & 91.270   & 91.223   & 91.235   & 91.240   & 91.240   & 91.240  \\
 \hline
 \Acp     &      5.68 &     6.13 &     8.05 &     6.29 &     6.58 &     7.94 &     5.95 &     6.50\\
 \hline
 \hline\hline
 \Acp     Corrected &      5.93 &     6.32 &     8.00 &     6.47 &     6.70 &     8.04 &     6.05 &     6.60\\
 \hline
 Statistical          &      0.53 &     0.90 &     2.26 &     1.00 &     0.97 &     3.70 &     0.59 &     1.20\\
 \hline
 Uncorrelated         &      0.24 &     0.23 &     1.25 &     0.53 &     0.25 &     2.40 &     0.37 &     0.49\\
 Correlated           &      0.36 &     0.17 &     0.49 &     0.27 &     0.04 &     0.49 &     0.32 &     0.23\\
 \hline
 Total Systematic     &      0.44 &     0.28 &     1.35 &     0.60 &     0.25 &     2.45 &     0.49 &     0.54\\
 \hline
 \hline
 $a( \Rb      )$ &    1.4318 &          &   2.8933 &  -2.3087 &          &   4.3200 &   4.1000 &         \\
 $\Rb     ^{\mathrm{used}}$ &    0.2172 &          &   0.2170 &   0.2164 &          &   0.2160 &   0.2155 &         \\
 \hline
 $a( \Rc      )$ &   -2.9383 &          &  -6.4736 &   5.4307 &          &  -6.7600 &  -3.8000 &         \\
 $\Rc     ^{\mathrm{used}}$ &    0.1720 &          &   0.1710 &   0.1671 &          &   0.1690 &   0.1720 &         \\
 \hline
 $a( \Abp     )$ &           &  -2.1333 &          &          &          &   6.4274 &          &         \\
 $\Abp    ^{\mathrm{used}}$ &           &     9.79 &          &          &          &     8.84 &          &         \\
 \hline
 $a( \Brbl    )$ &    1.8993 &          &   4.8529 &  -2.7618 &          &   3.5007 &   5.1094 &         \\
 $\Brbl   ^{\mathrm{used}}$ &     10.78 &          &    11.00 &    10.56 &          &    10.50 &    10.90 &         \\
 \hline
 $a( \Brbclp  )$ &   -1.0745 &          &  -3.9500 &   2.2786 &          &  -3.2917 &  -1.7660 &         \\
 $\Brbclp ^{\mathrm{used}}$ &      8.14 &          &     7.90 &     8.07 &          &     7.90 &     8.30 &         \\
 \hline
 $a( \Brcl    )$ &   -3.2732 &          &  -7.2520 &   4.8965 &          &  -6.5327 &  -3.9200 &         \\
 $\Brcl   ^{\mathrm{used}}$ &      9.80 &          &     9.80 &     9.90 &          &     9.80 &     9.80 &         \\
 \hline
 $a( \chiM    )$ &    0.0453 &          &          &   0.3852 &          &          &          &         \\
 $\chiM   ^{\mathrm{used}}$ &   0.12460 &          &          &  0.11770 &          &          &          &         \\
 \hline
 $a( \fDp     )$ &           &          &          &          &  -0.0221 &          &          &         \\
 $\fDp    ^{\mathrm{used}}$ &           &          &          &          &   0.2210 &          &          &         \\
 \hline
 $a( \fDs     )$ &           &          &          &          &   0.0788 &          &          &         \\
 $\fDs    ^{\mathrm{used}}$ &           &          &          &          &   0.1120 &          &          &         \\
 \hline
 $a( \fLc     )$ &           &          &          &          &   0.0115 &          &          &         \\
 $\fLc    ^{\mathrm{used}}$ &           &          &          &          &   0.0840 &          &          &         \\
 \hline
 \end{tabular}
 \end{center}
 \caption{The measurements of 
 \Acp    .   All numbers are given in \%.
}
 \label{tab:Acpinp}
 \end{table}
 \begin{table}[htb]
 \begin{center}
 \begin{sideways}
 \begin{minipage}[b]{\textheight}
 \begin{center}
 \begin{tabular}{|l||c|c|c|c|c|c|c|c|c|c|c|}
 \hline
           & \mca{4}{ALEPH} & \mca{3}{DELPHI} & \mca{1}{L3} & \mca{3}{OPAL} \\
 \hline
           &90-95 &90-95 &90-95 &91-95 &93-95\dag &92-95 &92-95 &90-95 &91-95 &90-95\dag &90-95\\
           &lepton &lepton &lepton &multi &lepton &$D$-meson &jet charge &lepton &jet charge &lepton &$D$-meson\\
           &\tmcite{ref:alasy} &\tmcite{ref:alasy} &\tmcite{ref:alasy} &\tmcite{ref:ajet} &\tmcite{ref:dlasy} &\tmcite{ref:ddasy} &\tmcite{ref:djasy} &\tmcite{ref:llasy} &\tmcite{ref:ojet} &\tmcite{ref:olasy} &\tmcite{ref:odsac}\\
 \hline\hline
 \roots\ (\GeV)       &  92.050   & 92.940   & 93.900   & 92.950   & 92.990   & 92.990   & 92.940   & 93.100   & 92.910   & 92.950   & 92.950  \\
 \hline
 \Abh     &      3.93 &    10.60 &     9.03 &     11.72 &    10.10 &     8.77 &    12.30 &    13.79 &    14.60 &    10.76 &    -3.10\\
 \hline
 \hline\hline
 \Abh     Corrected &       \mca{3}{10.03}
 &    11.71 &    10.05 &     8.72 &    12.30 &    13.63 &    14.63 &    10.75 &    -3.11\\
 \hline
 Statistical          &       \mca{3}{1.51}
 &     0.98 &     1.80 &     6.37 &     1.60 &     2.40 &     1.70 &     1.43 &     9.00\\
 \hline
 Uncorrelated         &      \mca{3}{0.14}
 &     0.12 &     0.14 &     0.97 &     0.24 &     0.34 &     0.64 &     0.25 &     2.03\\
 Correlated           &       \mca{3}{0.24}
 &     0.02 &     0.16 &     0.14 &     0.08 &     0.19 &     0.34 &     0.28 &     1.69\\
 \hline
 Total Systematic     &        \mca{3}{0.28}
 &     0.13 &     0.21 &     0.98 &     0.26 &     0.39 &     0.73 &     0.37 &     2.65\\
 \hline
 \hline
 $a( \Rb      )$ &   \mca{3}{-1.964} 
&  -9.5 &  -2.8859 &          &  -0.1962 &  -3.3756 & -12.9000 &  -0.8000 &         \\
 $\Rb     ^{\mathrm{used}}$ &      \mca{3}{0.2192} 
 &   0.2156 &   0.2164 &          &   0.2158 &   0.2170 &   0.2150 &   0.2155 &         \\
 \hline
 $a( \Rc      )$ &     \mca{3}{1.575}
&   0.3100 &   1.3577 &          &   1.2000 &   1.9869 &   0.6900 &   0.8000 &         \\
 $\Rc     ^{\mathrm{used}}$ &   \mca{3}{0.1710}
 &   0.1719 &   0.1671 &          &   0.1720 &   0.1734 &   0.1730 &   0.1720 &         \\
 \hline
 $a( \Ach     )$ &     \mca{3}{1.081}
&   1.2793 &          &          &          &   0.5206 &   1.3287 &          &         \\
 $\Ach    ^{\mathrm{used}}$ &     \mca{3}{12.51}
 &    12.42 &          &          &          &    12.39 &    12.08 &          &         \\
 \hline
 $a( \Brbl    )$ &   \mca{3}{-1.762}
 &          &  -2.3557 &          &          &  -2.0790 &          &  -1.3625 &         \\
 $\Brbl   ^{\mathrm{used}}$ &    \mca{3}{11.34}
&          &    10.56 &          &          &    10.50 &          &    10.90 &         \\
 \hline
 $a( \Brbclp  )$ &   \mca{3}{-0.2478}
 &          &  -0.7595 &          &          &  -1.1200 &          &   0.7064 &         \\
 $\Brbclp ^{\mathrm{used}}$ &    \mca{3}{7.86} 
&          &     8.07 &          &          &     8.00 &          &     8.30 &         \\
 \hline
 $a( \Brcl    )$ &    \mca{3}{1.524}
&          &   1.0703 &          &          &   1.9796 &          &   0.7840 &         \\
 $\Brcl   ^{\mathrm{used}}$ &    \mca{3}{9.80}
&          &     9.90 &          &          &     9.80 &          &     9.80 &         \\
 \hline
 $a( \chiM    )$ &    \mca{3}{6.584}
 &          &   1.6050 &          &          &          &          &          &         \\
 $\chiM   ^{\mathrm{used}}$ &    \mca{3}{0.12460}
 &          &  0.11770 &          &          &          &          &          &         \\
 \hline
 $a( \fDp     )$ &       \mca{3}{}     
  &          &          &   0.3978 &   0.4229 &          &          &          &         \\
 $\fDp    ^{\mathrm{used}}$ &   \mca{3}{}       
         &          &          &   0.2210 &   0.2330 &          &          &          &         \\
 \hline
 $a( \fDs     )$ &      \mca{3}{}      
         &          &          &  -0.0788 &   0.0211 &          &          &          &         \\
 $\fDs    ^{\mathrm{used}}$ &      \mca{3}{} 
         &          &          &   0.1120 &   0.1020 &          &          &          &         \\
 \hline
 $a( \fLc     )$ &        \mca{3}{} 
        &          &          &   0.0573 &  -0.0855 &          &          &          &         \\
 $\fLc    ^{\mathrm{used}}$ &    \mca{3}{}       
          &          &          &   0.0840 &   0.0630 &          &          &          &         \\
 \hline
 $a( \PcDst   )$ &    \mca{3}{}  &  -0.2800 &          &          &          &          &          &          &         \\
 $\PcDst  ^{\mathrm{used}}$ &      \mca{3}{}   &   0.1830 &          &          &          &          &          &          &         \\
 \hline
 \end{tabular}
 \end{center}
 \caption{The measurements of 
 \Abh    .   All numbers are given in \%.
}
 \label{tab:Abhinp}
 \end{minipage}
 \end{sideways}
 \end{center}
 \end{table}
 \begin{table}[htb]
 \begin{center}
 \begin{tabular}{|l||c|c|c|c|c|}
 \hline
           & \mca{1}{ALEPH} & \mca{2}{DELPHI} & \mca{2}{OPAL} \\
 \hline
           &91-95 &93-95\dag &92-95 &90-95\dag &90-95\\
           &$D$-meson &lepton &$D$-meson &lepton &$D$-meson\\
           &\tmcite{ref:adsac} &\tmcite{ref:dlasy} &\tmcite{ref:ddasy} &\tmcite{ref:olasy} &\tmcite{ref:odsac}\\
 \hline\hline
 \roots\ (\GeV)       &  92.960   & 92.990   & 92.990   & 92.950   & 92.950  \\
 \hline
 \Ach     &     10.82 &    10.50 &    11.78 &    15.62 &    16.50\\
 \hline
 \hline\hline
 \Ach     Corrected &     10.77 &    10.37 &    11.65 &    15.59 &    16.47\\
 \hline
 Statistical          &      3.30 &     2.90 &     3.20 &     2.02 &     4.10\\
 \hline
 Uncorrelated         &      0.79 &     0.41 &     0.52 &     0.57 &     0.92\\
 Correlated           &      0.18 &     0.28 &     0.07 &     0.62 &     0.43\\
 \hline
 Total Systematic     &      0.81 &     0.50 &     0.52 &     0.84 &     1.02\\
 \hline
 \hline
 $a( \Rb      )$ &           &  -4.0402 &          &   9.6000 &         \\
 $\Rb     ^{\mathrm{used}}$ &           &   0.2164 &          &   0.2155 &         \\
 \hline
 $a( \Rc      )$ &           &   7.5891 &          &  -8.9000 &         \\
 $\Rc     ^{\mathrm{used}}$ &           &   0.1671 &          &   0.1720 &         \\
 \hline
 $a( \Abh     )$ &   -2.6333 &          &          &          &         \\
 $\Abh    ^{\mathrm{used}}$ &     12.08 &          &          &          &         \\
 \hline
 $a( \Brbl    )$ &           &  -3.2492 &          &   9.5375 &         \\
 $\Brbl   ^{\mathrm{used}}$ &           &    10.56 &          &    10.90 &         \\
 \hline
 $a( \Brbclp  )$ &           &   1.5191 &          &  -1.5894 &         \\
 $\Brbclp ^{\mathrm{used}}$ &           &     8.07 &          &     8.30 &         \\
 \hline
 $a( \Brcl    )$ &           &   8.1341 &          &  -9.2120 &         \\
 $\Brcl   ^{\mathrm{used}}$ &           &     9.90 &          &     9.80 &         \\
 \hline
 $a( \chiM    )$ &           &  -0.2140 &          &          &         \\
 $\chiM   ^{\mathrm{used}}$ &           &  0.11770 &          &          &         \\
 \hline
 $a( \fDp     )$ &           &          &  -0.2984 &          &         \\
 $\fDp    ^{\mathrm{used}}$ &           &          &   0.2210 &          &         \\
 \hline
 $a( \fDs     )$ &           &          &   0.0539 &          &         \\
 $\fDs    ^{\mathrm{used}}$ &           &          &   0.1120 &          &         \\
 \hline
 $a( \fLc     )$ &           &          &   0.0764 &          &         \\
 $\fLc    ^{\mathrm{used}}$ &           &          &   0.0840 &          &         \\
 \hline
 \end{tabular}
 \end{center}
 \caption{The measurements of 
 \Ach    .   All numbers are given in \%.
}
 \label{tab:Achinp}
 \end{table}
 \begin{table}[htb]
 \begin{center}
 \begin{tabular}{|l||c|c|c|c|}
 \hline
           & \mca{4}{SLD} \\
 \hline
           &93-98\dag &93-98\dag &94-95\dag &96-98\dag\\
           &lepton &jet charge &$K^{\pm}$ &multi\\
           &\tmcite{ref:SLD_AQL} &\tmcite{ref:SLD_ABJ} &\tmcite{ref:SLD_ABK} &\tmcite{ref:SLD_vtxasy}\\
 \hline\hline
 \roots\ (\GeV)       &  91.280   & 91.280   & 91.280   & 91.280  \\
 \hline
 \cAb     &     0.924 &    0.907 &    0.855 &    0.921\\
 \hline
 Statistical          &     0.030 &    0.020 &    0.088 &    0.018\\
 \hline
 Uncorrelated         &     0.018 &    0.023 &    0.102 &    0.018\\
 Correlated           &     0.008 &    0.001 &    0.006 &    0.001\\
 \hline
 Total Systematic     &     0.020 &    0.023 &    0.102 &    0.018\\
 \hline
 \hline
 $a( \Rb      )$ &   -0.1237 &          &  -0.0139 &  -0.7283\\
 $\Rb     ^{\mathrm{used}}$ &    0.2164 &          &   0.2180 &   0.2158\\
 \hline
 $a( \Rc      )$ &    0.0308 &          &   0.0060 &   0.0359\\
 $\Rc     ^{\mathrm{used}}$ &    0.1674 &          &   0.1710 &   0.1722\\
 \hline
 $a( \cAc     )$ &    0.0534 &   0.0211 &  -0.0112 &   0.0095\\
 $\cAc    ^{\mathrm{used}}$ &     0.667 &    0.670 &    0.666 &    0.667\\
 \hline
 $a( \Brbl    )$ &   -0.1999 &          &          &         \\
 $\Brbl   ^{\mathrm{used}}$ &     10.62 &          &          &         \\
 \hline
 $a( \Brbclp  )$ &    0.0968 &          &          &         \\
 $\Brbclp ^{\mathrm{used}}$ &      8.07 &          &          &         \\
 \hline
 $a( \Brcl    )$ &    0.0369 &          &          &         \\
 $\Brcl   ^{\mathrm{used}}$ &      9.85 &          &          &         \\
 \hline
 $a( \chiM    )$ &    0.2951 &          &          &         \\
 $\chiM   ^{\mathrm{used}}$ &   0.11860 &          &          &         \\
 \hline
 \end{tabular}
 \end{center}
 \caption{The measurements of 
 \cAb    . }
 \label{tab:cAbinp}
 \end{table}
 \begin{table}[htb]
 \begin{center}
 \begin{tabular}{|l||c|c|c|}
 \hline
           & \mca{3}{SLD} \\
 \hline
           &93-98\dag &93-98\dag &96-98\dag\\
           &lepton &$D$-meson &K+vertex\\
           &\tmcite{ref:SLD_AQL} &\tmcite{ref:SLD_ACD} &\tmcite{ref:SLD_ACV}\\
 \hline\hline
 \roots\ (\GeV)       &  91.280   & 91.280   & 91.280  \\
 \hline
 \cAc     &     0.589 &    0.688 &    0.673\\
 \hline
 Statistical          &     0.055 &    0.035 &    0.029\\
 \hline
 Uncorrelated         &     0.045 &    0.020 &    0.024\\
 Correlated           &     0.021 &    0.003 &    0.002\\
 \hline
 Total Systematic     &     0.050 &    0.021 &    0.024\\
 \hline
 \hline
 $a( \Rb      )$ &    0.1855 &          &   0.5395\\
 $\Rb     ^{\mathrm{used}}$ &    0.2164 &          &   0.2158\\
 \hline
 $a( \Rc      )$ &   -0.4053 &          &  -0.0682\\
 $\Rc     ^{\mathrm{used}}$ &    0.1674 &          &   0.1722\\
 \hline
 $a( \cAb     )$ &    0.2137 &  -0.0673 &  -0.0187\\
 $\cAb    ^{\mathrm{used}}$ &     0.935 &    0.935 &    0.935\\
 \hline
 $a( \Brbl    )$ &    0.2874 &          &         \\
 $\Brbl   ^{\mathrm{used}}$ &     10.62 &          &         \\
 \hline
 $a( \Brbclp  )$ &   -0.1743 &          &         \\
 $\Brbclp ^{\mathrm{used}}$ &      8.07 &          &         \\
 \hline
 $a( \Brcl    )$ &   -0.3971 &          &         \\
 $\Brcl   ^{\mathrm{used}}$ &      9.85 &          &         \\
 \hline
 $a( \chiM    )$ &    0.0717 &          &         \\
 $\chiM   ^{\mathrm{used}}$ &   0.11860 &          &         \\
 \hline
 \end{tabular}
 \end{center}
 \caption{The measurements of 
 \cAc    . }
 \label{tab:cAcinp}
 \end{table}
 \begin{table}[htb]
 \begin{center}
 \begin{tabular}{|l||c|c|c|c|c|}
 \hline
           & \mca{1}{ALEPH} & \mca{1}{DELPHI} & \mca{2}{L3} & \mca{1}{OPAL} \\
 \hline
           &91-95\dag & 94-95\dag & 92    &94-95\dag &92-95\\
           &multi     & multi     &lepton &multi     &multi \\
           &\tmcite{ref:abl} &\tmcite{ref:dbl} &\tmcite{ref:lbl} &\tmcite{ref:lrbmixed} &\tmcite{ref:obl}\\
 \hline\hline
 \Brbl                &     10.70 &    10.70 &    10.68 &    10.22 &    10.85\\
 \hline                                                            
 Statistical          &      0.10 &     0.08 &     0.11 &     0.13 &     0.10\\
 \hline                                                            
 Uncorrelated         &      0.16 &     0.20 &     0.36 &     0.19 &     0.20\\
 Correlated           &      0.23 &     0.45 &     0.22 &     0.31 &     0.21\\
 \hline
 Total Systematic     &      0.28 &     0.49 &     0.42 &     0.36 &     0.29\\
 \hline
 \hline
 $a( \Rb      )$      &           &          &  -9.2571 &          &  -0.1808\\
 $\Rb     ^{\mathrm{used}}$                                        
                      &           &          &   0.2160 &          &   0.2169\\
 \hline                                                            
 $a( \Rc      )$      &           &          &          &   1.4450 &   0.4867\\
 $\Rc     ^{\mathrm{used}}$                                        
                      &           &          &          &   0.1734 &   0.1770\\
 \hline                                                            
 $a( \Brbclp  )$      &           &          &  -1.1700 &   0.1618 &         \\
 $\Brbclp ^{\mathrm{used}}$                                        
                      &           &          &     9.00 &     8.09 &         \\
 \hline                                                            
 $a( \Brcl    )$      &    -0.3078 &  -0.1960 &  -2.5480 &   0.9212 &         \\
 $\Brcl   ^{\mathrm{used}}$                                        
                      &      9.85 &     9.80 &     9.80 &     9.80 &         \\
 \hline                                                            
 $a( \chiM    )$      &    0.7683 &          &          &          &         \\
 $\chiM   ^{\mathrm{used}}$                                        
                      &    0.1178 &          &          &          &         \\
 \hline                                                            
 $a( \fDp     )$      &           &          &          &   0.5523 &   0.1445\\
 $\fDp    ^{\mathrm{used}}$                                        
                      &           &          &          &   0.2330 &   0.2380\\
 \hline                                                            
 $a( \fDs     )$      &           &          &          &   0.0213 &   0.0055\\
 $\fDs    ^{\mathrm{used}}$                                        
                      &           &          &          &   0.1030 &   0.1020\\
 \hline                                                            
 $a( \fLc     )$      &           &          &          &  -0.0427 &  -0.0157\\
 $\fLc    ^{\mathrm{used}}$                                        
                      &           &          &          &   0.0630 &   0.0650\\
 \hline
 \end{tabular}
 \end{center}
 \caption{The measurements of 
 \Brbl   .   All numbers are given in \%.
}
 \label{tab:Brblinp}
 \end{table}
 \begin{table}[htb]
 \begin{center}
 \begin{tabular}{|l||c|c|c|}
 \hline
           & \mca{1}{ALEPH} & \mca{1}{DELPHI} & \mca{1}{OPAL} \\
 \hline
           &91-95\dag &94-95\dag &92-95 \\
           &multi &multi &multi \\
           &\tmcite{ref:abl} &\tmcite{ref:dbl} &\tmcite{ref:obl} \\
 \hline\hline
 \Brbclp  &      8.18 &     7.98 &     8.41\\
 \hline
 Statistical          &      0.15 &     0.22 &     0.16\\
 \hline
 Uncorrelated         &      0.19 &     0.21 &     0.19\\
 Correlated           &      0.15 &     0.19 &     0.34\\
 \hline
 Total Systematic     &      0.24 &     0.28 &     0.39\\
 \hline
 \hline
 $a( \Rb      )$ &           &          &  -0.1808 \\
 $\Rb     ^{\mathrm{used}}$ &           &          &   0.2169 \\
 \hline
 $a( \Rc      )$ &    0.5026 &          &   0.3761 \\
 $\Rc     ^{\mathrm{used}}$ &    0.1709 &          &   0.1770 \\
 \hline
 $a( \Brcl    )$ &    0.3078 &          &                  \\
 $\Brcl   ^{\mathrm{used}}$ &      9.85 &          &                   \\
 \hline
 $a( \chiM    )$ &   -1.3884 &          &                  \\
 $\chiM   ^{\mathrm{used}}$ &   0.11940 &                    &         \\
 \hline
 $a( \fDp     )$ &           &          &   0.1190 \\
 $\fDp    ^{\mathrm{used}}$ &           &         &   0.2380\\
 \hline
 $a( \fDs     )$ &           &          &      0.0028\\
 $\fDs    ^{\mathrm{used}}$ &           &   &   0.1020\\
 \hline
 $a( \fLc     )$ &           &          &   -0.0110\\
 $\fLc    ^{\mathrm{used}}$ &           &    &   0.0660\\
 \hline
 \end{tabular}
 \end{center}
 \caption{The measurements of 
 \Brbclp .   All numbers are given in \%.
}
 \label{tab:Brbclpinp}
 \end{table}
 \begin{table}[htb]
 \begin{center}
 \begin{tabular}{|l||c|c|}
 \hline
           & \mca{1}{DELPHI} & \mca{1}{OPAL} \\
 \hline
           &92-95 &90-95\\
           &$D$+lepton &$D$+lepton\\
           &\tmcite{ref:drcd} &\tmcite{ref:ocl}\\
 \hline\hline
 $\Brcl$              &      9.64 &     9.58\\
 \hline
 Statistical          &      0.42 &     0.60\\
 \hline
 Uncorrelated         &      0.24 &     0.49\\
 Correlated           &      0.13 &     0.43\\
 \hline
 Total Systematic     &      0.27 &     0.65\\
 \hline
 \hline
 $a( \Brbl    )$ &   -0.5600 &  -1.4335\\
 $\Brbl   ^{\mathrm{used}}$ &     11.20 &    10.99\\
 \hline
 $a( \Brbclp  )$ &   -0.4100 &  -0.7800\\
 $\Brbclp ^{\mathrm{used}}$ &      8.20 &     7.80\\
 \hline
 \end{tabular}
 \end{center}
 \caption{The measurements of 
 $\Brcl$.   All numbers are given in \%.
}
 \label{tab:Brclinp}
 \end{table}
 \begin{table}[htb]
 \begin{center}
 \begin{tabular}{|l||c|c|c|c|}
 \hline
           & \mca{1}{ALEPH} & \mca{1}{DELPHI} & \mca{1}{L3} & \mca{1}{OPAL} \\
 \hline
           &90-95 &94-95\dag &90-95 &90-95\dag\\
           &multi &multi &lepton &lepton\\
           &\tmcite{ref:alasy} &\tmcite{ref:dbl} &\tmcite{ref:llasy} &\tmcite{ref:olasy}\\
 \hline\hline
 \chiM    &   0.12446 &  0.12700 &  0.11920 &  0.11380\\
 \hline
 Statistical          &   0.00515 &  0.01300 &  0.00680 &  0.00540\\
 \hline
 Uncorrelated         &   0.00252 &  0.00484 &  0.00214 &  0.00306\\
 Correlated           &   0.00394 &  0.00431 &  0.00252 &  0.00324\\
 \hline
 Total Systematic     &   0.00468 &  0.00648 &  0.00330 &  0.00445\\
 \hline
 \hline
 $a( \Rb      )$ &    0.0341 &          &          &         \\
 $\Rb     ^{\mathrm{used}}$ &    0.2192 &          &          &         \\
 \hline
 $a( \Rc      )$ &    0.0009 &          &   0.0004 &         \\
 $\Rc     ^{\mathrm{used}}$ &    0.1710 &          &   0.1734 &         \\
 \hline
 $a( \Brbl    )$ &    0.0524 &          &   0.0550 &   0.0170\\
 $\Brbl   ^{\mathrm{used}}$ &     11.34 &          &    10.50 &    10.90\\
 \hline
 $a( \Brbclp  )$ &   -0.0440 &          &  -0.0466 &  -0.0318\\
 $\Brbclp ^{\mathrm{used}}$ &      7.86 &          &     8.00 &     8.30\\
 \hline
 $a( \Brcl    )$ &    0.0035 &  -0.0020 &   0.0006 &   0.0039\\
 $\Brcl   ^{\mathrm{used}}$ &      9.80 &     9.80 &     9.80 &     9.80\\
 \hline
 \end{tabular}
 \end{center}
 \caption{The measurements of 
 \chiM   . }
 \label{tab:chiMinp}
 \end{table}
 \begin{table}[htb]
 \begin{center}
 \begin{tabular}{|l||c|c|}
 \hline
           & \mca{1}{DELPHI} & \mca{1}{OPAL} \\
 \hline
           &92-95 &90-95\\
           &$D$-meson &$D$-meson\\
           &\tmcite{ref:drcd} &\tmcite{ref:orcd}\\
 \hline\hline
 \PcDst   &    0.1740 &   0.1514\\
 \hline
 Statistical          &    0.0100 &   0.0096\\
 \hline
 Uncorrelated         &    0.0040 &   0.0088\\
 Correlated           &    0.0007 &   0.0011\\
 \hline
 Total Systematic     &    0.0041 &   0.0089\\
 \hline
 \hline
 $a( \Rb      )$ &    0.0293 &         \\
 $\Rb     ^{\mathrm{used}}$ &    0.2166 &         \\
 \hline
 $a( \Rc      )$ &   -0.0158 &         \\
 $\Rc     ^{\mathrm{used}}$ &    0.1735 &         \\
 \hline
 \end{tabular}
 \end{center}
 \caption{The measurements of 
 \PcDst  . }
 \label{tab:PcDstinp}
 \end{table}
 \clearpage
 \begin{table}[htb]
 \begin{center}
 \begin{tabular}{|l||c|c|c|}
 \hline
           & \mca{1}{ALEPH} & \mca{1}{DELPHI} & \mca{1}{OPAL} \\
 \hline
           &91-95 &92-95 &91-94\\
           &$D$ meson &$D$ meson &$D$ meson\\
           &\tmcite{ref:arcc} &\tmcite{ref:drcc} &\tmcite{ref:orcc}\\
 \hline\hline
 \RcfDp   &    0.0406 &   0.0384 &   0.0391\\
 \hline
 Statistical          &    0.0013 &   0.0013 &   0.0050\\
 \hline
 Uncorrelated         &    0.0014 &   0.0015 &   0.0042\\
 Correlated           &    0.0032 &   0.0025 &   0.0031\\
 \hline
 Total Systematic     &    0.0035 &   0.0030 &   0.0052\\
 \hline
 \hline
 $a( \fDp     )$ &           &   0.0008 &         \\
 $\fDp    ^{\mathrm{used}}$ &           &   0.2210 &         \\
 \hline
 $a( \fDs     )$ &           &  -0.0002 &         \\
 $\fDs    ^{\mathrm{used}}$ &           &   0.1120 &         \\
 \hline
 \end{tabular}
 \end{center}
 \caption{The measurements of 
 \RcfDp  . }
 \label{tab:RcfDpinp}
 \end{table}
 \begin{table}[htb]
 \begin{center}
 \begin{tabular}{|l||c|c|c|}
 \hline
           & \mca{1}{ALEPH} & \mca{1}{DELPHI} & \mca{1}{OPAL} \\
 \hline
           &91-95 &92-95 &91-94\\
           &$D$ meson &$D$ meson &$D$ meson\\
           &\tmcite{ref:arcc} &\tmcite{ref:drcc} &\tmcite{ref:orcc}\\
 \hline\hline
 \RcfDs   &    0.0207 &   0.0213 &   0.0160\\
 \hline
 Statistical          &    0.0033 &   0.0017 &   0.0042\\
 \hline
 Uncorrelated         &    0.0011 &   0.0010 &   0.0016\\
 Correlated           &    0.0053 &   0.0054 &   0.0043\\
 \hline
 Total Systematic     &    0.0054 &   0.0055 &   0.0046\\
 \hline
 \hline
 $a( \fDp     )$ &           &   0.0007 &         \\
 $\fDp    ^{\mathrm{used}}$ &           &   0.2210 &         \\
 \hline
 $a( \fDs     )$ &           &  -0.0009 &         \\
 $\fDs    ^{\mathrm{used}}$ &           &   0.1120 &         \\
 \hline
 $a( \fLc     )$ &           &  -0.0001 &         \\
 $\fLc    ^{\mathrm{used}}$ &           &   0.0840 &         \\
 \hline
 \end{tabular}
 \end{center}
 \caption{The measurements of 
 \RcfDs  . }
 \label{tab:RcfDsinp}
 \end{table}
 \begin{table}[htb]
 \begin{center}
 \begin{tabular}{|l||c|c|c|}
 \hline
           & \mca{1}{ALEPH} & \mca{1}{DELPHI} & \mca{1}{OPAL} \\
 \hline
           &91-95 &92-95 &91-94\\
           &$D$ meson &$D$ meson &$D$ meson\\
           &\tmcite{ref:arcc} &\tmcite{ref:drcc} &\tmcite{ref:orcc}\\
 \hline\hline
 \RcfLc   &    0.0157 &   0.0170 &   0.0091\\
 \hline
 Statistical          &    0.0016 &   0.0035 &   0.0050\\
 \hline
 Uncorrelated         &    0.0005 &   0.0016 &   0.0015\\
 Correlated           &    0.0044 &   0.0045 &   0.0035\\
 \hline
 Total Systematic     &    0.0045 &   0.0048 &   0.0038\\
 \hline
 \hline
 $a( \fDp     )$ &           &   0.0002 &         \\
 $\fDp    ^{\mathrm{used}}$ &           &   0.2210 &         \\
 \hline
 $a( \fDs     )$ &           &  -0.0001 &         \\
 $\fDs    ^{\mathrm{used}}$ &           &   0.1120 &         \\
 \hline
 $a( \fLc     )$ &           &  -0.0002 &         \\
 $\fLc    ^{\mathrm{used}}$ &           &   0.0840 &         \\
 \hline
 \end{tabular}
 \end{center}
 \caption{The measurements of 
 \RcfLc  . }
 \label{tab:RcfLcinp}
 \end{table}
 \begin{table}[htb]
 \begin{center}
 \begin{tabular}{|l||c|c|c|}
 \hline
           & \mca{1}{ALEPH} & \mca{1}{DELPHI} & \mca{1}{OPAL} \\
 \hline
           &91-95 &92-95 &91-94\\
           &$D$ meson &$D$ meson &$D$ meson\\
           &\tmcite{ref:arcc} &\tmcite{ref:drcc} &\tmcite{ref:orcc}\\
 \hline\hline
 \RcfDz   &    0.0965 &   0.0928 &   0.1000\\
 \hline
 Statistical          &    0.0029 &   0.0026 &   0.0070\\
 \hline
 Uncorrelated         &    0.0040 &   0.0038 &   0.0057\\
 Correlated           &    0.0045 &   0.0023 &   0.0041\\
 \hline
 Total Systematic     &    0.0060 &   0.0044 &   0.0070\\
 \hline
 \hline
 $a( \fDp     )$ &           &   0.0021 &         \\
 $\fDp    ^{\mathrm{used}}$ &           &   0.2210 &         \\
 \hline
 $a( \fDs     )$ &           &  -0.0004 &         \\
 $\fDs    ^{\mathrm{used}}$ &           &   0.1120 &         \\
 \hline
 $a( \fLc     )$ &           &  -0.0004 &         \\
 $\fLc    ^{\mathrm{used}}$ &           &   0.0840 &         \\
 \hline
 \end{tabular}
 \end{center}
 \caption{The measurements of 
 \RcfDz  . }
 \label{tab:RcfDzinp}
 \end{table}
 \begin{table}[htb]
 \begin{center}
 \begin{tabular}{|l||c|c|}
 \hline
           & \mca{1}{DELPHI} & \mca{1}{OPAL} \\
 \hline
           &92-95 &90-95\\
           &$D$ meson &$D$-meson\\
           &\tmcite{ref:drcc} &\tmcite{ref:orcd}\\
 \hline\hline
 \RcPcDst &    0.0282 &   0.0268\\
 \hline
 Statistical          &    0.0007 &   0.0005\\
 \hline
 Uncorrelated         &    0.0010 &   0.0010\\
 Correlated           &    0.0007 &   0.0009\\
 \hline
 Total Systematic     &    0.0012 &   0.0013\\
 \hline
 \hline
 $a( \fDp     )$ &    0.0006 &         \\
 $\fDp    ^{\mathrm{used}}$ &    0.2210 &         \\
 \hline
 $a( \fDs     )$ &   -0.0001 &         \\
 $\fDs    ^{\mathrm{used}}$ &    0.1120 &         \\
 \hline
 $a( \fLc     )$ &   -0.0004 &         \\
 $\fLc    ^{\mathrm{used}}$ &    0.0840 &         \\
 \hline
 \end{tabular}
 \end{center}
 \caption{The measurements of 
 \RcPcDst. }
 \label{tab:RcPcDstinp}
 \end{table}

\chapter{Heavy-Flavour Fit including Off-Peak Asymmetries}\label{app-HF-fit}
The full 18 parameter fit to the LEP and SLD data gave the following results:
\begin{eqnarray*}
  \Rbz    &=& 0.21647   \pm  0.00068\\
  \Rcz    &=& 0.1719    \pm  0.0031 \\
  \Abl    &=& 0.0508    \pm  0.0068 \\
  \Acl    &=&-0.035     \pm  0.017  \\
  \Abp    &=& 0.0975    \pm  0.0018 \\
  \Acp    &=& 0.0620    \pm  0.0036 \\
  \Abh    &=& 0.1150    \pm  0.0057 \\
  \Ach    &=& 0.130     \pm  0.013  \\
  \cAb    &=& 0.922     \pm  0.020  \\
  \cAc    &=& 0.670     \pm  0.026  \\
  \Brbl   &=& 0.1067    \pm  0.0021 \\
  \Brbclp &=& 0.0807    \pm  0.0017 \\
  \Brcl   &=& 0.0979    \pm  0.0031 \\
  \chiM   &=& 0.1195    \pm  0.0040 \\
  \fDp    &=& 0.234     \pm  0.016  \\
  \fDs    &=& 0.125     \pm  0.023  \\
  \fcb    &=& 0.096     \pm  0.023  \\
  \PcDst  &=& 0.1620    \pm  0.0048 \,
\end{eqnarray*}
with a $\chi^2/$d.o.f.{} of  $43/(99-18)$. The corresponding correlation
matrix is given in Table~\ref{tab:18parcor}.
The energy for the  peak$-$2, peak and peak+2 results are respectively
89.55 \GeV{}, 91.26 \GeV{} and 92.94 \GeV.
Note that the asymmetry results shown here are not the pole
asymmetries shown in Section~\ref{sec-HFSUM-LEP-SLD}.
The non-electroweak parameters do not depend on the treatment of the 
asymmetries.

\begin{table}[p]
\begin{center}
\begin{sideways}
\begin{minipage}[b]{\textheight}
\begin{center}
\footnotesize
\begin{tabular}{|l||rrrrrrrrrrrrrrrrrr|}
\hline
&\makebox[0.45cm]{$1)$}
&\makebox[0.45cm]{$2)$}
&\makebox[0.45cm]{$3)$}
&\makebox[0.45cm]{$4)$}
&\makebox[0.45cm]{$5)$}
&\makebox[0.45cm]{$6)$}
&\makebox[0.45cm]{$7)$}
&\makebox[0.45cm]{$8)$}
&\makebox[0.45cm]{$9)$}
&\makebox[0.45cm]{$10)$}
&\makebox[0.45cm]{$11)$}
&\makebox[0.45cm]{$12)$}
&\makebox[0.45cm]{$13)$}
&\makebox[0.45cm]{$14)$}
&\makebox[0.45cm]{$15)$}
&\makebox[0.45cm]{$16)$}
&\makebox[0.45cm]{$17)$}
&\makebox[0.45cm]{$18)$}\\
&\makebox[0.45cm]{\Rb}
&\makebox[0.45cm]{\Rc}
&\makebox[0.45cm]{$\Abb$}
&\makebox[0.45cm]{$\Acc$}
&\makebox[0.45cm]{$\Abb$}
&\makebox[0.45cm]{$\Acc$}
&\makebox[0.45cm]{$\Abb$}
&\makebox[0.45cm]{$\Acc$}
&\makebox[0.45cm]{\cAb}
&\makebox[0.45cm]{\cAc}
&\makebox[0.45cm]{BR}
&\makebox[0.45cm]{BR}
&\makebox[0.45cm]{BR}
&\makebox[0.45cm]{\chiM}
&\makebox[0.45cm]{$\fDp$}
&\makebox[0.45cm]{$\fDs$}
&\makebox[0.45cm]{$f(c_{bar.})$}
&\makebox[0.55cm]{PcDst}\\
&
&
&\makebox[0.45cm]{$(-2)$}
&\makebox[0.45cm]{$(-2)$}
&\makebox[0.45cm]{(pk)}
&\makebox[0.45cm]{(pk)}
&\makebox[0.45cm]{$(+2)$}
&\makebox[0.45cm]{$(+2)$}
&
&
&\makebox[0.45cm]{$(1)$}
&\makebox[0.45cm]{$(2)$}
&\makebox[0.45cm]{$(3)$}
&
&
&
&
&\\
\hline\hline
1)  &$  1.00$&$ -0.14$&$ -0.02$&$  0.00$&$ -0.07$&$  0.01$&$ -0.03$&$  0.00$&
     $ -0.08$&$  0.04$&$ -0.09$&$ -0.02$&$ -0.01$&$ -0.03$&$ -0.16$&$ -0.04$&
     $  0.13$&$  0.10$\\
2)  &$ -0.14$&$  1.00$&$  0.01$&$  0.01$&$  0.04$&$ -0.01$&$  0.02$&$ -0.01$&
     $  0.03$&$ -0.05$&$  0.06$&$ -0.03$&$ -0.30$&$  0.04$&$ -0.13$&$  0.17$&
     $  0.16$&$ -0.44$\\
3)  &$ -0.02$&$  0.01$&$  1.00$&$  0.16$&$  0.03$&$  0.01$&$  0.01$&$  0.00$&
     $  0.00$&$  0.00$&$  0.02$&$ -0.02$&$  0.00$&$  0.05$&$  0.00$&$  0.00$&
     $  0.00$&$ -0.01$\\
4)  &$  0.00$&$  0.01$&$  0.16$&$  1.00$&$  0.01$&$  0.01$&$  0.00$&$  0.00$&
     $  0.00$&$  0.00$&$  0.02$&$ -0.01$&$  0.02$&$  0.01$&$  0.00$&$  0.01$&
     $  0.00$&$  0.00$\\
5)  &$ -0.07$&$  0.04$&$  0.03$&$  0.01$&$  1.00$&$  0.15$&$  0.07$&$  0.01$&
     $  0.01$&$  0.00$&$  0.04$&$ -0.08$&$  0.00$&$  0.13$&$  0.01$&$  0.02$&
     $  0.00$&$ -0.03$\\
6)  &$  0.01$&$ -0.01$&$  0.01$&$  0.01$&$  0.15$&$  1.00$&$  0.01$&$  0.09$&
     $  0.00$&$  0.01$&$  0.13$&$ -0.14$&$ -0.08$&$  0.15$&$  0.00$&$  0.00$&
     $ -0.01$&$  0.00$\\
7)  &$ -0.03$&$  0.02$&$  0.01$&$  0.00$&$  0.07$&$  0.01$&$  1.00$&$  0.17$&
     $  0.01$&$  0.00$&$  0.01$&$ -0.03$&$  0.01$&$  0.06$&$  0.01$&$  0.01$&
     $ -0.01$&$ -0.01$\\
8)  &$  0.00$&$ -0.01$&$  0.00$&$  0.00$&$  0.01$&$  0.09$&$  0.17$&$  1.00$&
     $  0.00$&$  0.00$&$  0.02$&$ -0.04$&$ -0.03$&$  0.02$&$  0.00$&$ -0.01$&
     $  0.00$&$  0.01$\\
9)  &$ -0.08$&$  0.03$&$  0.00$&$  0.00$&$  0.01$&$  0.00$&$  0.01$&$  0.00$&
     $  1.00$&$  0.13$&$ -0.02$&$  0.01$&$  0.03$&$  0.06$&$  0.00$&$  0.00$&
     $  0.00$&$ -0.02$\\
10) &$  0.04$&$ -0.05$&$  0.00$&$  0.00$&$  0.00$&$  0.01$&$  0.00$&$  0.00$&
     $  0.13$&$  1.00$&$  0.02$&$ -0.04$&$ -0.02$&$  0.01$&$  0.00$&$  0.00$&
     $  0.00$&$  0.02$\\
11) &$ -0.09$&$  0.06$&$  0.02$&$  0.02$&$  0.04$&$  0.13$&$  0.01$&$  0.02$&
     $ -0.02$&$  0.02$&$  1.00$&$ -0.21$&$  0.01$&$  0.37$&$  0.03$&$  0.01$&
     $ -0.01$&$ -0.01$\\
12) &$ -0.02$&$ -0.03$&$ -0.02$&$ -0.01$&$ -0.08$&$ -0.14$&$ -0.03$&$ -0.04$&
     $  0.01$&$ -0.04$&$ -0.21$&$  1.00$&$  0.08$&$ -0.31$&$  0.02$&$ -0.01$&
     $ -0.01$&$  0.01$\\
13) &$ -0.01$&$ -0.30$&$  0.00$&$  0.02$&$  0.00$&$ -0.08$&$  0.01$&$ -0.03$&
     $  0.03$&$ -0.02$&$  0.01$&$  0.08$&$  1.00$&$  0.14$&$  0.01$&$ -0.03$&
     $ -0.02$&$  0.13$\\
14) &$ -0.03$&$  0.04$&$  0.05$&$  0.01$&$  0.13$&$  0.15$&$  0.06$&$  0.02$&
     $  0.06$&$  0.01$&$  0.37$&$ -0.31$&$  0.14$&$  1.00$&$  0.01$&$  0.01$&
     $  0.00$&$ -0.03$\\
15) &$ -0.16$&$ -0.13$&$  0.00$&$  0.00$&$  0.01$&$  0.00$&$  0.01$&$  0.00$&
     $  0.00$&$  0.00$&$  0.03$&$  0.02$&$  0.01$&$  0.01$&$  1.00$&$ -0.39$&
     $ -0.26$&$  0.10$\\
16) &$ -0.04$&$  0.17$&$  0.00$&$  0.01$&$  0.02$&$  0.00$&$  0.01$&$ -0.01$&
     $  0.00$&$  0.00$&$  0.01$&$ -0.01$&$ -0.03$&$  0.01$&$ -0.39$&$  1.00$&
     $ -0.50$&$ -0.08$\\
17) &$  0.13$&$  0.16$&$  0.00$&$  0.00$&$  0.00$&$ -0.01$&$ -0.01$&$  0.00$&
     $  0.00$&$  0.00$&$ -0.01$&$ -0.01$&$ -0.02$&$  0.00$&$ -0.26$&$ -0.50$&
     $  1.00$&$ -0.14$\\
18) &$  0.10$&$ -0.44$&$ -0.01$&$  0.00$&$ -0.03$&$  0.00$&$ -0.01$&$  0.01$&
     $ -0.02$&$  0.02$&$ -0.01$&$  0.01$&$  0.13$&$ -0.03$&$  0.10$&$ -0.08$&
     $ -0.14$&$  1.00$\\
\hline
\end{tabular}
\normalsize
\end{center}
\caption[]{
  The correlation matrix for the set of the 18 heavy flavour
  parameters. BR(1), BR(2) and BR(3) denote $\Brbl$, $\Brbclp$ and $\Brcl$
  respectively, PcDst denotes $\PcDst$.  }
\label{tab:18parcor}
\end{minipage}
\end{sideways}
\end{center}
\end{table}

\chapter{Detailed inputs and results on W-boson and four-fermion averages}
\label{4f_sec:appendix}

Tables~\ref{4f_tab:WWmeas}, \ref{4f_tab:WWcorr}, 
\ref{4f_tab:WWtheo}, \ref{4f_tab:rWWmeas},
\ref{4f_tab:Wbrmeas}, \ref{4f_tab:ZZmeas},  
\ref{4f_tab:WevTOTmeas} and~\ref{4f_tab:WevHADmeas}
give the details of the inputs and of the results
for the calculation of LEP averages
of the WW cross section, 
of the WW cross section ratio $\rww$,
of W decay branching fractions, 
of the ZZ cross section,
and of the total and hadronic single W cross sections.
For both inputs and results, whenever relevant,
the splitup of the errors into their various components
is given in the table.
For each measurement, 
the Collaborations %
provide
additional information which is necessary 
for the combination of LEP results,
such as the expected statistical error 
or the splitup of the systematic uncertainty 
into its correlated and uncorrelated components.

\begin{table}[hbtp]
\begin{center}
\begin{small}
\begin{tabular}{|c|ccccc|c|c|c|}
\cline{1-8}
& & & {\scriptsize (LCEC)} & {\scriptsize (LUEU)} & 
{\scriptsize (LUEC)} & & &
\multicolumn{1}{|r}{$\quad$} \\
$\sqrt{s}$ & $\sww$ & 
$\Delta\sww^\mathrm{stat}$ &
$\Delta\sww^\mathrm{syst}$ &
$\Delta\sww^\mathrm{syst}$ &
$\Delta\sww^\mathrm{syst}$ &
$\Delta\sww^\mathrm{syst}$ &
$\Delta\sww$ & 
\multicolumn{1}{|r}{$\quad$} \\
\cline{1-8}
\multicolumn{8}{|c|}
{\Aleph~\cite{4f_bib:aleww183,4f_bib:aleww189,
4f_bib:aleww1999,4f_bib:aleww2000}} &
\multicolumn{1}{|r}{$\quad$} \\
\cline{1-8}
182.7 \GeV & 
15.57 & $\pm$0.62 & $\pm$0.09 & $\pm$0.09 & $\pm$0.26 & $\pm$0.29& $\pm$0.68 &
\multicolumn{1}{|r}{$\quad$} \\
188.6 \GeV & 
15.71 & $\pm$0.34 & $\pm$0.05 & $\pm$0.09 & $\pm$0.15 & $\pm$0.18& $\pm$0.38 &
\multicolumn{1}{|r}{$\quad$} \\
191.6 \GeV & 
17.23 & $\pm$0.89 & $\pm$0.05 & $\pm$0.09 & $\pm$0.15 & $\pm$0.18& $\pm$0.91 &
\multicolumn{1}{|r}{$\quad$} \\
195.5 \GeV & 
17.00 & $\pm$0.54 & $\pm$0.05 & $\pm$0.09 & $\pm$0.15 & $\pm$0.18& $\pm$0.57 &
\multicolumn{1}{|r}{$\quad$} \\
199.5 \GeV & 
16.98 & $\pm$0.53 & $\pm$0.05 & $\pm$0.09 & $\pm$0.15 & $\pm$0.18& $\pm$0.56 &
\multicolumn{1}{|r}{$\quad$} \\
201.6 \GeV & 
16.16 & $\pm$0.74 & $\pm$0.05 & $\pm$0.09 & $\pm$0.15 & $\pm$0.18& $\pm$0.76 &
\multicolumn{1}{|r}{$\quad$} \\
204.9 \GeV & 
16.57 & $\pm$0.52 & $\pm$0.05 & $\pm$0.09 & $\pm$0.15 & $\pm$0.18& $\pm$0.55 &
\multicolumn{1}{|r}{$\quad$} \\
206.6 \GeV & 
17.32 & $\pm$0.41 & $\pm$0.05 & $\pm$0.09 & $\pm$0.15 & $\pm$0.18& $\pm$0.45 &
\multicolumn{1}{|r}{$\quad$} \\
\cline{1-8}
\multicolumn{8}{|c|}
{\Delphi~\cite{4f_bib:delww183,4f_bib:delww189,
4f_bib:delww1999,4f_bib:delww2000}} &
\multicolumn{1}{|r}{$\quad$} \\
\cline{1-8}
182.7 \GeV & 
15.86 & $\pm$0.69 & $\pm$0.09 & $\pm$0.07 & $\pm$0.24 & $\pm$0.27& $\pm$0.74 &
\multicolumn{1}{|r}{$\quad$} \\
188.6 \GeV & 
15.83 & $\pm$0.38 & $\pm$0.07 & $\pm$0.05 & $\pm$0.18 & $\pm$0.20& $\pm$0.43 &
\multicolumn{1}{|r}{$\quad$} \\
191.6 \GeV & 
16.90 & $\pm$1.00 & $\pm$0.07 & $\pm$0.06 & $\pm$0.20 & $\pm$0.22& $\pm$1.02 &
\multicolumn{1}{|r}{$\quad$} \\
195.5 \GeV & 
17.86 & $\pm$0.59 & $\pm$0.07 & $\pm$0.06 & $\pm$0.20 & $\pm$0.22& $\pm$0.63 &
\multicolumn{1}{|r}{$\quad$} \\
199.5 \GeV & 
17.35 & $\pm$0.56 & $\pm$0.07 & $\pm$0.06 & $\pm$0.20 & $\pm$0.22& $\pm$0.60 &
\multicolumn{1}{|r}{$\quad$} \\
201.6 \GeV & 
17.67 & $\pm$0.81 & $\pm$0.08 & $\pm$0.07 & $\pm$0.21 & $\pm$0.23& $\pm$0.84 &
\multicolumn{1}{|r}{$\quad$} \\
204.9 \GeV & 
17.44 & $\pm$0.60 & $\pm$0.06 & $\pm$0.05 & $\pm$0.21 & $\pm$0.22& $\pm$0.64 &
\multicolumn{1}{|r}{$\quad$} \\
206.6 \GeV & 
16.50 & $\pm$0.43 & $\pm$0.06 & $\pm$0.05 & $\pm$0.20 & $\pm$0.21& $\pm$0.48 & 
\multicolumn{1}{|r}{$\quad$} \\
\cline{1-8}
\multicolumn{8}{|c|}
{\Ltre~\cite{4f_bib:ltrww183,4f_bib:ltrww189,
4f_bib:ltrww1999,4f_bib:ltrww2000} } &
\multicolumn{1}{|r}{$\quad$} \\
\cline{1-8}
182.7 \GeV & 
16.53 & $\pm$0.67 & $\pm$0.08 & $\pm$0.14 & $\pm$0.21 & $\pm$0.26& $\pm$0.72 &
\multicolumn{1}{|r}{$\quad$} \\
188.6 \GeV & 
16.24 & $\pm$0.37 & $\pm$0.04 & $\pm$0.08 & $\pm$0.20 & $\pm$0.22& $\pm$0.43 &
\multicolumn{1}{|r}{$\quad$} \\
191.6 \GeV & 
16.39 & $\pm$0.90 & $\pm$0.08 & $\pm$0.08 & $\pm$0.21 & $\pm$0.24& $\pm$0.93 &
\multicolumn{1}{|r}{$\quad$} \\
195.5 \GeV & 
16.67 & $\pm$0.55 & $\pm$0.08 & $\pm$0.08 & $\pm$0.21 & $\pm$0.24& $\pm$0.60 &
\multicolumn{1}{|r}{$\quad$} \\
199.5 \GeV & 
16.94 & $\pm$0.57 & $\pm$0.08 & $\pm$0.08 & $\pm$0.21 & $\pm$0.24& $\pm$0.62 &
\multicolumn{1}{|r}{$\quad$} \\
201.6 \GeV & 
16.95 & $\pm$0.85 & $\pm$0.08 & $\pm$0.08 & $\pm$0.21 & $\pm$0.24& $\pm$0.88 &
\multicolumn{1}{|r}{$\quad$} \\
204.9 \GeV & 
17.35 & $\pm$0.59 & $\pm$0.08 & $\pm$0.08 & $\pm$0.21 & $\pm$0.24& $\pm$0.64 &
\multicolumn{1}{|r}{$\quad$} \\
206.6 \GeV & 
17.96 & $\pm$0.45 & $\pm$0.08 & $\pm$0.08 & $\pm$0.21 & $\pm$0.24& $\pm$0.51 &
\multicolumn{1}{|r}{$\quad$} \\
\cline{1-8}
\multicolumn{8}{|c|}
{\Opal~\cite{4f_bib:opaww183,4f_bib:opaww189,
4f_bib:opaww1999,4f_bib:opaww2000a,4f_bib:opaww2000b}} &
\multicolumn{1}{|r}{$\quad$} \\
\cline{1-8}
182.7 \GeV & 
15.43 & $\pm$0.61 & $\pm$0.14 & $\pm$0.00 & $\pm$0.22 & $\pm$0.26& $\pm$0.66 &
\multicolumn{1}{|r}{$\quad$} \\
188.6 \GeV & 
16.30 & $\pm$0.34 & $\pm$0.07 & $\pm$0.00 & $\pm$0.17 & $\pm$0.18& $\pm$0.38 &
\multicolumn{1}{|r}{$\quad$} \\
191.6 \GeV & 
16.60 & $\pm$0.88 & $\pm$0.12 & $\pm$0.00 & $\pm$0.40 & $\pm$0.42& $\pm$0.98 &
\multicolumn{1}{|r}{$\quad$} \\
195.5 \GeV & 
18.59 & $\pm$0.60 & $\pm$0.12 & $\pm$0.00 & $\pm$0.41 & $\pm$0.43& $\pm$0.74 &
\multicolumn{1}{|r}{$\quad$} \\
199.5 \GeV & 
16.32 & $\pm$0.54 & $\pm$0.10 & $\pm$0.00 & $\pm$0.37 & $\pm$0.38& $\pm$0.66 &
\multicolumn{1}{|r}{$\quad$} \\
201.6 \GeV & 
18.48 & $\pm$0.81 & $\pm$0.12 & $\pm$0.00 & $\pm$0.40 & $\pm$0.42& $\pm$0.91 &
\multicolumn{1}{|r}{$\quad$} \\
204.9 \GeV & 
15.97 & $\pm$0.52 & $\pm$0.10 & $\pm$0.00 & $\pm$0.36 & $\pm$0.37& $\pm$0.64 &
\multicolumn{1}{|r}{$\quad$} \\
206.6 \GeV & 
17.77 & $\pm$0.42 & $\pm$0.09 & $\pm$0.00 & $\pm$0.37 & $\pm$0.38& $\pm$0.57 &
\multicolumn{1}{|r}{$\quad$} \\
\cline{1-8}
\hline
\multicolumn{8}{|c|}{LEP Averages } & $\chi^2/\textrm{d.o.f.}$ \\
\hline
182.7 \GeV & 
15.79 & $\pm$0.32 & $\pm$0.10 & $\pm$0.04 & $\pm$0.11 & $\pm$0.15& $\pm$0.36 & 
 \multirow{8}{20.3mm}{$
   \hspace*{-0.3mm}
   \left\}
     \begin{array}[h]{rr}
       &\multirow{8}{8mm}{\hspace*{-4.2mm}27.42/24}\\
       &\\ &\\ &\\ &\\ &\\ &\\ &\\  
     \end{array}
   \right.
   $}\\
188.6 \GeV & 
16.00 & $\pm$0.18 & $\pm$0.05 & $\pm$0.03 & $\pm$0.08 & $\pm$0.10& $\pm$0.21 & \\
191.6 \GeV & 
16.72 & $\pm$0.46 & $\pm$0.07 & $\pm$0.03 & $\pm$0.11 & $\pm$0.13& $\pm$0.48 & \\
195.5 \GeV & 
17.43 & $\pm$0.29 & $\pm$0.07 & $\pm$0.04 & $\pm$0.10 & $\pm$0.13& $\pm$0.32 & \\
199.5 \GeV & 
16.84 & $\pm$0.28 & $\pm$0.07 & $\pm$0.04 & $\pm$0.10 & $\pm$0.13& $\pm$0.31 & \\
201.6 \GeV & 
17.23 & $\pm$0.40 & $\pm$0.07 & $\pm$0.04 & $\pm$0.10 & $\pm$0.13& $\pm$0.42 & \\
204.9 \GeV & 
16.71 & $\pm$0.28 & $\pm$0.07 & $\pm$0.04 & $\pm$0.10 & $\pm$0.13& $\pm$0.31 & \\
206.6 \GeV & 
17.33 & $\pm$0.22 & $\pm$0.06 & $\pm$0.04 & $\pm$0.10 & $\pm$0.12& $\pm$0.25 & \\
\hline
\end{tabular}
\end{small}
\caption{\small
W-pair production cross section (in pb) for different \CoM\ energies.
The first column contains the \CoM\ energy,
and the second, the measurements.
Observed statistical uncertainties are used in the fit
and are listed in the third column;
when asymmetric errors are quoted by the Collaborations,
the positive error is listed in the table and used in the fit.
The fourth, fifth and sixth columns contain
the components of the systematic errors,
as subdivided by the Collaborations into
LEP-correlated   energy-correlated   (LCEC),
LEP-uncorrelated energy-uncorrelated (LUEU),
LEP-uncorrelated energy-correlated   (LUEC).
The total systematic error is given in the seventh column,
the total error in the eighth.
For the LEP averages, the $\chi^2$ of the fit is also given
in the ninth column.}
\label{4f_tab:WWmeas} 
\end{center}
\end{table}

\begin{table}[hbtp]
\begin{center}
\hspace*{-0.3cm}
\renewcommand{\arraystretch}{1.2}
\begin{tabular}{|c|cccccccc|} 
\hline
\roots / \GeV 
      & 182.7 & 188.6 & 191.6 & 195.5 & 199.5 & 201.6 & 204.9 & 206.6 \\
\hline
182.7 & 1.000 & 0.197 & 0.113 & 0.169 & 0.169 & 0.128 & 0.166 & 0.201 \\
188.6 & 0.197 & 1.000 & 0.134 & 0.200 & 0.200 & 0.150 & 0.196 & 0.239 \\
191.6 & 0.113 & 0.134 & 1.000 & 0.119 & 0.119 & 0.090 & 0.118 & 0.143 \\
195.5 & 0.169 & 0.200 & 0.119 & 1.000 & 0.177 & 0.133 & 0.174 & 0.211 \\
199.5 & 0.169 & 0.200 & 0.119 & 0.177 & 1.000 & 0.133 & 0.175 & 0.212 \\
201.6 & 0.128 & 0.150 & 0.090 & 0.133 & 0.133 & 1.000 & 0.131 & 0.159 \\
204.9 & 0.166 & 0.196 & 0.118 & 0.174 & 0.175 & 0.131 & 1.000 & 0.209 \\
206.6 & 0.201 & 0.239 & 0.143 & 0.211 & 0.212 & 0.159 & 0.209 & 1.000 \\
\hline
\end{tabular}
\renewcommand{\arraystretch}{1.}
\caption{\small 
Correlation matrix for the LEP combined W-pair cross sections
listed at the bottom of Table~\protect\ref{4f_tab:WWmeas}.
Correlations are all positive and range from 9\% to 24\%.}
\label{4f_tab:WWcorr} 
\end{center}
\end{table}

\begin{table}[hbtp]
\begin{center}
\hspace*{-0.3cm}
\renewcommand{\arraystretch}{1.2}
\begin{tabular}{|c|c|c|c|c|} 
\hline
\roots & \multicolumn{4}{|c|}{WW cross section (pb)}                              \\
\cline{2-5} 
(GeV) & $\sww^{\footnotesize\Gentle}$ 
      & $\sww^{\footnotesize\KoralW}$   
      & $\sww^{\footnotesize\YFSWW}$    
      & $\sww^{\footnotesize\RacoonWW}$ \\
\hline
182.7 & $15.710\pm0.020$ & $15.619\pm0.002$ & $15.361\pm0.005$ & $15.368\pm0.008$ \\
188.6 & $16.647\pm0.020$ & $16.554\pm0.002$ & $16.266\pm0.005$ & $16.249\pm0.011$ \\
191.6 & $16.961\pm0.020$ & $16.865\pm0.002$ & $16.568\pm0.006$ & $16.519\pm0.009$ \\
195.5 & $17.262\pm0.020$ & $17.165\pm0.002$ & $16.841\pm0.006$ & $16.801\pm0.009$ \\
199.5 & $17.462\pm0.020$ & $17.361\pm0.002$ & $17.017\pm0.007$ & $16.979\pm0.009$ \\
201.6 & $17.532\pm0.020$ & $17.428\pm0.002$ & $17.076\pm0.006$ & $17.032\pm0.009$ \\
204.9 & $17.602\pm0.020$ & $17.497\pm0.002$ & $17.128\pm0.006$ & $17.079\pm0.009$ \\
206.6 & $17.621\pm0.020$ & $17.516\pm0.001$ & $17.145\pm0.006$ & $17.087\pm0.009$ \\
\hline
\end{tabular}
\renewcommand{\arraystretch}{1.}
\caption{\small 
W-pair cross section predictions (in pb) for different \CoM\ energies,
according to \Gentle~\protect\cite{4f_bib:gentle}, 
\KoralW~\protect\cite{4f_bib:koralw}, 
\YFSWW~\protect\cite{common_bib:yfsww} and 
\RacoonWW~\protect\cite{common_bib:racoonww},
for $\Mw=80.35$~GeV.
The errors listed in the table are only the statistical errors 
from the numerical integration of the cross section.}
\label{4f_tab:WWtheo} 
\end{center}
\end{table}

\begin{table}[hbtp]
\begin{center}
\begin{small}
\begin{tabular}{|c|cccccc|c|c|}
\hline
& & & {\scriptsize (LCEU)} & {\scriptsize (LCEC)} & 
{\scriptsize (LUEU)} & {\scriptsize (LUEC)} & & \\
$\sqrt{s}$ & $\rww$ & 
$\Delta\rww^\mathrm{stat}$ &
$\Delta\rww^\mathrm{syst}$ &
$\Delta\rww^\mathrm{syst}$ &
$\Delta\rww^\mathrm{syst}$ &
$\Delta\rww^\mathrm{syst}$ &
$\Delta\rww$ &
$\chi^2/\textrm{d.o.f.}$ \\
\hline
\hline
\multicolumn{9}{|c|}{\Gentle~\cite{4f_bib:gentle}}\\
\hline
182.7 \GeV & 
1.005 & $\pm$0.021 & $\pm$0.001 & $\pm$0.006 & $\pm$0.003& $\pm$0.007 & $\pm$0.023&
\multirow{8}{20.3mm}{$
  \hspace*{-0.3mm}
  \left\}
    \begin{array}[h]{rr}
      &\multirow{8}{6mm}{\hspace*{-4.2mm}27.42/24}\\
      &\\ &\\ &\\ &\\ &\\ &\\ &\\  
    \end{array}
  \right.
  $}\\
188.6 \GeV & 
0.961 & $\pm$0.011 & $\pm$0.001 & $\pm$0.003 & $\pm$0.002& $\pm$0.005 & $\pm$0.013&\\
191.6 \GeV & 
0.986 & $\pm$0.027 & $\pm$0.001 & $\pm$0.004 & $\pm$0.002& $\pm$0.006 & $\pm$0.028&\\
195.5 \GeV & 
1.010 & $\pm$0.017 & $\pm$0.001 & $\pm$0.004 & $\pm$0.002& $\pm$0.006 & $\pm$0.018&\\
199.5 \GeV & 
0.964 & $\pm$0.016 & $\pm$0.001 & $\pm$0.004 & $\pm$0.002& $\pm$0.006 & $\pm$0.018&\\
201.6 \GeV & 
0.983 & $\pm$0.023 & $\pm$0.001 & $\pm$0.004 & $\pm$0.002& $\pm$0.006 & $\pm$0.024&\\
204.9 \GeV & 
0.949 & $\pm$0.016 & $\pm$0.001 & $\pm$0.004 & $\pm$0.002& $\pm$0.006 & $\pm$0.018&\\
206.6 \GeV & 
0.984 & $\pm$0.012 & $\pm$0.001 & $\pm$0.004 & $\pm$0.002& $\pm$0.006 & $\pm$0.014&\\
\hline
Average & 
0.973 & $\pm$0.006 & $\pm$0.001 & $\pm$0.004 & $\pm$0.001& $\pm$0.006 & $\pm$0.009&
\hspace*{1.5mm}39.16/31\hspace*{-0.5mm}\\
\hline
\hline
\multicolumn{9}{|c|}{\KoralW~\cite{4f_bib:koralw}}\\
\hline
182.7 \GeV & 
1.011 & $\pm$0.021 & $\pm$0.000 & $\pm$0.006 & $\pm$0.003& $\pm$0.007 & $\pm$0.023&
\multirow{8}{20.3mm}{$
  \hspace*{-0.3mm}
  \left\}
    \begin{array}[h]{rr}
      &\multirow{8}{6mm}{\hspace*{-4.2mm}27.42/24}\\
      &\\ &\\ &\\ &\\ &\\ &\\ &\\  
    \end{array}
  \right.
  $}\\
188.6 \GeV & 
0.967 & $\pm$0.011 & $\pm$0.000 & $\pm$0.003 & $\pm$0.002& $\pm$0.005 & $\pm$0.013&\\
191.6 \GeV & 
0.991 & $\pm$0.027 & $\pm$0.000 & $\pm$0.004 & $\pm$0.002& $\pm$0.006 & $\pm$0.028&\\
195.5 \GeV & 
1.015 & $\pm$0.017 & $\pm$0.000 & $\pm$0.004 & $\pm$0.002& $\pm$0.006 & $\pm$0.018&\\
199.5 \GeV & 
0.970 & $\pm$0.016 & $\pm$0.000 & $\pm$0.004 & $\pm$0.002& $\pm$0.006 & $\pm$0.018&\\
201.6 \GeV & 
0.989 & $\pm$0.023 & $\pm$0.000 & $\pm$0.004 & $\pm$0.002& $\pm$0.006 & $\pm$0.024&\\
204.9 \GeV & 
0.955 & $\pm$0.016 & $\pm$0.000 & $\pm$0.004 & $\pm$0.002& $\pm$0.006 & $\pm$0.018&\\
206.6 \GeV & 
0.989 & $\pm$0.012 & $\pm$0.000 & $\pm$0.004 & $\pm$0.002& $\pm$0.006 & $\pm$0.014&\\
\hline
Average & 
0.979 & $\pm$0.006 & $\pm$0.000 & $\pm$0.004 & $\pm$0.001& $\pm$0.006 & $\pm$0.009&
\hspace*{1.5mm}39.20/31\hspace*{-0.5mm}\\
\hline
\hline
\multicolumn{9}{|c|}{\YFSWW~\cite{common_bib:yfsww}}\\
\hline
182.7 \GeV & 
1.028 & $\pm$0.021 & $\pm$0.000 & $\pm$0.006 & $\pm$0.003& $\pm$0.007 & $\pm$0.023&
\multirow{8}{20.3mm}{$
  \hspace*{-0.3mm}
  \left\}
    \begin{array}[h]{rr}
      &\multirow{8}{6mm}{\hspace*{-4.2mm}27.42/24}\\
      &\\ &\\ &\\ &\\ &\\ &\\ &\\  
    \end{array}
  \right.
  $}\\
188.6 \GeV & 
0.984 & $\pm$0.011 & $\pm$0.000 & $\pm$0.003 & $\pm$0.002& $\pm$0.005 & $\pm$0.013&\\
191.6 \GeV & 
1.009 & $\pm$0.028 & $\pm$0.000 & $\pm$0.004 & $\pm$0.002& $\pm$0.006 & $\pm$0.029&\\
195.5 \GeV & 
1.035 & $\pm$0.017 & $\pm$0.000 & $\pm$0.004 & $\pm$0.002& $\pm$0.006 & $\pm$0.019&\\
199.5 \GeV & 
0.990 & $\pm$0.016 & $\pm$0.000 & $\pm$0.004 & $\pm$0.002& $\pm$0.006 & $\pm$0.018&\\
201.6 \GeV & 
1.009 & $\pm$0.024 & $\pm$0.000 & $\pm$0.004 & $\pm$0.002& $\pm$0.006 & $\pm$0.025&\\
204.9 \GeV & 
0.976 & $\pm$0.016 & $\pm$0.000 & $\pm$0.004 & $\pm$0.002& $\pm$0.006 & $\pm$0.018&\\
206.6 \GeV & 
1.011 & $\pm$0.013 & $\pm$0.000 & $\pm$0.004 & $\pm$0.002& $\pm$0.006 & $\pm$0.015&\\
\hline
Average & 
0.998 & $\pm$0.006 & $\pm$0.000 & $\pm$0.004 & $\pm$0.001& $\pm$0.006 & $\pm$0.009&
\hspace*{1.5mm}39.04/31\hspace*{-0.5mm}\\
\hline
\hline
\multicolumn{9}{|c|}{\RacoonWW~\cite{common_bib:racoonww}}\\
\hline
182.7 \GeV & 
1.028 & $\pm$0.021 & $\pm$0.001 & $\pm$0.006 & $\pm$0.003& $\pm$0.007 & $\pm$0.023&
\multirow{8}{20.3mm}{$
  \hspace*{-0.3mm}
  \left\}
    \begin{array}[h]{rr}
      &\multirow{8}{6mm}{\hspace*{-4.2mm}27.42/24}\\
      &\\ &\\ &\\ &\\ &\\ &\\ &\\  
    \end{array}
  \right.
  $}\\
188.6 \GeV & 
0.985 & $\pm$0.011 & $\pm$0.001 & $\pm$0.003 & $\pm$0.002& $\pm$0.005 & $\pm$0.013&\\
191.6 \GeV & 
1.012 & $\pm$0.028 & $\pm$0.001 & $\pm$0.004 & $\pm$0.002& $\pm$0.006 & $\pm$0.029&\\
195.5 \GeV & 
1.037 & $\pm$0.017 & $\pm$0.001 & $\pm$0.004 & $\pm$0.002& $\pm$0.006 & $\pm$0.019&\\
199.5 \GeV & 
0.992 & $\pm$0.016 & $\pm$0.001 & $\pm$0.004 & $\pm$0.002& $\pm$0.006 & $\pm$0.018&\\
201.6 \GeV & 
1.012 & $\pm$0.024 & $\pm$0.001 & $\pm$0.004 & $\pm$0.002& $\pm$0.006 & $\pm$0.025&\\
204.9 \GeV & 
0.978 & $\pm$0.016 & $\pm$0.001 & $\pm$0.004 & $\pm$0.002& $\pm$0.006 & $\pm$0.018&\\
206.6 \GeV & 
1.014 & $\pm$0.013 & $\pm$0.001 & $\pm$0.004 & $\pm$0.002& $\pm$0.006 & $\pm$0.015&\\
\hline
Average & 
1.000 & $\pm$0.006 & $\pm$0.000 & $\pm$0.004 & $\pm$0.001& $\pm$0.006 & $\pm$0.009&
\hspace*{1.5mm}39.14/31\hspace*{-0.5mm}\\
\hline
\end{tabular}
\end{small}
\caption{\small
Ratios of LEP combined W-pair cross section measurements
to the expectations of the four theoretical models considered,
for different \CoM\ energies and for all energies combined.
The first column contains the \CoM\ energy,
the second the combined ratios,
the third the statistical errors.
The fourth, fifth, sixth and seventh columns contain
the sources of systematic errors that are considered as 
LEP-correlated   energy-uncorrelated (LCEU),
LEP-correlated   energy-correlated   (LCEC),
LEP-uncorrelated energy-uncorrelated (LUEU),
LEP-uncorrelated energy-correlated   (LUEC).
The total error is given in the eighth column.
The only LCEU systematic sources considered 
are the statistical errors on the cross section theoretical predictions,
while the LCEC, LUEU and LUEC sources are those coming from
the corresponding errors on the cross section measurements.}
\label{4f_tab:rWWmeas} 
\end{center}
\end{table}

\renewcommand{\arraystretch}{1.2}
\begin{table}[p]
\begin{center}
\begin{small}
\hspace*{-0.0cm}
\begin{tabular}{|l|cccc|c|c|c|}
\cline{1-8}
Decay & & & {\scriptsize (unc)} & {\scriptsize (cor)} & & & 
3$\times$3 correlation \\
channel & $\wwbr$ & 
$\Delta\wwbr^\mathrm{stat}$ &
$\Delta\wwbr^\mathrm{syst}$ &
$\Delta\wwbr^\mathrm{syst}$ &
$\Delta\wwbr^\mathrm{syst}$ &
$\Delta\wwbr$ & 
for $\Delta\wwbr$\\
\cline{1-8}
\multicolumn{8}{|c|}{\Aleph~\cite{4f_bib:aleww2000}}\\
\hline
\BWtoenu & 
10.95 & $\pm$0.27 & $\pm$0.15 & $\pm$0.04 & $\pm$0.16 & $\pm$0.31 &
\multirow{3}{47mm}{\mbox{$\Biggl(\negthickspace\negthickspace$
                     \begin{tabular}{ccc}
                      \phm1.000 &    -0.048 &    -0.271 \\
                         -0.048 & \phm1.000 &    -0.253 \\
                         -0.271 &    -0.253 & \phm1.000 \\
                     \end{tabular}
                     $\negthickspace\negthickspace\Biggr)$} } \\
\BWtomnu & 
11.11 & $\pm$0.25 & $\pm$0.14 & $\pm$0.04 & $\pm$0.15 & $\pm$0.29 & \\
\BWtotnu & 
10.57 & $\pm$0.32 & $\pm$0.20 & $\pm$0.04 & $\pm$0.20 & $\pm$0.38 & \\
\hline
\multicolumn{8}{c}{}\\

\cline{1-8}
\multicolumn{8}{|c|}{\Delphi~\cite{4f_bib:delww2000}}\\
\hline
\BWtoenu & 
10.36 & $\pm$0.30 & $\pm$0.15 & $\pm$0.05 & $\pm$0.16 & $\pm$0.34 &
\multirow{3}{47mm}{\mbox{$\Biggl(\negthickspace\negthickspace$
                     \begin{tabular}{ccc}
                      \phm1.000 &    -0.050 &    -0.330 \\
                         -0.050 & \phm1.000 &    -0.250 \\
                         -0.330 &    -0.250 & \phm1.000 \\
                     \end{tabular}
                     $\negthickspace\negthickspace\Biggr)$} } \\
\BWtomnu & 
10.62 & $\pm$0.26 & $\pm$0.09 & $\pm$0.05 & $\pm$0.10 & $\pm$0.28 & \\
\BWtotnu & 
10.99 & $\pm$0.39 & $\pm$0.26 & $\pm$0.03 & $\pm$0.26 & $\pm$0.47 & \\
\hline
\multicolumn{8}{c}{}\\

\cline{1-8}
\multicolumn{8}{|c|}{\Ltre~\cite{4f_bib:ltrww2000}}\\
\hline
\BWtoenu & 
10.40 & $\pm$0.26 & $\pm$0.13 & $\pm$0.06 & $\pm$0.14 & $\pm$0.30 & 
\multirow{3}{47mm}{\mbox{$\Biggl(\negthickspace\negthickspace$
                     \begin{tabular}{ccc}
                      \phm1.000 &    -0.016 &    -0.279 \\
                         -0.016 & \phm1.000 &    -0.295 \\
                         -0.279 &    -0.295 & \phm1.000 \\
                     \end{tabular}
                     $\negthickspace\negthickspace\Biggr)$} } \\
\BWtomnu & 
\phz9.72 & $\pm$0.27 & $\pm$0.14 & $\pm$0.06 & $\pm$0.15 & $\pm$0.31 & \\
\BWtotnu & 
11.78 & $\pm$0.38 & $\pm$0.20 & $\pm$0.06 & $\pm$0.21 & $\pm$0.43 & \\
\hline
\multicolumn{8}{c}{}\\

\cline{1-8}
\multicolumn{8}{|c|}{\Opal~\cite{4f_bib:opaww2000b}}\\
\hline
\BWtoenu & 
10.40 & $\pm$0.25 & $\pm$0.24 & $\pm$0.05 & $\pm$0.25 & $\pm$0.35 & 
\multirow{3}{47mm}{\mbox{$\Biggl(\negthickspace\negthickspace$
                     \begin{tabular}{ccc}
                      \phm1.000 &     0.141 &    -0.179 \\
                          0.141 & \phm1.000 &    -0.174 \\
                         -0.179 &    -0.174 & \phm1.000 \\
                     \end{tabular}
                     $\negthickspace\negthickspace\Biggr)$} } \\
\BWtomnu & 
10.61 & $\pm$0.25 & $\pm$0.23 & $\pm$0.06 & $\pm$0.24 & $\pm$0.35 & \\
\BWtotnu & 
11.18 & $\pm$0.31 & $\pm$0.37 & $\pm$0.05 & $\pm$0.37 & $\pm$0.48 & \\
\hline
\multicolumn{8}{c}{}\\

\cline{1-8}
\multicolumn{7}{|c}{LEP Average (without lepton universality assumption)}
&\multicolumn{1}{c|}{}\\
\hline
\BWtoenu & 
10.54 & $\pm$0.13 & $\pm$0.08 & $\pm$0.05 & $\pm$0.10 & $\pm$0.17 &
\multirow{3}{47mm}{\mbox{$\Biggl(\negthickspace\negthickspace$
                     \begin{tabular}{ccc}
                       \phm1.000 & \phm0.066 &    -0.214 \\
                       \phm0.066 & \phm1.000 &    -0.189 \\
                          -0.214 &    -0.189 & \phm1.000 \\
                     \end{tabular}
                     $\negthickspace\negthickspace\Biggr)$} } \\
\BWtomnu & 
10.54 & $\pm$0.13 & $\pm$0.08 & $\pm$0.05 & $\pm$0.09 & $\pm$0.16 & \\
\BWtotnu & 
11.09 & $\pm$0.17 & $\pm$0.13 & $\pm$0.04 & $\pm$0.14 & $\pm$0.22 & \\
\hline
$\chi^2/\textrm{d.o.f.}$ & \multicolumn{1}{|c|}{14.9/9} & 
\multicolumn{6}{c}{}\\
\cline{1-2} 
\multicolumn{8}{c}{}\\

\cline{1-7} 
\multicolumn{7}{|c|}{LEP Average (with lepton universality assumption)}
&\multicolumn{1}{c}{}\\
\cline{1-7} 
\BWtolnu & 
10.69 & $\pm$0.06 & $\pm$0.05 & $\pm$0.05 & $\pm$0.07 & $\pm$0.09 & 
\multicolumn{1}{c}{}\\
{\mbox{$\mathcal{B}(\mathrm{W}\rightarrow\mathrm{had.})$}}  & 
67.92 & $\pm$0.17 & $\pm$0.15 & $\pm$0.15 & $\pm$0.21 & $\pm$0.27 & 
\multicolumn{1}{c}{}\\
\cline{1-7} 
$\chi^2/\textrm{d.o.f.}$ & \multicolumn{1}{|c|}{18.8/11} &
\multicolumn{6}{c}{}\\
\cline{1-2} 
\end{tabular}
\vspace*{0.5cm}

\end{small}
\caption{\small
W branching fraction measurements (in \%).
The first column contains the decay channel, 
the second the measurements,
the third the statistical uncertainty.
The fourth and fifth column list 
the uncorrelated and correlated components
of the systematic errors,
as provided by the Collaborations.
The total systematic error is given in the sixth column and
the total error in the seventh. 
Correlation matrices
for the three leptonic branching fractions 
are given in the last column.
This table is identical to Table~7 of Ref.~\protect\cite{4f_bib:4f_m01},
because results are not updated with respect to those presented
for the winter 2001 conferences.}
\label{4f_tab:Wbrmeas} 
\end{center}
\end{table}
\renewcommand{\arraystretch}{1.}

\begin{table}[p]
\vspace*{-0.0cm}
\begin{center}
\begin{small}
\begin{tabular}{|c|cccc|c|c|c|}
\hline
$\sqrt{s}$ & $\szz$ & 
$\Delta\szz^\mathrm{stat}$ &
$\Delta\szz^\mathrm{syst\,(unc)}$ &
$\Delta\szz^\mathrm{syst\,(cor)}$ &
$\Delta\szz^\mathrm{syst}$ &
$\Delta\szz$ & 
$\Delta\szz^\mathrm{stat\,(exp)}$ \\
\hline
\multicolumn{8}{|c|}
{\Aleph~\cite{4f_bib:alezz189,
4f_bib:alezz1999,4f_bib:alezz2000}}\\
\hline
182.7 \GeV & 0.11 & $^{+0.16}_{-0.11}$ & 
$\pm$0.04 & $\pm$0.01 & $\pm$0.04 & $^{+0.16}_{-0.12}$ & $\pm$0.14 \\
188.6 \GeV & 0.67 & $^{+0.13}_{-0.12}$ & 
$\pm$0.04 & $\pm$0.01 & $\pm$0.04 & $^{+0.14}_{-0.13}$ & $\pm$0.13 \\
191.6 \GeV & 0.53 & $^{+0.34}_{-0.27}$ & 
$\pm$0.02 & $\pm$0.01 & $\pm$0.02 & $^{+0.34}_{-0.27}$ & $\pm$0.33 \\
195.5 \GeV & 0.69 & $^{+0.23}_{-0.20}$ & 
$\pm$0.03 & $\pm$0.01 & $\pm$0.03 & $^{+0.23}_{-0.20}$ & $\pm$0.23 \\
199.5 \GeV & 0.70 & $^{+0.22}_{-0.20}$ & 
$\pm$0.03 & $\pm$0.01 & $\pm$0.03 & $^{+0.22}_{-0.20}$ & $\pm$0.23 \\
201.6 \GeV & 0.70 & $^{+0.33}_{-0.28}$ & 
$\pm$0.02 & $\pm$0.01 & $\pm$0.02 & $^{+0.33}_{-0.28}$ & $\pm$0.35 \\
204.9 \GeV & 1.21 & $^{+0.26}_{-0.23}$ & 
$\pm$0.03 & $\pm$0.01 & $\pm$0.03 & $^{+0.26}_{-0.23}$ & $\pm$0.27 \\
206.6 \GeV & 1.01 & $^{+0.19}_{-0.17}$ & 
$\pm$0.02 & $\pm$0.01 & $\pm$0.02 & $^{+0.19}_{-0.17}$ & $\pm$0.18 \\
\hline
\multicolumn{8}{|c|}
{\Delphi~\cite
{4f_bib:delzz189,4f_bib:delzz1999a,
4f_bib:delzz1999b,4f_bib:delzz2000}}\\
\hline
182.7 \GeV & 0.38 & $\pm$0.18          & 
$\pm$0.04 & $\pm$0.01 & $\pm$0.04 & $\pm$0.18          & $\pm$0.15 \\
188.6 \GeV & 0.60 & $\pm$0.13          & 
$\pm$0.07 & $\pm$0.02 & $\pm$0.07 & $\pm$0.15          & $\pm$0.14 \\
191.6 \GeV & 0.55 & $\pm$0.33          & 
$\pm$0.08 & $\pm$0.02 & $\pm$0.08 & $\pm$0.34          & $\pm$0.40 \\
195.5 \GeV & 1.17 & $\pm$0.27          & 
$\pm$0.09 & $\pm$0.03 & $\pm$0.10 & $\pm$0.29          & $\pm$0.24 \\
199.5 \GeV & 1.08 & $\pm$0.24          & 
$\pm$0.10 & $\pm$0.03 & $\pm$0.11 & $\pm$0.26          & $\pm$0.23 \\
201.6 \GeV & 0.87 & $\pm$0.31          & 
$\pm$0.11 & $\pm$0.03 & $\pm$0.11 & $\pm$0.33          & $\pm$0.34 \\
204.9 \GeV & 1.05 & $\pm$0.23          & 
$\pm$0.12 & $\pm$0.04 & $\pm$0.12 & $\pm$0.26          & $\pm$0.23 \\
206.6 \GeV & 0.98 & $\pm$0.18          & 
$\pm$0.11 & $\pm$0.03 & $\pm$0.12 & $\pm$0.22          & $\pm$0.19 \\
\hline
\multicolumn{8}{|c|}
{\Ltre~\cite{common_bib:ltrzz183a,4f_bib:ltrzz183b,common_bib:ltrzz189,
4f_bib:ltrzz1999,common_bib:ltrzz2000new}} \\
\hline
182.7 \GeV & 0.31 & $^{+0.16}_{-0.15}$ & 
$\pm$0.05 & $\pm$0.01 & $\pm$0.05 & $^{+0.17}_{-0.15}$ & $\pm$0.16 \\
188.6 \GeV & 0.73 & $^{+0.15}_{-0.14}$ & 
$\pm$0.03 & $\pm$0.02 & $\pm$0.04 & $^{+0.15}_{-0.14}$ & $\pm$0.15 \\
191.6 \GeV & 0.29 & $\pm$0.22 & 
$\pm$0.01 & $\pm$0.02 & $\pm$0.02 & $\pm$0.22          & $\pm$0.34 \\
195.5 \GeV & 1.18 & $\pm$0.24          & 
$\pm$0.06 & $\pm$0.07 & $\pm$0.09 & $\pm$0.26          & $\pm$0.22 \\
199.5 \GeV & 1.25 & $\pm$0.25          & 
$\pm$0.06 & $\pm$0.07 & $\pm$0.09 & $\pm$0.27          & $\pm$0.24 \\
201.6 \GeV & 0.95 & $\pm$0.38 & 
$\pm$0.05 & $\pm$0.05 & $\pm$0.07 & $\pm$0.39          & $\pm$0.35 \\
204.9 \GeV & 0.84 & $\pm$0.22          & 
$\pm$0.05 & $\pm$0.05 & $\pm$0.07 & $\pm$0.23          & $\pm$0.23 \\
206.6 \GeV & 1.20 & $\pm$0.18          & 
$\pm$0.07 & $\pm$0.07 & $\pm$0.10 & $\pm$0.21          & $\pm$0.17 \\
\hline
\multicolumn{8}{|c|}
{\Opal~\cite{common_bib:opazz189,common_bib:opazz2000new}}\\
\hline
182.7 \GeV & 0.12 & $^{+0.20}_{-0.18}$ & 
$\pm$0.03 & $\pm$0.01 & $\pm$0.03 & $^{+0.20}_{-0.18}$ & $\pm$0.19 \\
188.6 \GeV & 0.80 & $^{+0.14}_{-0.13}$ & 
$\pm$0.06 & $\pm$0.02 & $\pm$0.06 & $^{+0.15}_{-0.14}$ & $\pm$0.14 \\
191.6 \GeV & 1.13 & $^{+0.46}_{-0.39}$ & 
$\pm$0.11 & $\pm$0.03 & $\pm$0.11 & $^{+0.47}_{-0.41}$ & $\pm$0.36 \\
195.5 \GeV & 1.19 & $^{+0.27}_{-0.24}$ & 
$\pm$0.09 & $\pm$0.03 & $\pm$0.09 & $^{+0.28}_{-0.26}$ & $\pm$0.25 \\
199.5 \GeV & 1.09 & $^{+0.25}_{-0.23}$ & 
$\pm$0.08 & $\pm$0.02 & $\pm$0.08 & $^{+0.26}_{-0.24}$ & $\pm$0.25 \\
201.6 \GeV & 0.94 & $^{+0.37}_{-0.32}$ & 
$\pm$0.07 & $\pm$0.03 & $\pm$0.08 & $^{+0.38}_{-0.33}$ & $\pm$0.37 \\
204.9 \GeV & 1.07 & $^{+0.26}_{-0.24}$ & 
$\pm$0.09 & $\pm$0.03 & $\pm$0.09 & $^{+0.28}_{-0.26}$ & $\pm$0.26 \\
206.6 \GeV & 1.07 & $^{+0.20}_{-0.19}$ & 
$\pm$0.07 & $\pm$0.03 & $\pm$0.08 & $^{+0.22}_{-0.21}$ & $\pm$0.21 \\
\hline
\hline
\multicolumn{7}{|c|}{LEP Averages } & $\chi^2/\textrm{d.o.f.}$ \\
\hline
182.7 \GeV & 0.23 & $\pm$0.08 & 
$\pm$0.02 & $\pm$0.01 & $\pm$0.02 & $\pm$0.08 & 2.28/3 \\
188.6 \GeV & 0.70 & $\pm$0.07 & 
$\pm$0.03 & $\pm$0.02 & $\pm$0.03 & $\pm$0.08 & 0.97/3 \\
191.6 \GeV & 0.60 & $\pm$0.18 & 
$\pm$0.03 & $\pm$0.02 & $\pm$0.04 & $\pm$0.18 & 2.88/3 \\
195.5 \GeV & 1.04 & $\pm$0.12 & 
$\pm$0.03 & $\pm$0.03 & $\pm$0.05 & $\pm$0.13 & 3.23/3 \\
199.5 \GeV & 1.01 & $\pm$0.12 & 
$\pm$0.04 & $\pm$0.03 & $\pm$0.05 & $\pm$0.13 & 2.80/3 \\
201.6 \GeV & 0.86 & $\pm$0.18 & 
$\pm$0.04 & $\pm$0.03 & $\pm$0.05 & $\pm$0.18 & 0.32/3 \\
204.9 \GeV & 1.03 & $\pm$0.12 & 
$\pm$0.04 & $\pm$0.03 & $\pm$0.05 & $\pm$0.13 & 1.11/3 \\
206.6 \GeV & 1.06 & $\pm$0.09 & 
$\pm$0.04 & $\pm$0.03 & $\pm$0.05 & $\pm$0.11 & 0.76/3 \\
\hline
\end{tabular}
\end{small}
\vspace*{-0.1cm}
\caption{\small
Z-pair production cross section (in pb) at different energies.
The first column contains the LEP \CoM\ energy,
the second the measurements and
the third the statistical uncertainty. 
The fourth and the fifth columns list 
the uncorrelated and correlated components 
of the systematic errors,
as provided by the Collaborations.
The total systematic error is given in the sixth column,
the total error in the seventh.
The eighth column lists,
for the four LEP measurements,
the symmetrized expected statistical error,
and for the LEP combined value,
the $\chi^2$ of the fit.}
\label{4f_tab:ZZmeas} 
\end{center}
\end{table}

\begin{table}[p]
\renewcommand{\arraystretch}{1.2}
\vspace*{-0.0cm}
\begin{center}
\begin{small}
\begin{tabular}{|c|ccc|c|c|}
\hline
$\sqrt{s}$ & $\swent$ & 
$\Delta\swent^\mathrm{stat}$ &
$\Delta\swent^\mathrm{syst}$ &
$\Delta\swent$ & 
$\Delta\swent^\mathrm{stat\,(exp)}$ \\
\hline
\multicolumn{6}{|c|}
{\Aleph~\cite{4f_bib:alesw183,4f_bib:alesw189,
4f_bib:alesw2000}}\\
\hline
182.7 \GeV & 0.61 & $\pm0.26$ & $\pm$0.06 & $\pm0.27$ & $\pm$0.25 \\
188.6 \GeV & 0.45 & $\pm0.14$ & $\pm$0.04 & $\pm0.15$ & $\pm$0.16 \\
191.6 \GeV & 1.31 & $\pm0.47$ & $\pm$0.11 & $\pm0.48$ & $\pm$0.40 \\
195.5 \GeV & 0.65 & $\pm0.24$ & $\pm$0.06 & $\pm0.25$ & $\pm$0.25 \\
199.5 \GeV & 0.99 & $\pm0.25$ & $\pm$0.10 & $\pm0.27$ & $\pm$0.24 \\
201.6 \GeV & 0.75 & $\pm0.35$ & $\pm$0.08 & $\pm0.36$ & $\pm$0.36 \\
204.9 \GeV & 0.78 & $\pm0.26$ & $\pm$0.07 & $\pm0.27$ & $\pm$0.26 \\
206.6 \GeV & 1.19 & $\pm0.22$ & $\pm$0.12 & $\pm0.25$ & $\pm$0.21 \\
\hline
\multicolumn{6}{|c|}
{\Delphi~\cite
{4f_bib:delsw189b,4f_bib:delsw2000}}\\
\hline
188.6 \GeV & 0.75 & $^{+0.29}_{-0.25}$ & 
$\pm$0.07 & $^{+0.30}_{-0.26}$ & $\pm$0.26 \\
191.6 \GeV & 0.17 & $^{+0.33}_{-0.17}$ & 
$\pm$0.07 & $^{+0.34}_{-0.18}$ & $\pm$0.61 \\
195.5 \GeV & 0.94 & $^{+0.40}_{-0.35}$ & 
$\pm$0.07 & $^{+0.41}_{-0.36}$ & $\pm$0.36 \\
199.5 \GeV & 0.51 & $^{+0.32}_{-0.31}$ & 
$\pm$0.07 & $^{+0.33}_{-0.32}$ & $\pm$0.30 \\
201.6 \GeV & 1.15 & $^{+0.55}_{-0.45}$ & 
$\pm$0.07 & $^{+0.55}_{-0.46}$ & $\pm$0.47 \\
204.9 \GeV & 0.56 & $^{+0.36}_{-0.31}$ & 
$\pm$0.06 & $^{+0.36}_{-0.32}$ & $\pm$0.35 \\
206.6 \GeV & 0.58 & $^{+0.25}_{-0.22}$ & 
$\pm$0.06 & $^{+0.26}_{-0.23}$ & $\pm$0.28 \\
\hline
\multicolumn{6}{|c|}
{\Ltre~\cite{4f_bib:ltrswdef,4f_bib:ltrsw183,4f_bib:ltrsw189,
4f_bib:ltrsw1999}} \\
\hline
182.7 \GeV & 0.80 & $^{+0.28}_{-0.25}$ & 
$\pm$0.05 & $^{+0.28}_{-0.25}$ & $\pm$0.26 \\
188.6 \GeV & 0.69 & $^{+0.16}_{-0.14}$ & 
$\pm$0.04 & $^{+0.16}_{-0.15}$ & $\pm$0.15 \\
191.6 \GeV & 1.06 & $^{+0.48}_{-0.41}$ & 
$\pm$0.09 & $^{+0.49}_{-0.42}$ & $\pm$0.45 \\
195.5 \GeV & 0.98 & $^{+0.27}_{-0.25}$ & 
$\pm$0.09 & $^{+0.28}_{-0.27}$ & $\pm$0.24 \\
199.5 \GeV & 0.79 & $^{+0.26}_{-0.23}$ & 
$\pm$0.06 & $^{+0.27}_{-0.24}$ & $\pm$0.25 \\
201.6 \GeV & 1.38 & $^{+0.45}_{-0.40}$ &
$\pm$0.13 & $^{+0.47}_{-0.42}$ & $\pm$0.38 \\
\hline
\multicolumn{6}{|c|}
{\Opal~\cite{4f_bib:opasw189}}\\
\hline
188.6 \GeV & 0.67 & $^{+0.16}_{-0.14}$ & 
$\pm$0.06 & $^{+0.17}_{-0.15}$ & $\pm$0.16 \\
\hline
\hline
\multicolumn{5}{|c|}{LEP Averages } & $\chi^2/\textrm{d.o.f.}$ \\
\hline
182.7 \GeV & 0.70 & $\pm$0.18 & $\pm$0.04 & $\pm$0.19 & 0.26/1 \\
188.6 \GeV & 0.62 & $\pm$0.09 & $\pm$0.03 & $\pm$0.09 & 1.60/3 \\
191.6 \GeV & 0.99 & $\pm$0.27 & $\pm$0.06 & $\pm$0.28 & 2.38/2 \\
195.5 \GeV & 0.84 & $\pm$0.16 & $\pm$0.05 & $\pm$0.16 & 0.92/2 \\
199.5 \GeV & 0.79 & $\pm$0.15 & $\pm$0.05 & $\pm$0.16 & 1.40/2 \\
201.6 \GeV & 1.06 & $\pm$0.23 & $\pm$0.06 & $\pm$0.24 & 1.38/2 \\
204.9 \GeV & 0.70 & $\pm$0.21 & $\pm$0.05 & $\pm$0.22 & 0.24/1 \\
206.6 \GeV & 0.94 & $\pm$0.17 & $\pm$0.08 & $\pm$0.18 & 2.71/1 \\
\hline
\end{tabular}
\end{small}
\vspace*{-0.1cm}
\caption{\small
Single-W total production cross section (in pb) at different energies.
The first column contains the LEP \CoM\ energy,
and the second the measurements. 
The third and fourth column list 
the statistical and systematic uncertainties,
and the fifth the total error.
The sixth column lists,
for the four LEP measurements,
the symmetrized expected statistical error,
and for the LEP combined value,
the $\chi^2$ of the fit.}
\label{4f_tab:WevTOTmeas} 
\end{center}
\renewcommand{\arraystretch}{1.}
\end{table}

\begin{table}[p]
\renewcommand{\arraystretch}{1.2}
\vspace*{-0.0cm}
\begin{center}
\begin{small}
\begin{tabular}{|c|ccc|c|c|}
\hline
$\sqrt{s}$ & $\swenh$ & 
$\Delta\swenh^\mathrm{stat}$ &
$\Delta\swenh^\mathrm{syst}$ &
$\Delta\swenh$ & 
$\Delta\swenh^\mathrm{stat\,(exp)}$ \\
\hline
\multicolumn{6}{|c|}
{\Aleph~\cite{4f_bib:alesw183,4f_bib:alesw189,
4f_bib:alesw2000}}\\
\hline
182.7 \GeV & 0.40 & $\pm0.23$ & $\pm$0.06 & $\pm0.24$ & $\pm$0.23 \\
188.6 \GeV & 0.31 & $\pm0.13$ & $\pm$0.04 & $\pm0.14$ & $\pm$0.14 \\
191.6 \GeV & 0.94 & $\pm0.43$ & $\pm$0.11 & $\pm0.44$ & $\pm$0.37 \\
195.5 \GeV & 0.45 & $\pm0.22$ & $\pm$0.06 & $\pm0.23$ & $\pm$0.23 \\
199.5 \GeV & 0.82 & $\pm0.24$ & $\pm$0.10 & $\pm0.26$ & $\pm$0.22 \\
201.6 \GeV & 0.68 & $\pm0.34$ & $\pm$0.08 & $\pm0.35$ & $\pm$0.33 \\
204.9 \GeV & 0.50 & $\pm0.24$ & $\pm$0.07 & $\pm0.25$ & $\pm$0.24 \\
206.6 \GeV & 0.95 & $\pm0.21$ & $\pm$0.12 & $\pm0.24$ & $\pm$0.19 \\
\hline
\multicolumn{6}{|c|}
{\Delphi~\cite
{4f_bib:delsw189b,4f_bib:delsw2000}}\\
\hline
188.6 \GeV & 0.44 & $^{+0.27}_{-0.24}$ & 
$\pm$0.07 & $^{+0.28}_{-0.25}$ & $\pm$0.25 \\
191.6 \GeV & 0.01 & $^{+0.18}_{-0.01}$ & 
$\pm$0.07 & $^{+0.19}_{-0.07}$ & $\pm$0.57 \\
195.5 \GeV & 0.78 & $^{+0.37}_{-0.33}$ & 
$\pm$0.07 & $^{+0.38}_{-0.34}$ & $\pm$0.33 \\
199.5 \GeV & 0.16 & $^{+0.28}_{-0.16}$ & 
$\pm$0.07 & $^{+0.29}_{-0.17}$ & $\pm$0.27 \\
201.6 \GeV & 0.55 & $^{+0.46}_{-0.39}$ & 
$\pm$0.07 & $^{+0.47}_{-0.40}$ & $\pm$0.43 \\
204.9 \GeV & 0.50 & $^{+0.34}_{-0.30}$ & 
$\pm$0.06 & $^{+0.35}_{-0.31}$ & $\pm$0.33 \\
206.6 \GeV & 0.37 & $^{+0.23}_{-0.20}$ & 
$\pm$0.06 & $^{+0.24}_{-0.21}$ & $\pm$0.26 \\
\hline
\multicolumn{6}{|c|}
{\Ltre~\cite{4f_bib:ltrswdef,4f_bib:ltrsw183,4f_bib:ltrsw189,
4f_bib:ltrsw1999}} \\
\hline
182.7 \GeV & 0.58 & $^{+0.23}_{-0.20}$ & 
$\pm$0.04 & $^{+0.23}_{-0.20}$ & $\pm$0.21 \\
188.6 \GeV & 0.52 & $^{+0.14}_{-0.13}$ & 
$\pm$0.03 & $^{+0.14}_{-0.13}$ & $\pm$0.14 \\
191.6 \GeV & 0.85 & $^{+0.45}_{-0.37}$ & 
$\pm$0.06 & $^{+0.45}_{-0.37}$ & $\pm$0.41 \\
195.5 \GeV & 0.66 & $^{+0.24}_{-0.22}$ & 
$\pm$0.05 & $^{+0.25}_{-0.23}$ & $\pm$0.21 \\
199.5 \GeV & 0.34 & $^{+0.23}_{-0.20}$ & 
$\pm$0.03 & $^{+0.23}_{-0.20}$ & $\pm$0.22 \\
201.6 \GeV & 1.09 & $^{+0.41}_{-0.36}$ &
$\pm$0.08 & $^{+0.42}_{-0.37}$ & $\pm$0.35 \\
\hline
\multicolumn{6}{|c|}
{\Opal~\cite{4f_bib:opasw189}}\\
\hline
188.6 \GeV & 0.53 & $^{+0.13}_{-0.12}$ & 
$\pm$0.05 & $^{+0.14}_{-0.13}$ & $\pm$0.13 \\
\hline
\hline
\multicolumn{5}{|c|}{LEP Averages } & $\chi^2/\textrm{d.o.f.}$ \\
\hline
182.7 \GeV & 0.50 & $\pm$0.16 & $\pm$0.04 & $\pm$0.16 & 0.31/1 \\
188.6 \GeV & 0.46 & $\pm$0.08 & $\pm$0.02 & $\pm$0.08 & 1.47/3 \\
191.6 \GeV & 0.73 & $\pm$0.25 & $\pm$0.06 & $\pm$0.25 & 1.94/2 \\
195.5 \GeV & 0.60 & $\pm$0.14 & $\pm$0.03 & $\pm$0.15 & 0.77/2 \\
199.5 \GeV & 0.46 & $\pm$0.14 & $\pm$0.04 & $\pm$0.14 & 3.60/2 \\
201.6 \GeV & 0.80 & $\pm$0.21 & $\pm$0.05 & $\pm$0.21 & 1.13/2 \\
204.9 \GeV & 0.50 & $\pm$0.20 & $\pm$0.05 & $\pm$0.20 & 0.00/1 \\
206.6 \GeV & 0.71 & $\pm$0.15 & $\pm$0.08 & $\pm$0.17 & 2.77/1 \\
\hline
\end{tabular}
\end{small}
\vspace*{-0.1cm}
\caption{\small
Single-W hadronic production cross section (in pb) at different energies.
The first column contains the LEP \CoM\ energy,
and the second the measurements. 
The third and fourth column list 
the statistical and systematic uncertainties,
and the fifth the total error.
The sixth column lists,
for the four LEP measurements,
the symmetrized expected statistical error,
and for the LEP combined value,
the $\chi^2$ of the fit.}
\label{4f_tab:WevHADmeas} 
\end{center}
\renewcommand{\arraystretch}{1.}
\end{table}

\end{appendix}

\clearpage

\bibliographystyle{lep2unsrt}
\bibliography{s01_ew,s01_hf,common,gg,ff,4f_s01,gc,mw}

\begin{thebibliography}{100}

\bibitem{bib-EWEP-00}
The LEP Collaborations ALEPH, DELPHI, L3, OPAL and the LEP Electroweak Working
  Group, and the SLD Heavy Flavour and Electroweak Groups, {\it A Combination
  of Preliminary Electroweak Measurements and Constraints on the Standard
  Model}, CERN--EP-2001-21, hep-ex/0103048.

\bibitem{LEPLS}
The LEP~Collaborations ALEPH, DELPHI, L3 and OPAL and the LEP Electroweak
  working group, {\it Combination procedure for the precise determination of Z
  boson parameters from results of the LEP experiments}, CERN--EP-2000-153,
  hep-ex/0101027.

\bibitem{ref:lephf}
The LEP Experiments: ALEPH, DELPHI, L3 and OPAL, Nucl.~Inst.~Meth. {\bf A378}
  (1996) 101.

\bibitem{ALEPHLS}
ALEPH Collaboration, D.~Decamp \etal, Z.~Phys. {\bf C48} (1990) 365;\\ ALEPH
  Collaboration, D.~Decamp \etal, Z.~Phys. {\bf C53} (1992) 1;\\ ALEPH
  Collaboration, D.~Buskulic \etal, Z.~Phys. {\bf C60} (1993) 71;\\ ALEPH
  Collaboration, D.~Buskulic \etal, Z.~Phys. {\bf C62} (1994) 539;\\ ALEPH
  Collaboration, R.~Barate \etal, Eur. Phys. J. {\bf C 14} (2000) 1.

\bibitem{DELPHILS}
DELPHI Collaboration, P.~Aarnio \etal, Nucl.~Phys. {\bf B367} (1991) 511;\\
  DELPHI Collaboration, P.~Abreu \etal, Nucl.~Phys. {\bf B417} (1994) 3;\\
  DELPHI Collaboration, P.~Abreu \etal, Nucl.~Phys. {\bf B418} (1994) 403;\\
  DELPHI Collaboration, P.~Abreu \etal, Eur. Phys. J. {\bf C 16} (2000) 371.

\bibitem{L3LS}
L3 Collaboration, B.~Adeva \etal, Z.~Phys. {\bf C51} (1991) 179;\\ L3
  Collaboration, O.~Adriani \etal, Phys.~Rep. {\bf 236} (1993) 1; \\ L3
  Collaboration, M.~Acciarri \etal, Z.~Phys. {\bf C62} (1994) 551;\\ L3
  Collaboration, M.~Acciarri \etal, Eur. Phys. J. {\bf C16} (2000) 1-40.

\bibitem{OPALLS}
OPAL Collaboration, G.~Alexander \etal, Z.~Phys. {\bf C52} (1991) 175;\\ OPAL
  Collaboration, P.D.~Acton \etal, Z.~Phys. {\bf C58} (1993) 219;\\ OPAL
  Collaboration, R.~Akers \etal, Z.~Phys. {\bf C61} (1994) 19;\\ OPAL
  Collaboration, G.~Abbiendi \etal, Eur. Phys. J. {\bf C14} (2000) 373;\\ OPAL
  Collaboration, G.~Abbiendi \etal, Eur. Phys. J. {\bf C19} (2001) 587.

\bibitem{ref:QEDCONV}
F.A.~Berends \etal, in {\em Z Physics at LEP 1, Vol.~1}, ed. {G.~Altarelli,
  R.~Kleiss and C.~Verzegnassi}, (CERN Report: CERN 89-08, 1989), p.~89.\\
  M.~B{\"{o}}hm \etal, in {\em Z Physics at LEP 1, Vol.~1}, ed. {G.~Altarelli,
  R.~Kleiss and C.~Verzegnassi}, (CERN Report: CERN 89-08, 1989), p. 203.

\bibitem{ref:consoli}
See, for example, M.~Consoli \etal, in {\em Z Physics at LEP 1, Vol.~1}, ed
  G.~Altarelli, R.~Kleiss and C.~Verzegnassi, (CERN Report: CERN 89-08, 1989),
  p.~7.

\bibitem{ref:Jadach91}
S.~Jadach, \etal, Phys.~Lett.~{\bf B257} (1991) 173.

\bibitem{ref:Skrzypek92}
M.~Skrzypek, Acta Phys.~Pol.~{\bf B23} (1992) 135.

\bibitem{ref:Montagna96}
G.~Montagna, \etal, Phys.~Lett.~{\bf B406} (1997) 243.

\bibitem{bib-lumthopal}
G. Montagna \etal, Nucl. Phys. {\bf B547} (1999) 39; \\ G. Montagna \etal,
  Phys. Rev. Lett. {\bf 459} (1999) 649.

\bibitem{bib-lumth99}
B.F.L Ward \etal, Phys. Lett. {\bf B450} (1999) 262.

\bibitem{bib-PCP99}
D.{} Bardin, M.{} Gr{\"u}newald and G.{} Passarino, {\it Precision Calculation
  Project Report}, hep-ph/9902452.

\bibitem{bib-ALEPHTAU}
ALEPH Collaboration, D.~Buskulic \etal, Zeit.~Phys. {\bf C69} (1996) 183;\\
  ALEPH Collaboration, A.~Heister \etal, CERN-EP/2001-027 and Eur. Phys. J.
  {\bf C20} (2001) 401.

\bibitem{bib-DELPHITAUnew}
DELPHI Collaboration, P.~Abreu \etal, Eur. Phys. J. {\bf C14} (2000) 585.

\bibitem{bib-L3TAUfin}
L3 Collaboration, M.~Acciarri \etal, Phys.~Lett.~{\bf B429} (1998) 387.

\bibitem{bib-OPALTAU}
OPAL Collaboration, G.~Abbiendi \etal, {\it Precision Neutral Current Asymmetry
  Parameter Measurements from the Tau Polarization at LEP}, CERN-EP-2001-023,
  Submitted to Eur. Phys. J. C.

\bibitem{bib-EWPPE187}
The LEP Collaborations ALEPH, DELPHI, L3, OPAL and the LEP Electroweak Working
  Group, {\it Combined Preliminary Data on $\Zzero$ Parameters from the LEP
  Experiments and Constraints on the Standard Model}, CERN--PPE/94-187.

\bibitem{ref:sld-s00}
SLD Collaboration, K.~Abe \etal, Phys.{} Rev.{} Lett.{} {\bf 84} (2000) 5945.

\bibitem{ref:sld-asym}
SLD Collaboration, K.~Abe \etal, SLAC-PUB-8618, (2000). Submitted to
  Phys.Rev.Lett.

\bibitem{ref:lephfnew}
The LEP Heavy Flavour Group, {\it Final input parameters for the LEP/SLD heavy
  flavour analyses,} LEPHF/01-01, \\
  http://www.cern.ch/LEPEWWG/heavy/lephf0101.ps.gz.

\bibitem{ref:alife}
ALEPH Collaboration, R.~Barate \etal, Physics Letters {\bf B 401} (1997) 150;\\
  ALEPH Collaboration, R.~Barate \etal, Physics Letters {\bf B 401} (1997) 163.

\bibitem{ref:drb}
DELPHI Collaboration, P.Abreu \etal, Eur. Phys. J. {\bf C10} (1999) 415.

\bibitem{ref:lrbmixed}
L3 Collaboration, M. Acciarri \etal, Eur. Phys. J. {\bf C13} (2000) 47.

\bibitem{ref:omixed}
OPAL Collaboration, G. Abbiendi \etal, Eur. Phys. J. {\bf C8} (1999) 217.

\bibitem{ref:SLD_R_B}
SLD Collaboration, K. Abe \etal, Phys. Rev. Lett. {\bf 80} (1998) 660; \\ see
  also~\cite{ref:groot2001}.

\bibitem{ref:arcd}
ALEPH Collaboration, R.~Barate {\em et~al.}, Eur. Phys. J. {\bf C4} (1998) 557.

\bibitem{ref:drcd}
DELPHI Collaboration, P.Abreu \etal, Eur. Phys. J. {\bf C12} (2000) 209.

\bibitem{ref:drcc}
DELPHI Collaboration, P.Abreu \etal, Eur. Phys. J. {\bf C12} (2000) 225.

\bibitem{ref:orcd}
OPAL Collaboration, K.~Ackerstaff \etal, Eur. Phys. J. {\bf C1} (1998) 439.

\bibitem{ref:arcc}
ALEPH Collaboration, R. Barate \etal, Eur. Phys. J. {\bf C16} (2000) 597.

\bibitem{ref:orcc}
OPAL Collaboration, G.~Alexander \etal, Z.~Phys.~{\bf C72} (1996) 1.

\bibitem{ref:SLD_R_C}
SLD Collaboration, {\it A Measurement of $R_c$ with the SLD Detector }
  SLAC--PUB--7880, contributed paper to ICHEP 98 Vancouver {\bf ICHEP'98 \#174
  };\\ see also~\cite{ref:groot2001}.

\bibitem{ref:afbqcd}
D.~Abbaneo {\em et~al.}, Eur. Phys. J. {\bf C4} (1998) 185.

\bibitem{ref:ZFITTER}
D.~Bardin \etal, Z.~Phys. {\bf C44} (1989) 493; Comp.~Phys.~Comm. {\bf 59}
  (1990) 303; Nucl.~Phys. {\bf B351}(1991) 1; Phys.~Lett. {\bf B255} (1991) 290
  and CERN-TH 6443/92 (May 1992); the most recent version of ZFITTER (6.21) is
  described in DESY 99-070, hep-ph/9908433 (Aug 1999) published in
  Comp.~Phys.~Comm. {\bf 133} (2001) 229.

\bibitem{ref:alasy}
ALEPH Collaboration, D.~Buskulic \etal, Phys.~Lett. {\bf B384} (1996)414;\\
  ALEPH Collaboration, {\it Measurement of the b and c forward-backward
  asymmetries using leptons} ALEPH 99-076 CONF 99-048, contributed paper to EPS
  99 Tampere {\bf HEP'99 \#6-65}.

\bibitem{ref:dlasy}
DELPHI Collaboration, P.Abreu \etal, Z.~Phys {\bf C65} (1995) 569;\\ DELPHI
  Collaboration, {\it Measurement of the Forward-Backward Asymmetries of
  $e^+e^-\ \rightarrow Z\ \rightarrow b \overline{b}$ and $e^+e^-\ \rightarrow
  Z\ \rightarrow c \overline{c}$ using prompt leptons } DELPHI 2000-101 CONF
  400, contributed paper to ICHEP 2000 Osaka {\bf ICHEP'00 \#377 }.\\ Delphi
  notes are available at http://wwwcn.cern.ch/\~{}pubxx/www/delsec/delnote/.

\bibitem{ref:llasy}
L3 Collaboration, O.~Adriani \etal, Phys.~Lett. {\bf B292 } (1992) 454; \\ L3
  Collaboration, {\it L3 Results on \Abb, \Acc and $\chi$ for the Glasgow
  Conference,} L3 Note 1624;\\ L3 Collaboration, M. Acciarri \etal, Phys. Lett.
  {\bf B448} (1999) 152.

\bibitem{ref:olasy}
OPAL Collaboration, G.~Alexander \etal, Z.~Phys. {\bf C70 } (1996) 357;\\ OPAL
  Collaboration, R.~Akers \etal, {\it Updated Measurement of the Heavy Quark
  Forward-Backward Asymmetries and Average B Mixing Using Leptons in
  Multihadronic Events}, OPAL Physics Note PN226 contributed paper to ICHEP96,
  Warsaw, 25-31 July 1996 {\bf PA05-007}\\ OPAL Collaboration, R.~Akers \etal,
  {\it QCD corrections to the bottom and charm forward-backward asymmetries}
  OPAL Physics Note PN284.

\bibitem{ref:ajet}
ALEPH Collaboration, A.~Heister \etal, {\it Measurement of $A_{FB}^b$ using
  Inclusive b-hadron Decays}, CERN-EP/2001-047 and Eur. Phys. J. C DOI
  10.1007/s100520100812.

\bibitem{ref:djasy}
DELPHI Collaboration, P.Abreu \etal, Eur. Phys. J. {\bf C9} (1999) 367.

\bibitem{ref:dnnasy}
DELPHI Collaboration, {\it Determination of $A_{FB}^b$ using inclusive charge
  reconstruction and lifetime tagging at LEP 1}, DELPHI 2001-027 CONF 468.

\bibitem{ref:ljet}
L3 Collaboration, M. Acciarri \etal, Phys. Lett. {\bf B439} (1998) 225.

\bibitem{ref:ojet}
OPAL Collaboration, K.Ackerstaff \etal, Z.~Phys. {\bf C75} (1997) 385.

\bibitem{ref:adsac}
ALEPH Collaboration, R.~Barate \etal, Phys. Lett. {\bf B434} (1998) 415.

\bibitem{ref:ddasy}
DELPHI Collaboration, P.Abreu \etal, Eur. Phys. J. {\bf C10} (1999) 219.

\bibitem{ref:odsac}
OPAL Collaboration, G.~Alexander \etal, Z.~Phys. {\bf C73} (1996) 379.

\bibitem{ref:SLD_AQL}
SLD Collaboration, K. Abe \etal, Phys. Rev. Lett. {\bf 30} (1999) 3384; SLD
  Collaboration, K. Abe \etal, SLAC-PUB-8951, to be submitted to Phys. Rev.
  Lett.

\bibitem{ref:SLD_ACD}
SLD Collaboration, K. Abe, \etal, Phys Rev. {\bf D63} 032005 (2001);\\ see
  also~\cite{ref:groot2001}.

\bibitem{ref:SLD_ABJ}
SLD Collaboration, K. Abe, \etal, Phys Rev. Lett. {\bf 81} 942 (1998);\\ see
  also~\cite{ref:groot2001}.

\bibitem{ref:SLD_ABK}
SLD Collaboration, K. Abe et al, Phys. Rev. Lett. {\bf 83}, 1902 (1999).

\bibitem{ref:SLD_vtxasy}
SLD Collaboration, {\it Direct measurement of ${\cal A}_b$ using charged kaons
  at the SLD detector}, SLAC-PUB-8200, contributed paper to EPS 99 Tampere {\bf
  HEP'99 \#6\_473 };\\ SLD Collaboration, {\it Direct measurement of Ab using
  charged vertices}, SLAC-PUB-8542, Contributed Paper to ICHEP2000 , {\bf \#
  743};\\ see also~\cite{ref:groot2001}.

\bibitem{ref:SLD_ACV}
SLD Collaboration, {\it Direct measurement of ${\cal A}_c$ using inclusive
  charm tagging at the SLD detector}, SLAC-PUB-8199, contributed paper to EPS
  99 Tampere {\bf HEP'99 \#6\_474 };\\ see also~\cite{ref:groot2001}.

\bibitem{ref:abl}
ALEPH Collaboration. , {\it Inclusive semileptonic branching ratios of b
  hadrons produced in Z decays}, CERN-EP/2001-057.

\bibitem{ref:dbl}
DELPHI Collaboration, {\it Measurement of the semileptonic b branching ratios
  in Z decays } DELPHI 99-111 CONF 298, contributed paper to EPS 99 Tampere
  {\bf HEP'99 \#5\_522 }.

\bibitem{ref:lbl}
L3 Collaboration, M. Acciarri \etal, Z Phys. {\bf C71} 379 (1996).

\bibitem{ref:obl}
OPAL Collaboration, G. Abbiendi \etal, Eur. Phys. J. {\bf C13} (2000) 225.

\bibitem{ref:ocl}
OPAL Collaboration, G.~Abbiendi \etal, Eur. Phys. J. {\bf C8} (1999) 573.

\bibitem{ALEPHcharge1996}
ALEPH Collaboration, D.~Buskulic \etal, Z.~Phys. {\bf C71} (1996) 357.

\bibitem{DELPHIcharge}
DELPHI Collaboration, P.~Abreu \etal, Phys.~Lett. {\bf B277} (1992) 371.

\bibitem{OPALcharge}
OPAL Collaboration, P.~D.~Acton \etal, Phys.~Lett. {\bf B294} (1992) 436.

\bibitem{JETSET}
T.~\hbox{Sj\"ostrand}, Comp.~Phys.~Comm. {\bf 82} (1994) 74.

\bibitem{HERWIG}
G.~Marchesini \etal, Comp.~Phys.~Comm. {\bf 67} (1992) 465.

\bibitem{gg:ref:LEPGG}
ALEPH Collab., ALEPH 2000-008 CONF 2000-005 and ref. therein;\\ DELPHI Collab.,
  DELPHI 2001-093 CONF 521 and ref. therein; \\ L3 Collab., L3 Note 2650 and
  ref. therein;\\ OPAL Collab. OPAL PN469.

\bibitem{gg:ref:drell}
S.~D. Drell, Ann. Phys. {\bf 4} (1958) 75.

\bibitem{gg:ref:low}
F.~E. Low, \PRL {\bf 14} (1965) 238.

\bibitem{gg:ref:eboli}
O.~J.~P. Eboli, A.~A. Natale, and S.~F. Novaes, \PL {\bf B271} (1991) 274.

\bibitem{gg:ref:ad}
K.~Agashe and N.~G. Deshpande, \PL {\bf B456} (1999) 60.

\bibitem{gg:ref:vachon}
B.~Vachon.
\newblock Excited electron contribution to the $\eegg$ cross-section.
\newblock hep-ph/0103132, 2001.

\bibitem{ff:ref:lepff-moriond2001}
LEPEWWG $\ff$ Subgroup, D.~Bourilkov {\it et. al.}, LEP2FF/01-01,ALEPH 2001-039
  PHYSIC 2000-013, DELPHI 2001-108 PHYS 896, L3 note 2663, OPAL TN690.

\bibitem{ff:ref:lepff-osaka}
LEPEWWG $\ff$ Subgroup, C.~Geweniger {\it et. al.}, LEP2FF/00-03,ALEPH 2000-088
  PHYSIC 2000-034, DELPHI 2000-168 PHYS 881, L3 note 2624, OPAL TN673.

\bibitem{ff:ref:lepff-moriond2000}
LEPEWWG $\ff$ Subgroup, D.~Bourilkov {\it et. al.}, LEP2FF/00-01,ALEPH 2000-026
  PHYSIC 2000-005, DELPHI 2000-046 PHYS 855, L3 note 2527, OPAL TN647.

\bibitem{ff:ref:lepff-tamp}
LEPEWWG $\ff$ Subgroup, D.~Bourilkov {\it et. al.}, LEP2FF/99-01,ALEPH 99-082
  PHYSIC 99-030, DELPHI 99-143 PHYS 829, L3 note 2443, OPAL TN616.

\bibitem{ff:ref:ffbar_web}
LEPEWWG $\ff$ subgroup: http://www.cern.ch/LEPEWWG/lep2/~.

\bibitem{ff:ref:expts}
ALEPH Collaboration ``Study of Fermion Pair Production in $\ee$ Collisions at
  130-183 GeV'', Eur. Phys. J. {\bf{C12}} (2000) 183; \\ ALEPH Collaboration,
  ``Fermion Pair Production in $\ee$ Collisions at 189 GeV and Limits on
  Physics Beyond the Standard Model'', ALEPH 99-018 CONF 99-013; \\ ALEPH
  Collaboration, ``Fermion Pair Production in $\ee$ Collisions from 192 to 202
  GeV'', ALEPH 2000-025 CONF 2000-021; \\ ALEPH Collaboration, ``Fermion Pair
  Production in $\ee$ Collisions at high energy and Limits on Physics beyond
  the Standard Model '', ALEPH 2001-019 CONF 2001-016; \\ DELPHI
  Collaboration,``Measurement and Interpretation of Fermion-Pair Production at
  LEP energies from 130 to 172 GeV'', Eur. Phys. J. {\bf{C11}} (1999), 383; \\
  DELPHI Collaboration, ``Measurement and Interpretation of Fermion-Pair
  Production at LEP Energies from 183 to 189 GeV'', Phys.Lett. {\bf{B485}}
  (2000), 45; \\ DELPHI Collaboration, ``Results on Fermion-Pair Production at
  LEP running from 192 to 202 GeV'', DELPHI 2000-128 OSAKA CONF 427; \\ DELPHI
  Collaboration, ``Results on Fermion-Pair Production at LEP running in 2000'',
  DELPHI 2001-094 CONF 522; \\ L3 Collaboration, ``Measurement of Hadron and
  Lepton-Pair Production at 161 GeV $< \sqrt{s} <$ 172 GeV at LEP'', Phys.
  Lett. {\bf{B407}} (1997) 361; \\ L3 Collaboration, ``Measurement of Hadron
  and Lepton-Pair Production at 130 GeV $< \sqrt{s} <$ 189 GeV at LEP'', Phys.
  Lett. {\bf{B479}} (2000), 101. \\ L3 Collaboration, ``Preliminary L3 Results
  on Fermion-Pair Production in 1999'', L3 note 2563; \\ L3
  Collaboration,``Preliminary L3 Results on Fermion-Pair Production in 2000'',
  L3 note 2648; \\ OPAL Collaboration, ``Tests of the Standard Model and
  Constraints on New Physics from Measurements of Fermion Pair Production at
  130 - 172 GeV at LEP'', Euro. Phys. J. {\bf C2} (1998) 441; \\ OPAL
  Collaboration, ``Tests of the Standard Model and Constraints on New Physics
  from Measurements of Fermion Pair Production at 183 GeV at LEP'', Euro. Phys.
  J. {\bf C6} (1999) 1; \\ OPAL Collaboration, ``Tests of the Standard Model
  and Constraints on New Physics from Measurements of Fermion Pair Production
  at 189 GeV at LEP'', Euro. Phys. J. {\bf C13} (2000) 553; \\ OPAL
  Collaboration, ``Tests of the Standard Model and Constraints on New Physics
  from Measurements of Fermion Pair Production at 192-202 GeV at LEP'', OPAL
  PN424 (2000); \\ OPAL Collaboration, ``Measurement of Standard Model
  Processes in e+e- Collisions at sqrt{s}~203-209 GeV'', OPAL PN469 (2001).

\bibitem{ff:ref:ZFITTER}
D.~Bardin {\it et al.}, CERN-TH 6443/92;
  http://www.ifh.de/$\sim$riemann/Zfitter/zf.html~.\\ Predictions are from
  ZFITTER versions 6.04 or later.\\ Definition 1 corresponds to the ZFITTER
  flags FINR=0 and INTF=0; definition 2 corresponds to FINR=0 and INTF=1 for
  hadrons, FINR=1 and INTF=1 for leptons.

\bibitem{ff:ref:TOPAZ0}
G. Montagna {\it et~al.}, Comput. Phys. Commun. {\bf 117} (1999) 278.

\bibitem{ff:ref:KK}
S.~Jadach {\it et al.}, http://home.cern.ch/$\sim$jadach/KKindex.html~.

\bibitem{ff:ref:lepffwrkshp}
M. Kobel {\it et al.}, ``Two-Fermion Production in Electron Positron
  Collisions'' in S. Jadach {\it et al.} [eds] , ``Reports of the Working
  Groups on Precision Calculations for LEP2 Physics: proceedings'' CERN
  2000-009, hep-ph/0007180.

\bibitem{common_bib:lyons}
L.~Lyons, D.~Gibaut and P.~Clifford, \NIMA{270}{1988}{110}.

\bibitem{ff:ref:hfpublished}
ALEPH Collaboration, Euro. Phys J. {\bf{C12}} (2000) 183; \\ DELPHI
  Collaboration, P.Abreu {\it et al.}, Euro. Phys J. {\bf{C11}}(1999); \\ L3
  Collaboration, M.Acciarri {\it et al.}, Phys.\ Lett.\ {\bf B485} (2000) 71;
  \\ OPAL Collaboration, G.Abbiendi {\it et al.}, Euro.\ Phys.\ J.\ {\bf C16}
  (2000) 41.

\bibitem{ff:ref:hfpreliminary}
ALEPH Collaboration, ALEPH 99-018 CONF 99-013; \\ ALEPH Collaboration, ALEPH
  2000-046 CONF 2000-029; \\ ALEPH Collaboration, ALEPH 2000-047 CONF 2000-030;
  \\ ALEPH Collaboration, ALEPH 2001-019 CONF 2001-016; \\ DELPHI
  Collaboration, DELPHI 2001-095 CONF 523; \\ L3 Collaboration, L3 Internal
  Note 2640, 1 March 2001.

\bibitem{ff:ref:hfconfnote}
LEPEWWG Heavy Flavour at LEP2 Subgroup, ``Combination of Heavy Flavour
  Measurements at LEP2'', LEP2FF/00-02.

\bibitem{ff:ref:hfzfit}
ZFITTER V6.23 is used.\\ D. Bardin {\it et al.}, Preprint hep-ph/9908433. \\
  Relevant ZFITTER settings used are FINR=0 and INTF=1.

\bibitem{ff:ref:hflep1-99}
DELPHI Collaboration, P.Abreu {\it et al.}, Euro Phys J. {\bf{C10}}(1999) 415.
  \\ The LEP collaborations {\it et al.}, CERN-EP/2000-016.

\bibitem{ff:ref:zprime-thry}
P. Langacker, R.W. Robinett and J.L. Rosner, Phys. Rev. {\bf D30} (1984) 1470;
  \\ D. London and J.L. Rosner, Phys. Rev. {\bf D34} (1986) 1530; \\ J.C. Pati
  and A. Salam, Phys. Rev. {\bf D10} (1974) 275; \\ R.N. Mohapatra and J.C.
  Pati, Phys. Rev. {\bf D11} (1975) 566.

\bibitem{ff:ref:sqsm}
G. Altarelli \etal, Z. Phys. {\bf{C45}} (1989) 109; \\ erratum Z. Phys.
  {\bf{C47}} (1990) 676.

\bibitem{ff:ref:lep1zprime}
DELPHI Collaboration, P. Abreu {\it et al.}, Zeit. Phys. {\bf{C65}} (1995) 603.

\bibitem{ff:ref:ELPthr}
E. Eichten, K. Lane, and M. Peskin, Phys. Rev. Lett. {\bf 50} (1983) 811.

\bibitem{ff:ref:Kroha}
H. Kroha, Phys. Rev. {\bf D46} (1992) 58.

\bibitem{4f_bib:4f_s01}
The LEP WW Working Group, LEPEWWG/XSEC/2001-03, note prepared for the summer
  2001 conferences, {\tt http://lepewwg.web.cern.ch/LEPEWWG/lepww/4f/Summer01}.

\bibitem{4f_bib:4f_s00}
The LEP WW Working Group, LEPEWWG/XSEC/2000-01, note prepared for the summer
  2000 conferences, {\tt http://lepewwg.web.cern.ch/LEPEWWG/lepww/4f/Summer00}.

\bibitem{4f_bib:4f_m01}
The LEP WW Working Group, LEPEWWG/XSEC/2001-01, note prepared for the winter
  2001 conferences, {\tt http://lepewwg.web.cern.ch/LEPEWWG/lepww/4f/Winter01}.

\bibitem{4f_bib:fourfrep}
M.W.~Gr{\"{u}}newald, G.~Passarino \etal, {\it ``Four fermion production in
  electron positron collisions''}, Four fermion working group report of the
  LEP2 \MC\ Workshop 1999/2000, in {\it ``Reports of the working groups on
  precision calculations for LEP2 Physics''} CERN 2000--009, {\tt
  http://arXiv.org/abs/hep-ph/0005309}.

\bibitem{common_bib:adloww161}
\Aleph\ Collaboration, R.~Barate \etal, Phys.~Lett.~{\bf B401} (1997) 347.\\
  \Delphi\ Collaboration, P.~Abreu \etal, Phys.~Lett.~{\bf B397} (1997) 158.\\
  \Ltre\ Collaboration, M.~Acciarri \etal, Phys.~Lett. {\bf B398} (1997) 223.\\
  \Opal\ Collaboration, K.~Ackerstaff \etal, Phys.~Lett. {\bf B389} (1996) 416.

\bibitem{common_bib:aleww172}
\Aleph\ Collaboration, R.~Barate \etal, Phys.~Lett.~{\bf B415} (1997) 435.

\bibitem{common_bib:delww172}
DELPHI Collaboration, P.~Abreu \etal, Eur.~Phys.~J. {\bf C2} (1998) 581.

\bibitem{common_bib:ltrww172}
L3 Collaboration, M.~Acciarri \etal, Phys. Lett. {\bf B407} (1997) 419.

\bibitem{common_bib:opaww172}
OPAL Collaboration, K.~Ackerstaff \etal, Eur.~Phys.~J. {\bf C1} (1998) 395.

\bibitem{4f_bib:aleww183}
\Aleph\ Collaboration, R.~Barate \etal, \PLB{453}{1999}{107}.

\bibitem{4f_bib:delww183}
\Delphi\ Collaboration, P.~Abreu \etal, \PLB{456}{1999}{310}.

\bibitem{4f_bib:ltrww183}
\Ltre\ Collaboration, M.~Acciarri \etal, \PLB{436}{1998}{437}.

\bibitem{4f_bib:opaww183}
\Opal\ Collaboration, G.~Abbiendi \etal, Eur. Phys. Jour. {\bf C8} (1999) 191.

\bibitem{4f_bib:aleww189}
\Aleph\ Collaboration, R.~Barate \etal, \PLB{484}{2000}{205}.

\bibitem{4f_bib:delww189}
\Delphi\ Collaboration, P.~Abreu \etal, \PLB{479}{2000}{89}.

\bibitem{4f_bib:ltrww189}
\Ltre\ Collaboration, M.~Acciarri \etal, \PLB{496}{2000}{19}.

\bibitem{4f_bib:opaww189}
\Opal\ Collaboration, G.~Abbiendi \etal, \PLB{493}{2000}{249}.

\bibitem{4f_bib:aleww1999}
\Aleph\ Collaboration, \Aleph\ 2000--005 \Conf\ 2000--002, ICHEP 2000 abstract
  288.

\bibitem{4f_bib:delww1999}
\Delphi\ Collaboration, \Delphi\ 2000--140 \Conf\ 439, ICHEP 2000 abstract 458.

\bibitem{4f_bib:opaww1999}
\Opal\ Collaboration, \Opal\ Physics Note PN420, ICHEP 2000 abstract 184. The
  W-pair production cross sections at $\roots=200$--202 GeV results have been
  updated in~\cite{4f_bib:opaww2000a}.

\bibitem{4f_bib:opaww2000a}
\Opal\ Collaboration, \Opal\ Physics Note PN437, ICHEP 2000 abstract 168.

\bibitem{4f_bib:ltrww1999}
\Ltre\ Collaboration, \L3\ Note 2514, ICHEP 2000 abstract 512.

\bibitem{4f_bib:aleww2000}
\Aleph\ Collaboration, \Aleph\ 2001--013 \Conf\ 2001--010, submitted to the
  Winter 2001 Conferences.

\bibitem{4f_bib:delww2000}
\Delphi\ Collaboration, \Delphi\ 2001--024 \Conf\ 465, submitted to the Winter
  2001 Conferences.

\bibitem{4f_bib:ltrww2000}
\Ltre\ Collaboration, \L3\ Note 2638, submitted to the Winter 2001 Conferences.

\bibitem{4f_bib:opaww2000b}
\Opal\ Collaboration, \Opal\ Physics Note PN469, submitted to the Winter 2001
  Conferences.

\bibitem{4f_bib:4f_pdg01}
The LEP WW Working Group, LEPEWWG/XSEC/2001-02, note prepared for the 2001
  update of the PDG Review, {\tt
  http://lepewwg.web.cern.ch/LEPEWWG/lepww/4f/PDG01}.

\bibitem{4f_bib:lepewwg97}
The LEP Collaborations \Aleph, \Delphi, \L3, \Opal, the LEP Electroweak Working
  Group and the SLD Heavy Flavour Working Group, {\it ``A Combination of
  Preliminary Electroweak Measurements and Constraints on the Standard
  Model''}, CERN--PPE/97--154.

\bibitem{4f_bib:lepewwg98}
The LEP Collaborations \Aleph, \Delphi, \L3, \Opal, the LEP Electroweak Working
  Group and the SLD Heavy Flavour Working Group, {\it ``A Combination of
  Preliminary Electroweak Measurements and Constraints on the Standard
  Model''}, CERN--EP/99--15.

\bibitem{common_bib:yfsww}
S. Jadach, W. P{\l}aczek, M. Skrzypek, B.F.L. Ward, Phys. Rev. {\bf D54} (1996)
  5434. \\ S. Jadach, W. P{\l}aczek, M. Skrzypek, B.F.L. Ward, Z. W\c{a}s,
  Phys. Lett. {\bf B417} (1998) 326. \\ S. Jadach, W. P{\l}aczek, M. Skrzypek,
  B.F.L. Ward, Z. W\c{a}s, Phys. Rev. {\bf D61} (2000) 113010; preprint
  CERN-TH-99-222, hep-ph/9907346.\\ S. Jadach, W. P{\l}aczek, M. Skrzypek,
  B.F.L. Ward, Z. W\c{a}s, preprint CERN-TH/2000-337, hep-ph/0007012; submitted
  to Phys. Lett. B.\\ S. Jadach, W. P{\l}aczek, M. Skrzypek, B.F.L. Ward, Z.
  W\c{a}s, {\it The Monte Carlo Event Generator {\tt YFSWW3} version {\tt 1.16}
  for W-Pair Production and Decay at LEP2/LC Energies}, preprint
  CERN-TH/2001-017, UTHEP-01-0101, hep-ph/0103163, accepted for publication by
  Comput. Phys. Commun.\\ The \YFSWW\ cross-sections at 155--215 GeV have been
  kindly provided by the authors.

\bibitem{common_bib:racoonww}
A.~Denner, S.~Dittmaier, M.~Roth and D.~Wackeroth, Nucl. Phys. {\bf B560}
  (1999) 33.\\ A.~Denner, S.~Dittmaier, M.~Roth and D.~Wackeroth, Nucl. Phys.
  {\bf B587} (2000) 67.\\ A.~Denner, S.~Dittmaier, M.~Roth and D.~Wackeroth,
  \PLB{475}{2000}{127}.\\ A.~Denner, S.~Dittmaier, M.~Roth and D.~Wackeroth,
  hep-ph/0101257.\\ The \RacoonWW\ cross-sections at 155--215 GeV have been
  kindly provided by the authors.

\bibitem{4f_bib:dpa}
See~\cite{4f_bib:fourfrep} and references therein for a discussion of complete
  $\oa$ radiative corrections to W-pair production in the LPA/DPA
  approximations.

\bibitem{4f_bib:dpaerr}
The theoretical uncertainty $\Delta\sigma/\sigma$ on the W-pair production
  cross section calculated in the LPA/DPA above 170 GeV can be parametrised as
  $\Delta\sigma/\sigma=0.4\oplus0.072\cdot t_1\cdot t_2$, where
  $t_1=(200-2\cdot\Mw)/(\roots-2\cdot\Mw)$ and $t_2=(1-(\frac{2\cdot
  M_{\mathrm{W}}}{200})^2)/ (1-(\frac{2\cdot M_{\mathrm{W}}}{\sqrt{s}})^2)$. In
  the threshold region, a 2\% uncertainty is assigned. Private communication
  from the authors of ~\cite{common_bib:racoonww,common_bib:yfsww}, March 2001.

\bibitem{4f_bib:gentle}
D.~Bardin, J.~Biebel, D.~Lehner, A.~Leike, A.~Olchevski and T.~Riemann,
  \CPC{104}{1997}{161}. See also~\cite{4f_bib:fourfrep}.\\ The \Gentle\
  cross-sections at 155--215 GeV have been kindly provided by E.~Lan\c{c}on and
  A.~Ealet.

\bibitem{4f_bib:koralw}
M. Skrzypek, S. Jadach, M.~Martinez, W.~P{\l}aczek, Z.~W\c{a}s, Phys. Lett.
  {\bf B372} (1996) 289. \\ S. Jadach, W. P{\l}aczek, M. Skrzypek, Z. W\c{a}s,
  Comput. Phys. Commun. {\bf 94} (1996) 216. \\ S. Jadach, W. P{\l}aczek, M.
  Skrzypek, B.F.L. Ward, Z. W\c{a}s, Comput. Phys. Commun. {\bf 119} (1999)
  272. \\ S. Jadach, W. P{\l}aczek, M. Skrzypek, B.F.L. Ward, Z. W\c{a}s,
  preprint hep-ph/0104049, submitted to Comput. Phys. Commun.\\ The ``\KoralW''
  cross-sections at 155--215 GeV have been kindly provided by the authors. They
  have actually been computed using \YFSWW~\cite{common_bib:yfsww}, switching
  off non-leading $\oa$ radiative corrections and the screening of the Coulomb
  correction, to reproduce the calculation from \KoralW.

\bibitem{4f_bib:screening}
A.~P.~Chapovsky and V.~A.~Khoze, Eur. Phys. J. {\bf C9} (1999) 449.

\bibitem{4f_bib:melnikov}
K.~Melnikov and O.~Yakovlev, Phys. Lett. {\bf B324} (1994) 217; \\ V.~S.~Fadin,
  V.~A.~Khoze and A.~D.~Martin, Phys. Rev. {\bf D49} (1994) 2247.

\bibitem{4f_bib:yellowreport}
W.~Beenakker \etal, {\it ``WW Cross-Sections and Distributions''}, in {\it
  ``Physics at LEP2''}, eds.~G.~Altarelli \etal, CERN 96-01.

\bibitem{common_bib:pdg2000}
Particle Data Group, D.E.~Groom \etal, \EPJ{15}{2000}{1}.

\bibitem{4f_bib:alezz189}
\Aleph\ Collaboration, R.~Barate \etal, \PLB{469}{1999}{287}.

\bibitem{4f_bib:delzz189}
\Delphi\ Collaboration, P.~Abreu \etal, \PLB{497}{2001}{199}.

\bibitem{common_bib:ltrzz183a}
\Ltre\ Collaboration, M.~Acciarri \etal, \PLB{450}{1999}{281}. The Z-pair
  cross-section at 183 GeV therein follows the \Ltre\ definition: the
  corresponding {\sc NC02} cross-section is given in~\cite{4f_bib:ltrzz183b}.

\bibitem{4f_bib:ltrzz183b}
\L3\ Collaboration, \L3\ Note 2366, submitted to the Winter 1999 Conferences.

\bibitem{common_bib:ltrzz189}
\Ltre\ Collaboration, M.~Acciarri \etal, \PLB{465}{1999}{363}.

\bibitem{common_bib:opazz189}
\Opal\ Collaboration, G.~Abbiendi \etal, \PLB{476}{2000}{256}.

\bibitem{4f_bib:ltrzz1999}
\Ltre\ Collaboration, M.~Acciarri \etal, \PLB{497}{2001}{23}.

\bibitem{common_bib:opazz2000new}
\Opal\ Collaboration, \Opal\ Physics Note PN482, submitted to the Summer 2001
  Conferences.

\bibitem{4f_bib:opazz1999}
\Opal\ Collaboration, \Opal\ Physics Note PN423, ICHEP 2000 abstract 154.

\bibitem{4f_bib:alezz1999}
\Aleph\ Collaboration, \Aleph\ 2000--004 \Conf\ 2000--001, ICHEP 2000 abstract
  287. Systematic errors on the measurement were not quoted in this note and
  have only been made available in~\cite{4f_bib:alezz2000}.

\bibitem{4f_bib:alezz2000}
\Aleph\ Collaboration, \Aleph\ 2001--006 \Conf\ 2001--003, submitted to the
  Winter 2001 Conferences.

\bibitem{4f_bib:delzz1999a}
\Delphi\ Collaboration, \Delphi\ 2000--145 \Conf\ 444, ICHEP 2000 abstract 659.
  Systematic errors on the ZZ measurement at 192--202 GeV, not quoted in this
  note, are scaled linearly~\cite{4f_bib:delzz1999b} from those estimated at
  189 GeV.

\bibitem{4f_bib:delzz1999b}
\Delphi\ Collaboration, private communication by P.~Bambade, March 2001.

\bibitem{4f_bib:delzz2000}
\Delphi\ Collaboration, \Delphi\ 2001-015 \Conf\ 456, submitted to the Winter
  2001 Conferences.

\bibitem{common_bib:ltrzz2000new}
\Ltre\ Collaboration, \L3\ Note 2696, submitted to the Summer 2001 Conferences.

\bibitem{4f_bib:yfszz}
S. Jadach, W. P{\l}aczek, B.F.L. Ward, \PRD{56}{1997}{6939}.

\bibitem{4f_bib:zzto}
G.~Passarino, in~\cite{4f_bib:fourfrep}.

\bibitem{4f_bib:alesw2000}
\Aleph\ Collaboration, \Aleph\ 2001--017 \Conf\ 2001--014, submitted to the
  Winter 2001 Conferences.

\bibitem{4f_bib:delsw2000}
\Delphi\ Collaboration, \Delphi\ 2001--099 \Conf\ 527, submitted to the Summer
  2001 Conferences.

\bibitem{4f_bib:ltrsw189}
\Ltre\ Collaboration, M.~Acciarri \etal, \PLB{487}{2000}{229}. See
  also~\cite{4f_bib:ltrswdef}.

\bibitem{4f_bib:alesw183}
\Aleph\ Collaboration, R.~Barate \etal, \PLB{462}{1999}{389}. The single W
  cross section at 183 GeV therein follows the \Aleph\ definition: the
  corresponding value using the common LEP definition is given
  in~\cite{4f_bib:alesw189}.

\bibitem{4f_bib:ltrswdef}
\Ltre\ Collaboration, private communication by S.~Schmidt-Kaerst, March 2001.
  The single W cross-sections at 183, 189 and 192--202~GeV
  in~\cite{4f_bib:ltrsw183,4f_bib:ltrsw189,4f_bib:ltrsw1999} all follow the
  \Ltre\ definition. The corresponding values using the common LEP definition,
  computed by applying a conversion factor determined for each result with
  \Grace~\cite{4f_bib:grace}, have all been provided in private communications.

\bibitem{4f_bib:ltrsw183}
\Ltre\ Collaboration, M.~Acciarri \etal, \PLB{436}{1998}{417}. See
  also~\cite{4f_bib:ltrswdef}.

\bibitem{4f_bib:alesw189}
\Aleph\ Collaboration, \Aleph\ 2000--022 \Conf\ 2000--019, submitted to the
  Winter 2000 Conferences.

\bibitem{4f_bib:delsw189a}
\Delphi\ Collaboration, \Delphi\ 2000-047 \Conf\ 362, submitted to the Winter
  2000 Conferences. The decays of the W boson to taus are not included in the
  total single W cross section at 189--202~GeV. A new definition has been
  adopted in~\cite{4f_bib:delsw189b}.

\bibitem{4f_bib:delsw189b}
\Delphi\ Collaboration, \Delphi\ 2000-143 \Conf\ 442/2, contributed paper for
  ICHEP 2000. The decays of the W boson to taus are here included in the total
  single W cross section at 189--202~GeV. A different definition was used
  in~\cite{4f_bib:delsw189a}.

\bibitem{4f_bib:ltrsw1999}
\Ltre\ Collaboration, \L3\ Note 2518, submitted to the Winter 2000 Conferences.
  See also~\cite{4f_bib:ltrswdef}.

\bibitem{4f_bib:opasw189}
\Opal\ Collaboration, \Opal\ Physics Note PN427, March 2000.

\bibitem{4f_bib:swmor00}
The LEP WW Working Group, LEPEWWG/WW/00--02, {\it LEP Single W Cross Section
  for Winter 2000 Conferences},\\ {\tt
  http://lepewwg.web.cern.ch/LEPEWWG/lepww/4f/Winter00/ww-0002.ps}.

\bibitem{4f_bib:ltrsw172}
\Ltre\ Collaboration, M.~Acciarri \etal, \PLB{403}{1997}{168}. This measurement
  of the single W cross-section at 161--172~GeV is superseded by that published
  for 130--183~GeV~\cite{4f_bib:ltrsw183}.

\bibitem{4f_bib:wto}
G.~Passarino, \NPB{578}{2000}{3}.\\ G.~Passarino, \NPB{574}{2000}{451}.\\ The
  \WTO\ cross-sections at 160--210 GeV have been kindly provided by the author.

\bibitem{4f_bib:wphact}
E.~Accomando and A.~Ballestrero, \CPC{99}{1999}{270}.\\ The \WPHACT\
  cross-sections at 160--210 GeV have been kindly provided by A.~Ballestrero.

\bibitem{4f_bib:grace}
J.~Fujimoto \etal, \CPC{100}{1997}{74}.\\ Y.~Kurihara \etal,
  Prog.~Theor.~Phys.~{\bf 103}~(2000)~1199.\\ The \Grace\ cross-sections at
  160--210 GeV have been kindly computed by R.~Tanaka.

\bibitem{4f_bib:swap}
G.~Montagna \etal, {\it ``Higher--order QED corrections to single W production
  in electron--positron collisions''}, FNT/T--2000/08, {\tt
  http://arXiv.org/abs/hep-ph/0006307}.

\bibitem{gc_bib:LEP2YR}
G. Gounaris \etal, in {\em Physics at LEP 2}, Report CERN 96-01 (1996), eds G.
  Altarelli, T. Sj{\"o}strand, F. Zwirner, Vol. 1, p. 525.

\bibitem{gc_bib:ALEPH-cTGC-Oalpha}
ALEPH Collaboration, {\em Measurement of Triple Gauge-Boson Couplings in
  $e^+e^-$ collisions up to 208GeV}, ALEPH 2001-027 CONF 2001-021.

\bibitem{Montagna:2001ej}
G.~Montagna, M.~Moretti, O.~Nicrosini, M.~Osmo, and F.~Piccinini, Phys. Lett.
  {\bf B515} (2001) 197--205.

\bibitem{Hagiwara:1987vm}
K.~Hagiwara, R.~D. Peccei, D.~Zeppenfeld, and K.~Hikasa, Nucl. Phys. {\bf B282}
  (1987) 253.

\bibitem{Gounaris:2000tb}
G.~J. Gounaris, J.~Layssac, and F.~M. Renard, Phys. Rev. {\bf D62} (2000)
  073013.

\bibitem{Belanger:1992qi}
G.~Belanger and F.~Boudjema, Phys. Lett. {\bf B288} (1992) 210--220.

\bibitem{Stirling:1999ek}
W.~J. Stirling and A.~Werthenbach, Eur. Phys. J. {\bf C14} (2000) 103--110.

\bibitem{Stirling:1999xa}
W.~J. Stirling and A.~Werthenbach, Phys. Lett. {\bf B466} (1999) 369.

\bibitem{Belanger:1999aw}
G.~Belanger, F.~Boudjema, Y.~Kurihara, D.~Perret-Gallix, and A.~Semenov, Eur.
  Phys. J. {\bf C13} (2000) 283--293.

\bibitem{gc_bib:ALEPH-nTGC}
ALEPH Collaboration, {\em Limits on anomalous neutral gauge couplings using
  data from ZZ and Z$\gamma$ production between 183-208 GeV}, ALEPH 2001-061
  (July 2001) CONF 2001-041.

\bibitem{gc_bib:DELPHI-nTGC}
DELPHI Collaboration, {\em Study of Trilinear Gauge Boson Couplings ZZZ,
  $ZZ\gamma$ and $Z\gamma\gamma$}, DELPHI 2001-097 (July 2001) CONF 525.

\bibitem{gc_bib:L3-hTGC}
L3 Collaboration, M. Acciari \etal, Phys. Lett. {\bf B 436} (1999) 187;\\ L3
  Collaboration, M. Acciari \etal, Phys. Lett. {\bf B 489} (2000) 55.\\ L3
  Collaboration, {\em Search for anomalous ZZg and Zgg couplings in the process
  ee$\rightarrow$Zg at LEP}, L3 Note 2672 (July 2001).

\bibitem{gc_bib:OPAL-hTGC}
OPAL Collaboration, G. Abbiendi \etal, Eur. Phys. J. {\bf C 17} (2000) 13.

\bibitem{gc_bib:L3-fTGC}
See references~\cite{common_bib:ltrzz183a,
  common_bib:ltrzz189,common_bib:ltrzz2000new}.

\bibitem{gc_bib:OPAL-fTGC}
See references~\cite{common_bib:opazz189,common_bib:opazz2000new}.

\bibitem{gc_bib:ALEPH-QGC}
ALEPH Collaboration, {\em Constraints on Anomalous Quartic Gauge Boson
  Couplings}, ALEPH 2001-069 CONF 2001-049.

\bibitem{gc_bib:L3-QGC}
  Collaboration, M.Acciarri \etal, Phys. Lett. {\bf B 490} (2000) 187; \\ L3
  Collaboration, {\em Measurement of the $W^+W^-$g Cross Section and Direct
  Limits on Anomalous Quartic Gauge Boson Couplings at LEP}, L3 Note 2675 (June
  2001).

\bibitem{gc_bib:OPAL-QGC}
OPAL Collaboration, G. Abbiendi \etal, Phys. Lett. {\bf B 471} (1999) 293.

\bibitem{gc_bib:moriond01}
The LEP-TGC combination group, LEPEWWG/TGC/2001-01, March 2001.

\bibitem{mw:bib:A-mw183}
ALEPH Collaboration, R.~Barate et~al., Phys. Lett. {\bf B453} (1999) 121--137.

\bibitem{mw:bib:A-mw189}
ALEPH Collaboration, R.~Barate et~al., Eur. Phys. J. {\bf C17} (2000) 241--261.

\bibitem{mw:bib:A-mw20x}
ALEPH Collaboration, {\it Measurement of the W Mass and Width in $\epem$
  Collisions at $\roots\sim 192-208~\GeV$}, ALEPH note 2001-020 CONF 2001-017.

\bibitem{mw:bib:D-mw183}
DELPHI Collaboration, P.~Abreu et~al., Phys. Lett. {\bf B462} (1999) 410--424.

\bibitem{mw:bib:D-mw189}
DELPHI Collaboration, P.~Abreu et~al., Phys. Lett. {\bf B511} (2001) 159--177.

\bibitem{mw:bib:D-mw20X}
DELPHI Collaboration, {\it{Measurement of the mass and width of the W Boson in
  $\epem$ collisions at $\roots = 192-209~\GeV$}}, DELPHI 2001-103 CONF 531.

\bibitem{mw:bib:L-mw183}
L3 Collaboration, M.~Acciarri et~al., Phys. Lett. {\bf B454} (1999) 386--398.

\bibitem{mw:bib:L-mw189}
L3 Collaboration, {\it{Preliminary Results on the Measurement of Mass and Width
  of the W Boson at LEP}}, L3 Note 2377, March 1999.

\bibitem{mw:bib:L-mw19X}
L3 Collaboration, {\it{Preliminary Results on the Measurement of Mass and Width
  of the W Boson at LEP}}, L3 Note 2575, July 2000.

\bibitem{mw:bib:L-mw20X}
L3 Collaboration, {\it{Preliminary Results on the Measurement of Mass and Width
  of the W Boson at LEP}}, L3 Note 2637, February 2001.

\bibitem{mw:bib:O-mw183}
OPAL Collaboration, G.~Abbiendi et~al., Phys. Lett. {\bf B453} (1999) 138--152.

\bibitem{mw:bib:O-mw189}
OPAL Collaboration, G. Abbiendi \etal, \PLB{507}{2001}{29}.

\bibitem{mw:bib:O-mw19X}
OPAL Collaboration, {\it{Measurement of the Mass of the W Boson in \epem\
  annihilations at 192-202 GeV}}, OPAL Physics Note PN422 (updated July 2000).

\bibitem{mw:bib:O-mwlvlv}
OPAL Collaboration, {\it{Determination of the W mass in the fully leptonic
  channel using an unbinned maximum likelihood fit}}, OPAL Physics Note PN480,
  July 2001.

\bibitem{mw:bib:energy}
LEP Energy Working Group, LEPEWG 01/01, March 2001.

\bibitem{mw:bib:ski}
Torbjorn Sjostrand and Valery~A. Khoze, Z. Phys. {\bf C62} (1994) 281--310.

\bibitem{ref:CHARMIIgn}
CHARM II Collaboration, P.~Vilain \etal, Phys.~Lett. {\bf B335} (1994) 246.

\bibitem{bib-UA2MW}
UA2 Collaboration, J.~Alitti \etal, Phys.~Lett. {\bf B276} (1992) 354.

\bibitem{bib-CDFMW1}
CDF Collaboration, F.~Abe \etal, Phys.~Rev.~Lett. {\bf 65} (1990) 2243;\\ CDF
  Collaboration, F.~Abe \etal, Phys.~Rev. {\bf D43} (1991) 2070.

\bibitem{bib-CDFMW2}
CDF Collaboration, F.~Abe \etal, Phys.~Rev.~Lett. {\bf 75} (1995) 11;\\ CDF
  Collaboration, F.~Abe \etal, Phys.~Rev. {\bf D52} (1995) 4784.\\ A.~Gordon,
  talk presented at XXXIInd Rencontres de Moriond, Les Arcs, 16-22 March 1997,
  to appear in the proceedings.

\bibitem{bib-D0MW}
D\O\ Collaboration, S.~Abachi \etal, \PRL{84}{2000}{222}.

\bibitem{bib-MWAVE-00}
{\tt http://www-cdf.fnal.gov/physics/ewk/wmass\_new.htm}.

\bibitem{bib-topCDF}
CDF Collaboration, W.~Yao, {\it t Mass at CDF}, talk presented at ICHEP 98,
  Vancouver, B.C., Canada, 23-29 July, 1998.

\bibitem{bib-topD0}
D\O{} Collaboration, B.~Abbott \etal, \PRL{84}{2000}{222}.

\bibitem{bib-Tevatop}
The Top Averaging Group, L.\,Demortier \etal, for the CDF and D\O{}
  Collaborations, FERMILAB-TM-2084 (1999).

\bibitem{bib-CCFRnn}
CCFR/NuTeV Collaboration, K.~McFarland \etal, Eur.~Phys.~Jour.~{\bf C1} (1998)
  509.

\bibitem{bib-NuTeV}
NuTeV Collaboration, K.~McFarland, talk presented at the XXXIIIth Rencontres de
  Moriond, Les Arcs, France, 15-21 March, 1998, hep-ex/9806013. The preliminary
  result quoted is a combination of the NuTeV and CCFR results.

\bibitem{QWCs:exp:1}
C.~S. Wood et~al., Science {\bf 275} (1997) 1759.

\bibitem{QWCs:exp:2}
S.~C. Bennett and C.~E. Wieman, Phys. Rev. Lett. {\bf 82} (1999) 2484--2487.

\bibitem{QWCs:theo:2000}
A.~Derevianko, Phys. Rev. Lett. {\bf 85} (2000) 1618.

\bibitem{QWCs:theo:2001}
S.~G.~Porsev M.~G.~Kozlov and I.~I. Tupitsyn, Phys. Rev. Lett. {\bf 86} (2001)
  3260.

\bibitem{bib-NuTeV-final}
NuTeV Collaboration, G.P. Zeller \etal, {\em A Precise Determination of
  Electroweak Parameters in Neutrino-Nucleon Scattering}, preprint
  hep-ex/0110059.

\bibitem{bib-Gmu}
T.{} van Ritbergen and R.G. Stuart, Phys.{} Rev.{} Lett.{} {\bf 82} (1999) 488.

\bibitem{bib-BP01}
H.~Burkhardt and B.~Pietrzyk, Phys.~Lett. {\bf B513} (2001) 46.

\bibitem{ref:sld-s99}
SLD Collaboration, J.~Brau, {\it Electroweak Precision Measurements with
  Leptons}, talk presented at EPS-HEP-99, Tampere, Finland, 15-21 July 1999.

\bibitem{bib-PCLI}
{\it Reports of the working group on precision calculations for the Z
  resonance}, eds.~D.~Bardin, W.~Hollik and G.~Passarino, CERN Yellow Report
  95-03, Geneva, 31 March 1995.

\bibitem{bib-twoloop}
G.~Degrassi, S.~Fanchiotti and A.~Sirlin, Nucl.~Phys.~{\bf B351} (1991) 49;\\
  G.~Degrassi and A.~Sirlin, Nucl.~Phys.~{\bf B352} (1991) 342;\\ G.~Degrassi,
  P.~Gambino and A.~Vicini, Phys.~Lett.~{\bf B383} (1996) 219;\\ G.~Degrassi,
  P.~Gambino and A.~Sirlin, Phys.~Lett.~{\bf B394} (1997) 188;\\ G.~Degrassi
  and P.~Gambino, Nucl.~Phys.~{\bf B567} (2000) 3.

\bibitem{bib-QCDEW}
A.~Czarnecki and J.~K{\"u}hn, Phys.~Rev.~Lett.~{\bf 77} (1996) 3955;\\
  R.~Harlander, T.~Seidensticker and M.~Steinhauser, Phys.~Lett.~{\bf B426}
  (1998) 125.

\bibitem{bib-SMNEW}
Electroweak libraries:\\ ZFITTER: see Reference~\citen{ref:ZFITTER};\\ BHM
  (G.~Burgers, W.~Hollik and M.~Martinez): W.~Hollik, Fortschr.~Phys. {\bf38}
  (1990) 3, 165; M.~Consoli, W.~Hollik and F.~Jegerlehner: Proceedings of the
  Workshop on Z physics at LEP I, CERN Report 89-08 Vol.I,7 and G.~Burgers,
  F.~Jegerlehner, B.~Kniehl and J.~K{\"u}hn: the same proceedings, CERN Report
  89-08 Vol.I, 55; \\ TOPAZ0 Version 4.0i: G.~Montagna, O.~Nicrosini,
  G.~Passarino, F.~Piccinni and R.~Pittau, Nucl.~Phys. {\bf B401} (1993) 3;
  Comp.~Phys.~Comm. {\bf 76} (1993) 328.\\ These computer codes have upgraded
  by including the results of~\cite{bib-PCLI} and references therein. ZFITTER
  and TOPAZ0 have been further updated using the results of
  references~\citen{bib-twoloop} and~\citen{bib-QCDEW}. See, D.~Bardin and
  G.~Passarino, {\it Upgrading of Precision Calculations for Electroweak
  Observables}, CERN-TH/98-92, hep-ph/9803425.

\bibitem{FHWW-f2l-MW}
A. Freitas, W. Hollik, W. Walter and G. Weiglein, Phys. Lett. {\bf B495} (2000)
  338.

\bibitem{BGPWprivate}
D.~Bardin, P.~Gambino, G.~Passarino, G.~Weiglein, private communication, spring
  2001.

\bibitem{bib-SMALFAS}
T.~Hebbeker, M.~Martinez, G.~Passarino and G.~Quast, Phys.~Lett. {\bf B331}
  (1994) 165;\\ P.A.~Raczka and A.~Szymacha, Phys. Rev. {\bf D54} (1996)
  3073;\\ D.E.~Soper and L.R.~Surguladze, Phys. Rev. {\bf D54} (1996) 4566.

\bibitem{bib-alphalept}
M.~Steinhauser, Phys.~Lett.~{\bf B429} (1998) 158.

\bibitem{bib-JEG2}
S.~Eidelmann and F.~Jegerlehner, Z.~Phys. {\bf C67} (1995) 585.

\bibitem{BES_01}
The BES Collaboration, J.Z.Bai \etal, {\it Measurements of the Cross Section
  for $e^+e^-\rightarrow $ hadrons at Center-of-Mass Energies from 2 to 5 GeV},
  hep-ex/0102003.

\bibitem{bib-Swartz}
M.~L.~Swartz, Phys.~Rev. {\bf D53} (1996) 5268.

\bibitem{bib-Zeppe}
A.D.~Martin and D.~Zeppenfeld, Phys.~Lett. {\bf B345} (1994) 558.

\bibitem{bib-Alemany}
R.~Alemany, \etal, Eur.~Phys.~J.~{\bf C2} (1998) 123.

\bibitem{bib-Davier}
M.~Davier and A.~H{\"o}cker, Phys.~Lett.~{\bf B419} (1998) 419.

\bibitem{bib-alphaKuhn}
J.H.~K{\"u}hn and M.~Steinhauser, Phys.{} Lett.{} {\bf B437} (1998) 425.

\bibitem{bib-Erler}
J.~Erler, Phys.~Rev.~{\bf D59}, (1999) 054008.

\bibitem{bib-ADMartin}
A.~D.~Martin, J.~Outhwaite and M.~G.~Ryskin, Phys.~Lett. {\bf B492} (2000) 69.

\bibitem{bib-jeger99}
F. Jegerlehner, {\it Hadronic Effects in (g-2) and $\alpha_{QED}$(M$_Z$) :
  Status and Perspectives}, Proc. of Int. Symp. on Radiative Corrections,
  Barcelona, Sept. 1998, page 75.

\bibitem{Siggi-Bethke-alpha-s}
S.~Bethke, hep-ex/0004021, J. Phys. G26 (2000) R27.

\bibitem{MINUIT}
F.~James and M.~Roos, Comput. Phys. Commun. {\bf 10} (1975) 343.

\bibitem{ref:TOPAZ0}
G.~Montagna \etal, Comput. Phys. Commun. {\bf 117} (1999) 278;\\ {\tt
  http://www.to.infn.it/$\sim$giampier/topaz0.html~}.

\bibitem{ref:HiggsOsaka}
K.{}Hoffman, {\it Year 2000 update for OPAL and LEP HIGGSWG results"}, talk
  presented at ICHEP2000, Osaka, July 27 - August 2,2000, to appear in the
  proceedings.

\bibitem{ref:groot2001}
N. De Groot {\it Electroweak results from SLD}, talk presented at XXXVIth
  Rencontres de Moriond, Electroweak Interactions and Unified Theories, Les
  Arcs, March 2001, hep-ex/0105058;\\ Serbo V. Presented at Int. Europhys.
  Conf. High Energy Phys., July 2001, Budapest, Hungary.

\end{thebibliography}

\vfill

\section*{Links to LEP results on the World Wide Web}
The physics notes describing the preliminary results of the four LEP
experiments submitted to the summer 2001 conferences, as well as
additional documentation from the LEP electroweak working
group are available on the World Wide Web at: \\[2mm]
\begin{tabular}{ll}
  ALEPH:   &
  {\tt http://alephwww.cern.ch/ALPUB/oldconf/summer01/summer.html}\\
  DELPHI:  &
  {\tt http://delphiwww.cern.ch/\~{}pubxx/www/delsec/conferences/summer01/}\\
  L3:      &
  {\tt http://l3www.cern.ch/conferences/Budapest2001/}\\
  OPAL:    &
  {\tt http://opal.web.cern.ch/Opal/physnote.html}\\     
  LEP-EWWG: &
  {\tt http://lepewwg.web.cern.ch/LEPEWWG/}\\
\end{tabular}

\end{document}